\documentclass[a4paper,12pt]{article}
\usepackage[dvipdfmx]{graphicx}
\usepackage{amsfonts,amsthm,amsmath,amssymb,graphicx,float,mathtools}
\numberwithin{equation}{section}
\usepackage{subfigure}
\usepackage{color}
\usepackage{bbm} %for \mathbbm{1} etc
\usepackage{tensor} % for left subscript
\usepackage{verbatim} % for comments
\usepackage[
	  pagebackref=false,
	  colorlinks=true,
      linkcolor=blue,
      urlcolor=blue,
      filecolor=black,
      citecolor=red,
      pdfstartview=FitV,
      pdftitle={},
        pdfauthor={},
        pdfsubject={},
        pdfkeywords={},
        pdfpagemode=None,
        bookmarksopen=true
      ]{hyperref}
\usepackage{ascmac}
\usepackage{tikz-cd}% for graphs
\usepackage{caption}% for captions
\DeclareMathOperator{\sech}{sech}

\begin{document}

\newtheorem{definition}{Definition}[section]
\newcommand{\be}{\begin{equation}}
\newcommand{\ee}{\end{equation}}
\newcommand{\bea}{\begin{eqnarray}}
\newcommand{\eea}{\end{eqnarray}}
\newcommand{\LE}{\left[}
\newcommand{\R}{\right]}
\newcommand{\nn}{\nonumber}
\newcommand{\Tr}{\text{Tr}}
\newcommand{\N}{\mathcal{N}}
\newcommand{\G}{\Gamma}
\newcommand{\vf}{\varphi}
\newcommand{\LL}{\mathcal{L}}
\newcommand{\Op}{\mathcal{O}} 
\newcommand{\HH}{\mathcal{H}}
\newcommand{\arctanh}{\text{arctanh}}
\newcommand{\up}{\uparrow}
\newcommand{\down}{\downarrow}
\newcommand{\ket}[1]{\left| #1 \right>}
\newcommand{\bra}[1]{\left< #1 \right|}
\newcommand{\ketbra}[1]{\left|#1\right>\left<#1\right|}
\newcommand{\rd}{\partial}
\newcommand{\de}{\partial}
\newcommand{\ba}{\begin{eqnarray}}
\newcommand{\ea}{\end{eqnarray}}
\newcommand{\db}{\bar{\partial}}
\newcommand{\we}{\wedge}
\newcommand{\ca}{\mathcal}
\newcommand{\lr}{\leftrightarrow}
\newcommand{\f}{\frac}
\newcommand{\s}{\sqrt}
\newcommand{\vp}{\varphi}
\newcommand{\hvp}{\hat{\varphi}}
\newcommand{\tvp}{\tilde{\varphi}}
\newcommand{\tp}{\tilde{\phi}}
\newcommand{\ti}{\tilde}
\newcommand{\ap}{\alpha}
\newcommand{\pr}{\propto}
\newcommand{\mb}{\mathbf}
\newcommand{\ddd}{\cdot\cdot\cdot}
\newcommand{\no}{\nonumber \\}
\newcommand{\la}{\langle}
\newcommand{\lb}{\rangle}
\newcommand{\ep}{\epsilon}
 \def\we{\wedge}
 \def\lr{\leftrightarrow}
 \def\f {\frac}
 \def\ti{\tilde}
 \def\ap{\alpha}
 \def\pr{\propto}
 \def\mb{\mathbf}
 \def\ddd{\cdot\cdot\cdot}
 \def\no{\nonumber \\}
 \def\la{\langle}
 \def\lb{\rangle}
 \def\ep{\epsilon}
\newcommand{\mcl}{\mathcal}
 \def\g{\gamma}
\def\Tr{\text{tr}}

\begin{titlepage}
\thispagestyle{empty}

\begin{flushright}

\end{flushright}
\bigskip

\begin{center}
  \noindent{\large \textbf{Local operator quench induced by two-dimensional inhomogeneous and homogeneous CFT Hamiltonians}}\\
\vspace{2cm}

\vspace{1cm}
\renewcommand\thefootnote{\mbox{$\fnsymbol{footnote}$}}
Weibo Mao \footnote{maoweibo21@mails.ucas.ac.cn}${}^{1}$,
Masahiro Nozaki\footnote{mnozaki@ucas.ac.cn}${}^{1,2}$,
 Kotaro Tamaoka\footnote{tamaoka.kotaro@nihon-u.ac.jp}${}^{3}$ and Mao Tian Tan\footnote{maotian.tan@apctp.org}${}^{4}$\\

\vspace{1cm}
${}^{1}${\small \sl Kavli Institute for Theoretical Sciences, University of Chinese Academy of Sciences,
Beijing 100190, China}\\
${}^{2}${\small \sl RIKEN Interdisciplinary Theoretical and Mathematical Sciences (iTHEMS), \\Wako, Saitama 351-0198, Japan}\\
${}^{3}${\small \sl Department of Physics, College of Humanities and Sciences, Nihon University, \\Sakura-josui, Tokyo 156-8550, Japan}\\
${}^{4}${\small \sl Asia Pacific Center for Theoretical Physics, Pohang, Gyeongbuk, 37673, Korea}\\

\vskip 4em
\end{center}
%%%%%%%%%%%%%%%%%%%%%%%%%%%%%
\begin{abstract}
We explore non-equilibrium processes in two-dimensional conformal field theories (2d CFTs) due to the growth of operators induced by inhomogeneous and homogeneous Hamiltonians by investigating the time dependence of the partition function, energy density, and entanglement entropy.
The non-equilibrium processes considered in this paper are constructed out of the Lorentzian and Euclidean time evolution governed by different Hamiltonians.
We explore the effect of the time ordering on entanglement dynamics so that we find that in a free boson CFT and RCFTs, this time ordering does not affect the entanglement entropy, while in the holographic CFTs, it does. 
Our main finding is that in the holographic CFTs, the non-unitary time evolution induced by the inhomogeneous Hamiltonian can retain the initial state information longer than in the unitary time evolution.
%Consequently, we found that if we regulate the state after the operator grows with time, the time dependence of these quantities exhibits behaviors that have not been found so far. Furthermore, we found the gravity dual of the systems considered. 
\end{abstract}
%%%%%%%%%%%%%%%%%%%%%%%%
\end{titlepage} 
%%%%%%%%%%%%%%%%%%%%%%%%%%%%%%%
\tableofcontents

%%%%%%%%%%%%%%%%%%%%%%%%%%%%%%%%%%%%%%%
\section{Introduction and summary}
%%%%%%%%%%%%%%%%%%%%%%%%%%%%%%%%%%%%%%%

%\subsection*{General background}

% In the last few years, it has emerged that holographic conformal field theories (holographic CFTs), the theories could be described by the gravity, could have chaotic properties.
% To explore the mechanism behind holography, a vast of research studies the their chaotic property \cite{PhysRevA.43.2046,PhysRevE.50.888,2008Natur.452..854R,2011RvMP...83..863P,2005JSMTE..04..010C,2019Natur.567...61L,2020PhRvL.124x0505J,2021PhRvX..11b1010B,2016PhRvA..94d0302S,PhysRevA.94.062329,2016arXiv160701801Y,2017PhRvA..95a2120Y,2018PhRvA..97d2105Y,2017PhRvE..95f2127C,2017arXiv171003363Y,2017NatPh..13..781G,2016arXiv161205249W,PhysRevX.7.031011,2017arXiv170506714M}.
% One of the profound chaotic phenomena induced by the holographic CFT is the operator growth, where the ``size" of local operator grows during time evolution.
% During the time evolution induced by the strong chaotic Hamiltonians such as the holographic CFT Hamiltonians, the size of the local operator could exponentially grow with times \cite{Shenker:2013pqa,Maldacena:2015waa,Roberts:2014ifa}.

In recent years, it has come to light that holographic conformal field theories (CFTs), which can be described by semiclassical gravity, possess chaotic properties. To elucidate the mechanism behind holography, a vast literature examining their chaotic properties has been accumulated\cite{PhysRevA.43.2046,PhysRevE.50.888,2008Natur.452..854R,2011RvMP...83..863P,2005JSMTE..04..010C,2019Natur.567...61L,2020PhRvL.124x0505J,2021PhRvX..11b1010B,2016PhRvA..94d0302S,PhysRevA.94.062329,2016arXiv160701801Y,2017PhRvA..95a2120Y,2018PhRvA..97d2105Y,2017PhRvE..95f2127C,2017arXiv171003363Y,2017NatPh..13..781G,2016arXiv161205249W,PhysRevX.7.031011,2017arXiv170506714M}.
One of the main chaotic phenomena observed in holographic CFTs, in which the ``size" of local operators grow during time evolution, is known as operator growth. In fact, under time evolution by these maximally chaotic Hamiltonians, the size of these local operators can grow exponentially in time \cite{Shenker:2013pqa,Maldacena:2015waa,Roberts:2014ifa}.

Some pioneering works on operator growth explored the time dependence of entanglement entropy after a vacuum state has been excited by a local operator before undergoing regular time evolution \cite{2014arXiv1405.5946C,Nozaki:2014uaa,PhysRevD.90.041701,Nozaki:2014hna,2015JHEP...02..171A}.
In these papers, we similarly start from the vacuum state with the insertion of a single local operator and evolve the system with two-dimensional CFT Hamiltonians. Unlike previous works, we will use spatially inhomogeneous Hamiltonians to either time evolve the system or to regulate the state.

% Partitioning the space into two complementary regions $A$ and $\overline{A}$, the density matrix associated with region $A$ is $\rho_{A}=\Tr_{\overline{A}}\rho$, where $\rho$ is the density operator of the total system. The von Neumann entropy of subsystem $A$ is given in terms of this reduced density matrix as
% \be
% S_A=-\Tr_A\left(\rho_A \log{\rho_A}\right).
% \ee
In non-holographic theories such as RCFTs or the free boson CFT, the time dependence of the entanglement entropy of a subsystem $A$, which we denote $S_A$, is well-described by the propagation of excitations called quasiparticles. When the subsystem $A$ is half of an infinite system, the entanglement entropy $S_A$ increases by a constant determined by the local operator inserted, at a time that is determined by the motion of these quasiparticles. On the other hand, in $2$d holographic CFTs, the entanglement entropy $S_A$ grows logarithmically at late times and so the dependence of $S_A$ on the local operator is hidden by this logarithmic growth. This suggests that the scrambling effect of $2$d holographic Hamiltonians hides the information of the local operator inserted\footnote{In this paper, we will treat the heavy operators as local operator quench in holographic CFTs. If we consider R\'enyi entropy and treat the light local operator, we can keep some information even at the leading order. See \cite{Kusuki:2019gjs}, for example.}. These studies have been generalized in different ways. In \cite {2015JHEP...08..011C,Caputa:2014eta}, the initial states were respectively chosen to be the thermofield double state and the thermal state instead of the vacuum state. In \cite{Kusuki:2019avm,Caputa:2019avh}, the authors considered the time evolution with two local operators inserted instead of one. %The pseudo entropy of local operator quenches have been computed in \cite{He2023,PhysRevD.109.025014,Guo2022}. 
Some other studies of local operator quenches can be found in \cite{Guo2018,Chen2015,Guo2015}. In this paper, we will explore this scrambling process by using spatially inhomogeneous Hamiltonians instead of the usual uniform Hamiltonian.

The inhomogenous Hamiltonians considered in this paper were originally introduced as inhomogeneous deformations of spin systems with the intention of reducing the systems' dependence on  boundary conditions \cite{Gendiar01102009,Gendiar01022010,Hikihara_2011,2011PhRvA..83e2118G}. We consider a family of M\"{o}bius Hamiltonians that are parametrized by a parameter $\theta$ that controls the spatial inhomogeneity. Setting $\theta=0$ gives the usual uniform Hamiltonian. On the other hand, sending $\theta \rightarrow\infty$ gives the SSD Hamiltonian which is the most inhomogeneous Hamiltonian we consider. These inhomogeneous deformations were subsequently introduced in $2$d CFTs \cite{Okunishi_2016,2016PhRvB..93w5119W} which is tantamount to placing these CFTs on curved spacetimes \cite{Caputa:2020mgb}. Currently, these inhomogeneous deformations are used to analytically explore the properties of quantum systems following global quenches and periodic Floquet driving \cite{PhysRevB.97.184309,2019JPhA...52X5401M,2021arXiv211214388G, PhysRevLett.118.260602,2018arXiv180500031W,2020PhRvX..10c1036F,Han_2020,2021PhRvR...3b3044W,2020arXiv201109491F,2021arXiv210910923W,PhysRevB.103.224303,PhysRevResearch.2.023085,Moosavi2021,PhysRevLett.122.020201,10.21468/SciPostPhys.3.3.019,2024arXiv240216555B}. In addition, the non-equilibrium processes induced by these inhomogeneous quenches can be used to cool systems to the ground state \cite{2016arXiv161104591Z,2020PhRvR...2c3347R,HM,2018PhRvL.120u0604A,2019PhRvB..99j4308M,2022arXiv221100040W} which produces non-local correlations in conformal field theories \cite{2023arXiv230208009G,2023arXiv231019376N}. This production of non-local correlations was further explored in spin chains in \cite{2023arXiv230501019G}. The generalization of Floquet driving in CFTs with dimensions greater than 2 have been studied in \cite{Das:2023xaw}. The relationship between local quenches and M\"obius quenches were explored in \cite{2023arXiv230904665K}, while the entanglement entropy for quantum quenches in generic inhomogeneous CFTs with open boundary conditions was derived in \cite{Liu:2023tiq}.

In this paper, we consider a Lorentzian time evolution induced by a Hamiltonian $H_1$, followed by a subsequent Euclidean time evolution induced by a Hamiltonian $H_2$, and vice versa, in order to explore the effect of the ordering of time evolutions with different signatures on the non-equilibrium process. We also studied the growth of operators induced by homogeneous and inhomogeneous Hamiltonians in both Euclidean and Lorentzian signature by investigating the time dependence of entanglement entropy and the energy density. The time-evolved states considered in this paper are 
\be \label{eq:state-considered}
\begin{split}
&\ket{\Phi_a (\tilde{t})}=\mathcal{N}_ae^{-i H_1 \tilde{t}}e^{-\epsilon H_2} \mathcal{O} \ket{0},\\
&\ket{\Phi_b (\tilde{t})}=\mathcal{N}_be^{-\epsilon H_2} e^{-i H_1 \tilde{t}}\mathcal{O}\ket{0},\\
\end{split}
\ee
where $\epsilon$ denotes the time duration of the Euclidean evolution, $\tilde{t}$ denotes the time duration of the Lorentzian evolution, $\mathcal{N}_{a,b}$ guarantees that the norms of the states are unity, and the Hamiltonians $H_i$ considered in this paper are the undeformed, sine-square deformed, and M\"obius Hamiltonians in two-dimensional conformal field theories ($2$d CFTs) on the circle with circumference $L$. We assume that the ground state for the undeformed Hamiltonian is equal to that of the deformed Hamiltonian and we denote this ground state by $\ket{0}$, so $\ket{0}$ is the ground state of both $H_1$ and $H_2$. When $H_1$ is not equal to $H_2$, the state $\ket{\Phi_a (\tilde{t})}$ is different from the state $\ket{\Phi_b (\tilde{t})}$. 

\begin{figure}
    \centering
\includegraphics[keepaspectratio, scale=0.2]{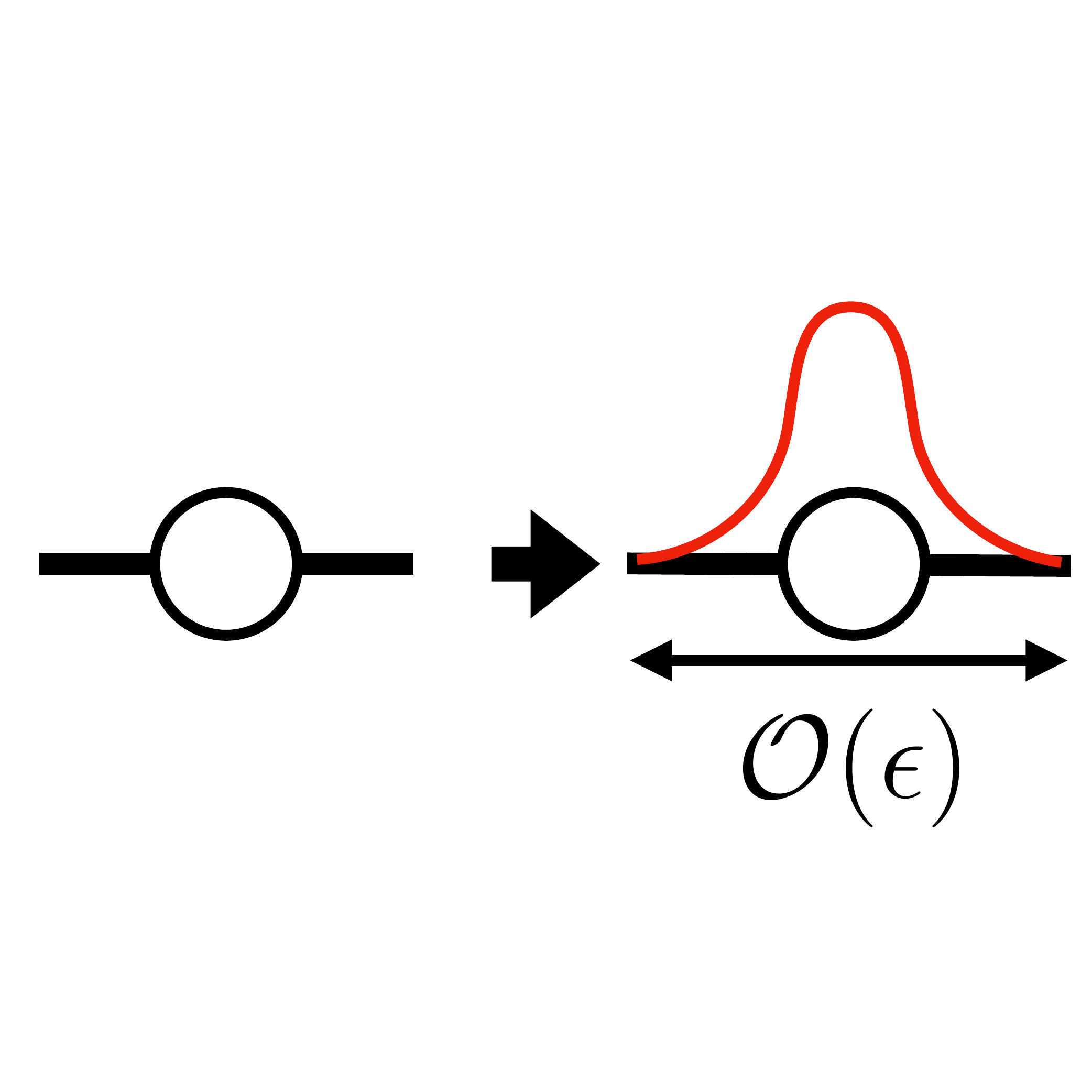}
    \caption{A sketch depicting the smearing of a local operator. After acting $e^{-\epsilon H_2}$ on a state with an insertion of a local operator, the local operator is smeared within a region of size $\mathcal{O}(\epsilon)$. The circle and red curve illustrate the local and smeared local operators respectively.}
\label{smearing-process}
\end{figure}

The role of the Euclidean time evolution is to reduce the effect of the local operators on the non-equilibrium processes. To see this, note that the Euclidean time evolution operator can be expanded in terms of the eigenstates of $H_2$ as $e^{-\epsilon H_2} =\sum_{a=1}^{M}e^{-\epsilon E_{a}} \ket{a} \bra{a}$, where $E_a$ and $\ket{a}$ denote the eigenvalues and eigenstates of $H_2$ respectively. Here, $a$ labels the eigenstates while $E_M$ is the maximum eigenvalue. Since exponential function $e^{-x}$ is negligible for $x\gg 1$, we approximate the Euclidean time evolution by
\be
e^{-\epsilon H_2} =\sum_{a=1}^{M}e^{-\epsilon E_{a}} \ket{a} \bra{a} \approx \sum_{a=1}^{\tilde{M}}e^{-\epsilon E_{a}} \ket{a} \bra{a},
\ee
where $\tilde{M}$ labels the largest eigenvalue such that $\epsilon E_{\tilde{M}}=\mathcal{O}(1)$. 
Thus, the Euclidean time evolution approximately removes contributions from the high-energy modes with $E_h>E_{\tilde{M}}$. In other words, the Euclidean time evolution effectively reduces the dimension of the Hilbert space of states contributing to the non-equilibrium process. For the states in \eqref{eq:state-considered}, the terms in the expansion in the energy eigenbasis of $H_2$ with $\bra{E_h}\mathcal{O}\ket{0}$ in $\ket{\Phi_a (\tilde{t})}$ and $\bra{E_h}e^{-iH_1\tilde{t}}\mathcal{O}\ket{0}$ in $\ket{\Phi_b (\tilde{t})}$ are negligible. Therefore, the local operator is smeared within a region of size $\mathcal{O}(\epsilon)$ as in Fig. \ref{smearing-process} and the Euclidean time evolution has reduced the locality of the system.

\begin{figure}[htbp]
    \begin{tabular}{cc}
      \begin{minipage}[t]{0.5\hsize}
        \centering
        \includegraphics[keepaspectratio, scale=0.2]{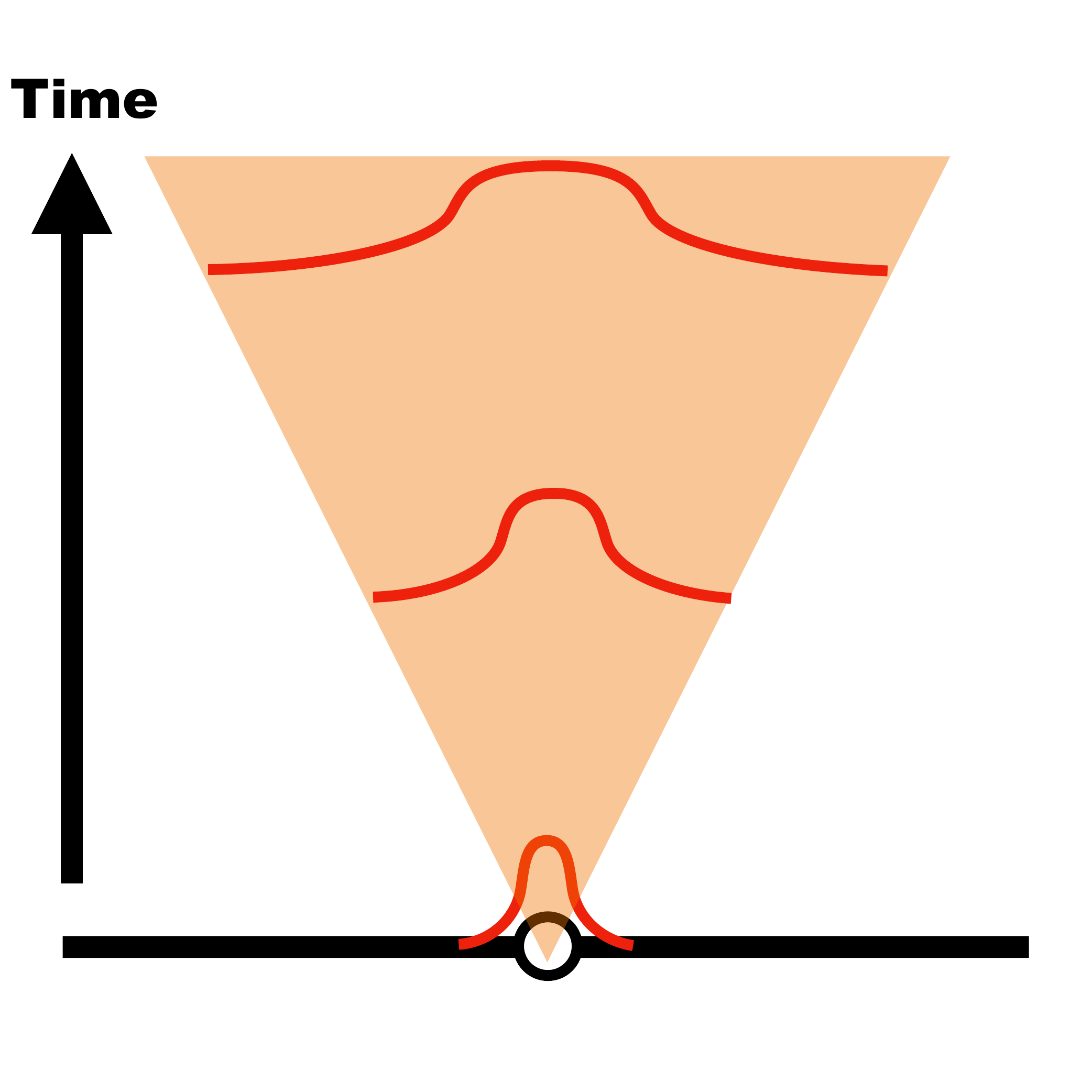}
        
    [a] Unitary process    
    
      \end{minipage}
      &\begin{minipage}[t]{0.5\hsize}
        \centering
        \includegraphics[keepaspectratio, scale=0.2]{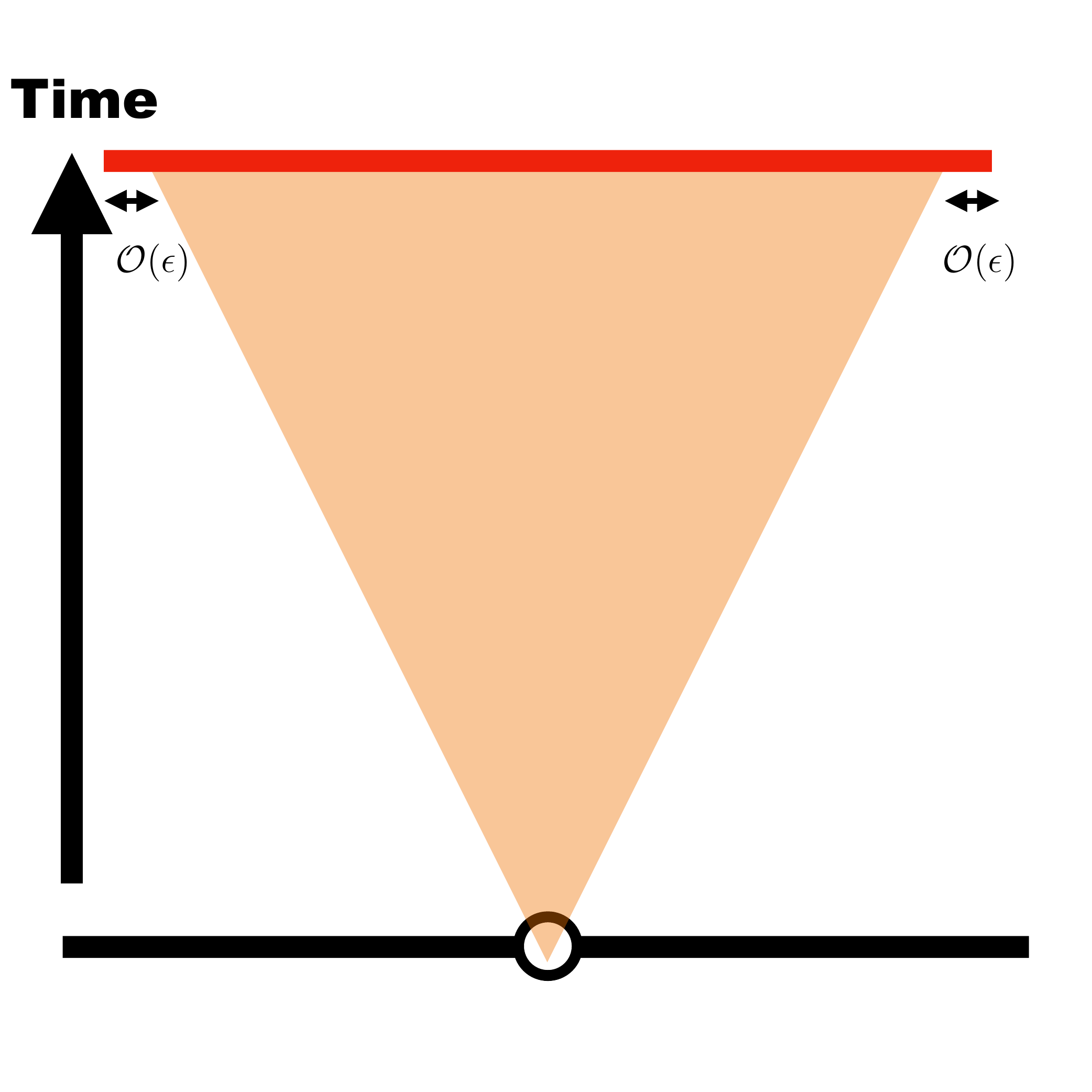}
        
          [b] Non-unitary process

      \end{minipage} 

    \end{tabular}
      \caption{Unitary and non-unitary processes. In [a], we smear the local operator, then evolve the system in real time. In [b], we evolve the system in real time before smearing the local operator. The orange region illustrates the space-time region where the local operator is delocalized by the dynamics. In the case of $\ket{\Phi_a}$, %``information" introduced by the smearing 
      the effect of smearing is delocalized during the real time evolution. In the case of $\ket{\Phi_b}$, the %information introduced by the 
      effect of smearing is spatially localized.}
        \label{Fig:unitary-non-unitary-process}
  \end{figure}

%, and the dynamical structure should be changed.
%To study how the Euclidean evolution hides the local structure of the system, we consider the systems in (\ref{eq:state-considered}). 
Once the Lorentzian time evolution is included, the order in which the Lorentzian and Euclidean time evolutions are performed potentially lead to different physics. In $\ket{\Phi_a (\tilde{t})}$, the local operator state first evolves under Euclidean time evolution before a subsequent Lorentzian time evolution with leads to operator growth. On the other hand, in $\ket{\Phi_b (\tilde{t})}$, the order of time evolution is swapped; the operator grows during the initial Lorentzian time evolution before it undergoes a subsequent Euclidean time evolution.
In the former case, the initial Euclidean time evolution approximately removes the contribution of high-energy modes where the fine-grained information of the local operator might be encoded before the smeared local operator grows under the final Lorentzian time evolution (see the left panel of Fig. \ref{Fig:unitary-non-unitary-process}). Contrast this with the latter case where the local operator grows during the Lorentzian time evolution before the Euclidean time evolution operator removes the high-energy modes.
In this case, during the initial Lorentzian time evolution, the information of the local operator which is encoded in the high-energy modes may spread to the low-energy modes.
Consequently, even after the final Euclidean time evolution, the fine-grained information of the local operator might still remain (see the right panel of Fig. \ref{Fig:unitary-non-unitary-process}).

\subsection*{Summary}
%%%%%%%%%%%%%%%%%%%%%%%%%%%%%%%%%%%%%%%
Let us summarize the main results of this paper below.

%%%%%%%%%%%%%%%%%%%%%%%%%%%%%%%%%%%%%%%
\subsubsection*{Time dependence of the partition function}
%%%%%%%%%%%%%%%%%%%%%%%%%%%%%%%%%%%%%%%
Since the time dependence of the partition function is determined by that of the two point function for the vacuum state, it is independent of the details of the CFTs.
For the M\"obius Hamiltonians, the partition functions in the cases considered in this paper are finite. 
For SSD Hamiltonian, when the excitations created by the local operator hit a spatial point where the Hamiltonian density vanishes, the partition function becomes infinite. This suggests that at the spatial position where the Hamiltonian density vanishes, the corresponding Euclidean time evolution operator does not reduce the Hilbert space to a subspace that only contains the low energy modes.
% This suggests that at the point where the density of Hamiltonian, inducing the Euclidean time evolution, vanishes, this Euclidean time evolution operator does not reduce the Hilbert space to that spanned by the low energy modes.
%%%%%%%%%%%%%%%%%%%%%%%%%%%%%%%%%%%%%%%

%%%%%%%%%%%%%%%%%%%%%%%%%%%%%%%%%%%%%%%
\subsubsection*{Time dependence of energy density}
%%%%%%%%%%%%%%%%%%%%%%%%%%%%%%%%%%%%%%%
We also explored the time dependence of the energy density which is a CFT-independent quantity as well. During both unitary and non-unitary Lorentzian time evolutions, the time dependence of the energy density follows the propagation of excitations with velocities determined by the Hamiltonians. An exception occurs for the non-unitary Lorentzian time evolution. 
For example, during non-unitary process induced by the SSD time evolution, the energy density at the origin grows quartically in time.
% Only during the non-unitary Lorentzian time evolution induced by the M\"obius Hamiltonian with $\theta=\infty$, so-called SSD Hamiltonian, for large times, the energy density at the special point quaternically grows with time. 

%%%%%%%%%%%%%%%%%%%%%%%%%%%%%%%%%%%%%%%
\subsubsection*{Time dependence of entanglement entropy in two-dimensional conformal field theories}
%%%%%%%%%%%%%%%%%%%%%%%%%%%%%%%%%%%%%%%
Unlike the partition function and the energy density, the entanglement entropy of a single interval in $2$d CFTs depends on the theory under consideration.
% Then, we explored the time dependence of the entanglement entropy for the single interval
% in $2$d CFTs. This depends on the details of $2$d CFTs.

%%%%%%%%%%%%%%%%%%%%%%%%%%%%%%%%%%%%%%%
{\bf Two-dimensional free bosons and rational CFTs}:
%%%%%%%%%%%%%%%%%%%%%%%%%%%%%%%%%%%%%%%
During the inhomogeneous time evolutions in $2$d free bosons and rational CFTs (RCFTs), the time ordering of Lorentzian and Euclidean time evolution does not affect the growth of entanglement entropy when it is well-behaved.
The time evolution of the entanglement entropy follows the propagation of quasiparticles created by the insertion of the local operators. 
In the time interval where entanglement entropy deviates from the vacuum entanglement entropy, its value is determined by the local operator that is inserted. For a particular Hermitian sum of vertex operators in the free boson CFT, this value is $\log2$. For general RCFTs, this value is given by the logarithm of the quantum dimension of the local operator.

%%%%%%%%%%%%%%%%%%%%%%%%%%%%%%%%%%%%%%%
{\bf Two-dimensional holographic CFTs}:
%%%%%%%%%%%%%%%%%%%%%%%%%%%%%%%%%%%%%%%
In contrast to the 2d free boson CFT and RCFTs, the ordering of the Lorentzian and Euclidean time evolution does affect the growth of entanglement entropy.
% During the non-equilibrium process in $2$d holographic CFTs, contrary to $2$d free n and RCFs, the time ordering of Lorentzian and Euclidean time evolution does affect the growth of entanglement entropy.
For M\"obius time evolution, the time dependence of entanglement entropy exhibits quantum revivals.
In other words, the value of entanglement entropy oscillates periodically in time.
During the Lorentzian time evolution induced by the M\"obius Hamiltonian, the period of quantum revivals is $L \cosh{2\theta}$, while during the time evolution induced by the uniform Hamiltonian, the period is $L$. Here, $L$ is the system size and $\theta$ is the parameter determining the inhomogeneity of the M\"obius Hamiltonian. When the entanglement entropy becomes larger than the vacuum entanglement entropy, the deviation between these two quantities depends on the ordering of the Lorentzian and Euclidean time evolution (see Section \ref{sec:EE-for-Mobius} for the details of the analysis.).
% In the time intervals where the value of the entanglement entropy is larger than the vacuum entanglement entropy, it depends on the time ordering of Lorentzian and Euclidean time evolution (See Section \ref{sec:EE-for-Mobius} for the details of the analysis.).

Finally, we summarize our findings on the SSD limit, where $\theta \rightarrow \infty$.
% Subsequently, we will report on our findings in the SSD limit, where $\theta \rightarrow \infty$.
During the real time evolution induced by the uniform Hamiltonian, the time evolution of the entanglement entropy is similar to the M\"{o}bius case. There are a few main differences between the M\"obius Hamiltonians and the SSD Hamiltonian. Firstly, during unitary time evolution with the uniform Hamiltonian, if the operator is inserted at the origin where the SSD Hamiltonian density is zero, the entanglement entropy becomes infinite in the SSD case. Secondly, during the non-unitary time evolution with the uniform Hamiltonian, the entanglement entropy becomes infinite when the excitations created by the local operator hit the origin.
% The differences from the M\"obius Hamiltonian with $\theta \neq \infty$ are: during the unitary time evolution, if the operator is inserted at the special point where the SSD Hamiltonian density is zero, the entanglement entropy becomes infinite; during the non-unitary time evolution, when the excitations created by the local operator hit the special point, the entanglement entropy becomes infinite.
Together with the divergences observed in the partition function, this suggests that the SSD euclidean time evolution does not reduce the dimension of the local Hilbert space at the origin. During real time evolution induced by the SSD Hamiltonian, the entanglement entropy remains constant if the local operator is inserted at the origin where the SSD Hamiltonian density vanishes but changes in time if the local operator is inserted somewhere else.
% During the real time evolution induced by the SSD Hamiltonian, if the local operator is inserted at the point where the SSD Hamiltonian density vanishes, the entanglement entropy does not depend on the time, while when the operator is not inserted at this special point, the entanglement entropy does. 
Yet another difference with inhomogeneous Hamiltonian with $\theta \neq \infty$ is the absence of quantum revivals in the entanglement entropy.
% However, contrary to the inhomogeneous Hamiltonian with $\theta \neq \infty$, the time dependence of the entanglement entropy does not exhibit the quantum revivals.
The time dependence of entanglement entropy is well described by the propagation of quasiparticles created by the local operator (see Section \ref{sec:quasiparticle-picture}.), although this quasiparticle picture can not determine the value of the entanglement entropy. 
If one of the boundaries of the subsystem is at the origin where the SSD Hamiltonian density is zero, the information of the local operator survives longer under the non-unitary real time evolution than during the unitary real time evolution.
% during the non-unitary real time evolution survives longer than that during the unitary real time evolution.

%%%%%%%%%%%%%%%%%%%%%%%%%%%%%%%%%%%%%%%
\subsection*{Organization of this paper}
%%%%%%%%%%%%%%%%%%%%%%%%%%%%%%%%%%%%%%%
In Section \ref{sec:Preliminary}, we present the details of the systems considered in this paper, define entanglement entropy, and present the time evolution of local operators in the Heisenberg picture.
In Section \ref{sec:universal}, the time evolution of the CFT-independent quantities, namely the partition function and the energy density, are presented.
In Section \ref{sec:ee-in-integrable}, we report the time dependence of entanglement entropy in $2$d free boson CFT and general RCFTs, and then propose an effective picture describing the time dependence of entanglement entropy in such non-holographic theories.
In Section \ref{sec:e-e-holographic}, the time dependence of entanglement entropy in $2$d holographic CFTs is presented. We conclude the paper in Section \ref{sec:discussion} where we discuss our findings in this paper.
%%%%%%%%%%%%%%%%%%%%%%%%%%%%%%%%%%%%%%%
%\subsubsection*{Two-dimensional holographic conformal field theories ($2$d holographic CFTs)}
%%%%%%%%%%%%%%%%%%%%%%%%%%%%%%%%%%%%%%%
%In the following, we will report on the entanglement dynamics in $2$d holographic CFTs, the CFTs possess the graity duals.

%{\bf Survival of information about the local operator:} Let us focus on the case where the Lorentzian time evolution is induced by the sine-square deformed (SSD) Hamiltonian. 
%If the local operator is smeared by the Euclidean evolution induced by the homogeneous Hamiltonian, and subsequently we evolve the system with the SSD Hamiltonian, then the entanglement entropy logarithmically grows with time. 
%This behavior is independent of the local operator inserted into the state.
%In the different time-ordered non-equilibrium process where the local operator grows during the SSD time evolution, and subsequently the Euclidean evolution smears the local operator, the entanglement entropy saturates to the constant value that depends on the local operator inserted.
%\textcolor{red}{\bf MN: I should explain the following explanation more:}We can see from this time dependence that the growth of smearing hides the information of the local operator.

%\textcolor{red}{\bf MN: We should add other results here.}
%%%%%%%%%%%%%%%%%%%%%%%%%%%%%%%%%%%%%%%
\section{Preliminary \label{sec:Preliminary}}
%%%%%%%%%%%%%%%%%%%%%%%%%%%%%%%%%%%%%%%
In this section, we will explain the details of the inhomogeneous Hamiltonians and systems considered, how to compute the entanglement entropy in the path-integral formalism, and discuss the real time trajectory, induced by the inhomogeneous Hamiltonians, of the local operator.

%%%%%%%%%%%%%%%%%%%%%%%%%%%%%%%%%%%%%%%
\subsection{Inhomogeneous Hamiltonians}
%%%%%%%%%%%%%%%%%%%%%%%%%%%%%%%%%%%%%%%
Here, we will explain the Hamiltonians considered in this paper.
The CFT Hamiltonians considered in this paper are undeformed, M\"obius, and SSD Hamiltonians defined as 
\be
\begin{small}
    H_{\text{inh}}=\int dx f(x)h(x), 
\end{small}
\ee
where the envelope function are defined as 
\be\label{EnvelopeFunction}
\begin{small}
    f(x)=\begin{cases}
        1 & \text{for~} H \\
        1-\tanh{\left(2\theta\right)}\cos{\left(\f{2\pi x}{L}\right)}& \text{for~} H_{\text{M\"obius}} \\
        2\sin^2{\left(\f{\pi x}{L}\right)} & \text{for~} H_{\text{SSD}} \\
    \end{cases}.
\end{small}
\ee
For $\theta=0$, $H_{\text{M\"obius}}$ reduces to the undeformed Hamiltonian, $H$, while for $\theta=\infty$, $H_{\text{M\"obius}}$ reduces to the SSD Hamiltonian $H_{\text{SSD}}$. In the SSD limit, where $\theta=\infty$, the deformed Hamiltonian density, $\overline{h}(x)=f(x)h(x)$ vanishes at $x=0$, while at $x=L/4$ and $x=3L/4$, the deformed density reduces to the undeformed one. %\textcolor{red}{, up to a proportionality constant}.
Let us defined the Virasoro generator as 
\be
L_n=\oint \f{dz}{2\pi i}z^{n+1} T(z), \overline{L}_n=\oint \f{d\overline{z}}{2\pi i}\overline{z}^{n+1} \overline{T}(\overline{z}),
\ee
where $n$ are integer numbers, $(z, \overline{z})=(e^{\f{2\pi(ix+\tau)}{L}},e^{\f{2\pi(-ix+\tau)}{L}})=(e^{\f{2\pi w}{L}},e^{\f{2\pi\overline{w}}{L}})$, and the chiral and anti-chiral parts of the energy-momentum tensor is defined by 
$h(x)=T(w)+\overline{T}(\overline{w})$. In terms of Virasoro generators, the inhomogeneous Hamiltonian are given by
\be \label{eq:Hamiltonian-virasoro}
\begin{split}
    &H_0=\f{2\pi}{L}\left(L_0+\overline{L}_0\right),\\
    &H_{\text{M\"obius}}=\f{2\pi}{L}\left[L_0+\overline{L}_0-\f{\tanh{2\theta}}{2}\left(L_{1}+L_{-1}+\overline{L}_{1}+\overline{L}_{-1}\right)\right], \\
    &H_{\text{SSD}}=\f{2\pi}{L}\left[L_0+\overline{L}_0-\f{1}{2}\left(L_{1}+L_{-1}+\overline{L}_{1}+\overline{L}_{-1}\right)\right].
\end{split}
\ee
%%%%%%%%%%%%%%%%%%%%%%%%%%%%%%%%%%%%%%%
\subsection{The systems considered in this paper}
%%%%%%%%%%%%%%%%%%%%%%%%%%%%%%%%%%%%%%%
We explain the system considered in this paper.
We start from the vacuum state with the insertion at the spatial position $x$ of the local operator, and then evolve the systems according to the evolution constructed of the Euclidean and Lorentzian time evolution.
More precisely, the systems considered in this paper are
\be \label{Local_states}
\begin{split}
    &\ket{\Psi_1}=\mathcal{N}_1e^{-i H_{\text{M$\ddot{o}$bius}}t_1}e^{-\epsilon H_0}\mathcal{O}(x)\ket{0},\ket{\Psi_2}=\mathcal{N}_2e^{-\epsilon H_0}e^{-i H_{\text{M$\ddot{o}$bius}}t_1}\mathcal{O}(x)\ket{0},\\
    &\ket{\Psi_3}=\mathcal{N}_3e^{-i H_{0}t_0}e^{-\epsilon H_{\text{M$\ddot{o}$bius}}}\mathcal{O}(x)\ket{0},\ket{\Psi_4}=\mathcal{N}_4 e^{-\epsilon H_{\text{M$\ddot{o}$bius}}}e^{-i H_{0}t_0}\mathcal{O}(x)\ket{0},\\
\end{split}
\ee
where the normalization constants satisfy with
\be \label{rdm_real}
\begin{split}
&\mathcal{N}_1^{-2} = \bra{0}\mathcal{O}^{\dagger}(x)e^{-2\epsilon H_0}\mathcal{O}(x)\ket{0}, \mathcal{N}_2^{-2} =  \bra{0}\mathcal{O}^{\dagger}(x)e^{i H_{\text{M$\ddot{o}$bius}}t_1}e^{-2\epsilon H_0}e^{-i H_{\text{M$\ddot{o}$bius}}t_1}\mathcal{O}(x)\ket{0}\\
 &\mathcal{N}_3^{-2} = \bra{0}\mathcal{O}^{\dagger}(x)e^{-2\epsilon H_{\text{M$\ddot{o}$bius}}}\mathcal{O}(x)\ket{0}, \mathcal{N}_4^{-2} =  \bra{0}\mathcal{O}^{\dagger}(x)e^{i H_{0}t_0}e^{-2\epsilon H_{\text{M$\ddot{o}$bius}}}e^{-i H_{0}t_0}\mathcal{O}(x)\ket{0}.\\
\end{split}
\ee
The inverse normalization constants, $\mathcal{N}^{-2}_{1}$ ($\mathcal{N}^{-2}_{3}$), are independent of $t_1$ ($t_0$), while $\mathcal{N}^{-2}_{2}$ ($\mathcal{N}^{-2}_{4}$) varies with times under the evolution induced by $H_{\text{M$\ddot{o}$bius}}$ ($H_0$).  
Thus, since the normalization constants for $i=1,3$ are constant and those for $i=2,4$, we call the time evolution for $i=1,3$ the unitary one, while we call the time evolution for $i=2,4$ the non-unitary one.
The systems considered are defined on the spatial circle with the circumference $L$.
The difference between $\ket{\Psi_1}$ ($\ket{\Psi_3}$) and $\ket{\Psi_2}$ ($\ket{\Psi_4}$) is the time ordering of the Euclidean and Lorentzian time evolution.
In $\ket{\Psi_{i=1,3}}$, the systems initially follow the Euclidean time evolution, and then undergo Lorentzian time evolution, while in $\ket{\Psi_{i=2,4}}$, the ordering of the time evolution is opposite. For $i=1,2$, the time evolution is induced by the inhomogeneous Hamiltonian, while for $i=3,4$, it is induced by the uniform Hamiltonian.
The density operators associated with the systems in (\ref{Local_states}) are given by %\textcolor{red}{MT: I think the first two lines should have $t_1$.}
\be \label{eq:density-operator-in-L}
\begin{split}
   & \rho_{1}=\mathcal{N}^2_1e^{-i H_{\text{M$\ddot{o}$bius}}t_1}e^{-\epsilon H_0}\mathcal{O}(x)\ket{0} \bra{0}\mathcal{O}^{\dagger}(x)e^{-\epsilon H_0}e^{i H_{\text{M$\ddot{o}$bius}}t_1},\\ 
   & \rho_{2}=\mathcal{N}^2_2e^{-\epsilon H_0}e^{-i H_{\text{M$\ddot{o}$bius}}t_1}\mathcal{O}(x)\ket{0} \bra{0}\mathcal{O}^{\dagger}(x)e^{i H_{\text{M$\ddot{o}$bius}}t_1}e^{-\epsilon H_0},\\
   & \rho_{3}=\mathcal{N}^2_3e^{-i t_0 H_0}e^{-\epsilon H_{\text{M$\ddot{o}$bius}}}\mathcal{O}(x)\ket{0} \bra{0}\mathcal{O}^{\dagger}(x)e^{-\epsilon H_{\text{M$\ddot{o}$bius}}}e^{i H_0t_0},\\ 
    & \rho_{4}=\mathcal{N}^2_4e^{-\epsilon H_{\text{M$\ddot{o}$bius}}}e^{-i t_0 H_0}\mathcal{O}(x)\ket{0} \bra{0}\mathcal{O}^{\dagger}(x)e^{i H_0t_0}e^{-\epsilon H_{\text{M$\ddot{o}$bius}}},\\ 
\end{split}
\ee

%%%%%%%%%%%%%%%%%%%%%%%%%%%%%%%%%%%%%%%
\subsection{Entanglement entropy}
%%%%%%%%%%%%%%%%%%%%%%%%%%%%%%%%%%%%%%%
In this paper, we will mainly explore the effect on the entanglement dynamics of the time ordering of Euclidean and Lorentzian time evolution by using entanglement entropy.
Therefore, let us define the entanglement entropy.
To this end, we divide the circle into a single interval, $A$, and a complement subsystem, $\overline{A}$, and then define a reduced density matrix associated to $A$ by $\rho_{A,i}=\Tr_{\overline{A}}\rho_{i}$.
Endpoints of $A$ are $X_1$ and $X_2$, and we assume that $X_1>X_2>0$.
We will consider the time-dependence of $S_{A,i}$ in only three cases in the following section (See Fig.\ \ref{Fig:subsystems} ):
\be
\begin{split}
    A=\begin{cases}
        \left\{x\big{|}0 \le x\le X_2, X_1\le x \le L\right\},~\text{where}~X_2>L-X_1>0~\text{and}~X_1>\f{L}{2}>X_2>0~\text{for}~(a)\\
        \left\{x\big{|}X_2 \le x\le X_1\right\}, ~\text{where}~ 0<X_2<\f{L}{2}<X_1~\text{and}~\f{L}{2}-X_2>X_1-\f{L}{2}  ~\text{for}~(b)\\
         \left\{x\big{|}X_2 \le x\le X_1\right\}, ~\text{where}~ 0<X_2<X_1<\f{L}{2}  ~\text{for}~(c)\\
    \end{cases}.
\end{split}
\ee
\begin{center}
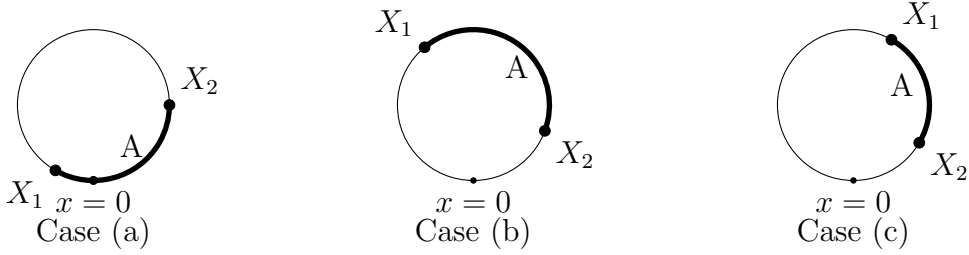
 
\begin{tikzpicture}
    \begin{scope}[shift={(0,0)}]
        % Figure a
        \draw (0,0) circle (1cm);
        \filldraw (00:1) circle (2pt) node[above right] {$X_2$};
        \filldraw (240:1) circle (2pt) node[below left] {$X_1$};
        \draw[line width=2pt] (240:1) arc[start angle=240, end angle=360, radius=1] node[midway, above] {A};
        \node at (0,-1.7) {Case (a)};
        \filldraw (0,-1) circle (1.5pt) node[below] {$x=0$};
    \end{scope}
    
    \begin{scope}[shift={(5,0)}]\label{Fig:cases}
        % Figure b
        \draw (0,0) circle (1cm);
        \filldraw (130:1) circle (2pt) node[above left] {$X_1$};
        \filldraw (-20:1) circle (2pt) node[below right] {$X_2$};
        \draw[line width=2pt] (130:1) arc[start angle=130, end angle=-20, radius=1] node[midway, below] {A};
        \node at (0,-1.7) {Case (b)};
         \filldraw (0,-1) circle (1pt) node[below] {$x=0$};
    \end{scope}
    
    \begin{scope}[shift={(10,0)}]
        % Figure c
        \draw (0,0) circle (1cm);
        \filldraw (60:1) circle (2pt) node[above right] {$X_1$};
        \filldraw (-30:1) circle (2pt) node[below right] {$X_2$};
        \draw[line width=2pt] (-30:1) arc[start angle=-30, end angle=60, radius=1] node[midway, left] {A};
        \node at (0,-1.7) {Case (c)};
         \filldraw (0,-1) circle (1pt) node[below] {$x=0$};
    \end{scope}
\end{tikzpicture}
\captionof{figure}{The subsystems considered in this paper. In (a), the subsystem includes only the origin. In (b), the subsystem includes only $x=\f{L}{2}$. In (c), the subsystem does not contain both the origin and $\f{L}{2}$. The origin and $\f{L}{2}$ are the fixed points during the SSD time evolution, as explained in Section \ref{Sec:trajectory-of-LO}.  \label{Fig:subsystems}}
\end{center}

%We define the entanglement entropies as Von Neumann entropy for the reduced density matrices,
%\be
%\begin{split}
 %   S_A=\lim_{n\rightarrow 1}S^{(n)}_{A}=-\Tr_A \rho_A \log{\rho_A}.
%\end{split}
%\ee
%In addition to it, 
We define the $n$-th moment of the R\'enyi entanglement entropy as 
\be
\begin{split}
    S^{(n)}_A=\f{1}{1-n}\log{\left[\Tr_A\rho^{(n)}_A\right]}.
\end{split}
\ee
Then,  we define entanglement entropy as the Von Neumann limit, where $n\rightarrow 1$, of $S^{(n)}_A$,
\be
S_A=\lim_{n\rightarrow 1}S^{(n)}_{A}=-\Tr_A\left(\rho_A \log{\rho_A}\right).
\ee

%\textcolor{red}{MT: There is no need to write (2.9) and (2.11) twice. I suggest removing (2.9) and defining R\'{e}nyi EE in (2.10) first then explain that von Neumann is obtained by taking a limit as in (2.11).}
%%%%%%%%%%%%%%%%%%%%%%%%%%%%%%%%%%%%%%%
\subsubsection{Euclidean path integral}
%%%%%%%%%%%%%%%%%%%%%%%%%%%%%%%%%%%%%%%
In this paper, we will employ the twist operator formalism suitable for the analytic computation \cite{Calabrese_2004,Calabrese_2009}.
To do so, we first define the Euclidean R\'enyi entanglement entropy, a cousin of the R\'enyi entanglement entropy, and then compute it in the path-integral formalism. 
Subsequently, we will perform the analytic continuation from Euclidean time to Lorentzian time. 
Consequently, we explore the time dependence of the entanglement entropy for a single interval.
Define the Euclidean density operators as a Euclidean counterpart of the density operators in (\ref{eq:density-operator-in-L}),
\be
\begin{split}
   & \rho_{1,E}=\mathcal{N}^2_{E,1}e^{- H_{\text{M$\ddot{o}$bius}}\tau_1}e^{-\epsilon H_0}\mathcal{O}(x)\ket{0} \bra{0}\mathcal{O}^{\dagger}(x)e^{-\epsilon H_0}e^{ H_{\text{M$\ddot{o}$bius}}\tau_1}=\mathcal{N}^2_{E,1}U^E_{1,\epsilon}\mathcal{O}(x)\ket{0} \bra{0}\mathcal{O}^{\dagger}(x)\tilde{U}^E_{1,\epsilon},\\
    & \rho_{2,E}=\mathcal{N}^2_{E,2}e^{-\epsilon H_0}e^{- H_{\text{M$\ddot{o}$bius}}\tau_1}\mathcal{O}(x)\ket{0} \bra{0}\mathcal{O}^{\dagger}(x)e^{ H_{\text{M$\ddot{o}$bius}}\tau_1}e^{-\epsilon H_0}=\mathcal{N}^2_{E,2}U^E_{2,\epsilon}\mathcal{O}(x)\ket{0} \bra{0}\mathcal{O}^{\dagger}(x)\tilde{U}^E_{2,\epsilon},\\
    & \rho_{3,E}=\mathcal{N}^2_{E,3}e^{-\tau_0 H_0}e^{-\epsilon H_{\text{M$\ddot{o}$bius}}}\mathcal{O}(x)\ket{0} \bra{0}\mathcal{O}^{\dagger}(x)e^{-\epsilon H_{\text{M$\ddot{o}$bius}}}e^{ H_0\tau_0}=\mathcal{N}^2_{E,3}U^E_{3,\epsilon}\mathcal{O}(x)\ket{0} \bra{0}\mathcal{O}^{\dagger}(x)\tilde{U}^E_{3,\epsilon},\\ 
    & \rho_{4,E}=\mathcal{N}^2_{E,4}e^{-\epsilon H_{\text{M$\ddot{o}$bius}}}e^{-\tau_0 H_0}\mathcal{O}(x)\ket{0} \bra{0}\mathcal{O}^{\dagger}(x)e^{\tau_0 H_0}e^{-\epsilon H_{\text{M$\ddot{o}$bius}}}=\mathcal{N}^2_{E,4}U^E_{4,\epsilon}\mathcal{O}(x)\ket{0} \bra{0}\mathcal{O}^{\dagger}(x)\tilde{U}^E_{4,\epsilon},\\ 
\end{split}
\ee
where $\mathcal{N}^2_{E,i=1 \sim 4}$ guarantee that $\Tr \rho_{i,E}=1$, and they are given by
\be
\begin{split}
    &\mathcal{N}^2_{E,1}=\f{1}{\bra{0}\mathcal{O}^{\dagger}(x)e^{-2\epsilon H_0}\mathcal{O}(x)\ket{0}},~\mathcal{N}^2_{E,2}=\f{1}{\bra{0}\mathcal{O}^{\dagger}(x)e^{ H_{\text{M$\ddot{o}$bius}}\tau_1}e^{-2\epsilon H_0}e^{- H_{\text{M$\ddot{o}$bius}}\tau_1}\mathcal{O}(x)\ket{0}},\\
    &\mathcal{N}^2_{E,3} = \f{1}{\bra{0}\mathcal{O}^{\dagger}(x)e^{-2\epsilon H_{\text{M$\ddot{o}$bius}}}\mathcal{O}(x)\ket{0}},~ \mathcal{N}_{E,4}^{2} =  \f{1}{\bra{0}\mathcal{O}^{\dagger}(x)e^{ H_{0}\tau_0}e^{-2\epsilon H_{\text{M$\ddot{o}$bius}}}e^{- H_{0}\tau_0}\mathcal{O}(x)\ket{0}}.\\
\end{split}
\ee
In this paper, $\mathcal{O}$ is a primary operator with the conformal dimension $(h_{\mathcal{O}},h_{\mathcal{O}})$. 
Define an Euclidean reduced density matrix by $\rho_{A,i,E}=\Tr_{\overline{A}}\rho_{i,E}$ and then define the Euclidean $n$-th R$\acute{e}$nyi and Von Neumann entanglement entropies  by
\be
S^{(n)}_{A,i,E}=\f{1}{1-n}\log{\left[\Tr_A\rho^n_{A,i,E}\right]},~~S_{A,i,E}=\lim_{n\rightarrow 1} S^{(n)}_{A,i,E}
\ee

In the twist operator formalism \cite{2004JSMTE..06..002C,2015JHEP...02..171A}, the Euclidean $n$-th R$\acute{e}$nyi entanglement entropy for single interval $A$ is given by
%\be
%S^{(n)}_{A,i}=\f{1}{1-n}\log{\left[\mathcal{N}^{2n}_{E,i=1\sim 4}\cdot\bra{0}\mathcal{O}_n^{\dagger}(x)\tilde{U}_{i,E}\sigma_{n}(X_1)\overline{\sigma}_n(X_2){U}_{i,E}\mathcal{O}_n(x)\ket{0}\right]},
%\ee
\be
S^{(n)}_{A,i}=\f{1}{1-n}\log{\left[\f{\left \langle\mathcal{O}_n^{\dagger}(x)\tilde{U}^E_{i,\epsilon}\sigma_{n}(X_1)\overline{\sigma}_n(X_2){U}^E_{i,\epsilon}\mathcal{O}_n(x)\right \rangle}{\left \langle\mathcal{O}_n^{\dagger}(x)\tilde{U}^E_{i,\epsilon}{U}^E_{i,\epsilon}\mathcal{O}_n(x)\right \rangle^n}\right]},
\ee
where $\langle \cdot \rangle$ denotes the vacuum expectation value, $\sigma_n$ and $\overline{\sigma}_n$ are the twist and anti-operators with the conformal dimension $\left(h_n, \overline{h}_n\right)=\left(\f{c(n^2-1)}{24n},\f{c(n^2-1)}{24n}\right)$, and $\mathcal{O}_n$ is a primary operator with the conformal dimensions $(nh_{\mathcal{O}},nh_{\mathcal{O}})$ as in \cite{2015JHEP...02..171A}. %\footnote{\textcolor{red}{\bf MN: Kotaro, the formula I wrote down is correct?}}. 
In terms of Heisenberg picture, $S^{n}_{A,i,E}$ reduces to 
%\textcolor{red}{MT: This is a minor point but I think $\mathcal{O}$ and $\mathcal{O}^\dagger$ should be swapped although it does not affect the result.}
\be
\begin{split}
    S^{(n)}_{A,i,E} &=\f{1}{1-n}\log\bigg{[}\left \langle \mathcal{O}^{\dagger,H}_{n,i,\epsilon}(x)\sigma_{n}(X_1)\overline{\sigma}_n(X_2)\mathcal{O}^H_{n,i,-\epsilon}(x)\right \rangle\bigg{]}-\f{n}{1-n}\log\bigg{[}\left \langle  \mathcal{O}^{\dagger,H}_{i,\epsilon}(x)\mathcal{O}^H_{i,-\epsilon}(x)\right \rangle\bigg{]},\\
   % S^{(n)}_{A,2,E} &=\f{1}{1-n}\log\bigg{[}\bra{0}\mathcal{N}^2_{E,2}e^{\epsilon H_0}e^{-H_{\text{M$\ddot{o}$bius}}\tau_1}\mathcal{O}(x)e^{H_{\text{M$\ddot{o}$bius}}\tau_1}e^{-\epsilon H_0}\sigma_{n}(X_1)\overline{\sigma}_n(X_2)\\
    %&~~~~~~~~~~~~~~~~~\times e^{-\epsilon H_0}e^{-H_{\text{M$\ddot{o}$bius}}\tau_1}\mathcal{O}^{\dagger}(x)e^{H_{\text{M$\ddot{o}$bius}}\tau_1}e^{\epsilon H_0}\ket{0}\bigg{]}
\end{split}
\ee
%\footnote{\textcolor{red}{\bf MN: We should check whether or not this is correct.}}
where $ \mathcal{O}^H_{n,i,\pm\epsilon}(x)$ denote the local operator in the Heisenberg picture. Here, we use the facts that $U_{i,E,\pm \epsilon}\ket{0}=\ket{0}$ and $U^E_{i,\epsilon}\tilde{U}^E_{i,-\epsilon}=U^E_{i,-\epsilon}\tilde{U}^E_{i,\epsilon}=\tilde{U}^E_{i,\epsilon}U^E_{i,-\epsilon}=\tilde{U}^E_{i,-\epsilon}U^E_{i,\epsilon}={\bf 1}$.
The primary operators in the Heisenberg picture are given by
\be \label{eq: op-in-Heisenberg}
 \mathcal{O}^H_{n,i,\epsilon}(x)=U^E_{i,-\epsilon}\mathcal{O}(w_x,\overline{w}_x)\tilde{U}^E_{i,\epsilon} =\left|\f{dw^{\text{New},i}_{\epsilon}}{dw_x}\right|^{2nh_{\mathcal{O}}}\mathcal{O}(w^{\text{New},i}_{\epsilon},\overline{w}^{\text{New},i}_{\epsilon}),
\ee
where complex coordinates $(w, \overline{w})$ are defined by $(w, \overline{w})=(\tau+ix,\tau-ix)$, and $(w_x, \overline{w}_x)=(ix,-ix)$.
The details of computation on (\ref{eq: op-in-Heisenberg}) and the trajectories of local operators during the evolution considered are reported in Appendix \ref{App:trajectory-H-t-C}.
Consequently, in terms of $(w^{\text{New},i}_{\pm \epsilon}, \overline{w}^{\text{New},i}_{\pm \epsilon})$, the $n$-th moment of R\'enyi entanglement entropy is given by
\be \label{eq:entanglement-entropy-fomula-1}
\begin{split}
    S^{(n)}_{A,i,E} &=\f{1}{1-n}\log{\left[\f{\left \langle \mathcal{O}_n^{\dagger}(w^{\text{New},i}_{\epsilon},\overline{w}^{\text{New},i}_{\epsilon})\sigma_{n}(w_{X_1},\overline{w}_{X_1})\overline{\sigma}_n(w_{X_2},\overline{w}_{X_2})\mathcal{O}_n(w^{\text{New},i}_{-\epsilon},\overline{w}^{\text{New},i}_{-\epsilon})\right \rangle}{\left \langle \mathcal{O}^{\dagger}(w^{\text{New},i}_{\epsilon},\overline{w}^{\text{New},i}_{\epsilon})\mathcal{O}(w^{\text{New},i}_{-\epsilon},\overline{w}^{\text{New},i}_{-\epsilon})\right \rangle^n}\right]}\\
    %&-\f{n}{1-n}\log\bigg{[}\left \langle \mathcal{O}^{\dagger}(w^{\text{New},i}_{\epsilon},\overline{w}^{\text{New},i}_{\epsilon})\mathcal{O}(w^{\text{New},i}_{-\epsilon},\overline{w}^{\text{New},i}_{-\epsilon})\right \rangle\bigg{]}.\\
   % S^{(n)}_{A,2,E} &=\f{1}{1-n}\log\bigg{[}\bra{0}\mathcal{N}^2_{E,2}e^{\epsilon H_0}e^{-H_{\text{M$\ddot{o}$bius}}\tau_1}\mathcal{O}(x)e^{H_{\text{M$\ddot{o}$bius}}\t·u_1}e^{-\epsilon H_0}\sigma_{n}(X_1)\overline{\sigma}_n(X_2)\\
    %&~~~~~~~~~~~~~~~~~\times e^{-\epsilon H_0}e^{-H_{\text{M$\ddot{o}$bius}}\tau_1}\mathcal{O}^{\dagger}(x)e^{H_{\text{M$\ddot{o}$bius}}\tau_1}e^{\epsilon H_0}\ket{0}\bigg{]}
\end{split}
\ee
Thus, the contributions from the conformal factors are canceled out.
In the following sections, we consider the time dependence of the entanglement entropy in various $2$d CFTs.
%%%%%%%%%%%%%%%%%%%%%%%%%
\subsection{Time trajectory of the local operators \label{Sec:trajectory-of-LO}}
\begin{figure}[htbp]
    \begin{tabular}{cc}
     \begin{minipage}[t]{0.5\hsize}
       \centering
       \includegraphics[keepaspectratio, scale=0.5]{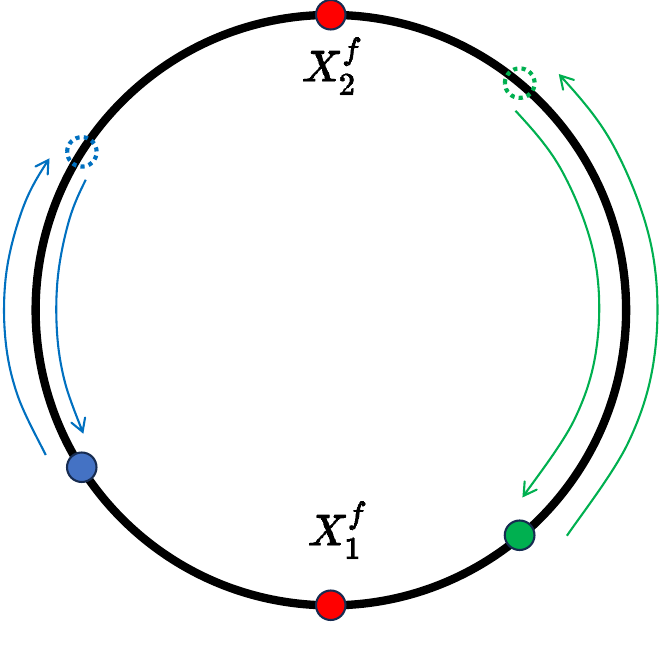}
        
   [a]  Möbius evolution
    
     \end{minipage}
     &\begin{minipage}[t]{0.5\hsize}
       \centering
        \includegraphics[keepaspectratio, scale=0.5]{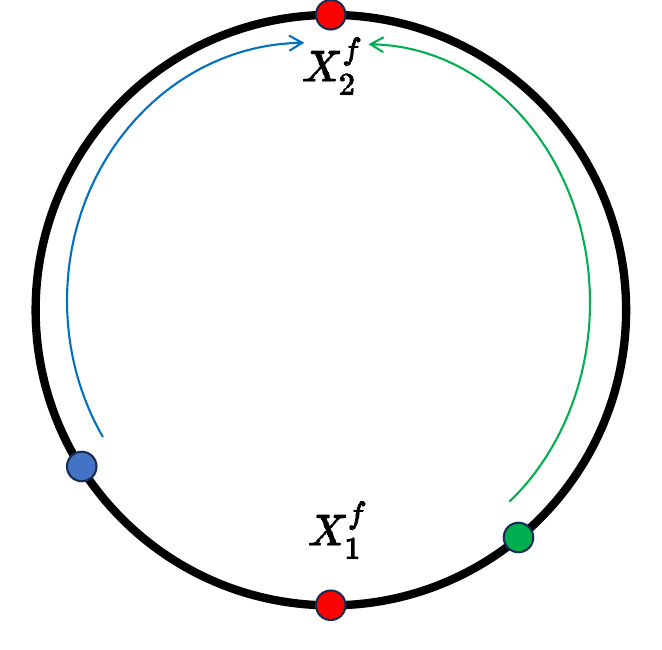}
        
         [b] SSD evolution

      \end{minipage} 

    \end{tabular}
      \caption{The picture illustrating the spatial trajectory of the operator during the time evolution. Panel [a] illustrates the time evolution of the local operator during the Möbius time evolution, while [b] does it during the SSD time evolution. The initial insertion points of the operator are marked by blue and green, and their trajectories are denoted by blue and green curves. The two fixed points, $X_1^f=0$ and $X_2^f=L / 2$, are marked in red.}
        \label{Fig:trajectory}
  \end{figure}

%%%%%%%%%%%%%%%%%%%%%%%%%
We will close this section by exploring the time trajectory, induced by the M\"obius and SSD Hamiltonian, of the local operator after performing the analytic continuation.
Before performing the analytic continuation, we define the spatial position of the local operator by
\be
X^{\text{New},i=1,2}_{\pm \epsilon}=\f{w^{\text{New},i=1,2}_{\pm \epsilon}-\overline{w}^{\text{New},i=1,2}_{\pm \epsilon}}{2i}.
\ee
Then, we perform the analytic continuation, $\tau_1=it_1$.
Let $X^f_1$ and $X^{f}_2$ denote $0$ and $\f{L}{2}$. 
We assume that at $t_1=0$, the local operator is inserted at $x$.
During the time evolution induced by the M\"obius Hamiltonian, the local operator for $X_1^f<x<X_2^f$ periodically oscillates with time between $X_1^f$ and $X_2^f$, while for $X_2^f<x<L$ the one does between $L$ and $X_2^f$ (See Fig. \ref{Fig:trajectory}).
During the time evolution induced by $H_{\text{SSD}}$, if the local operator is inserted at either $X_1^f$ or $X_2^f$, the local operator does not spatially move.
Therefore, we call them fixed points.
In addition to them, during the SSD time evolution, the local operator spatially moves to $X_2^f$, and consequently accumulates around $X_2^f$.

%%%%%%%%%%%%%%%%%%%%%%%%%%%%%%%%%%%%%%%
\section{Universal quantities \label{sec:universal}}
%%%%%%%%%%%%%%%%%%%%%%%%%%%%%%%%%%%%%%%
%\textcolor{red}{\bf MN: Before submitting the draft to arXiv, we should do cross-check the computations here again.}

In this section, we will explore the time dependence of universal quantities which are independent of the details of the $2$d CFTs. First, we present the time dependence of the partition function, and then we will report on the time dependence of the expectation value of the energy density.
%%%%%%%%%%%%%%%%%%%%%%%%%%%%%%%%%%%%%%%
\subsection{Time-dependence of partition function}
%%%%%%%%%%%%%%%%%%%%%%%%%%%%%%%%%%%%%%%
%Before reporting the time dependence of entanglement entropy, we will study the time dependence of partition function for the states considered in this paper.
Define the partition function as 
\be
\begin{split}
    Z_{i=1,\cdots,4}= \mathcal{N}^{-2}_{E,i=1,\cdots,4} %= 
   % Z^0 \left \langle \mathcal{O}^{\dagger}(w^{\text{New},i}_{\epsilon},\overline{w}^{\text{New},i}_{\epsilon})\mathcal{O}(w^{\text{New},i}_{-\epsilon},\overline{w}^{\text{New},i}_{-\epsilon})\right \rangle,
\end{split}
\ee
%\textcolor{red}{MT: I suggest dropping the $\mathcal{N}^{2}_{i=1,\cdots,4}$ because the way is defined here appears to be the inverse of (2.6). Also, would there be an additional conformal factor?}
where $Z_0$ is the partition function for the vacuum state.
Then, we perform the analytic continuation, $\tau_{i=0,1}=it_{0,1}$, and explore the time dependence of the partition functions.
Since $Z_0$ is time-independent and we are focusing on the time dependent piece of the partition function, we redefine the partition function as $\tilde{Z}_i=\f{Z_i}{Z_0}$.
%%%%%%%%%%%%%%%%%%%%%%%%%%%%%%%%%%%%%%%
\subsubsection{M\"obius case}
%%%%%%%%%%%%%%%%%%%%%%%%%%%%%%%%%%%%%%%
Let us consider the time dependence of the redefined partition function for the four density operators in (\ref{eq:density-operator-in-L}). 
For general $\theta \neq \infty$, their time dependence is determined by
\be
\begin{split}
    &\tilde{Z}_1=\left(\f{L}{\pi \sinh{\left(\f{2\pi \epsilon}{L}\right)}}\right)^{4h_{\mathcal{O}}}\underset{\f{\epsilon}{L} \ll 1}{\approx} \f{1}{\left(2\epsilon\right)^{4h_{\mathcal{O}}}}=:\tilde{Z}^{(1)}_1,\\
     % &\tilde{Z}_2=\tilde{Z}_1 \times \left[\f{1}{\left(\chi_1^2-\chi_2^2\right)^2}\right]^{h_{\mathcal{O}}}\underset{\f{\epsilon}{L} \ll 1}{\approx} \tilde{Z}^{(1)}_1 \times \left[\f{1}{\left(\chi_1^2-\chi_2^2\right)^2}\right]^{h_{\mathcal{O}}},\\
     &\tilde{Z}_2=\tilde{Z}_1 \times  \left[\frac{1}{\chi \overline{\chi}}\right]^{2 h_{\mathcal{O}}}\underset{\f{\epsilon}{L} \ll 1}{\approx} \tilde{Z}^{(1)}_1 \times \left[\frac{1}{\chi \overline{\chi}}\right]^{2 h_{\mathcal{O}}},\\
    &\tilde{Z}_3 =\left[\f{\pi}{L\sinh{\left(\f{2\pi \epsilon}{L_{\text{eff}}}\right)}\cosh{2\theta}\left( 1-\cos{\left(\frac{2 \pi x}{L}\right)}\tanh{2\theta}\right)}\right]^{4h_{\mathcal{O}}}\underset{\f{\epsilon}{L} \ll 1}{\approx}\tilde{Z}^{(1)}_1 \times\left(\frac{1}{ \left( 1-\cos(\frac{2 \pi x}{L}) \tanh (2 \theta)  \right)}\right)^{4 h_\mathcal{O}},\\
    &\tilde{Z}_4 =\left[\f{\pi}{L\cosh{2\theta}\sinh{\left(\f{2\pi \epsilon}{L_{\text{eff}}}\right)}}\right]^{4h_{\mathcal{O}}}\times\left[\f{1}{\left( 1-\cos(\frac{2 \pi (t_0+x)}{L}) \tanh (2 \theta)  \right)\left( 1-\cos(\frac{2 \pi (t_0-x)}{L}) \tanh (2 \theta)  \right)}\right]^{2h_{\mathcal{O}}}\\
    &\underset{\f{\epsilon}{L} \ll 1}{\approx}\tilde{Z}^{(1)}_1 \times\left[\f{1}{\left( 1-\cos(\frac{2 \pi (t_0+x)}{L}) \tanh (2 \theta)  \right)\left( 1-\cos(\frac{2 \pi (t_0-x)}{L}) \tanh (2 \theta)  \right)}\right]^{2h_{\mathcal{O}}}
    %%%%%%%%%%%%%%%
\end{split}
\ee
where the $\chi$ are defined as
% \be
% \begin{split}
%     \chi_1=1+\left(\cosh{4\theta}-\sinh{4\theta}\cos{\left(\f{2\pi x}{L}\right)}-1\right)\left(\f{1-\cos{\left(\f{2\pi t_1}{L_{\text{eff}}}\right)}}{2}\right),~\chi_2=\sinh{2\theta}\sin{\left(\f{2\pi t_1}{L_{\text{eff}}}\right)}\sin{\left(\f{2\pi x}{L}\right)}
% \end{split}
% \ee
\be
\begin{split}
    \chi=\cos ^2\left(\frac{\pi  t_1}{L_{\text{eff}}}\right)+\sin ^2\left(\frac{\pi  t_1}{L_{\text{eff}}}\right) \left(\cosh (2 \theta )+\sinh (2 \theta ) \left(-\cos \left(\frac{2 \pi  x}{L}\right)+i \sin \left(\frac{2 \pi  x}{L}\right)\right)\right)^2
\end{split}
\ee
\if[0]
\be
\begin{split}
&\tilde{Z}_1\approx \f{1}{\left(2\epsilon\right)^{4h_{\mathcal{O}}}},\tilde{Z}_3 \approx\tilde{Z}_1 \times\left(\frac{1}{ \left( \cos(\frac{2 \pi x}{L}) \tanh (2 \theta)-1  \right)}\right)^{4 h_\mathcal{O}},\\
    %%%%%%%%%%%%%%%
&\tilde{Z}_2\approx \tilde{Z}_1 \times \left( \frac{\text{sech}^4(2 \theta ) \left(-\tanh (2 \theta ) \sin \left(\frac{\pi  t_1}{L}\right)+\text{sech}(2 \theta ) e^{\frac{2 i \pi  x}{L}} \cos \left(\frac{\pi  t_1}{L}\right)+e^{\frac{2 i \pi  x}{L}} \sin \left(\frac{\pi  t_1}{L}\right)\right)}{ \left(\tanh (2 \theta ) \sin \left(\frac{\pi  t_1}{L}\right)+\text{sech}(2 \theta ) e^{\frac{2 i \pi  x}{L}} \cos \left(\frac{\pi  t_1}{L}\right)-e^{\frac{2 i \pi  x}{L}} \sin \left(\frac{\pi  t_1}{L}\right)\right)} \right)^{2 h_{\mathcal{O}}} \\
     %%%%%%%%%%%%%%%%%%%%%%
     &\times\left(\f{1} {\left(\text{sech}(2 \theta ) \cos \left(\frac{\pi  t_1}{L}\right)-\sin \left(\frac{\pi  t_1}{L}\right) \left(\tanh (2 \theta ) e^{\frac{i \pi  x}{L}}-e^{-\frac{i \pi  x}{L}}\right)\right)^3}\right)^{2 h_{\mathcal{O}}} \\
     %%%%%%%%%%%%%%%%%%%%%%%%%%%%
    &\times\left(\f{1}{\left(\text{sech}(2 \theta ) \cos \left(\frac{\pi  t_1}{L}\right)+\sin \left(\frac{\pi  t_1}{L}\right) \left(\tanh (2 \theta ) e^{\frac{i \pi  x}{L}}-e^{-\frac{i \pi  x}{L}}\right)\right)} \right)^{2 h_{\mathcal{O}}},\\
    %%%%%%%%%%%
    &\tilde{Z}_4 \approx\tilde{Z}_1 \times\left(
\frac{e^{\frac{4 i \pi x}{L}}}{\left(-1+\cos \left(\frac{2 \pi(x-t_0)}{L}\right) \tanh (2 \theta)\right)\left(-1+\cos \left(\frac{2 \pi(x+t_0)}{L}\right) \tanh (2 \theta)\right)}\right)^{2 h_\mathcal{O}}.
\end{split}
\ee
\fi
Here, since for $\theta \neq \infty$, $\left|1- \cos{(A)}\tanh{2\theta}\right|>0$%\footnote{\textcolor{red}{We should check if this is true for $i=2$.}} 
, where $A$ is assumed to be a real parameter, $\tilde{Z}_{i=3,4}$ do not diverge.
%\footnote{\textcolor{red}{MN: Weibo Mao, please confirm $\tilde{Z}_{I=2}$ does not diverge during the time evolution.\textcolor{green}{MWB: Confirmed, and added the reason.}}} 
 Additionally, since $\left|\chi\right|>0$ for any $t_1, x, \theta \in \mathbb{R}$ , %we have $\left(\chi \overline{\chi}\right)>0$.  it implies that
 $\tilde{Z}_{i=2}$ does not diverge.
Therefore, these partition functions for the M\"obius case are well-defined and the Euclidean time evolution induced by the M\"obius Hamiltonian. 
The redefined partition functions, $\tilde{Z}_{i=1,3}$, are independent of time, while $Z_{i=2,4}$ depends on the time.
The partition function, $\tilde{Z}_{i=2}$, is a periodic function with the period $L_{\text{eff}}$, while $\tilde{Z}_{i=4}$ is the periodic one with $L$.%\footnote{\textcolor{red}{We should check if this is correct. Also, we should check if $Z_{i=2}$ is the periodic function and what is the period.}}.
%%%%%%%%%%%%%%%%%%%%%%%%%%%%%%%%%%%%%%%
\subsubsection{SSD limit}
%%%%%%%%%%%%%%%%%%%%%%%%%%%%%%%%%%%%%%%
Now, we look closely at the time dependence of the partition functions for the states considered in the SSD limit, where $\theta \rightarrow \infty$.
Since $\tilde{Z}_{i=1}$ is independent of $\theta$, $\tilde{Z}_{i=1}$ in the SSD limit is the same as that for finite $\theta$.
In the SSD limit, the redefined partition function for $i=3$ reduces to
\be
\begin{split}
   &%\tilde{Z}_1\approx\f{1}{\left(2\epsilon\right)^{4h_{\mathcal{O}}}}~,
   \tilde{Z}_3= \left(\frac{\csc ^2\left(\frac{\pi  x}{L}\right)}{4 \epsilon }\right)^{4h_{\mathcal{O}}}.\\
   \end{split}
\ee
Thus, $\tilde{Z}_3$ depends on the insertion point of the local operator. 
If the local operator is inserted at $x=0$, $\tilde{Z}_3$ becomes infinite.
Since the Hamiltonian density of $H_{\text{SSD}}$ vanishes at $x=0$, and $\mathcal{O}(x=0)$ commutes with $e^{-\epsilon H_{\text{SSD}}}$, this damping functor, $e^{-\epsilon H_{\text{SSD}}}$, can not keep the state finite.
In other words, this suggests that at $x=0$, the Euclidean time evolution induced by $H_{\text{SSD}}$ does not play as a regulator.
%The divergence of $Z_3$ suggests that the Euclidean time evolution, $e^{-\epsilon H_{\text{SSD}}}$, can not inhibit the UV divergence of the system.
%This may be because the Hamiltonian density at $x=0$ of $H_{\text{SSD}}$ commutes with the local operator at the same spatial point. 
In the SSD limit, the time dependence of the partition functions for $i=2,4$ at the leading other of the small $\epsilon$ reduces to 
\be
\begin{split}
       \tilde{Z}_2&=\tilde{Z}_1 \times \left[\f{1}{\left[16\sin^4{\left(\f{\pi x}{L}\right)}\left(\f{\pi t_1}{L}\right)^4-8\sin^2{\left(\f{\pi x}{L}\right)}\cos{\left(\f{2\pi x}{L}\right)}\left(\f{\pi t_1}{L}\right)^2+1\right]^2}\right]^{h_{\mathcal{O}}}\\
       &\underset{\f{\epsilon}{L} \ll 1}{\approx} \tilde{Z}^{(1)}_1 \times \left[\f{1}{\left[16\sin^4{\left(\f{\pi x}{L}\right)}\left(\f{\pi t_1}{L}\right)^4-8\sin^2{\left(\f{\pi x}{L}\right)}\cos{\left(\f{2\pi x}{L}\right)}\left(\f{\pi t_1}{L}\right)^2+1\right]^2}\right]^{h_{\mathcal{O}}},\\
  \tilde{Z}_4&= \left(\frac{1}{2 \epsilon \left[\cos{\left(\f{2\pi x}{L}\right)}-\cos{\left(\f{2\pi t_0}{L}\right)}\right]} \right)^{4h_{\mathcal{O}}},
\end{split}
\ee
where at $t_0=n L\pm x$, where $n$ is an integer, $\tilde{Z}_4$ diverges. %\footnote{\textcolor{red}{\bf MN: We should check it is in the case even without taking the small $\epsilon$ limit.}}. 
Here, $\tilde{Z}_4$ is a periodic function of $t_0$ with a period $L$.
Our interpretation is that when the excitations created by the local operator hit $X=0$, the spatial point where the Hamiltonian density of $H_{\text{SSD}}$ vanishes, the partition function becomes infinite. A possible explanation is that the Euclidean time evolution induced by $H_{\text{SSD}}$ does not work as a regulator at the origin.
In the large $t_1$-regime, the $t_1$ dependence of $Z_2$ is approximately given by
\be
\begin{split}
    \tilde{Z_2}\approx \begin{cases}
        \left(\frac{L^8 \csc ^8\left(\frac{\pi  x}{L}\right)}{4096 \pi ^8 t_1^8 \epsilon ^4}\right)^{h_{\mathcal{O}}} & x\neq 0\\
        \frac{1}{(2 \epsilon)^{4h_{\mathcal{O}}}} & x=0
    \end{cases}.
\end{split}
\ee
Thus, when the operator is inserted at $x\neq 0$, the partition function decays with $t_1$, while when the operator is inserted at $x=0$, the partition function is independent of time.
This suggests that during the SSD time evolution, the local operator at $x=0$ does not change the entanglement structure.  
%%%%%%%%%%%%%%%%%%%%%%%%%%%%%%%%%%%%%%%
\subsection{Non-unitarity}
%%%%%%%%%%%%%%%%%%%%%%%%%%%%%%%%%%%%%%%
Let us define the current as the derivative of the partition with respect to time,
\be
{\bf j}_i= \f{d \left \langle \Psi_i(t)\big{|}\Psi_i(t) \right \rangle}{dt_{a=0,1}},
\ee
where $i$ labels the states. 
The symbol $t_1$ is for $i=1,2$, while $t_0$ is for $i=3,4$.
For $i=1,3$, ${\bf j}_i$ vanishes, which means that the time evolution is unitary because the norm of state is conserved.
For $i=2,4$, ${\bf j}_i$ does not vanish. 
The currents for $i=2,4$ are determined by
\be \label{eq:current-Mobius}
\begin{split}
&{\bf j}_2=i\left(\f{2\pi}{L}\right)\sinh{\left(\f{2\pi \epsilon}{L}\right)}\tanh{2\theta}\bra{\Psi_2(t)}(L_1-L_{-1}+\overline{L}_1-\overline{L}_{-1})\ket{\Psi_2(t)},\\
&{\bf j}_4=i\left(\f{2\pi}{L}\right)\left(\f{2\tanh{2\theta}}{\text{csch}{4\theta}}\right)^{\f{1}{2}}\sinh{\left(\f{2\pi \epsilon}{L}\sqrt{2\tanh{2\theta}\text{csch}{4\theta}}\right)}\bra{\Psi_4(t)}(L_1-L_{-1}+\overline{L}_1-\overline{L}_{-1})\ket{\Psi_4(t)},\\
\end{split}
\ee
We can see from (\ref{eq:current-Mobius}) that for both $\ket{\Psi}_2$ and $\ket{\Psi}_4$, the dissipation of the partition function is determined by $L_{1}-L_{-1}+\overline{L}_1-\overline{L}_{-1}$.
In SSD limit, the equations for the current reduce to 
\be
\begin{split}
    &{\bf j}_2=i\left(\f{2\pi}{L}\right)\sinh{\left(\f{2\pi \epsilon}{L}\right)}\bra{\Psi_2(t)}(L_1-L_{-1}+\overline{L}_1-\overline{L}_{-1})\ket{\Psi_2(t)},\\
     &{\bf j}_4=2\epsilon i \left(\f{2\pi}{L}\right)^2 \bra{\Psi_4(t)}(L_1-L_{-1}+\overline{L}_1-\overline{L}_{-1})\ket{\Psi_4(t)}.
\end{split}
\ee
Non-unitarity of the dynamics for $i=2,4$ is due to the time evolution of expectation value of $L_{1}-L_{-1}+\overline{L}_1-\overline{L}_{-1}$.% \footnote{\textcolor{red}{\bf MN: If we can give some interpretation on $L_{1}-L_{-1}+\overline{L}_1-\overline{L}_{-1}$, it is better. If no one has good interpretations of this, I will remove this footnote.}}.

%%%%%%%%%%%%%%%%%%%%%%%%%%%%%%%%%%%%%%%
\subsection{Energy density}
%%%%%%%%%%%%%%%%%%%%%%%%%%%%%%%%%%%%%%%
Above, we explored the global properties of the systems by using the (redefined) partition functions.
Here, we will explore the time dependence of a local universal quantity, the energy density.
In our calculation, we begin with the calculation of the energy density in the Euclidean path integral, and then perform the analytic continuation to real time.
Define the expectation values of chiral and anti-chiral parts of the energy density as 
\be
\begin{split}
&\left \langle T_{ww} (w_X,\overline{w}_X)\right \rangle_{i,E} =\Tr \left( \rho_{i,E} T_{ww}(w_X,\overline{w}_X) \right)= \text{\footnotesize{$\f{\left\langle \mathcal{O}^{\dagger}\left(w^{\text{New},i}_{\epsilon}, \overline{w}^{\text{New},i}_{\epsilon}\right) T_{ww}(w_X,\overline{w}_X) \mathcal{O}\left(w^{\text{New},i}_{-\epsilon}, \overline{w}^{\text{New},i}_{-\epsilon}\right)\right \rangle }{\left\langle \mathcal{O}^{\dagger}\left(w^{\text{New},i}_{\epsilon},\overline{w}^{\text{New},i}_{\epsilon}\right)\mathcal{O}\left(w^{\text{New},i}_{-\epsilon}, \overline{w}^{\text{New},i}_{-\epsilon}\right)\right \rangle}$}},\\
&\left \langle T_{\overline{w}\overline{w}}(w_X,\overline{w}_X) \right \rangle_{i,E} =\Tr \left( \rho_{i,E} T_{\overline{w}\overline{w}}(w_X,\overline{w}_X) \right)= \text{\footnotesize{$\f{\left\langle \mathcal{O}^{\dagger}\left(w^{\text{New},i}_{\epsilon}, \overline{w}^{\text{New},i}_{\epsilon}\right) T_{\overline{w}\overline{w}}(w_X,\overline{w}_X) \mathcal{O}\left(w^{\text{New},i}_{-\epsilon}, \overline{w}^{\text{New},i}_{-\epsilon}\right)\right \rangle }{\left\langle \mathcal{O}^{\dagger}\left(w^{\text{New},i}_{\epsilon}, \overline{w}^{\text{New},i}_{\epsilon}\right)\mathcal{O}\left(w^{\text{New},i}_{-\epsilon}, \overline{w}^{\text{New},i}_{-\epsilon}\right)\right \rangle}$}},\\
\end{split}
\ee
where the energy density is assumed to be defined as the linear combination of $T_{ww}$ and the chiral and anti-chiral parts of the energy density, $T=T_{ww}+T_{\overline{w}\overline{w}}$. 
We call $\left\langle T_{ww} \right \rangle_i$ and $\left\langle T_{\overline{w}\overline{w}}\right \rangle_i$ chiral and anti-chiral energy densities, respectively.
By performing the conformal map from cylinder to the flat space and using the Ward-Takahashi identity, the the expectation values of energy densities \footnote{Here, the Ward-Takahashi identity is \be\begin{split}
    &\left\langle T_{zz}(z) \mathcal{O}(z_1,\overline{z}_1)\mathcal{O}(z_2,\overline{z}_2) \right \rangle=\sum_{i=1}^2\left[\f{h_{\mathcal{O}}}{\left(z-z_i\right)^2}+\f{1}{z-z_i}\partial_{z_i}\right]\left\langle \mathcal{O}(z_1,\overline{z}_1)\mathcal{O}(z_2,\overline{z}_2) \right \rangle,\\
    &\left\langle T_{\overline{z}\overline{z}}(\overline{z}) \mathcal{O}(z_1,\overline{z}_1)\mathcal{O}(z_2,\overline{z}_2) \right \rangle=\sum_{i=1}^2\left[\f{h_{\mathcal{O}}}{\left(\overline{z}-\overline{z}_i\right)^2}+\f{1}{z-z_i}\partial_{\overline{z}_i}\right]\left\langle \mathcal{O}(z_1,\overline{z}_1)\mathcal{O}(z_2,\overline{z}_2) \right \rangle.\\
\end{split}\ee} are given by
\be
\begin{split}
    \left \langle T_{ww} (w_X,\overline{w}_X)\right \rangle_{i,E} &=-\f{c}{24}\left(\f{2\pi}{L}\right)^2+h_{\mathcal{O}}\left(\f{dz_{X}}{dw_X}\right)^2\left[\f{1}{z_X-z^{\text{New},i}_{\epsilon}}-\f{1}{z_X-z^{\text{New},i}_{-\epsilon}}\right]^2,\\
    %&-\left(\f{4}{\left(z_{X}-z^{\text{New},i}_{\epsilon}\right)\left(z_{X}-z^{\text{New},i}_{-\epsilon}\right)\left(z^{\text{New},i}_{\epsilon}-z^{\text{New},i}_{-\epsilon}\right)^{2h_{\mathcal{O}}}}\right)\Bigg{)},\\
    \left \langle T_{\overline{w}\overline{w}} (w_X,\overline{w}_X)\right \rangle_{i,E} &=-\f{c}{24}\left(\f{2\pi}{L}\right)^2+h_{\mathcal{O}}\left(\f{d\overline{z}_{X}}{d\overline{w}_X}\right)^2\left[\f{1}{\overline{z}_X-\overline{z}^{\text{New},i}_{\epsilon}}-\f{1}{\overline{z}_X-\overline{z}^{\text{New},i}_{-\epsilon}}\right]^2.\\
    %&-\left(\f{4}{\left(\overline{z}_{X}-\overline{z}^{\text{New},i}_{\epsilon}\right)\left(\overline{z}_{X}-\overline{z}^{\text{New},i}_{-\epsilon}\right)\left(\overline{z}^{\text{New},i}_{\epsilon}-\overline{z}^{\text{New},i}_{-\epsilon}\right)^{2h_{\mathcal{O}}}}\right)\Bigg{)}.\\
\end{split}
\ee

%%%%%%%%%%%%%%%%%%%%%%%%%%%%%%%%%%%%%%%
\subsubsection{M\"obius case}
%%%%%%%%%%%%%%%%%%%%%%%%%%%%%%%%%%%%%%%
First, we will closely look at the time dependence of the chiral and anti-chiral energy densities during the M\"obius time evolution.
After performing analytic continuations, $\tau_{i=0,1}=it_{0,1}$, at the second order of the small $\epsilon$ expansion, they are determined by
\be
\begin{split}
    &\left \langle T_{ww}\left(w_X\right) \right \rangle_{i=1,\cdot,4} \approx \left(\f{2\pi}{L}\right)^2\left[-\f{c}{24}+\epsilon^2\mathcal{T}^{\theta}_{i=1,\cdot,4}\right],
    %%%%%%%%%%%%%%%%%%%%%%%%%%%%%%%
    ~\left \langle T_{\overline{w}\overline{w}}\left(\overline{w}_X\right) \right \rangle_{i=1,\cdot,4}\approx \left(\f{2\pi}{L}\right)^2\left[-\f{c}{24}+\epsilon^2\overline{\mathcal{T}}^{\theta}_{i=1,\cdot,4}\right],\\
    %%%%%%%%%%%%%%%%%%%%%%%%%%%%%%%%%%%%%%%%%%%%%%%%%%%%%
   % & \left \langle T_{ww}\left(w_X\right) \right \rangle_{2} \approx \left(\f{2\pi}{L}\right)^2\left[-\f{c}{24}+\epsilon^2\mathcal{T}^{\theta}_{i=2}\right],
    %%%%%%%%%%%%%%%%%%%%%%%%%%%%%%%%%%%%%%%%%%%%%%%%%%%%%
 %   ~ \left \langle T_{\overline{w}\overline{w}}\left(\overline{w}_X\right) \right \rangle_{2} \approx \left(\f{2\pi}{L}\right)^2\left[-\f{c}{24}+\epsilon^2 \overline{\mathcal{T}}^{\theta}_{i=2}\right],\\
    %%%%%%%%%%%%%%%%%%%%%%%%%%%%%%%%%%%%%%%%%%%%%%%%%%%%%
  % & \left \langle T_{ww}\left(w_X\right) \right \rangle_{3} \approx \left(\f{2\pi}{L}\right)^2\left[-\f{c}{24}+\epsilon^2 \mathcal{T}^{\theta}_{i=3}\right],\\
    %%%%%%%%%%%%%%%%%%%%%%%%%%%%%%%%%%%%%%%%%%%%%%%%%%%%%
%   & \left \langle T_{\overline{w}\overline{w}}\left(\overline{w}_X\right) \right \rangle_{3} \approx \left(\f{2\pi}{L}\right)^2\left[-\f{c}{24}+\epsilon^2\overline{\mathcal{T}}^{\theta}_{i=3}\right],\\
    %%%%%%%%%%%%%%%%%%%%%%%%%%%%%%%%%%%%%%%%%%%%%%%%%%%%%
%    & \left \langle T_{ww}\left(w_X\right) \right \rangle_{4} \approx \left(\f{2\pi}{L}\right)^2\left[-\f{c}{24}+\epsilon^2 \mathcal{T}^{\theta}_{i=4}\right],\\
    %%%%%%%%%%%%%%%%%%%%%%%%%%%%%%%%%%%%%%%%%%%%%%%%%%%%%
 %  & \left \langle T_{\overline{w}\overline{w}}\left(\overline{w}_X\right) \right \rangle_{4} \approx  \left(\f{2\pi}{L}\right)^2\left[-\f{c}{24}+\epsilon^2 \overline{\mathcal{T}}^{\theta}_{i=4}\right],\\
    %%%%%%%%%%%%%%%%%%%%%%%%%%%%%%%%%%%%%%%%%%%%%%%%%%%%%
\end{split}
\ee
where functions, $\mathcal{T}^{\theta}_{i=1,\cdots, 4}$, $\overline{\mathcal{T}}^{\theta}_{i=1,\cdots, 4}$, $D_{i=1,2,\pm}$ and $N_{\pm}$, are defined as 
\be
\begin{split}
    &\mathcal{T}^{\theta}_{i=1}=\f{\pi^2h_{\mathcal{O}}}{L^2 D_{+}},~\overline{\mathcal{T}}^{\theta}_{i=1}=\f{\pi^2h_{\mathcal{O}}}{L^2 D_{-}},~\mathcal{T}^{\theta}_{i=2}=\f{\pi^2 h_{\mathcal{O}} N_+}{L^2 D_+},~\overline{\mathcal{T}}^{\theta}_{i=2}=\f{\pi^2h_{\mathcal{O}} N_-}{L^2 D_-},\\
    %%%%%%%%%%%%%%%%%%%%%%%%%%%%%%%%%%%%%%%%%%%%%%%%%%%%%
    &\mathcal{T}^{\theta}_{i=3}=\f{\pi^2 \epsilon^2 h_{\mathcal{O}}\left(\tanh{2\theta}\cos{\left(\f{2\pi x}{L}\right)}-1\right)^2}{L^2\sin^4{\left[\f{\pi(t_0+X-x)}{L}\right]}},~\overline{\mathcal{T}}^{\theta}_{i=3}=\f{\pi^2  h_{\mathcal{O}}\left(\tanh{2\theta}\cos{\left(\f{2\pi x}{L}\right)}-1\right)^2}{L^2\sin^4{\left[\f{\pi(t_0-X+x)}{L}\right]}},\\
    %%%%%%%%%%%%%%%%%%%%%%%%%%%%%%%%%%%%%%%%%%%%%%%%%%%%%
    &\mathcal{T}^{\theta}_{i=4}=\f{\pi^2  h_{\mathcal{O}}\left(\tanh{2\theta}\cos{\left(\f{2\pi (t_0-x)}{L}\right)}-1\right)^2}{L^2\sin^4{\left[\f{\pi(t_0+X-x)}{L}\right]}},~
    \overline{\mathcal{T}}^{\theta}_{i=4}=\f{\pi^2 h_{\mathcal{O}}\left(\tanh{2\theta}\cos{\left(\f{2\pi (t_0+x)}{L}\right)}-1\right)^2}{L^2\sin^4{\left[\f{\pi(t_0-X+x)}{L}\right]}},\\
    %%%%%%%%%%%%%%%%%%%%%%%%%%%%%%%%%%%%%%%%%%%%%%%%%%%%%%
    % &D_1=\bigg{[}\sin{\left[\f{\pi\left(t_1 \text{sech}(2\theta)+x-X\right)}{L}\right]}\left(1-\cosh{2\theta}\right)-\sin{\left[\f{\pi\left(t_1 \text{sech}(2\theta)-x+X\right)}{L}\right]}\left(\cosh{2\theta}+1\right)\\
    % &+\sinh{2\theta}\left(\sin{\left[\f{\pi\left(t_1 \text{sech}(2\theta)-x-X\right)}{L}\right]}+\sin{\left[\f{\pi\left(t_1 \text{sech}(2\theta)+x+X\right)}{L}\right]}\right)\bigg{]}^4,\\
    % &D_2=\bigg{[}\sin{\left[\f{\pi\left(t_1 \text{sech}(2\theta)-x+X\right)}{L}\right]}\left(1-\cosh{2\theta}\right)-\sin{\left[\f{\pi\left(t_1 \text{sech}(2\theta)+x-X\right)}{L}\right]}\left(\cosh{2\theta}+1\right)\\
    % &+\sinh{2\theta}\left(\sin{\left[\f{\pi\left(t_1 \text{sech}(2\theta)-x-X\right)}{L}\right]}+\sin{\left[\f{\pi\left(t_1 \text{sech}(2\theta)+x+X\right)}{L}\right]}\right)\bigg{]}^4,\\
    &D_{\pm}=\bigg{[}\frac{1}{2} \left(\sinh ^2(2 \theta ) \cos ^2\left(\frac{\pi  (x+X)}{L}\right)+\cosh ^2(2 \theta ) \cos ^2\left(\frac{\pi  (x-X)}{L}\right)\right)\\
    &-\left(\sinh (4 \theta ) \cos \left(\frac{\pi  (x-X)}{L}\right) \cos \left(\frac{\pi  (x+X)}{L}\right)+\sin ^2\left(\frac{\pi  (x-X)}{L}\right)\right)\\
    &+\frac{1}{2} \cos \left(\frac{2 \pi  t_1}{L_{\text{eff}}}\right) \left(\cosh ^2(2 \theta ) \left(-\cos ^2\left(\frac{\pi  (x-X)}{L}\right)\right)-\sinh ^2(2 \theta ) \cos ^2\left(\frac{\pi  (x+X)}{L}\right)\right)\\
    &+\frac{1}{2} \cos \left(\frac{2 \pi  t_1}{L_{\text{eff}}}\right) \left(\sinh (4 \theta ) \cos \left(\frac{\pi  (x-X)}{L}\right) \cos \left(\frac{\pi  (x+X)}{L}\right)+\sin ^2\left(\frac{\pi  (x-X)}{L}\right)\right)\\
    &\pm \sin \left(\frac{2 \pi  t_1}{L_{\text{eff}}}\right) \sin \left(\frac{\pi  (x-X)}{L}\right) \left(\sinh (2 \theta ) \cos \left(\frac{\pi  (x+X)}{L}\right)-\cosh (2 \theta ) \cos \left(\frac{\pi  (x-X)}{L}\right)\right)\bigg{]}^2,\\
    &N_{\pm}=\text{\small{$\bigg{[}\cosh ^2(2 \theta ) \left(1-\tanh (2 \theta ) \cos \left(\frac{2 \pi  x}{L}\right)\right)+\frac{1}{2} \sinh (4 \theta ) \cos \left(\frac{2 \pi  t_1}{L_{\text{eff}}}\right) \left(\cos \left(\frac{2 \pi  x}{L}\right)-\tanh (2 \theta )\right)$}}\\
    &\text{\footnotesize{$\pm \sinh (2 \theta ) \sin \left(\frac{2 \pi  x}{L}\right) \sin \left(\frac{2 \pi  t_1}{L_{\text{eff}}}\right)\bigg{]}^2$}},
\end{split}
\ee

where the effective system size is defined as $L_{\text{eff}}=L \cosh{2\theta}$.
Thus, the leading order behaviors of these densities in the small $\epsilon$ expansion are $\mathcal{O}(1)$ and independent of time and the location of the local operator, while the next-to-leading order terms %\textcolor{red}{(I think "next-to-leading order term" sounds better)} 
are $\mathcal{O}(\epsilon^2)$, and the insertion of the local operator contributes to these behaviors. 
The second order terms of energy densities for $i=3,4$ are periodic functions with $t_0$, and their periods are $L$
% \footnote{\textcolor{red}{\bf MN: I guess that the second orders of energy densities for $i=1,2$ are periodic function with $t_1$ and their periods are $L_{\text{eff}}$. We should check this.}}.
For $i=3,4$, this small $\epsilon$ expansion becomes invalid at $t_0=\pm (X-x)+nL$, where $n$ are integers because the second order terms of energy densities diverges.
This suggests that the local operator at $x$ induces localized energies and they propagate to left and right at the speed of light
% \footnote{\textcolor{red}{We should check the time dependence of the energy densities for $i=1,2$.}}. 
When these local excitations reach $X$, the energy densities at $X$ drastically grow.
%%%%%%%%%%%%%%%%%%%%%%%%%%%%%%%%%%%%%%%
\subsubsection{SSD limit}
%%%%%%%%%%%%%%%%%%%%%%%%%%%%%%%%%%%%%%%
% After taking the SSD limit and then performing analytic continuation, we will explore the time dependence of the energy densities.
Next, we explore the time dependence of the energy density after taking the SSD limit and then performing analytic continuation. %Because they are complicated, we postpone their details to Appendix \ref{App:energy-densities-for-WO}. 
Because the details are complicated, we relegate them to Appendix \ref{App:energy-densities-for-WO}. 
To simplify them and explore their properties, we take the small $\epsilon$ limit and closely look at the first and second order terms of the energy densities in this limit.
In this limit, the energy densities for four states are approximated by 
\be
\begin{split}\label{eq:ee-with-small-epsilon-limit}
    &\text{\small{$\left \langle T_{ww}\left(w_X\right) \right \rangle_{i=1,\dots,4} \approx \left(\f{2\pi}{L}\right)^2 \left[ -\f{c}{24}+\epsilon^2\mathcal{T}^{\theta=\infty}_{i=1,\dots,4}\right]$}},
   % &+\f{2h_{\mathcal{O}}e^{\f{i2\pi X}{L}}\left(2\pi t_1 \sin{\left[\f{\pi x}{L}\right]}+Le^{-i\f{\pi x}{L}}\right)^2}{\left(4\pi t_1 \sin{\left[\f{\pi x}{L}\right]}\sin{\left[\f{\pi X}{L}\right]}+2 L \sin{\left[\f{\pi (X-x)}{L}\right]}\right)^2} \left(-1+\f{2\left[2\pi t_1\sin{\left(\f{\pi x}{L}\right)}+Le^{\f{-i\pi x}{L}}\right]^{4h_{\mathcal{O}}}}{(4\pi L\epsilon)^{2h_{\mathcal{O}}}}\right)\bigg{]},\\
    %%%%%%%%%%%%%%%%%%%%%%%%%%%%%%%%%%%%%%%%%%%%%%%%%%%%%
   \text{\small{$ \left \langle T_{\overline{w}\overline{w}}\left(\overline{w}_X\right) \right \rangle_{i=1,\dots,4} \approx\left(\f{2\pi}{L}\right)^2 \left[ -\f{c}{24}+\epsilon^2\overline{\mathcal{T}}^{\theta=\infty}_{i=1,\dots,4}\right]$}},\\
    %%%%%%%%%%%%%%%%%%%%%%%%%%%%%%%%%%%%%%%%%%%%%%%%%%%%%
  %  & \left \langle T_{ww}\left(w_X\right) \right \rangle_{2} \approx \left(\f{2\pi}{L}\right)^2 \left[ -\f{c}{24}+\epsilon^2 \mathcal{T}_{i=2}\right],
    %%%%%%%%%%%%%%%%%%%%%%%%%%%%%%%%%%%%%%%%%%%%%%%%%%%%%
 %   \left \langle T_{\overline{w}\overline{w}}\left(\overline{w}_X\right) \right \rangle_{2} \approx \left(\f{2\pi}{L}\right)^2 \left[ -\f{c}{24}+\epsilon^2\overline{\mathcal{T}}_{i=2}\right],\\
    %%%%%%%%%%%%%%%%%%%%%%%%%%%%%%%%%%%%%%%%%%%%%%%%%%%%%
 %  & \left \langle T_{ww}\left(w_X\right) \right \rangle_{3} \approx \left(\f{2\pi}{L}\right)^2\left[-\f{c}{24}+\epsilon^2\mathcal{T}_{i=3}\right],
    %%%%%%%%%%%%%%%%%%%%%%%%%%%%%%%%%%%%%%%%%%%%%%%%%%%%%
 %  \left \langle T_{\overline{w}\overline{w}}\left(\overline{w}_X\right) \right \rangle_{3} \approx \left(\f{2\pi}{L}\right)^2\left[-\f{c}{24}+\epsilon^2\overline{\mathcal{T}}_{i=3}\right],\\
    %%%%%%%%%%%%%%%%%%%%%%%%%%%%%%%%%%%%%%%%%%%%%%%%%%%%%
  %  & \left \langle T_{ww}\left(w_X\right) \right \rangle_{4} \approx \left(\f{2\pi}{L}\right)^2\left[-\f{c}{24}+\epsilon^2\mathcal{T}_{i=4}\right],
    %%%%%%%%%%%%%%%%%%%%%%%%%%%%%%%%%%%%%%%%%%%%%%%%%%%%%
   % \left \langle T_{\overline{w}\overline{w}}\left(\overline{w}_X\right) \right \rangle_{4} \approx \left(\f{2\pi}{L}\right)^2\left[-\f{c}{24}+\epsilon^2\overline{\mathcal{T}}_{i=4}\right],\\
    %%%%%%%%%%%%%%%%%%%%%%%%%%%%%%%%%%%%%%%%%%%%%%%%%%%%%
\end{split}
\ee
where the functions, $\mathcal{T}^{\theta=\infty}_{i=1,\cdots, 4}$ and $\overline{\mathcal{T}}^{\theta=\infty}_{i=1,\cdots, 4}$, are defined as
\be
\begin{split}
    &\mathcal{T}^{\theta=\infty}_{i=1}=\f{4h_{\mathcal{O}}\pi^2L^2}{\left[2\pi t_1 \sin{\left[\f{\pi X}{L}\right]}\sin{\left[\f{\pi x}{L}\right]}+L\sin{\left[\f{\pi(X-x)}{L}\right]}\right]^4},\\
    %%%%%%%%%%%%%%%%%%%%%%%%%%%%%%%%%%%%%%%%%%%%%%%%%%%%%
    &\overline{\mathcal{T}}^{\theta=\infty}_{i=1}=\f{4h_{\mathcal{O}}\pi^2L^2}{\left[2\pi t_1 \sin{\left[\f{\pi X}{L}\right]}\sin{\left[\f{\pi x}{L}\right]}+L\sin{\left[\f{\pi(x-X)}{L}\right]}\right]^4},\\
    %%%%%%%%%%%%%%%%%%%%%%%%%%%%%%%%%%%%%%%%%%%%%%%%%%%%%
    &\mathcal{T}^{\theta=\infty}_{i=2}=\f{h_{\mathcal{O}}\pi^2\left[L^2-4\pi^2t_1^2\sin^2{\left[\f{\pi x}{L}\right]}+4\pi t_1 L\sin{\left[\f{\pi x}{L}\right]}\cos{\left[\f{\pi x}{L}\right]}\right]^2}{L^2\left[2\pi t_1 \sin{\left[\f{\pi X}{L}\right]}\sin{\left[\f{\pi x}{L}\right]}+L\sin{\left[\f{\pi(X-x)}{L}\right]}\right]^4},\\
    %%%%%%%%%%%%%%%%%%%%%%%%%%%%%%%%%%%%%%%%%%%%%%%%%%%%%
    &\overline{\mathcal{T}}^{\theta=\infty}_{i=2}=\f{h_{\mathcal{O}}\pi^2\left[L^2-4\pi^2t_1^2\sin^2{\left[\f{\pi x}{L}\right]}-4\pi t_1 L\sin{\left[\f{\pi x}{L}\right]}\cos{\left[\f{\pi x}{L}\right]}\right]^2}{L^2\left[2\pi t_1 \sin{\left[\f{\pi X}{L}\right]}\sin{\left[\f{\pi x}{L}\right]}+L\sin{\left[\f{\pi(x-X)}{L}\right]}\right]^4},\\
    %%%%%%%%%%%%%%%%%%%%%%%%%%%%%%%%%%%%%%%%%%%%%%%%%%%%%
    &\mathcal{T}^{\theta=\infty}_{i=3}=\f{4\pi  h_{\mathcal{O}}\sin^4{\left[\f{\pi x}{L}\right]}}{L^2\sin^4{\left[\f{\pi(t_0-x+X)}{L}\right]}},~\overline{\mathcal{T}}^{\theta=\infty}_{i=3}=\f{4\pi h_{\mathcal{O}}\sin^4{\left[\f{\pi x}{L}\right]}}{L^2\sin^4{\left[\f{\pi(t_0+x-X)}{L}\right]}},\\
    &\mathcal{T}^{\theta=\infty}_{i=4}=\f{4\pi  h_{\mathcal{O}}\sin^4{\left[\f{\pi (t_0-x)}{L}\right]}}{L^2\sin^4{\left[\f{\pi(t_0-x+X)}{L}\right]}},~\overline{\mathcal{T}}^{\theta=\infty}_{i=4}=\f{4\pi  h_{\mathcal{O}}\sin^4{\left[\f{\pi (t_0+x)}{L}\right]}}{L^2\sin^4{\left[\f{\pi(t_0+x-X)}{L}\right]}}.
\end{split}
\ee
For $i=1,2$, unlike the M\"obius time evolution, $\mathcal{T}^{\theta=\infty}_{i=1,2}$ and $\overline{\mathcal{T}}^{\theta=\infty}_{i=1,2}$ becomes non-periodic functions.
Since the functions, $\mathcal{T}^{\theta=\infty}_{i=1,2}$ and $\overline{\mathcal{T}}^{\theta=\infty}_{i=1,2}$, diverge at $t_1=\f{L}{2\pi}\left[\cot{\left(\f{\pi X}{L}\right)}-\cot{\left(\f{\pi x}{L}\right)}\right]$ and $t_1=\f{-L}{2\pi}\left[\cot{\left(\f{\pi X}{L}\right)}-\cot{\left(\f{\pi x}{L}\right)}\right]$, this small $\epsilon$ expansion becomes invalid.   
%This suggests that the local exciations are created by the local operator at $x$, and then propagates to left and right with the velocity $v=\2\sin^2{\left(\f{\pi X}{L}\right)$
%for $i=1,2$, the chiral (anti-chiral) energy density is localized around $t_1=\f{L}{2\pi}\left[\cot{\left(\f{\pi X}{L}\right)}-\cot{\left(\f{\pi x}{L}\right)}\right]$ ($t_1=\f{-L}{2\pi}\left[\cot{\left(\f{\pi X}{L}\right)}-\cot{\left(\f{\pi x}{L}\right)}\right]$). 
One possible interpretation of this energy divergence is that the local operator at the spatial location $x$ creates two local excitations called quasiparticles, and they propagate to left and right with velocities $v=\pm 2\sin^2{\left[\f{\pi x}{L}\right]}$. 
%These quasiparticles can be used to described the $t_1$-dependence of $S^{(2)}_A$ in the $2$d rational CFTs (See Section \ref{}). 
When this local excitations hit the point where we measure the energy densities, they grow drastically.
%The locations of energy localization are determined by that of these local excitations.
For $i=1$, the chiral and anti-chiral energy densities at $X=0$ are independent of $t_1$. 
For $i=2$, the energy densities at $X=0$ depends on $t_1$, and for  large $t_1$, the time dependence is approximated by
\be
 \left \langle T_{ww}\left(w_X\right) \right \rangle_{2}=\left \langle T_{\overline{w}\overline{w}}\left(\overline{w}_X\right) \right \rangle_{2} \approx \left(\f{2\pi}{L}\right)^2\left[-\f{c}{24}+\f{16\pi^6h_{\mathcal{O}}\epsilon^2 t^4_1}{L^6}\right].
\ee
Thus, in the late time regime, the energy is localized at the origin regardless of the insertion position of the local operator.
This accumulation of energy around $X=0$ can not be described by the propagation of quasiparticles with $v=\pm 2\sin^2{\left[\f{\pi x}{L}\right]}$ because the time for quasiparticles to arrive at the origin is infinite. 
Thus, this goes beyond the quasiparticle picture. 
%One possible interpretation of 
For the late time regime, in the spatial regions where the energy is not localized, the chiral and anti-chiral energy densities are independent of the insertion location of the local operator, and it is given by 
\be
\left \langle T_{ww}\left(w_X\right) \right \rangle_{2}=\left \langle T_{\overline{w}\overline{w}}\left(\overline{w}_X\right) \right \rangle_{2} \approx \left(\f{2\pi}{L}\right)^2\left[-\f{c}{24}+\f{\pi^2h_{\mathcal{O}}\epsilon^2 }{L^2 \sin^4{\left[\f{\pi X}{L}\right]}}\right].
\ee

For $i=3,4$, $\mathcal{T}^{\theta=\infty}_{i=3,4}$ and $\overline{\mathcal{T}}^{\theta=\infty}_{i=3,4}$ are the periodic functions of $t_0$ with the period, $L$.
The energy densities at $X$ diverge at the times, $t_0=\pm (x-X)+n L$.
As in for $i=1,2$, their time dependence can be described by the quasiparticles that are created by the local operator which propagate to left and right at the speed of light. 
%For $i=3,4$, the $t_0$-dependence of the chiral (anti-chiral) energy density exhibits the peaks of it at $t_0=\pm (x-X)+n L$ where $n$ is integer.
%These peaks follows the quasiparticle picture that describes $t_0$ dependence of $S^{(2)}_A$ in the $2$d rational CFTs.
%In this picture, two quasiparticles are created at the insertion point of the local operator.
%One of them propagates left, and the other propagates right at the speed of light.
%The chiral and anti-chial energies localize around the points where the quasiparticles exist.
For $i=4$, the $t_0$-dependence of numerators of $\mathcal{T}^{\theta=\infty}_{i=4}$ and $\overline{\mathcal{T}}^{\theta=\infty}_{i=4}$ exhibit the dynamical behavior that is not explained by this quasiparticle picture.
For simplicity, focus on the chiral energy density in the middle time regime, when $L \gg t_0\gg x$ and $L \gg t_0 \gg x-X$.
In this time regime, the chiral energy density is approximated by
\be \label{eq:energy-density-4}
\left \langle T_{ww}\left(w_X\right) \right \rangle_{4}\approx \left(\f{2\pi}{L}\right)^2\left[-\f{c}{24}+\f{4\pi \epsilon^2 h_{\mathcal{O}}}{L^2}\right].\\
\ee
In the time regime where  $L \gg t_0 \gg X-x$, the anti-chiral energy density is approximated by (\ref{eq:energy-density-4}).
% This suggests that even when the time passes enough after the local excitation passes the spatial point at $X$, 
This suggests that long after the local excitation has passed the spatial point $X$, the energy density can deviate from the vacuum one.
The energy densities for $i=3$ do not have such time regimes where the term at order $\mathcal{O}(\epsilon^2)$ is approximated by a constant. 
Thus, during the non-unitary time evolution, some properties of the energy densities go beyond the quasiparticle picture.

\section{Entanglement entropy in Integrable Theories \label{sec:ee-in-integrable}}

%%%%%%%%%%%%%%%%%%%%%%%%%%%%%%%%%%%
In the following sections, we will explore the time dependence of a quantity that depends on the operator contents of the $2$d CFT, namely the entanglement entropy.
In this section, we focus on the time-dependent part of the entanglement entropy in $2$d free bosons and rational CFTs (RCFTs). We collectively refer to these theories as integrable theories as opposed to the maximally chaotic holographic CFTs.
Let us define the change in the $n$-th R\'{e}nyi entanglement entropy as \cite{Numasawa2016}
\begin{equation}
    \Delta S_{A,_i}^{(n)}(t_i) = \frac{1}{1-n} \log \frac{\Tr_A(\rho_A)^n}{(\Tr_A\rho_A)^n} - \frac{1}{1-n} \log \frac{\Tr_A(\rho_A^{\text{g.s.}})^n}{(\Tr_A\rho_A^{\text{g.s.}})^n}
\end{equation}
% \be
% \Delta S^{(n)}_{A,i}(t)= S^{(n)}_{A,i}(t) -\Delta S^{(n)}_{A,i}(t=0).
% \ee
where $\rho_A$ and $\rho_A^{\text{g.s.}}$ are the density matrices for the local operator excited state and the ground state respectively, and $t_i = t_1$ for $i=1,2$ and $t_i = t_0$ for $i=3,4$. This change, $\Delta S^{(n)}_{A,i}(t_i)$, is independent of the UV-cutoff because this is defined by subtracting the $n$-th R\'{e}nyi entanglement entropy for the vacuum state.
Then, in this section, we consider the change in the second R\'{e}nyi entanglement entropy after the inhomogeneous and homogeneous local operator quenches. The change in the entanglement entropy after a single local operator quench can be written in terms of a path integral over a n-sheeted Riemann surface with two copies of the local operator on each Riemann sheet \cite{Numasawa2016,Nozaki:2014hna,2014arXiv1405.5946C,Nozaki:2014uaa, PhysRevD.90.041701,Caputa2015}
\begin{align}\label{ChangeEntropyLocalQuench}
    &\Delta S_{A,i}^{(n)}(t_i) \nonumber \\
    =& \frac{1}{1-n}\log \left[\frac{\left\langle \mathcal{O}^\dagger(w_1^{\text{New},i},\overline{w}_1^{\text{New},i})\mathcal{O}(w_2^{\text{New},i},\overline{w}_2^{\text{New},i})\ldots \mathcal{O}^\dagger(w_{2n-1}^{\text{New},i},\overline{w}_{2n-1}^{\text{New},i})\mathcal{O}(w_{2n}^{\text{New},i},\overline{w}_{2n}^{\text{New},i}) \right\rangle_{C^n}}{\left\langle
    \mathcal{O}^\dagger(w_\epsilon^{\text{New},i},\overline{w}_\epsilon^{\text{New},i})\mathcal{O}(w_{-\epsilon}^{\text{New},i},\overline{w}_{-\epsilon}^{\text{New},i})
    \right\rangle_C^n}\right]
\end{align}
where
\begin{equation}
  w_1^{\text{New},i}=w_\epsilon^{\text{New},i} ,\hspace{1cm} w_2^{\text{New},i}=w_{-\epsilon}^{\text{New},i}
\end{equation}
where $C$ and $C^n$ are the cylinder and n-replicated cylinder and $w_{2j-1}^{\text{New},i}$ and $w_{2j}^{\text{New},i}$ are the corresponding coordinates of $w_{1}^{\text{New},i}$ and $w_{2}^{\text{New},i}$ on the $j^{\text{th}}$ sheet for $j=1,\ldots,n$. The same is true for the anti-holomorphic coordinates. The branch cut on each Riemann surface runs from $w = iX_2$ to $w = iX_1$ and from $\bar{w} = -iX_2$ to $\bar{w} = -iX_1$. We assume that the local operator is spinless with conformal dimension $h_\mathcal{O}$.

Perform a cylinder-to-plane conformal transformation $z = e^{\frac{2\pi w}{L}}$ and $\bar{z} = e^{\frac{2\pi \bar{w}}{L}}$. The branch cut is now given by an arc on the unit circle running from $z_2 = e^{i\frac{2\pi X_2}{L}}$ to $z_1 = e^{i\frac{2\pi X_1}{L}}$ and from $\bar{z}_2 = e^{-i\frac{2\pi X_2}{L}}$ to $\bar{z}_1 = e^{-i\frac{2\pi X_1}{L}}$. Next, we uniformize this Riemann surface with a suitable conformal transformation. To orient the branch cut along arbitrary directions, consider the conformal transformation 
\begin{align}\label{UniformizationMap}
    \zeta^n =& \frac{z-e^{i\frac{2\pi X_2}{L}}}{e^{i\frac{2\pi X_1}{L}}-z} e^{i\psi}=-e^{i\psi+i\frac{\pi(X_2-X_1)}{L}}\frac{\sinh{\frac{\pi(w-iX_2)}{L}}}{\sinh{\frac{\pi(w-iX_1)}{L}}}, \nonumber\\
    \overline{\zeta}^n =& \frac{\overline{z}-e^{-i\frac{2\pi X_2}{L}}}{e^{-i\frac{2\pi X_1}{L}}-\overline{z}} e^{-i\psi}=-e^{-i\psi+i\frac{\pi(X_1-X_2)}{L}}\frac{\sinh{\frac{\pi(\overline{w}+iX_2)}{L}}}{\sinh{\frac{\pi(\overline{w}+iX_1)}{L}}}
\end{align} 
where $\psi\in\mathbb{R}$ is an arbitrary parameter that allows us to orient the branch cut so that the branch cut is oriented along the rays with arguments $\psi-\frac{\pi(X_1-X_2)}{L}$ and $\frac{\pi(X_1-X_2)}{L}-\psi$ on the $\zeta^n$ and $\Bar{\zeta}^n$ complex planes respectively. Note that it does not matter whether or not $X_2<X_1$ or $X_1<X_2$. In both cases, the branch cut lies along the same direction.  

\subsection{Second R\'{e}nyi Entropy}
For simplicity, set $n=2$. The uniformized coordinates are related by $\zeta_1^{\text{New},i}=-\zeta_3^{\text{New},i}$ and $\zeta_2^{\text{New},i}=-\zeta_4^{\text{New},i}$ and $\overline{\zeta}_1^{\text{New},i}=-\overline{\zeta}_3^{\text{New},i}$ and $\overline{\zeta}_2^{\text{New},i}=-\overline{\zeta}_4^{\text{New},i}$. 
% Arbitrary four-point functions on the complex plane with coordinates $Z$ and $\Bar{Z}$ can be written in terms of the conformal block by performing the conformal transformation \textcolor{red}{KT:``in terms of conformal block'' sounds a bit confusing. How about e.g. By performing the conformal transformation (XXX equation 4.5 XXX), arbitrary four-point functions on the complex plane can be written as (XXX equation 4.6 XXX)?}
By performing a conformal transformation
\begin{equation}
    \eta = \frac{(Z-Z_1)(Z_3-Z_4)}{(Z-Z_4)(Z_3-Z_1)},\hspace{1cm} \overline{\eta} = \frac{(\overline{Z}-\overline{Z}_1)(\overline{Z}_3-\overline{Z}_4)}{(\overline{Z}-\overline{Z}_4)(\overline{Z}_3-\overline{Z}_1)},
\end{equation}
arbitrary four-point functions on the complex plane can be written as
% so that 
\begin{align}\label{FourPointFunctionComplexPlane}
    &\left\langle \mathcal{O}^\dagger(Z_1,\overline{Z}_1)\mathcal{O}(Z_2,\overline{Z}_2)\mathcal{O}^\dagger(Z_3,\overline{Z}_3)\mathcal{O}(Z_4,\overline{Z}_4) \right\rangle_\mathbb{C} \\ \nonumber
    =&\frac{1}{Z_{13}^{2h_\mathcal{O}}Z_{24}^{2h_\mathcal{O}}}\frac{1}{\overline{Z}_{13}^{2\overline{h}_\mathcal{O}}\overline{Z}_{24}^{2\overline{h}_\mathcal{O}}} \lim_{\eta_4,\overline{\eta}_4\rightarrow \infty} \eta_4^{2h_\mathcal{O}}\overline{\eta}_4^{2\overline{h}_\mathcal{O}} \left\langle \mathcal{O}^\dagger(0)\mathcal{O}(\eta_2,\overline{\eta}_2)\mathcal{O}^\dagger(1)\mathcal{O}(\infty) \right\rangle_\mathbb{C}
\end{align}
where $\eta_2,\overline{\eta}_2$ are the cross-ratios and $Z_{ij}=Z_i-Z_j$. These cross ratios are defined as
\be
\eta_2=\f{Z_{12}Z_{34}}{Z_{13}Z_{24}},\hspace{1cm} \overline{\eta}_2=\f{\overline{Z}_{12}\overline{Z}_{34}}{\overline{Z}_{13}\overline{Z}_{24}}.
\ee

Therefore, the ratio of correlation functions in \eqref{ChangeEntropyLocalQuench} can be written as a function of the cross-ratio alone
\begin{align}\label{NormalizedCorrelator}
    &\frac{\left\langle \mathcal{O}^\dagger(w_1^{\text{New},i},\overline{w}_1^{\text{New},i})\mathcal{O}(w_2^{\text{New},i},\overline{w}_2^{\text{New},i})\mathcal{O}^\dagger(w_{3}^{\text{New},i},\overline{w}_{3}^{\text{New},i})\mathcal{O}(w_{4}^{\text{New},i},\overline{w}_{4}^{\text{New},i}) \right\rangle_{C^2}}{\left\langle
    \mathcal{O}^\dagger(w_\epsilon^{\text{New},i},\overline{w}_\epsilon^{\text{New},i})\mathcal{O}(w_{-\epsilon}^{\text{New},i},\overline{w}_{-\epsilon}^{\text{New},i})
    \right\rangle_C^2} \nonumber \\
    =& \left[\eta_2(1-\eta_2)\bar{\eta}_2(1-\bar{\eta}_2)\right]^{2h_\mathcal{O}} G(\eta_2,\Bar{\eta}_2)
\end{align}
where 
\begin{equation}\label{FourPointFunction}
    G(x,\Bar{x}) = \lim_{\eta,\bar{\eta}\rightarrow\infty} \eta^{2h_\mathcal{O}}\Bar{\eta}^{2h_\mathcal{O}} \left\langle\mathcal{O}^\dagger(0)\mathcal{O}(x,\Bar{x})\mathcal{O}^\dagger(1)\mathcal{O}(\infty)
    \right\rangle_\mathbb{C}
\end{equation}
and the cross-ratio is
\begin{equation}\label{CrossRatio}
    \eta_2 = \frac{\left(\zeta_1-\zeta_2\right)\left(\zeta_3-\zeta_4\right)}{\left(\zeta_1-\zeta_3\right)\left(\zeta_2-\zeta_4\right)},\hspace{1cm}\bar{\eta}_2 = \frac{\left(\bar{\zeta}_1-\bar{\zeta}_2\right)\left(\bar{\zeta}_3-\bar{\zeta}_4\right)}{\left(\bar{\zeta}_1-\bar{\zeta}_3\right)\left(\bar{\zeta}_2-\bar{\zeta}_4\right)}
\end{equation}
These formulas \eqref{NormalizedCorrelator}, \eqref{FourPointFunction} and \eqref{CrossRatio} hold for arbitrary choices of the branch cut direction $\psi$ as one would expect. 
The change in the second R\'{e}nyi entropy for a quench by the local operator $\mathcal{O}$ in any 2d CFT is 
\begin{equation}
    \Delta S_{A,i}^{(2)}(t_i) = -\log \left[\eta_2(1-\eta_2)\overline{\eta}_2(1-\overline{\eta}_2)\right]^{2h_\mathcal{O}} G(\eta_2,\overline{\eta}_2)
\end{equation}

\subsubsection*{Free Boson}
To obtain non-trivial entanglement dynamics under the local operator quench, we simply choose the local operator to be a sum of vertex operators \cite{Nozaki:2014hna,2014arXiv1405.5946C,Nozaki:2014uaa}
\begin{equation}\label{SumVertexOperator}
    \mathcal{O}= \frac{e^{\frac{i}{2}\phi}+e^{-\frac{i}{2}\phi}}{\sqrt{2}}
\end{equation}
where $\phi$ is a $c=1$ free boson. This local operator has conformal dimensions $h_{\mathcal{O}}=\overline{h}_{\mathcal{O}}=\frac{1}{8}$ and the four-point function \eqref{FourPointFunction} is 
\begin{equation}
    G(z,\bar{z}) = \frac{1+|z|+|1-z|}{2\sqrt{|z(1-z)|}}
\end{equation}
The change in entanglement entropy for the sum of vertex operator \eqref{SumVertexOperator} is \cite{Nozaki:2014hna}
\begin{equation}\label{SecondRenyiEntropyFreeBoson}
    \Delta S_{A,i}^{(2)}(t_i) = \log \frac{2}{1+|\eta_2|+|1-\eta_2|}.
\end{equation}
just as in \cite{Caputa2015}. As we will see later, when $\frac{\epsilon}{L}\rightarrow0$, the cross ratios $\eta_2,\overline{\eta}_2\rightarrow0,1$. For these four possibilities, the second R\'{e}nyi entropy \eqref{SecondRenyiEntropyFreeBoson} to leading order in $\frac{\epsilon}{L}$ is
\begin{equation}
    \Delta S_{A,i}^{(2)}(t_i) \xrightarrow{\epsilon/L\rightarrow 0}\begin{cases}
        0,& \text{if}\qquad \eta_2,\overline{\eta}_2\rightarrow0 \qquad\text{or}\qquad\eta_2,\overline{\eta}_2\rightarrow1 \\
        \log 2,& \text{if} \qquad\eta_2\rightarrow0,\overline{\eta}_2\rightarrow1  \qquad\text{or}\qquad\eta_2\rightarrow1,\overline{\eta}_2\rightarrow0 
    \end{cases}
\end{equation}

\subsubsection*{General RCFTs}
For general RCFTs, the four-point function can be written as a sum over intermediate primaries as \cite{BELAVIN1984333,PhysRevD.90.041701,Numasawa2016}
\begin{equation}
    G(z,\overline{z})= \sum_b (C_{\mathcal{O}\mathcal{O}}^b)^2 F_\mathcal{O}(b|z) \overline{F}_\mathcal{O}(b|\overline{z})
\end{equation}
where $b$ runs over the primaries  and the conformal blocks are normalized so that $F_\mathcal{O}(b|z)= z^{h_b-2h_\mathcal{O}}(1+\mathcal{O}(z))$. %The two-point function has been normalized such that $C_{\mathcal{O}\mathcal{O}}^0$=1 where $0$ represents the vacuum. 
By using a fusion transformation, the cross ratios of the conformal blocks can be changed as
\begin{equation}\label{FusionTransform}
    F_\mathcal{O}(b|1-z) = \sum_c F_{bc}[\mathcal{O}] F_\mathcal{O}(c|z)
\end{equation}
For RCFTs,
\begin{equation}
    F_{00}[\mathcal{O}] = \frac{1}{d_\mathcal{O}}
\end{equation}
where $0$ is the vacuum and $d_\mathcal{O}$ is the quantum dimension of the primary operator $\mathcal{O}$. When the cross ratios $\eta_2,\overline{\eta}_2\rightarrow0,1$,
applying the asymptotic form of the conformal blocks as well as the fusion transformation \eqref{FusionTransform} and the crossing symmetry gives
% As we will see later, when $\frac{\epsilon}{L}\rightarrow0$, the cross ratios $\eta_2,\overline{\eta}_2\rightarrow0,1$. For these four possibilities, applying the asymptotic form of the conformal blocks as well as the fusion transformation \eqref{FusionTransform} and the crossing symmetry gives
\begin{equation}
    \Delta S_{A,i}^{(2)}(t_i) \xrightarrow{\epsilon/L\rightarrow 0}\begin{cases}
        0,& \text{if}\qquad \eta_2,\overline{\eta}_2\rightarrow0 \qquad\text{or}\qquad\eta_2,\overline{\eta}_2\rightarrow1 \\
        \log d_\mathcal{O},& \text{if} \qquad\eta_2\rightarrow0,\overline{\eta}_2\rightarrow1  \qquad\text{or}\qquad\eta_2\rightarrow1,\overline{\eta}_2\rightarrow0 
    \end{cases}
\end{equation}
This is exactly the same behaviour as for the free boson \eqref{SecondRenyiEntropyFreeBoson}, but with $\log d_\mathcal{O}$ instead of $\log2$. Therefore, the results of the local operator quench in the RCFT can be obtained from that of the free boson by simply replacing $\log 2$ with $\log d_\mathcal{O}$.

\subsubsection*{Cross ratio}
The cross ratio does not depend on the operator content of the $2$d CFTs and is explicitly given by 
\begin{equation}\label{SecondRenyiCrossRatio}
    \eta_2= \frac{1}{2}\left[1-\frac{\left(\zeta_{+\epsilon}^{\text{New},i}\right)^2+\left(\zeta_{-\epsilon}^{\text{New},i}\right)^2}{2\zeta_{+\epsilon}^{\text{New},i}\zeta_{-\epsilon}^{\text{New},i}}\right].
\end{equation}
The same expression holds for the anti-holomorphic components by replacing the holomorphic quantities with the anti-holomorphic quantities. These expressions \eqref{SecondRenyiEntropyFreeBoson} and \eqref{SecondRenyiCrossRatio} are valid for arbitrary directions of the uniformization branch cut $\psi$ as expected. The only subtlety in the computation of the cross ratio \eqref{SecondRenyiCrossRatio} lies in the computation of the denominator. In what follows, let $\hat{\boldsymbol{\cdot}}$ denote both holomorphic and anti-holomorphic coordinates and $\sigma=+1/-1$ for holomorphic/anti-holomorphic coordinates. For example, $\hat{\zeta}_{\rho\epsilon}$ denote both holomorphic and anti-holomorphic coordinates with $\rho=\pm1$ indicating their Euclidean time positions. Assume that to second order in the small $\epsilon$ expansion, where $\delta= \frac{2\pi\epsilon}{L} \ll 1$, these coordinates take the form
\begin{equation}\label{UniformizationCoordinateExpansion}
    \hat{\zeta}^2_{\rho\epsilon} = \xi_0(t_i,L,x,X_1,X_2,\theta,\sigma) \left[1+i\delta\rho \xi_1(t_i,L,x,X_1,X_2,\theta,\sigma)+\delta^2 \xi_2(t_i,L,x,X_1,X_2,\theta,\sigma)
    \right]
\end{equation}
where $\xi_0,\xi_1,\xi_2$ are arbitrary real functions that are independent of $\rho$ and $x$ is the spatial location of the local operator insertion. To first order in $\delta \ll 1$, 
\begin{equation}
    \hat{\zeta}^2_{\rho\epsilon}=\xi_0(t_i,L,x,X_1,X_2,\theta,\sigma)+i\delta\rho \xi_0(t_i,L,x,X_1,X_2,\theta,\sigma)\xi_1(t_i,L,x,X_1,X_2,\theta,\sigma).
\end{equation}
Assume that $\xi_0(t_i,L,x,X_1,X_2,\theta,\sigma),\xi_1(t_i,L,x,X_1,X_2,\theta,\sigma)\in\mathbb{R},$ are finite and 
\begin{equation}
    \xi_0(t_i,L,x,X_1,X_2,\theta,\sigma), \xi_1(t_i,L,x,X_1,X_2,\theta,\sigma)\neq0.
\end{equation}
This means that $\xi_0(t_i,L,x,X_1,X_2,\theta,\sigma)\xi_1(t_i,L,x,X_1,X_2,\theta,\sigma)$ is finite and non-zero. Therefore, to linear order in $\delta$, $\hat{\zeta}^2_{\pm\epsilon}$ are two numbers that lie slightly off the positive or negative real axis on the complex $\zeta, \overline{\zeta}$ planes. Following \cite{PhysRevD.90.041701}, we choose the phase $\psi$ in the uniformization map \eqref{UniformizationMap} such that the branch cut lies on the negative real axis on the complex $\zeta^n$, $\overline{\zeta}^n$ plane. If $\xi_0(t_i,L,x,X_1,X_2,\theta,\sigma)>0$, then $\hat{\zeta}^2_{\rho\epsilon}$ lies slightly off the positive real axis for both $\rho=\pm1$ and in the $\delta\rightarrow 0$,
\begin{equation}
    \hat{\zeta}_{\rho\epsilon}=\sqrt{\xi_0}\sqrt{1+i\delta\rho \xi_1+\delta^2 \xi_2}
\end{equation}
so the denominator of the cross ratio \eqref{SecondRenyiCrossRatio} is
\begin{equation}
    \hat{\zeta}_{+\epsilon}\hat{\zeta}_{-\epsilon}= \xi_0\left[1+\frac{1}{2}\delta^2(\xi_1^2+2\xi_2)\right]
\end{equation}
On the other hand, if $\xi_0(t_i,L,x,X_1,X_2,\theta,\sigma)<0$, these coordinates $\hat{\zeta}^2_{+\epsilon}$ will lie slightly off the negative real axis where the branch cut is located. In order to take the $\delta\rightarrow 0$ limit while avoiding the branch cut, we can rotate them as such
\begin{equation}
    \hat{\zeta}^2_{\rho_1\epsilon} = e^{i\pi}\left(-\hat{\zeta}^2_{\rho_1\epsilon}\right), \hspace{1cm} \hat{\zeta}^2_{\rho_2\epsilon} = e^{-i\pi}\left(-\hat{\zeta}^2_{\rho_2\epsilon}\right)
\end{equation}
where $(\rho_1,\rho_2)=(1,-1)$ or $(\rho_1,\rho_2)=(-1,1)$ depending on the sign of $\xi_0\xi_1$. The function $-\hat{\zeta}^2_{\pm\epsilon}$ now lie slightly off the positive real axis and we can safely take the $\delta\rightarrow 0$ limit. The denominator in the cross ratio when $\xi_0<0$ is approximately given by
\begin{equation}
    \hat{\zeta}^2_{+\epsilon}\hat{\zeta}^2_{-\epsilon} = \sqrt{|\xi_0|\left[1+i\delta \xi_1+\delta^2 \xi_2\right]}\sqrt{|\xi_0|\left[1-i\delta \xi_1+\delta^2 \xi_2\right]}=|\xi_0|\left[1+\frac{1}{2}\delta^2(\xi_1^2+2\xi_2)\right]
\end{equation}
Therefore, for both sign of $\xi_0$, as long as $\xi_0\neq0$ and is finite, 
\begin{equation}
    \hat{\zeta}_{+\epsilon}\hat{\zeta}_{-\epsilon} = |\xi_0|\left[1+\frac{1}{2}\delta^2(\xi_1^2+2\xi_2)\right]
\end{equation}
Putting this back into the cross ratio \eqref{SecondRenyiCrossRatio}, we find that to quadratic order in $\delta$, the cross ratio is given by
\begin{equation}\label{CrossRatioSecondOrder}
    \hat{\eta}_2 = \frac{1}{2}\left\{1-\text{sgn}(\xi_0)\left[1-\frac{1}{2}\delta^2\xi_1^2\right]\right\}
\end{equation}
which does not depend on $\xi_2$. Of course, we require that $\xi_2$ does not depend on $\rho=\pm1$ and does not diverge and that both $\xi_0,\xi_1$ are finite and non-zero. It does not matter if $\xi_2$ vanishes or not.

\subsubsection*{Simple Example}
The results for the change in the second R\'{e}nyi entropy for these integrable theories for various choices of subsystems and two different local operator insertion positions are listed in Appendix \ref{IntegrableTheoryAppendix}. Let us list the result for one such case here. Consider a subsystem that contains the origin as in case (a) and place the local operator at $X_2^f$. For Lorentzian time evolution with the SSD Hamiltonian, $i=1,2$, the cross ratio to second order in $\frac{\epsilon}{L}$ is 
\begin{equation}
    \hat{\eta}_2 = \frac{1}{2}\left\{1-\text{sgn}\left[\frac{t_1-\frac{\sigma L}{2\pi}\cot{\frac{\pi X_2}{L}}}{t_1-\frac{\sigma L}{2\pi}\cot{\frac{\pi X_1}{L}}}\right]\left[1-\left(\frac{\epsilon}{L}\right)^2\times \text{positive number}\right]\right\}
\end{equation}
so $\eta_2,\overline{\eta}_2\rightarrow0,1$ as $\frac{\epsilon}{L}\rightarrow 0$ as mentioned earlier. The change in the second R\'{e}nyi entropy for $i=1,2$ becomes
\begin{equation}
    \Delta S_{A,i}^{(2)}(t_1) = \begin{cases}
        \log 2, &i=1,2, \frac{L}{2\pi}\cot{\frac{\pi X_2}{L}}<t_1< \frac{L}{2\pi}|\cot{\frac{\pi X_1}{L}}|\\
        0&i=1,2,0<t_1< \frac{L}{2\pi}\cot{\frac{\pi X_2}{L}}\quad \text{and}\quad \frac{L}{2\pi}|\cot{\frac{\pi X_1}{L}}|<t_1 
    \end{cases}
\end{equation}
in the free boson CFT and the $\log 2$ gets replaced with $\log d_\mathcal{O}$ for generic RCFTs. This is well-described by the quasiparticle picture which is detailed in the following subsection.

\subsection{Quasiparticle Picture \label{sec:quasiparticle-picture}}

It turns out that the behavior of the change in the second R\'{e}nyi entropy is well-described by a quasiparticle picture. At the initial time $t_i=0$, for $i=0,1$, a Bell pair of quasiparticles, one left-moving and the other right-moving, are created at the position of the local operator $\mathcal{O}$. These quasiparticles propagate with speed given by the envelope function $f(x)$ defined in \eqref{EnvelopeFunction}.

For finite values of $\theta$, a quasiparticle that begins at position $x$ arrives at position $X$ at time $t_i$ given by
\begin{equation}\label{QuasiparticleTrajectoryGeneralTheta}
    \tan\frac{\pi t_i}{L\cosh{2\theta}}=\mu e^{2\theta} \frac{\tan\frac{\pi X}{L}-\tan\frac{\pi x}{L}}{1+e^{4\theta}\tan\frac{\pi X}{L}\tan\frac{\pi x}{L}},
\end{equation}
where $\mu=+1/-1$ for right/left moving quasiparticles as found previously in \cite{2021arXiv211214388G,2023arXiv230208009G}. For the uniform Hamiltonian, setting $\theta=0$ gives $X = (x +\mu t_0) \, \text{mod}\, L$  which is the trajectory for quasiparticles moving with uniform unit speed. Since we think of the local operators at $x$ as the source of the quasiparticles and the endpoints of the subsystems $X_1$ and $X_2$ as the target $X$, by looking at the corresponding factors in the coordinates $\hat{\zeta}^n_{\rho\epsilon}$, we see that the holomorphic coordinates correspond to the left moving quasiparticles and the anti-holomorphic coordinates correspond to the right moving ones.

In the SSD limit, a quasiparticle that starts off at position $x$ at $t_1=0$ winds up at position $X$ at time $t_1$ that is determined by
\begin{equation}
    \pm t_1 = \frac{L}{2\pi}\left(\cot{\frac{\pi x}{L}} -\cot{\frac{\pi X}{L}}\right)
\end{equation}
where $+/-$ corresponds to right/left moving quasiparticles. The factor of $\cot{\frac{\pi x}{L}}$ diverges at $X_1^f$ since the quasiparticles that start off at that position don't move. To get non-trivial entanglement dynamics, place the local operator at the other fixed point $X_2^f$.

The second R\'{e}nyi entropy of a subsystem in the local operator excited state is the same as that of the vacuum state unless the subsystem contains exactly one member of the entangled bell pair of quasiparticles in which case the second R\'{e}nyi entropy increases by an amount that is determined by the local operator $\mathcal{O}$.

\subsection{Summary for integrable theories}\label{SummaryIntegrableTheories}

While the calculation for the change in the second R\'{e}nyi entropy for the local operator excited state in integrable theories is involved, the final result is simple and easy to understand. Therefore, we summarize the key physical result here and relegate the details of the calculation to Appendix \ref{IntegrableTheoryAppendix}.

The entanglement entropy is described by a pair of bell pairs that are created at the location of the local operator insertion as first described in \cite{PhysRevD.90.041701} with the qubits propagating with a speed given by the M\"{o}bius/SSD envelope function. When either member of the bell pair, but not both, is inside the subsystem, the second R\'{e}nyi entropy increases by $\log 2$ for Hermitian sum of vertex operators \eqref{SumVertexOperator} in the free boson theory and by $\log d_\mathcal{O}$ in general RCFTs for a primary $\mathcal{O}$. 

The entanglement entropy is completely determined by the unitary time evolution operator and does not depend on the regulator Hamiltonian except for certain choices of $x$ which causes the regulator to vanish. This is because the coordinates $\zeta^2_{\pm\epsilon}$ have a real part that at the leading $\mathcal{O}\left(\left(\frac{\epsilon}{L}\right)^0\right)$ order is given by the unitary operator while the imaginary part at the leading $\mathcal{O}\left(\left(\frac{\epsilon}{L}\right)\right)$ order is dependent on the regulator Hamiltonian. These coordinates therefore lie slightly off the real axis of the complex $\zeta^2$, $\overline{\zeta}^2$ planes with the real part determined by the unitary time evolution while the regulator Hamiltonian only serves to introduce a small separation between the two operators $\mathcal{O}$ and $\mathcal{O}^\dagger$. The jumps in the second R\'{e}nyi entanglement entropy are determined by the location of the operators along the real axis and hence only depend on the unitary time evolution. This also explains why swapping the order of the Lorentzian and Euclidean time evolution operators has no effect on the change in the second R\'{e}nyi entropy to leading order in $\frac{\epsilon}{L}$ when $\Delta S_{A,i}^{(2)}(t_i)$ is finite for the three choices of subsystems (a), (b) and (c) considered in this paper, including the case where the interval $A=[0,X_1]$, $0<X_1<\frac{L}{2}$, ends on the fixed point $X_1^f$. Since the leading order term in $\frac{\epsilon}{L}$ in the coordinates and hence the cross ratios are independent of the ordering of the Lorentzian and Euclidean evolution, so is the change in the second R\'{e}nyi entropy $\Delta S_{A,i}^{(2)}(t_i)$ to leading order. However, the order of the Lorentzian and Euclidean time evolution does affect the times at which $\Delta S_{A,i}^{(2)}(t_i)$ diverges which only occurs when the Euclidean time evolution is generated by the SSD Hamiltonian. When the local operator is inserted at $X_1^f$, $\Delta S_{A,3}^{(2)}(t_0)$ diverges for all $t_0$ while $\Delta S_{A,4}^{(2)}(t_0)$ diverges only for $t_0 = nL$ where $n\in \mathbb{Z}$. On the other hand, when the local operator is inserted at $X_2^f$, $\Delta S_{A,3}^{(2)}(t_0)$ remains finite for all $t_0$ while $\Delta S_{A,4}^{(2)}(t_0)$ diverges when $t_0 = \left(n+\frac{1}{2}\right)L$ where $n\in \mathbb{Z}$.

%%%%%%%%%%%%%%%%%%%%%%%%%%%%%%%%%%%%%%%%%%%%%%%%%%%%%%%%%%%%%%%
%%%%%%%%%%%%%%%%%%%%%%%%%%%%%%%%%%%%%%%
\section{Entanglement entropy in two-dimensional Holographic CFTs ($2$d holographic CFTs) \label{sec:e-e-holographic}}
%%%%%%%%%%%%%%%%%%%%%%%%%%%%%%%%%%%%%%%
Here, we explore the time dependence of $S_{A,i}$ during the M\"obius/SSD time evolution.
We begin by the Euclidean R\'enyi entanglement entropy in (\ref{eq:entanglement-entropy-fomula-1}), and then map from the cylinder to the complex plane, $(z,\overline{z})=(e^{\f{2\pi w}{L}},e^{\f{2\pi \overline{w}}{L}})$.
Then, $S^{(n)}_{A,i,E} $ is given by
\be \label{eq:entanglement-entropy-fomula-2}
\begin{split}
    S^{(n)}_{A,i,E} &=\f{1}{1-n}\log{\left[\f{\left \langle \mathcal{O}_n^{\dagger}(z^{\text{New},i}_{\epsilon},\overline{z}^{\text{New},i}_{\epsilon})\sigma_{n}(z_{X_1},\overline{z}_{X_1})\overline{\sigma}_n(z_{X_2},\overline{z}_{X_2})\mathcal{O}_n(z^{\text{New},i}_{-\epsilon},\overline{z}^{\text{New},i}_{-\epsilon})\right \rangle}{\left \langle \mathcal{O}^{\dagger}(z^{\text{New},i}_{\epsilon},\overline{z}^{\text{New},i}_{\epsilon})\mathcal{O}(z^{\text{New},i}_{-\epsilon},\overline{z}^{\text{New},i}_{-\epsilon})\right \rangle^n}\right]}\\
    &-\f{c(1+n)}{24n}\log{\left[\prod_{i=1,2}\left|\f{dz_{X_i}}{dw_{X_i}}\right|\right]},
\end{split}
\ee
where $z^{\text{New},i}_{\pm \epsilon}= e^{\f{2\pi w^{\text{New},i}_{\pm \epsilon}}{L}}$, $z_{X_{i=1,2}}=e^{\f{2\pi w_{X_{i=1,2}}}{L}}$ and $\overline{z}$ is the complex conjugate of $z$. 
Furthermore, to simplify the form of $S^{(n)}_{A,i,E}$, we perform a conformal map,
\be
\tilde{z}(z)=\f{\left(z^{\text{New},i}_{\epsilon}-z\right)\left(z_{X_2}-z^{\text{New},i}_{-\epsilon}\right)}{\left(z-z^{\text{New},i}_{-\epsilon}\right)\left(z^{\text{New},i}_{\epsilon}-z_{X_2}\right)}, \overline{\tilde{z}}(\overline{z})=\f{\left(\overline{z}^{\text{New},i}_{\epsilon}-\overline{z}\right)\left(\overline{z}_{X_2}-\overline{z}^{\text{New},i}_{-\epsilon}\right)}{\left(\overline{z}-\overline{z}^{\text{New},i}_{-\epsilon}\right)\left(\overline{z}^{\text{New},i}_{\epsilon}-\overline{z}_{X_2}\right)}.
\ee
Then, (\ref{eq:entanglement-entropy-fomula-2}) reduces to the function %, called conformal blocks, of the cross ratios,
of the cross ratios $(z_{c,i},\overline{z}_{c,i})$,
\be\label{eq:entanglement-entropy-fomula-3}
\begin{split}
    S^{(n)}_{A,i,E}=\f{1}{1-n}\log{\left[\left|z_{X_1}-z_{X_2}\right|^{-4nh_n}\left|1-z_{c,i}\right|^{4nh_n}G_n(z_{c,i},\overline{z}_{c,i})\right]}-\f{c(n+1)}{6n}\log{\left(\f{2\pi}{L}\right)},
\end{split}
\ee
where $(z_{c,i},\overline{z}_{c,i})$ are defined as 
\be
\left(z_{c,i},\overline{z}_{c,i}\right)=\left(\f{\left(z^{\text{New},i}_{\epsilon}-z_{X_1}\right)\left(z_{X_2}-z^{\text{New},i}_{-\epsilon}\right)}{\left(z_{X_1}-z^{\text{New},i}_{-\epsilon}\right)\left(z^{\text{New},i}_{\epsilon}-z_{X_2}\right)}, \f{\left(\overline{z}^{\text{New},i}_{\epsilon}-\overline{z}_{X_1}\right)\left(\overline{z}_{X_2}-\overline{z}^{\text{New},i}_{-\epsilon}\right)}{\left(\overline{z}_{X_1}-\overline{z}^{\text{New},i}_{-\epsilon}\right)\left(\overline{z}^{\text{New},i}_{\epsilon}-\overline{z}_{X_2}\right)}\right).
\ee
%\footnote{\textcolor{red}{MN: I will add $\lim_{n\rightarrow 1}  G_n$ here.}}
Here, we assume that $z_{c,i}, \overline{z}_{c,i} \approx 1$, and then we perform the OPE at $z_{c,i}, \overline{z}_{c,i} \approx 1$.
Subsequently, we closely look at the behavior of $\log{G_n}(z_{c,i},\overline{z}_{c,i)}$ at the leading order of the small $n$ limit, where $n\approx 1$.
The leading behavior of $\log{G_n}(z_{c,i},\overline{z}_{c,i)}$ reduces to 
\be
\log{G_n}(z_{c,i}, \overline{z}_{c,i}) \approx \f{c(1-n)}{6}\log{\left[\f{z^{\f{(1-\alpha_{\mathcal{O}})}{2}}_{c,i}\overline{z}^{\f{(1-\overline{\alpha}_{\mathcal{O}})}{2}}_{c,i}(1-z_{c,i}^{\alpha_{\mathcal{O}}})(1-\overline{z}_{c,i}^{\overline{\alpha}_{\mathcal{O}}})}{\alpha_{\mathcal{O}}\overline{\alpha}_{\mathcal{O}}}\right]},
\ee
where $\alpha_{\mathcal{O}}=\overline{\alpha}_{\mathcal{O}}=\sqrt{1-24h_{\mathcal{O}}/c}$. 
Consequently, in the Von Neumann limit, $n\rightarrow 1$, (\ref{eq:entanglement-entropy-fomula-3}) reduces to
\be \label{eq:eeS-cblock}
\begin{split}
    S_{A,i,E}=&\f{c}{6}\log{\left[\f{z_{c,i}^{\f{1-\alpha_{\mathcal{O}}}{2}}\overline{z}_{c,i}^{\f{1-\overline{\alpha}_{\mathcal{O}}}{2}}(1-z_{c,i}^{\alpha_{\mathcal{O}}})(1-\overline{z}_{c,i}^{\overline{\alpha}_{\mathcal{O}}})}{\alpha_{\mathcal{O}}\overline{\alpha}_{\mathcal{O}}}\right]}\\
    &+\f{c}{6}\log{\left[\left|z_{X_1}-z_{X_2}\right|^2\right]}-\f{c}{6}\log{\left[\left|1-z_{c,i}\right|^2\right]}-\f{c}{3}\log{\left(\f{2\pi}{L}\right)}.
\end{split}
\ee
%Here, we assume that $z_{c,i}, \overline{z}_{c,i} \approx 1$,  we perform the OPE at $z_{c,i}, \overline{z}_{c,i} \approx 1$, and then obtain $\lim_{n\rightarrow 1 }\log{G_n}(z_{c,i}, \overline{z}_{c,i})$,
%\be
%\lim_{n\rightarrow 1 }\log{G_n}(z_{c,i}, \overline{z}_{c,i}) \approx \f{c(1-n)}{6}\log{\left[\f{z^{\f{(1-\alpha_{\mathcal{O}})}{2}}_{c,i}}{\alpha_{\mathcal{O}}\overline{\alpha}_{\mathcal{O}}}\right]}
%\ee
Finally, take the analytic continuation, $\tau_1=it_1$ and $\tau_0=it_0$.
We will explore the time evolution of $S_{A,i}$ in the three cases shown in Fig. \ref{Fig:subsystems}
In this section, the insertions of local operator are at $x=X_{a=1,2}^f$. %The time dependence of $S_A$ for more general cases are reported in Appendix \ref{App:geneal}.
We will explore the characteristic properties of the dynamics for the state, $\ket{\Psi}_i$, by investigating the time dependence of entanglement entropy.
%%%%%%%%%%%%%%%%%%%%%%%%%%%%%%%%%%%%%%%
\subsection{M\"obius Hamiltonian \label{sec:EE-for-Mobius}}
%%%%%%%%%%%%%%%%%%%%%%%%%%%%%%%%%%%%%%%
Here, we will investigate the time dependence of $S_{A,i}$ for finite $\theta$.
First, we will explore the time dependence of $S_{A,i}$ when the local operator is inserted at $x=X_1^f=0$, and then we will explore the time dependence for the insertion at $x=X_2^f=L/2$ of the local operator.
%%%%%%%%%%%%%%%%%%%%%%%%%%%%%%%%%%%%%%%
\subsubsection{When the local operator is inserted at $x=X^f_1=0$.}
%%%%%%%%%%%%%%%%%%%%%%%%%%%%%%%%%%%%%%%
Let us present the time dependence of the cross ratios for $i=1\sim 4$ when the local operator is inserted at $x=X^f_1=0$.
To explore the properties of dynamics irrelevant to the regulator as possible, we investigate the time dependence of the cross ratios and $S_{A,i}$ in the small $\epsilon$ expansion, $\f{\epsilon}{L} \ll 1$.
When the local operator is inserted at $x$, the small $\epsilon$ expansion of the cross ratios analytically-continued to the real time is approximated by
\be \label{eq:small-cross-Xf1}
\begin{split}
    &z_{c,i} \approx 1+ i \epsilon f_i(T_i,x, \theta)\\
    &\overline{z}_{c,i} \approx 1+ i \epsilon g_i(T_i,x, \theta),
\end{split}
\ee
where $T_{i=1,2}=t_1$, $T_{i=3,4}=t_0$, and $x$ denotes the insertion point of the local operator. 
For all the subsystems and the locations, considered in this paper, of the local operator, in the small $\epsilon$ limit, the next-to-leading terms of $z_{c,i}$ and $\overline{z}_{c,i}$ are $\mathcal{O}(\epsilon)$, and they are pure imaginary.
We define $f_i(T_i,x, \theta)$ and $g_i(T_i,x, \theta)$ as the coefficient of $i\epsilon$ of $z_{c,i}$ and $\overline{z}_{c,i}$.
The details of the cross ratios, when the local operator is inserted at $X^{f}_{1}$, are reported in Appendix \ref{App:crossratiosX1}. In (\ref{eq:small-cross-Xf1}), the $\theta$ in $f_{i}(T_{i},x,\theta)$ or $g_{i}(T_{i},x,\theta)$ is a finite number. We use
$f_{i}(T_{i},x,\infty)$ or $g_{i}(T_{i},x,\infty)$ to represent the limit as $\theta \rightarrow \infty$.
We represent the time dependence of the cross ratio by the second order in the small $\epsilon$ limit.
Define $t_x$ by 
\be
t_x= \f{L_{\text{eff}}}{\pi}\tan^{-1}{\left[e^{2\theta}\tan{\left(\f{\pi x}{L}\right)}\right]}.
\ee
%where $L_{\text{eff}}=L \cosh{2\theta}$.
%In the cases of (a), (b), and (c) 
For $i=1,2$, $f_{i=1,2}(T_{i=1,2},X^f_1,\theta)$ ($g_{i=1,2}(T_{i=1,2},X^f_1,\theta)$) is positive in the time intervals, $n L_{\text{eff}}+t_{L-X_2}>t_1>n L_{\text{eff}}+t_{L-X_1}$ ($(n+1)L_{\text{eff}}+t_{X_2}>t_1>nL_{\text{eff}}+t_{X_1}$), where $n$ is an non-negative integer. It is negative in the time intervals, $(n+1) L_{\text{eff}}+t_{L-X_1}>t_1>n L_{\text{eff}}+t_{L-X_2}$ ($nL_{\text{eff}}+t_{X_1}>t_1>nL_{\text{eff}}+t_{X_2}$). 
%In the case of (c), the second order of $z_{c,i}$ ($\overline{z}_{c,i}$) is negative in the time intervals, $(n+1)L_{\text{eff}}+t_{L-X_1}>t_1>n L_{\text{eff}}+t_{L-X_2}$ ($nL_{\text{eff}}+t_{X_1}>t_1>nL_{\text{eff}}+t_{X_2}$), while that is positive in the time intervals, $n L_{\text{eff}}+t_{L-X_2}>t_1>n L_{\text{eff}}+t_{L-X_1}$ ($(n+1) L_{\text{eff}}+t_{ X_2}>t_1>nL_{\text{eff}}+t_{ X_1}$).
Since the denominators of $f_{i=1,2}(T_{i=1,2},X^f_1,\theta)$ ($g_{i=1,2}(T_{i=1,2},X^f_1,\theta)$) vanishes around the times, $t_1=nL_{\text{eff}} +t_{L-X_1}$ or $t_1=nL_{\text{eff}} +t_{L-X_2}$ ($t_1=nL_{\text{eff}} +t_{ X_1}$ or $t_1=nL_{\text{eff}} +t_{ X_2}$), the small $\epsilon$ expansion in (\ref{eq:small-cross-Xf1}) breaks down.
For $i=3,4$, $f_{i=3,4}(T_{i=1,2},X^f_1,\theta)$ ($g_{i=3,4}(T_{i=1,2},X^f_1,\theta)$) is positive in the time intervals, $nL-X_2>t_0>nL-X_1$ ($(n+1)L+X_2>t_0>nL+X_1$), while it is negative in the time intervals, $(n+1)L-X_1>t_0>nL-X_2$ ($nL+X_2>t_0>nL+X_1$). Since $f_{i=1,2}(T_{i=1,2},X^f_1,\theta)$ ($g_{i=1,2}(T_{i=1,2},X^f_1,\theta)$) vanishes around the times $t_0=nL -X_2$ or $t_0=nL -X_1$ ($t_0=nL +X_2$ or $t_0=nL +X_1$), the small $\epsilon$ expansion in (\ref{eq:small-cross-Xf1}) breaks down. As in \cite{2015JHEP...02..171A}, this suggests that we need to choose a different branch %\textcolor{red}{(MT: I think you mean choose a different branch since the branch cut should be fixed.)}
from that before the trajectory of the cross ratio encircles the origin,
\be
z_{c,i} \rightarrow e^{\pm  2i\pi }z_{c,i}, \overline{z}_{c,i} \rightarrow e^{\pm 2i \pi}\overline{z}_{c,i}.
\ee
where $\pm$ is determined by how the trajectory encircles the origin. 
For example in the case of $z_{c,3}$, when $L-X_1>t_0>0$, the location of $z_{c,3}$ is infinitesimally negative along the imaginary direction.
% of $z_{c,3}$ in $L-X_2>t_0>L-X_1$, while that is infinitesimally positive in $L-X_2>t_0>L-X_1$.
Conversely, within the interval $L-X_2>t_0>L-X_1$, the location becomes infinitesimally positive.
This suggests that around $t_0=L-X_1$, the trajectory of $z_{c,3}$ encircles the origin clockwise, so that $z_{c,3} \rightarrow e^{-  2i\pi }z_{c,3}$.
The values of (\ref{eq:eeS-cblock}) depends on the branches%cuts \textcolor{red}{(branch or branch cut?)}
, and the entanglement entropy in $2$d holographic CFTs should be given by the geodesics length, \cite{2006PhRvL..96r1602R,2006JHEP...08..045R}. 
In this section, we assume that the trajectories of cross ratios respect causality, and we choose the branches %\textcolor{red}{(branch or branch cut?)} 
such that the value of (\ref{eq:eeS-cblock}) is minimized.
Thus, we determine the time dependence of $S_{A,i}$.

Before reporting the detailed time evolution of $S_{A,i}$, we will present the common behavior of $S_{A,i}$ for the M\"obius case.
The outlined time dependence of $S_{A,i}$ follows the propagation of the quasiparticles as in the integrable theories explained above.
In other words, the local operator produces an entangled pair at its insertion point, and the quasiparticles of this pair propagate left and right at the velocity determined by the envelope function.
The time dependence of $S_{A,i=1,2}$ ($S_{A,i=3,4}$) are given by the periodic functions of $t_1$ ($t_0$) with the period, $L_{\text{eff}}$ ($L$).
Unlike the integrable theories, the detailed time dependence of entanglement entropy for $i=1~ (i=3)$ is different from that for $i=2~ (i=4)$.
In other words, in the time intervals where the value of entanglement entropy is larger than that of the vacuum one, the value of $S_{A, i=1}$ ($S_{A, i=3}$) is different from that of $S_{A, i=2}$ ($S_{A, i=4}$).
Therefore, the time dependence of $S_{A,i}$ in $2$d CFTs depends on the time ordering of the Euclidean and real time evolution.  
Since we can see from the time dependence of $S_{A,i}$ for (a) that the characteristic behavior of entanglement dynamics for the M\"obius case, we present $S_{A,i}$ only for (a) here, and report on that for (b) and (c) in Appendix \ref{App:HEE-MO-x0}. The time dependence of $S_{A,i}$ is given by %\textcolor{red}{(MT: Maybe I missed it but I don't see the specific form of $f_i$ and $g_i$ written anywhere apart from their definition in \eqref{eq:small-cross-Xf1}.)}
\be
\begin{split}\nonumber
%%%%%%%%%%%%%%%%%%%%%%%%%%%%%%%%%
     &S_{A,1} \approx \frac{c}{3}  \log \left[\frac{L }{\pi } \sin \left(\frac{\pi  \left(X_1-X_2\right)}{L}\right)\right]   
    \\ &+ \begin{cases}
0 &  t_{L-X_1}>t_1>0\\
\frac{c}{6} \log \left[\frac{2 \sin \left[\pi \alpha_{\mathcal{O}}\right]}{ \alpha_{\mathcal{O}}}\right]-\frac{c}{6} \log \left[ \epsilon  f_1(t_1,X^{f}_1,\theta)\right] &   nL_{\text{eff}}+t_{1,+}^t>t_1>nL_{\text{eff}}+t_{L-X_1}\\
\frac{c}{6} \log \left[\frac{2 \sin \left[\pi \alpha_{\mathcal{O}}\right]}{ \alpha_{\mathcal{O}}}\right]-\frac{c}{6} \log \left[ \epsilon  g_1(t_1,X^{f}_1,\theta)\right]  &   nL_{\text{eff}}+t_{X_2}>t_1>nL_{\text{eff}}+t_{1,+}^t\\
0 &   nL_{\text{eff}}+t_{L-X_2}>t_1>nL_{\text{eff}}+t_{X_2}\\
\frac{c}{6} \log \left[\frac{2 \sin \left[\pi \alpha_{\mathcal{O}}\right]}{ \alpha_{\mathcal{O}}}\right]-\frac{c}{6} \log \left[ -\epsilon  f_1(t_1,X^{f}_1,\theta)\right] &  nL_{\text{eff}}+ t_{1,-}^t>t_1>nL_{\text{eff}}+t_{L-X_2}\\
\frac{c}{6} \log \left[\frac{2 \sin \left[\pi \alpha_{\mathcal{O}}\right]}{ \alpha_{\mathcal{O}}}\right]-\frac{c}{6} \log \left[ - \epsilon  g_1(t_1,X^{f}_1,\theta)\right] &   nL_{\text{eff}}+t_{X_1}>t_1>nL_{\text{eff}}+ t_{1,-}^t\\
0 & (n+1) L_{\text{eff}}+t_{L-X_1}>t_1>nL_{\text{eff}}+t_{X_1}\\
     \end{cases},\\
%%%%%%%%%%%%%%%%%%%%%%%%%%%%%%%%%
     &S_{A,2} \approx  \frac{c}{3}  \log \left[\frac{L }{\pi } \sin \left(\frac{\pi  \left(X_1-X_2\right)}{L}\right)\right]   
    \\ &+ \begin{cases}
0 &  t_{L-X_1}>t_1>0\\
\frac{c}{6} \log \left[\frac{2 \sin \left[\pi \alpha_{\mathcal{O}}\right]}{ \alpha_{\mathcal{O}}}\right]-\frac{c}{6} \log \left[ \epsilon  f_2(t_1,X^{f}_1,\theta)\right] &  nL_{\text{eff}}+ t_{1,+}^t>t_1>nL_{\text{eff}}+t_{L-X_1}\\
\frac{c}{6} \log \left[\frac{2 \sin \left[\pi \alpha_{\mathcal{O}}\right]}{ \alpha_{\mathcal{O}}}\right]-\frac{c}{6} \log \left[ \epsilon  g_2(t_1,X^{f}_1,\theta)\right]  &   nL_{\text{eff}}+t_{X_2}>t_1>nL_{\text{eff}}+t_{1,+}^t\\
0 &   nL_{\text{eff}}+t_{L-X_2}>t_1>nL_{\text{eff}}+t_{X_2}\\
\frac{c}{6} \log \left[\frac{2 \sin \left[\pi \alpha_{\mathcal{O}}\right]}{ \alpha_{\mathcal{O}}}\right]-\frac{c}{6} \log \left[ -\epsilon  f_2(t_1,X^{f}_1,\theta)\right] &  nL_{\text{eff}}+ t_{1,-}^t>t_1>nL_{\text{eff}}+t_{L-X_2}\\
\frac{c}{6} \log \left[\frac{2 \sin \left[\pi \alpha_{\mathcal{O}}\right]}{ \alpha_{\mathcal{O}}}\right]-\frac{c}{6} \log \left[ - \epsilon  g_2(t_1,X^{f}_1,\theta)\right] &   nL_{\text{eff}}+t_{X_1}>t_1>nL_{\text{eff}}+ t_{1,-}^t\\
0 & (n+1) L_{\text{eff}}+t_{L-X_1}>t_1>nL_{\text{eff}}+t_{X_1}\\
     \end{cases},\\
%%%%%%%%%%%%%%%%%%%%%%%%%%%%%%%%%
      &S_{A,3} \approx \frac{c}{3}  \log \left[\frac{L }{\pi } \sin \left(\frac{\pi  \left(X_1-X_2\right)}{L}\right)\right]  
    \\ &+ \begin{cases}
0 & L-X_1>t_0>0\\
\frac{c}{6} \log \left[\frac{2 \sin \left[\pi \alpha_{\mathcal{O}}\right]}{ \alpha_{\mathcal{O}}}\right]-\frac{c}{6} \log \left[ \epsilon  f_3(t_0,X^{f}_1,\theta)\right] &nL+t^{t}_{0, +}>t_0>(n+1)L-X_1\\
\frac{c}{6} \log \left[\frac{2 \sin \left[\pi \alpha_{\mathcal{O}}\right]}{ \alpha_{\mathcal{O}}}\right]-\frac{c}{6} \log \left[ \epsilon  g_3(t_0,X^{f}_1,\theta)\right] &nL+X_2>t_0>nL+t^{t}_{0, +}\\ 
0&(n+1)L-X_2>t_0>nL+X_2\\
\frac{c}{6} \log \left[\frac{2 \sin \left[\pi \alpha_{\mathcal{O}}\right]}{ \alpha_{\mathcal{O}}}\right]-\frac{c}{6} \log \left[ -\epsilon  f_3(t_0,X^{f}_1,\theta)\right] &nL+t^{t}_{0, -}>t_0>(n+1)L-X_2\\
\frac{c}{6} \log \left[\frac{2 \sin \left[\pi \alpha_{\mathcal{O}}\right]}{ \alpha_{\mathcal{O}}}\right]-\frac{c}{6} \log \left[ -\epsilon  g_3(t_0,X^{f}_1,\theta)\right] &nL+X_1>t_0>nL+t^{t}_{0, -}\\
0 &(n+2)L-X_1>t_0>nL+X_1\\
    \end{cases},\\
\end{split}
\ee
     \be\label{EE-for-Mobius-0-a}
\begin{split}
    &S_{A,4}\approx \frac{c}{3}  \log \left[\frac{L }{\pi } \sin \left(\frac{\pi  \left(X_1-X_2\right)}{L}\right)\right]  
    \\ &+ \begin{cases}
0 & L-X_1>t_0>0\\
\frac{c}{6} \log \left[\frac{2 \sin \left[\pi \alpha_{\mathcal{O}}\right]}{ \alpha_{\mathcal{O}}}\right]-\frac{c}{6} \log \left[ \epsilon  f_4(t_0,X^{f}_1,\theta)\right] &nL+t^{t}_{0, +}>t_0>(n+1)L-X_1\\
\frac{c}{6} \log \left[\frac{2 \sin \left[\pi \alpha_{\mathcal{O}}\right]}{ \alpha_{\mathcal{O}}}\right]-\frac{c}{6} \log \left[ \epsilon  g_4(t_0,X^{f}_1,\theta)\right] &nL+X_2>t_0>nL+t^{t}_{0, +}\\ 
0&(n+1)L-X_2>t_0>nL+X_2\\
\frac{c}{6} \log \left[\frac{2 \sin \left[\pi \alpha_{\mathcal{O}}\right]}{ \alpha_{\mathcal{O}}}\right]-\frac{c}{6} \log \left[ -\epsilon  f_4(t_0,X^{f}_1,\theta)\right] &nL+t^{t}_{0, -}>t_0>(n+1)L-X_2\\
\frac{c}{6} \log \left[\frac{2 \sin \left[\pi \alpha_{\mathcal{O}}\right]}{ \alpha_{\mathcal{O}}}\right]-\frac{c}{6} \log \left[ -\epsilon  g_4(t_0,X^{f}_1,\theta)\right] &nL+X_1>t_0>nL+t^{t}_{0, -}\\
0 &(n+2)L-X_1>t_0>nL+X_1\\
    \end{cases}.
\end{split}
\ee
%%%%%%%%%%%%%%%%%%%%%%%%%%
Where $n$ is an integer greater than or equal to 0 , and $t^{t}_{1, \pm}, t^{t}_{0, \pm}$ are positive. we define $t^{t}_{1, \pm}, t^{t}_{0, \pm}$ as follows:
\be\label{eq:ttpm}
t_{1,\pm}^t= \f{L_{\text{eff}}}{\pi} \tan^{-1} \left[\pm \sqrt{e^{4 \theta} \tan{\f{\pi \left( L-X_1 \right)}{L}} \tan{\f{\pi X_2}{L}} }\right] 
, \cos \left(\frac{\pi t^{t}_{0, \pm}}{L}\right)= \pm \sqrt{\frac{\prod_{i=1}^2 \cos \left[\frac{\pi X_i}{L}\right]}{\cos \left[\frac{\pi\left(X_1+X_2\right)}{L}\right]}}
\ee
%%%%%%%%%%%%%%%%%%%%%%%%%%
We also studied the time dependence of $S_{A,i}$ when the local operator is inserted at $x=\f{L}{2}$.
However, the characteristic behavior of $S_{A,i}$ for the insertion at $x=\f{L}{2}$ of the local operator is the same as that for $x=0$.
Therefore, we postpone $S_{A,i}$ for $x=\f{L}{2}$ to Appendix \ref{App:HEE-MO-xL2}. 
%%%%%%%%%%%%%%%%%%%%%%%%%%%%%%%%%%%%%%%
\subsection{SSD limit}
%%%%%%%%%%%%%%%%%%%%%%%%%%%%%%%%%%%%%%%
Now, let us consider the time dependence of $S_{A,i}$ in the SSD limit, $\theta \rightarrow \infty$.
In this limit, $L_{\text{eff}}$ will become infinite, while $L$ will remain finite. 
This suggests that the time dependence of $S_{A,i=1,2}$ in the SSD limit may not be periodic, while for $i=3,4$ that may remain periodic.
%%%%%%%%%%%%%%%%%%%%%%%%%%%%%%%%%%%%%%%
\subsubsection{When the insertion of $\mathcal{O}$ is at $x=X_1^f$}
%%%%%%%%%%%%%%%%%%%%%%%%%%%%%%%%%%%%%%%
We consider the time evolution of $S_A$ when the local operator $\mathcal{O}$ is the inserted at $x=X_1^f$.
In this case, $t_1$-dependence of $S_{A,i=1,2}$ is given by
\be
\begin{split}
S_{A,1}= S_{A,2}= \f{c}{3}\log{\left[\f{L}{\pi}\sin{\left[\f{\pi(X_1-X_2)}{L}\right]}\right]}.
\end{split}
\ee

%\footnote{\textcolor{red}{\bf MN: We should check id this is correct even without taking $\epsilon$ to be small.}\textcolor{green}{\bf MWB: Some calculation could be find at \ref{App:WeiBoep}}}
We can see from $t_1$-dependence of $S_{A,i=1,2}$ that the local operator at $x=X_1^f$ does not grow with time during the evolution induced by the SSD Hamiltonian.
This is because the Hamiltonian density of $H_{\text{SSD}}$ is zero at the origin, so that this evolution operator trivially acts on the local operator.
For $\ket{\Psi_3}$, the cross ratios are exactly unity, so that
$S_{A,i=3}$ becomes infinite.
This is because the local operators inserted coincide with each other since $H_{\text{SSD}}$ does not play as the regulator.
%\footnote{\textcolor{red}{\bf We should consider this more carefully.}}.
For $\ket{\Psi_4}$, the cross ratios are approximated by
\if[0]
\be
\begin{split}
    z_{c,4} \approx 1+\f{4i\pi \epsilon \sin^2{\left[\f{\pi t_0}{L}\right]}\sin{\left[\f{\pi (X_1-X_2)}{L}\right]}}{L\prod_{i=1,2}\sin{\left[\f{\pi (t_0+X_i)}{L}\right]}} +\mathcal{O}(\epsilon^2),~\overline{z}_{c,4} \approx 1-\f{4i\pi \epsilon \sin^2{\left[\f{\pi t_0}{L}\right]}\sin{\left[\f{\pi (X_1-X_2)}{L}\right]}}{L\prod_{i=1,2}\sin{\left[\f{\pi (t_0-X_i)}{L}\right]}} +\mathcal{O}(\epsilon^2),
\end{split}
\ee
\fi
\be
\begin{split}
    z_{c,4} \approx 1-\f{4i\pi \epsilon \sin^2{\left[\f{\pi t_0}{L}\right]}\sin{\left[\f{\pi (X_1-X_2)}{L}\right]}}{L\prod_{i=1,2}\sin{\left[\f{\pi (t_0+X_i)}{L}\right]}} +\mathcal{O}(\epsilon^2),~\overline{z}_{c,4} \approx 1+\f{4i\pi \epsilon \sin^2{\left[\f{\pi t_0}{L}\right]}\sin{\left[\f{\pi (X_1-X_2)}{L}\right]}}{L\prod_{i=1,2}\sin{\left[\f{\pi (t_0-X_i)}{L}\right]}} +\mathcal{O}(\epsilon^2),
\end{split}
\ee
where around $t_0=t_{i=1,2,n,\pm}$, these expansions are invalid.
The definition of $t_{i=1,2,n,\pm}$ are
\be
t_{i=1,2,n,\pm} = n L \pm X_{i=1,2},
\ee
where $n$ is an integer number. 
The time dependence of the denominator of $z_{c,4}$ and $\overline{z}_{c,4}$ in this case is the same as that for M\"obius case when the local operator is inserted at $x=X_1^f$. 
This suggests we may have the candidates, corresponding to the branches%\textcolor{red}{(corresponding to different branches?)}
, of geodesics. 
The time dependence of entanglement entropy is determined by the minimal one of these candidates.
% \textcolor{green}{\bf MWB: Minimal or Maximal?}
Here, we report on the time dependence of $S_{A,4}$ for only (a) here, while that for (b) and (c) are presented in Appendix \ref{app:EE-SSD-Xf1}.
The time dependence of $S_{A,4}$ in (a) is determined by
     \be
\begin{split}
    &S_{A,4}\approx \frac{c}{3}  \log \left[\frac{L }{\pi } \sin \left(\frac{\pi  \left(X_1-X_2\right)}{L}\right)\right]  
    \\ &+ \begin{cases}
0 & L-X_1>t_0>0\\
\frac{c}{6} \log \left[\frac{2 \sin \left[\pi \alpha_{\mathcal{O}}\right]}{ \alpha_{\mathcal{O}}}\right]-\frac{c}{6} \log \left[ \epsilon  f_4(t_0,X^{f}_1,\infty)\right] &nL+t^{t}_{0, +}>t_0>(n+1)L-X_1\\
\frac{c}{6} \log \left[\frac{2 \sin \left[\pi \alpha_{\mathcal{O}}\right]}{ \alpha_{\mathcal{O}}}\right]-\frac{c}{6} \log \left[ \epsilon  g_4(t_0,X^{f}_1,\infty)\right] &nL+X_2>t_0>nL+t^{t}_{0, +}\\ 
0&(n+1)L-X_2>t_0>nL+X_2\\
\frac{c}{6} \log \left[\frac{2 \sin \left[\pi \alpha_{\mathcal{O}}\right]}{ \alpha_{\mathcal{O}}}\right]-\frac{c}{6} \log \left[ -\epsilon  f_4(t_0,X^{f}_1,\infty)\right] &nL+t^{t}_{0, -}>t_0>(n+1)L-X_2\\
\frac{c}{6} \log \left[\frac{2 \sin \left[\pi \alpha_{\mathcal{O}}\right]}{ \alpha_{\mathcal{O}}}\right]-\frac{c}{6} \log \left[ -\epsilon  g_4(t_0,X^{f}_1,\infty)\right] &nL+X_1>t_0>nL+t^{t}_{0, -}\\
0 &(n+2)L-X_1>t_0>nL+X_1\\
    \end{cases}.
\end{split}
\ee
where the $t^{t}_{0,\pm}$ is given by (\ref{eq:ttpm}).

%%%%%%%%%%%%%%%%%%%%%%%%%%%%%%%%%%%%%%%
\subsubsection*{Large $L$ limit}
%%%%%%%%%%%%%%%%%%%%%%%%%%%%%%%%%%%%%%%
Now, let us take the large $L$ limit to check if the time dependence of $S_{A,4}$ is consistent with that in \cite{2015JHEP...02..171A}. In this limit, $z_{c,4}$ and $\overline{z}_{c,4}$ are approximated by
\be
\begin{split}
    z_{c,4} \approx 1- \f{4i\pi^2 \epsilon (X_1-X_2) t_0^2}{L^2(t_0+X_1)(t_0+X_2)},  \overline{z}_{c,4} \approx 1+\f{4i\pi^2 \epsilon (X_1-X_2) t_0^2}{L^2(t_0-X_1)(t_0-X_2)}
\end{split}
\ee
 For simplicity, let us assume that $\f{L}{2}>X_1>X_2>0$.
Consequently, the $t_0$-dependence of $S_{A,4}$ is given by
\be
\begin{split}
    S_{A,4} \approx \begin{cases}
\f{c}{3}\log{\left[(X_1-X_2)\right]} & X_2>t_0\\
\f{c}{6}\log{\left[\f{\sin{\left(\pi \alpha_{\mathcal{O}}\right)}}{\pi \alpha_{\mathcal{O}}}\cdot \f{L^2(X_1-t_0)(t_0-X_2)(X_1-X_2)}{2\pi \epsilon t_0^2}\right]}& X_1>t_0>X_2 \\
\f{c}{3}\log{\left[(X_1-X_2)\right]} & t_0>X_1
    \end{cases}
\end{split}
\ee
Thus, this is consistent with that in \cite{2015JHEP...02..171A}.
%%%%%%%%%%%%%%%%%%%%%%%%%%%%%%%%%%%%%%%
\subsubsection{When the insertion of $\mathcal{O}$ is at $x=X_2^f$}
%%%%%%%%%%%%%%%%%%%%%%%%%%%%%%%%%%%%%%%
%\textcolor{red}{\bf MN: I will edit the file from here.}
Now, let us consider the case where the local operator is inserted at the other fixed point, $x=X_2^f=L/2$.
To second order in the small $\epsilon$ expansion, the analytic-continued cross ratios are given by
\be \label{expansion-of-corss-ratios-1}
\begin{split}
    &z_{c,1}\approx 1-\frac{2 i \pi  L \epsilon  \sin \left(\frac{\pi  (X_1-X_2)}{L}\right)}{\left(L \cos \left(\frac{\pi  X_1}{L}\right)-2\pi t_1 \sin \left(\frac{\pi  X_1}{L}\right)\right) \left(L \cos \left(\frac{\pi  X_2}{L}\right)-2 \pi t_1 \sin \left(\frac{\pi  X_2}{L}\right)\right)}+ \mathcal{O}(\epsilon^2),\\
    &\overline{z}_{c,1}\approx 1+\frac{2 i \pi  L \epsilon  \sin \left(\frac{\pi  (X_1-X_2)}{L}\right)}{\left(2 \pi t_1 \sin \left(\frac{\pi  X_1}{L}\right)+L \cos \left(\frac{\pi  X_1}{L}\right)\right) \left(2 \pi t_1 \sin \left(\frac{\pi  X_2}{L}\right)+L \cos \left(\frac{\pi  X_2}{L}\right)\right)}+ \mathcal{O}(\epsilon^2),\\
    %%%%%%%%%%%%%%%%%%%%%%%%%%%%%%%%%%%%
    &z_{c,2}\approx 1-\frac{2i\pi  \epsilon  \left(L^2+4 \pi ^2 t_1^2\right) \sin{\left[\f{\pi (X_1-X_2)}{L}\right]}}{L \left(L \cos \left(\frac{\pi  X_1}{L}\right)-2 \pi  t_1 \sin \left(\frac{\pi  X_1}{L}\right)\right) \left(L \cos \left(\frac{\pi  X_2}{L}\right)-2 \pi  t_1 \sin \left(\frac{\pi  X_2}{L}\right)\right)}+ \mathcal{O}(\epsilon^2),\\
    &\overline{z}_{c,2}\approx1+\frac{2i\pi  \epsilon  \left(L^2+4 \pi ^2 t_1^2\right)  \sin{\left[\f{\pi (X_1-X_2)}{L}\right]}}{L \left(2 \pi  t_1 \sin \left(\frac{\pi  X_1}{L}\right)+L \cos \left(\frac{\pi  X_1}{L}\right)\right) \left(2 \pi  t_1 \sin \left(\frac{\pi  X_2}{L}\right)+L \cos \left(\frac{\pi  X_2}{L}\right)\right)}+ \mathcal{O}(\epsilon^2),\\
    %%%%%%%%%%%%%%%%%%%%%%%%%%%%%%%%%%%%
    &z_{c,3}\approx 1-\f{4i\pi \epsilon\sin{\left[\f{\pi(X_1-X_2)}{L}\right]}}{L \cos{\left[\f{\pi(t_0+X_1)}{L}\right]}\cos{\left[\f{\pi(t_0+X_2)}{L}\right]}}+ \mathcal{O}(\epsilon^2),~
    \overline{z}_{c,3}\approx1+\f{4i\pi \epsilon\sin{\left[\f{\pi(X_1-X_2)}{L}\right]}}{L \cos{\left[\f{\pi(t_0-X_1)}{L}\right]}\cos{\left[\f{\pi(t_0-X_2)}{L}\right]}}+ \mathcal{O}(\epsilon^2),\\
    %%%%%%%%%%%%%%%%%%%%%%%%%%%%%%%%%%%%
    &z_{c,4}\approx 1-\f{4i\pi \epsilon \cos^2{\left(\f{\pi t_0}{L}\right)}\sin{\left[\f{\pi(X_1-X_2)}{L}\right]}}{L \cos{\left[\f{\pi(t_0+X_1)}{L}\right]}\cos{\left[\f{\pi(t_0+X_2)}{L}\right]}}+ \mathcal{O}(\epsilon^2),~
    \overline{z}_{c,4}\approx1+\f{4i\pi \cos^2{\left(\f{\pi t_0}{L}\right)}\sin{\left[\f{\pi(X_1-X_2)}{L}\right]}}{L \cos{\left[\f{\pi(t_0-X_1)}{L}\right]}\cos{\left[\f{\pi(t_0-X_2)}{L}\right]}}+ \mathcal{O}(\epsilon^2).\\
\end{split}
\ee
%\footnote{\textcolor{green}{ MWB: The sign of imaginary part of $z_{c, 3}$ $\overline{z}_{c, 3}$ $z_{c, 4}$ $\overline{z}_{c, 4}$ in (5.17) do not match with SSD limitation of (D.2).}}
For $i=1,2$, define the characteristic time scales by
\be
\tilde{t}_{i=1,2,\pm} =\f{L}{2\pi}\tan{\left[\f{\pi}{L}\left(\f{L}{2} \pm X_i\right)\right]}.
\ee
The second order of cross ratios, $z_{c,i=1,2}$ ($\overline{z}_{c,i=1,2}$), is positive in the time interval, $\tilde{t}_{2,-}>t_1>\tilde{t}_{1,-}$ ($\tilde{t}_{2,+}>t_1$ or $t_1>\tilde{t}_{1,+}$), while it is negative in $\tilde{t}_{1,-}>t_1$ or $t_1>\tilde{t}_{2,-}$ ($\tilde{t}_{1,+}>t_1>\tilde{t}_{2,+}$).
Around $t_1=\tilde{t}_{i,-}$ ($t_1=\tilde{t}_{i,+}$), the small $\epsilon$ expansion breaks down because the coefficient of $\epsilon^2$ becomes drastically large.
Thus, unlike the M\"obius case, the cross ratios in the SSD limit do not periodically behave. 
For $i=3,4$, the time intervals determining the sign of the cross ratios and the times for the small $\epsilon$ expansion to be invalid are the same as those for the case of M\"obius Hamiltonian. %Therefore, the choices of the branch cuts depending on these time intervals should  be the same as s those for the case of M\"obius Hamiltonian. 
%Thus, when the location of the local operator inserted is at $x=X_2^f$, the time dependence of $S_{A,i}$ remains finite.
However, while during the unitary time evolution corresponding to $i=3$, the value of $S_{A,i=3}$ is finite, during the non-unitary one corresponding to $i=4$, $S_{A,i=4}$ diverges at $t_0=L\cdot\left(\f{1}{2}+n\right)$ because the cross ratios becomes unity for any $\epsilon$.%\footnote{\textcolor{red}{MN: We should check if this is correct even without taking the small $\epsilon$ limit.}}.
This suggests that $e^{-\epsilon H_{\text{SSD}}}$ does not work as a regulator at $x=X_2^f$.
As in the M\"obius case, the outlined time evolution of $S_{A,i}$ can be described by the quasiparticle picture.
As explained in the M\"obius case, the entangled pair is induced at the insertion point of the local operator, and then the quasiparticles of this pair moves to left and right at the velocity determined by the envelope functions: for $i=1,2$ the velocity is $|2\sin^2{\left(\f{\pi x}{L}\right)}|$; for $i=3,4$ the velocity is the speed of light.
When only one of pair is in the subsystems considered, the entanglement between this pair contributes to $S_{A,i}$.
Since we can see the properties of entanglement dynamics in the SSD limit from $S_{A,i}$ for (a), we only present it.
The readers interested in (b) and (c) should look at Appendix \ref{app:EE-SSD-Xf2}.

%By taking  the time evolution of $S_{A,i}$ is given as follows. 
%Let us consider the $t_1$-dependence of $S_{A,i}$ in case (a). 
In the AdS/CFT correspondence, the time dependence of $S_{A,i}$ in (a) is given by %\textcolor{blue}{KT: maybe better to use holographic CFT rather than AdS/CFT}%\footnote{\textcolor{red}{\bf MN: I should check (a) again }} 
\be\label{EE-for-SSD-L/2-a}
\begin{split}
   & S_{A,1}\approx \f{c}{3}\log{\left[\f{L}{\pi}\sin{\left[\f{\pi(X_1-X_2)}{L}\right]}\right]} \\ &+\begin{cases}
         0 &\tilde{t}_{2,-}>t_1>0\\
        %%%%%%%%%%%%%%%%%%%%%%%%%%%%%%%%
\frac{c}{6} \log \left[\frac{2 \sin \left[\pi \alpha_{\mathcal{O}}\right]}{ \alpha_{\mathcal{O}}}\right]-\frac{c}{6} \log \left[- \epsilon  f_1(t_1,X^{f}_2,\infty)\right] & \tilde{t}^{t}_{1}>t_1>\tilde{t}_{2,-}\\
%%%%%%%%%%%%%%%%%%%%%%%%%%%%%
\frac{c}{6} \log \left[\frac{2 \sin \left[\pi \alpha_{\mathcal{O}}\right]}{ \alpha_{\mathcal{O}}}\right]-\frac{c}{6} \log \left[- \epsilon  g_1(t_1,X^{f}_2,\infty)\right]  & \tilde{t}_{1,+}>t_1>\tilde{t}^{t}_{1}\\
%%%%%%%%%%%%%%%%%%%%%%%%%%%%%%
0 & t_1>\tilde{t}_{1,+}\\
    \end{cases},\\
    %%%%%%%%%%%%%%%%%%%%%
    %%%%%%%%%%%%%%%%%%%%%
       & S_{A,2}\approx \f{c}{3}\log{\left[\f{L}{\pi}\sin{\left[\f{\pi(X_1-X_2)}{L}\right]}\right]} \\ &+\begin{cases}
         0 &\tilde{t}_{2,-}>t_1>0\\
        %%%%%%%%%%%%%%%%%%%%%%%%%%%%%%%%
\frac{c}{6} \log \left[\frac{2 \sin \left[\pi \alpha_{\mathcal{O}}\right]}{ \alpha_{\mathcal{O}}}\right]-\frac{c}{6} \log \left[-\epsilon  f_2(t_1,X^{f}_2,\infty)\right]  & \tilde{t}^{t}_{1}>t_1>\tilde{t}_{2,-}\\
%%%%%%%%%%%%%%%%%%%%%%%%%%%%%
\frac{c}{6} \log \left[\frac{2 \sin \left[\pi \alpha_{\mathcal{O}}\right]}{ \alpha_{\mathcal{O}}}\right]-\frac{c}{6} \log \left[-\epsilon  g_2(t_1,X^{f}_2,\infty)\right]  & \tilde{t}_{1,+}>t_1>\tilde{t}^{t}_{1}\\
%%%%%%%%%%%%%%%%%%%%%%%%%%%%%%
0 & t_1>\tilde{t}_{1,+}\\
    \end{cases},\\
%%%%%%%%%%%%%%%%%%%%%%%%%%%%%%
%%%%%%%%%%%%%%%%%%%%%%%%%%%%%%
       & S_{A,3}\approx \f{c}{3}\log{\left[\f{L}{\pi}\sin{\left[\f{\pi(X_1-X_2)}{L}\right]}\right]}\\ &+\begin{cases}
         0 &\f{L}{2}-X_2>t_0>0\\
        %%%%%%%%%%%%%%%%%%%%%%%%%%%%%%%%
\frac{c}{6} \log \left[\frac{2 \sin \left[\pi \alpha_{\mathcal{O}}\right]}{ \alpha_{\mathcal{O}}}\right]-\frac{c}{6} \log \left[- \epsilon  f_3(t_0,X^{f}_2,\infty)\right]  &nL+ \tilde{t}^{t}_{0,+}>t_1>\left(\f{1}{2}+n\right)L-X_2\\
%%%%%%%%%%%%%%%%%%%%%%%%%%%%%
\frac{c}{6} \log \left[\frac{2 \sin \left[\pi \alpha_{\mathcal{O}}\right]}{ \alpha_{\mathcal{O}}}\right]-\frac{c}{6} \log \left[- \epsilon  g_3(t_0,X^{f}_2,\infty)\right]  &\left(n-\f{1}{2}\right)L+X_1>t_1>nL+ \tilde{t}^{t}_{0,+}\\
%%%%%%%%%%%%%%%%%%%%%%%%%%%%%
0 &\left(n+\f{3}{2}\right)L-X_1>t_0>\left(n-\f{1}{2}\right)L+X_1\\
%%%%%%%%%%%%%%%%%%%%%%%%%%%%%%
\frac{c}{6} \log \left[\frac{2 \sin \left[\pi \alpha_{\mathcal{O}}\right]}{ \alpha_{\mathcal{O}}}\right]-\frac{c}{6} \log \left[\epsilon  f_3(t_0,X^{f}_2,\infty)\right]  & nL+\tilde{t}^t_{0,-}>t_0>\left(n+\f{3}{2}\right)L-X_1\\
%%%%%%%%%%%%%%%%%%%%%%%%%%%%%
\frac{c}{6} \log \left[\frac{2 \sin \left[\pi \alpha_{\mathcal{O}}\right]}{ \alpha_{\mathcal{O}}}\right]-\frac{c}{6} \log \left[ \epsilon  g_3(t_0,X^{f}_2,\infty)\right]  & \left(n+\f{1}{2}\right)L+X_2>t_0>nL+\tilde{t}^t_{0,-}\\
%%%%%%%%%%%%%%%%%%%%%%%%%%%%%
0&\left(\f{3}{2}+n\right)L-X_2>t_0>\left(\f{1}{2}+n\right)L+X_2
    \end{cases},\\
    %%%%%%%%%%%%%%%%%%%%%
    %%%%%%%%%%%%%%%%%%%%%
      & S_{A,4}\approx \f{c}{3}\log{\left[\f{L}{\pi}\sin{\left[\f{\pi(X_1-X_2)}{L}\right]}\right]}\\ &+\begin{cases}
                0 &\f{L}{2}-X_2>t_0>0\\
        %%%%%%%%%%%%%%%%%%%%%%%%%%%%%%%%
\frac{c}{6} \log \left[\frac{2 \sin \left[\pi \alpha_{\mathcal{O}}\right]}{ \alpha_{\mathcal{O}}}\right]-\frac{c}{6} \log \left[- \epsilon  f_4(t_0,X^{f}_2,\infty)\right]  &nL+ \tilde{t}^{t}_{0,+}>t_1>\left(\f{1}{2}+n\right)L-X_2\\
%%%%%%%%%%%%%%%%%%%%%%%%%%%%%
\frac{c}{6} \log \left[\frac{2 \sin \left[\pi \alpha_{\mathcal{O}}\right]}{ \alpha_{\mathcal{O}}}\right]-\frac{c}{6} \log \left[- \epsilon  g_4(t_0,X^{f}_2,\infty)\right]  &\left(n-\f{1}{2}\right)L+X_1>t_1>nL+ \tilde{t}^{t}_{0,+}\\
%%%%%%%%%%%%%%%%%%%%%%%%%%%%%
0 &\left(n+\f{3}{2}\right)L-X_1>t_0>\left(n-\f{1}{2}\right)L+X_1\\
%%%%%%%%%%%%%%%%%%%%%%%%%%%%%%
\frac{c}{6} \log \left[\frac{2 \sin \left[\pi \alpha_{\mathcal{O}}\right]}{ \alpha_{\mathcal{O}}}\right]-\frac{c}{6} \log \left[\epsilon  f_4(t_0,X^{f}_2,\infty)\right]  & nL+\tilde{t}^t_{0,-}>t_0>\left(n+\f{3}{2}\right)L-X_1\\
%%%%%%%%%%%%%%%%%%%%%%%%%%%%%
\frac{c}{6} \log \left[\frac{2 \sin \left[\pi \alpha_{\mathcal{O}}\right]}{ \alpha_{\mathcal{O}}}\right]-\frac{c}{6} \log \left[ \epsilon  g_4(t_0,X^{f}_2,\infty)\right]  & \left(n+\f{1}{2}\right)L+X_2>t_0>nL+\tilde{t}^t_{0,-}\\
%%%%%%%%%%%%%%%%%%%%%%%%%%%%%
0&\left(\f{3}{2}+n\right)L-X_2>t_0>\left(\f{1}{2}+n\right)L+X_2
    \end{cases}.\\
\end{split}
\ee
where $n$ is an integer greater than or equal to $0$, and $\tilde{t}_{i,\pm}$ are positive.  
The characteristic parameters are given by 
\be
\begin{split}
    % &f(X_1,X_2,t_1)= \f{-\prod_{i=1}^2\left[2\pi t_1 \sin{\left(\f{\pi X_i}{L}\right)}-L\cos{\left(\f{\pi X_i}{L}\right)}\right]}{2\pi L \epsilon \sin{\left[\f{\pi (X_1-X_2)}{L}\right]}},~F(X_1,X_2,t_1)= \f{f(X_1,X_2,t_1)}{(L^2+4\pi^2t_1^2)},\\
    % &g(X_1,X_2,t_1)= \f{\prod_{i=1}^2\left[2\pi t_1 \sin{\left(\f{\pi X_i}{L}\right)}+L\cos{\left(\f{\pi X_i}{L}\right)}\right]}{2\pi L \epsilon \sin{\left[\f{\pi (X_1-X_2)}{L}\right]}},~G(X_1,X_2,t_1)= \f{g(X_1,X_2,t_1)}{(L^2+4\pi^2t_1^2)},\\
    % &p(X_1,X_2,t_1)=\f{L \prod_{i=1}^2\cos{\left[\f{\pi (t_0+X_i)}{L}\right]}}{4\pi \epsilon \sin{\left[\f{\pi (X_1-X_2)}{L}\right]}},~P(X_1,X_2,t_1)=\f{p(X_1,X_2,t_1)}{\cos^2{\left(\f{\pi t_0}{L}\right)}},\\
    % &q(X_1,X_2,t_1)=-\f{L \prod_{i=1}^2\cos{\left[\f{\pi (t_0-X_i)}{L}\right]}}{4\pi \epsilon \sin{\left[\f{\pi (X_1-X_2)}{L}\right]}},~Q(X_1,X_2,t_1)=-\f{q(X_1,X_2,t_1)}{\cos^2{\left(\f{\pi t_0}{L}\right)}},\\
   &\tilde{t}_1^{t}=\sqrt{\tilde{t}_{1,+}\tilde{t}_{2,-}}, ~\cos{\left(\f{\pi \tilde{t}_{0, \pm }}{L}\right)}= \pm \sqrt{\f{-\prod_{i=1}^2\sin{\left[\f{\pi X_i}{L}\right]}}{\cos{\left[\f{\pi (X_1+X_2)}{L}\right]}}}
\end{split}
\ee

For general $x$, except for $x\neq 0$, the time dependence of $S_{A,i}$ is finite because the local operators inserted do not coincide with each other (See Appendix \ref{App:HEE-SSD-generalx} for the details of $S_{A,i}$). 

%Note that the computation on $S_{A,4}$ breaks down at $t_0=L\left(n+\f{1}{2}\right)$ because $z_{c,4}$ and $\overline{z}_{c,4}$ exactly becomes unity. %\footnote{\textcolor{red}{\bf We should consider the physical meaning of it.}}. 
%Since the Hamiltonian density at $x=X_1^f$ of $H_{\text{SSD}}$ is zero, the damping factor, $e^{-\epsilon H_{\text{SSD}}}$, can not tame the high-energy mode there adequately.
%At $t_0=L\left(n+\f{1}{2}\right)$, the excitation created by the local operator at $x=X_2^f$ could arrive at $x=X_1^f$, so that $S_{A,4}$ becomes diverge.
%For large times, $S_{A,i=1,2}$ is approximated by 
%\be \label{eq:asymp-ee1}
%\begin{split}
%&S_{A,1}\approx \f{c}{3}\log{\left(\f{t_1}{\epsilon}\right)}+ \f{c}{3}\log{\left(\f{t_1}{L}\right)},\\
%&S_{A,2}\approx \f{c}{3}\log{\left[\f{\sin{\left[\pi \alpha_{\mathcal{O}}\right]}}{\pi \alpha_{\mathcal{O}}}\right]}+\f{c}{3}\log{\left[\f{\pi L \sin{\left(\f{\pi X_1}{L}\right)}\sin{\left(\f{\pi X_2}{L}\right)}}{\epsilon}\right]}.
%\end{split}
%\ee
%Thus, $S_{A,1}$ logarithmically grows with $t_1$, while $S_{A,2}$ depends on the details of local operator, not $t_1$.
%For $\ket{\Psi_2}$, the information about the local operator remains even for large $t_1$.
%%%%%%%%%%%%%%%%%%%%%%%%%%%%%%%%%%%%%%%%%%%%%%%%%%
\subsection{Information survival}
%%%%%%%%%%%%%%%%%%%%%%%%%%%%%%%%%%%%%%%%%%%%%%%%%%
We close this section by exploring the contribution from the time ordering of the Euclidean and Lorentzian time evolution to the information scrambling.
Let us assume that the time dependence of entanglement entropy follows the propagation of quasiparticles.
During the evolution induced by the SSD Hamiltonian, the quasiparticles move to and then accumulate around $x=X^f_1$.
Therefore, if the boundary of $A$ is at $x=X^f_1$, the quantum entanglement of quasiparticles keeps contributing to the time dependence of $S_A$. 
To check whether or not the information of the local operator such as the dependence of $S_A$ on the local operator conformal dimension survives at late times, we will explore the time dependence of $S_{A,i=1,2}$ in the SSD limit. 
We assume that the insertion point of the local operator is $X_2^f$, and take $A$ to be 
\be
A=\left\{x \bigg{|} 0\le x \le X_2 < \f{L}{2}\right\}.
\ee
%\textcolor{blue}{MT: To get non-trivial dynamics, insert the local operator at $X_2^f$ and not $X_1^f$.} 
We closely look at the leading order of $S_{A,i=1,2}$ in the small $\epsilon$ limit.
Then, the time dependence of $S_{A,i=1,2}$ in the SSD limit is given by 
\be
\begin{split}
    &S_{A,i=1}\approx \f{c}{3}\log{\left[\f{L}{\pi}\sin{\left(\f{\pi X_1}{L}\right)}\right]}\\ &+\begin{cases}
        0 & \f{L}{2\pi \tan{\left(\f{\pi X_1}{L}\right)}}>t_1>0\\
        \f{c}{6} \log{\left[\f{2 \sin{\left[\pi\alpha_{\mathcal{O}}\right]}}{\alpha_{\mathcal{O}}}\right]}+\f{c}{6}\log{\left[\f{t_1-\f{L}{2\pi \tan{\left(\f{\pi X_1}{L}\right)}}}{\epsilon}\right]} & t_1>\f{L}{2\pi \tan{\left(\f{\pi X_1}{L}\right)}}>0\\
    \end{cases},\\
    &S_{A,i=2}\approx \f{c}{3}\log{\left[\f{L}{\pi}\sin{\left(\f{\pi X_1}{L}\right)}\right]}\\ &+\begin{cases}
        0 & \f{L}{2\pi \tan{\left(\f{\pi X_1}{L}\right)}}>t_1>0\\
        \f{c}{6} \log{\left[\f{2 \sin{\left[\pi\alpha_{\mathcal{O}}\right]}}{\alpha_{\mathcal{O}}}\right]}+\f{c}{6}\log{\left[\f{L^2\left(t_1-\f{L}{2\pi \tan{\left(\f{\pi X_1}{L}\right)}}\right)}{\epsilon\left(L^2+4\pi^2 t_1^2\right)}\right]} & t_1>\f{L}{2\pi \tan{\left(\f{\pi X_1}{L}\right)}}>0\\
    \end{cases},\\
\end{split}
\ee
Thus, as we expected, $S_{A,i=1,2}$ is not the vacuum entanglement entropy.
Subsequently, we consider the time dependence of $S_{A,i=1,2}$ in the late time regime $t_1 \gg \f{L}{2\pi \tan{\left(\f{\pi X_1}{L}\right)}}, \f{L}{2\pi}$.
In this late time regime, the time dependence of $S_{A,i=1}$ is approximately %determined by
\be \label{eq:S1-late}
S_{A,i=1}\approx \f{c}{3}\log{\left[\f{L}{\pi}\sin{\left(\f{\pi X_1}{L}\right)}\right]}+ \f{c}{6} \log{\left[\f{2 \sin{\left[\pi\alpha_{\mathcal{O}}\right]}}{\alpha_{\mathcal{O}}}\right]}+\f{c}{6}\log{\left[\f{t_1}{\epsilon}\right]},
\ee
where since the last term of (\ref{eq:S1-late}) grows logarithmically with $t_1$ and becomes larger than the term depending on $\alpha_{\mathcal{O}}$, the information of the local operator inserted is locally hidden by this logarithmic growth.
Then, we consider the late-time dependence of $S_{A,i=2}$.
In the late-time interval, $\f{L^2}{\epsilon}\gg t_1 \gg \f{L}{2\pi \tan{\left(\f{\pi X_1}{L}\right)}}, \f{L}{2\pi}$, the time dependence of $S_{A,i=2}$ is approximately %determined by 
\be \label{eq:S2-late}
S_{A,i=2}\approx \f{c}{3}\log{\left[\f{L}{\pi}\sin{\left(\f{\pi X_1}{L}\right)}\right]}+ \f{c}{6} \log{\left[\f{2 \sin{\left[\pi\alpha_{\mathcal{O}}\right]}}{\alpha_{\mathcal{O}}}\right]}-\f{c}{6}\log{\left[\f{4\pi^2 \epsilon t_1}{L^2}\right]}.
\ee
The last term of (\ref{eq:S2-late})  contributes positively to $S_{A,i=2}$, while it deceases with $t_1$.
Therefore, in this time interval, the dependence on $\alpha_{\mathcal{O}}$ of $S_{A,i=2}$ can still remain. In the very late-time regime, $t_1 > \f{L^2}{\epsilon}$, the small $\epsilon$ expansion in (\ref{expansion-of-corss-ratios-1}) breaks down because the second order behavior of the cross ratios overcomes the leading one.
By taking the late time limit, $t_1 \gg 1$, and then taking the small $\epsilon$ limit, let us explore the late time behavior of the cross ratios.
In this limit, the leading order of the cross ratios is approximated by $-1$, so we cannot use the conformal block in (\ref{eq:eeS-cblock}). In the time regime, $\f{L^2}{\epsilon}\gg t_1 \gg \f{L}{2\pi \tan{\left(\f{\pi X_1}{L}\right)}}, \f{L}{2\pi}$, for $i=1$, the information of the local operators is locally hidden, while $i=2$, it cannot be hidden.
This suggests that the initial state information for $i=2$ survives longer than that for $i=1$ during the SSD holographic time evolution.
It would be interesting to explore the time dependence of $S_{A,i=2}$ in the very late-time region $t_1 \gg \f{L^2}{\epsilon}$ to check whether or not the information of the local operator remains during the $2$d holographic time evolution forever. We leave it as an open problem.
%%%%%%%%%%%%%%%%%%%%%%%%%%%%%%%%%%%%%%%
%\section{Gravity dual of local operator quenched states in (\ref{Local_states}) \label{section:gravity-dual}}
%%%%%%%%%%%%%%%%%%%%%%%%%%%%%%%%%%%%%%%
%Here, we consider the gravity dual to the states in (\ref{Local_states}).
%As in \cite {1999AIPC..484..147B,2012JHEP...12..027R}, the three dimensional gravity corresponds to $2$d holographic CFT can be obtained by computing the expectation value of energy density.
%Therefore, let us compute the expectation value of energy density.

\section{Discussions \label{sec:discussion}}
%%%%%%%%%%%%%%%%%%%%%%%%%%%%%%%%%%%%%%%%%%%%%%%%%%%%%%%%%%%%%%%%
In this paper, we explored the time dependence of the entanglement entropy for the excited states induced by the insertion of the local operators in $2$d free, rational, and holographic CFTs.
The time evolution considered is constructed of Euclidean and Lorentzian time evolution.
For $i=1,3$, the systems undergo the Euclidean evolution first and then undergo the Lorentzian time evolution, while the order of the time evolution for $i=2,4$ is the opposite.  
Since the norm of the state for $i=2,4$ is not invariant under the real time evolution, the evolution for $i=2,4$ is non-unitary.
Our main findings in this paper are threefold.
%%%%%%%%%%Regulator%%%%%%%%%%%%%%%%%%%%%%%%
The first one is about the Euclidean and Lorentzian time evolution induced by the SSD Hamiltonian.
Since the Hamiltonian density of the SSD Hamiltonian at $x=X_1^f$ is zero, $H_{\text{SSD}}$ is expected to commute with $\mathcal{O}(x=X_1^f)$. 
When the local operator is at $x=X_1^f$, for $i=1,2$, the entanglement entropy, energy density, and partition function do not depend on the times, while for $i=3,4$, they are divergent, as we expected. 
This suggests that in the continuum limit, the Euclidean and real time evolution of $H_{\text{SSD}}$ does not play as the regulator and time evolution operator at $x=X_1^f$, respectively.%\textcolor{red}{(MT: I suggest saying "This suggests that in the continuum limit, the Euclidean time evolution of $H_{\text{SSD}}$ does not work as a regulator at $x=X_1^f$")}
%%%%%%%%%%%Free-and-holography%%%%%%%%%%%%%%%%%%%%%%%
The second main result is that in free and rational CFTs, the time dependence of the entanglement entropy does not depend on the time ordering of the Euclidean and Lorentzian time evolution, while in the holographic CFTs, the ordering of these time evolutions with different signatures matters.
%%%%%%%%%%%Information-Survival%%%%%%%%%%%%%%%%%%%%%%%
The third one is about the local operator information survival for $i=1,2$. 
We considered the time dependence of the entanglement entropy when $X^f_1$ is put on the boundary of the subsystems.
In $2$d holographic CFTs, during the unitary time evolution (for $i=1$), in the late time regime, the entanglement entropy grows logarithmically in time, so that this logarithmic growth hides the dependence of the entanglement entropy on the conformal dimension.
During the non-unitary time evolution (for $i=2$), the dependence of the entanglement entropy on the local operator remains for a longer time than that for $i=1$.
However, we could not determine if for $i=2$, the information about the local operator inserted remains forever. 
If we first take the late time limit, $t_1 \gg 1$, and subsequently take the small $\epsilon$ limit, then the cross ratios approaches to $-1$.
Therefore, the approximation of the conformal blocks around $z, \overline{z} \approx 1$ is invalid.
% In this paper, we could not consider a situation where the quantum entanglement of quasiparticles contributes forever to the entanglement entropy during the evolution induced by the holographic CFT Hamiltonian because the inhomogeneous Hamiltonians are defined on the compact space.
In this paper, we could not consider a situation where the quasiparticles contribute forever to the entanglement entropy during the evolution induced by the holographic CFT Hamiltonian because the inhomogeneous Hamiltonians are defined on a compact space.
It would be interesting to explore if the information of the local operator inserted remains during a non-unitary process on a non-compact space that is more similar to the setup considered in \cite{2015JHEP...02..171A,2014arXiv1405.5946C} .
For example, it would be interesting to explore the system during the time evolution constructed out of the Euclidean evolution induced by the Rindler Hamiltonian and the real time evolution induced by the homogeneous time evolution.

%%%%%%%%%%%%%%%%%%%%%%%%%%%%%%%%%%%%%%%%%%%%%%%
\section*{Acknowledgements}
%%%%%%%%%%%%%%%%%%%%%%%%%%%%%%%%%%%%%%%%%%%%%%%
We thank useful discussions with Tadashi Takayanagi, Chen Bai, and Farzad Omidi. 
%K.G.~is supported by JSPS KAKENHI Grant-in-Aid for Early-Career Scientists (21K13930) and Research Fellowships of Japan Society for the Promotion of Science for Young Scientists (22J00663).
M.N.~is supported by funds from the University of Chinese Academy of Sciences (UCAS), funds from the Kavli
Institute for Theoretical Sciences (KITS).
K.T.~is supported by JSPS KAKENHI Grant No.~21K13920 and MEXT KAKENHI Grant No.~22H05265. M.T. is supported by an
appointment to the YST Program at the APCTP through the Science and Technology Promotion Fund and Lottery Fund of the Korean Government, as well as the Korean Local Governments -
Gyeongsangbuk-do Province and Pohang City.
%%%%%%%%%%%%%%%%%%%%%%%%%%%%%%%%%%%%%%%%
%\subsection*{Future directions}
%%%%%%%%%%%%%%%%%%%%%%%%%%%%%%%%%%%%%%%%%%%%%%%%%%%%%%%%%%%%%%%%

\appendix

%%%%%%%%%%%%%%%%%%%%%%%%
\section{Trajectory of the local operator }\label{App:trajectory}
%%%%%%%%%%%%%%%%%%%%%%%%
Here, we will explain how to compute the trajectories of the local operator and present the details.
%%%%%%%%%%%%%%%%%%%%%%%%
\subsection{How to compute the local operator trajectories \label{App:trajectory-H-t-C}}
%%%%%%%%%%%%%%%%%%%%%%%%
Here, we derive the transformation of the primary operator during the unitary and non-unitary time evolution.
The Euclidean time evolution operators considered here are %$U^{E}_{i,\epsilon}$ and $\tilde{U}^{E}_{i,\epsilon}$,
\be
\begin{split}
    &U^{E}_{i,\epsilon}=\begin{cases}
        e^{-\tau_1H_{\text{M\"obius}}}e^{-\epsilon H_0} &~\text{for}~ i=1\\
        e^{-\epsilon H_0}e^{-\tau_1H_{\text{M\"obius}}}  &~\text{for}~ i=2\\
        e^{-\tau_0 H_0}e^{-\epsilon H_{\text{M\"obius}}}  &~\text{for}~ i=3\\
        e^{-\epsilon H_{\text{M\"obius}}}e^{-\tau_0 H_0}  &~\text{for}~ i=4\\
    \end{cases},~
    \tilde{U}^{E}_{i,\epsilon}=\begin{cases}
        e^{-\epsilon H_0}e^{\tau_1H_{\text{M\"obius}}} &~\text{for}~ i=1\\
        e^{\tau_1H_{\text{M\"obius}}}e^{-\epsilon H_0}  &~\text{for}~ i=2\\
        e^{-\epsilon H_{\text{M\"obius}}}e^{\tau_0 H_0}  &~\text{for}~ i=3\\
        e^{\tau_0 H_0} e^{-\epsilon H_{\text{M\"obius}}} &~\text{for}~ i=4\\
    \end{cases}.\\
\end{split}
\ee
First, we focus on $H_0$.
In the complex coordinates, $(w,\overline{w})$, it is given by
\be
\begin{split}
    H_0=\f{1}{2i\pi}\oint dw T(w)+\f{1}{2i\pi}\oint d\overline{w} \overline{T}(\overline{w})
\end{split}
\ee
Then, by performing the conformal transformation, $(z,\overline{z})=(e^{\f{2\pi w}{L}}, e^{\f{2\pi \overline{w}}{L}})$, the energy densities transform as 
\be
\begin{split}
    T(w)=\left(\f{dz}{dw}\right)^2 T(z)+\f{c}{12}\left\{z;w\right\}, ~\overline{T}(\overline{w})=\left(\f{d\overline{z}}{d\overline{w}}\right)^2 \overline{T}(\overline{z})+\f{c}{12}\left\{\overline{z};\overline{w}\right\},
\end{split}
\ee
where Schwarzian derivatives are defined as 
\be
\begin{split}
   &\left\{z;w\right\}= \left(\f{\f{d^3z}{dw^3}}{\f{dz}{dw}}\right)-\f{3}{2}\left(\f{\f{d^2z}{dw^2}}{\f{dz}{dw}}\right)^2, ~\left\{\overline{z};\overline{w}\right\}= \left(\f{\f{d^3\overline{z}}{d\overline{w}^3}}{\f{d\overline{z}}{d\overline{w}}}\right)-\f{3}{2}\left(\f{\f{d^2\overline{z}}{d\overline{w}^2}}{\f{d\overline{z}}{d\overline{w}}}\right)^2.
\end{split}
\ee
Consequently, $H_0$ reduces to
\be
\begin{split}
    H_0=\f{2\pi}{L}\cdot \left[\f{1}{2i\pi}\oint dz z T(z)+\f{1}{2i\pi}\oint d\overline{z} \overline{z}\overline{T}(\overline{z})-\f{c}{12}\right]=\f{2\pi}{L}\left[L^{z}_0+\overline{L}^{\overline{z}}_0-\f{c}{12}\right],
\end{split}
\ee
where the Virasoro generators are defined as 
\be
L^{z}_n=\f{1}{2i\pi}\oint dz z^{n+1} T(z),~ \overline{L}^{\overline{z}}_n=\f{1}{2i\pi}\oint d\overline{z} \overline{z}^{n+1} \overline{T}(\overline{z}).
\ee
%\textcolor{red}{MT: I believe that the factor of $i$ in (A.2), (A.5), and (A.6) should be in the denominator.}
The transformation, induced by $H_0$, of the primary operator, $\mathcal{O}$, with the conformal factor $(h_{\mathcal{O}},h_\mathcal{O})$ is given by
\be
e^{aH_0}\mathcal{O}(z,\overline{z})e^{-aH_0}=e^{\f{2\pi a}{L}\left(L^{z}_0+\overline{L}^{\overline{z}}_0\right)}\mathcal{O}(z,\overline{z})e^{-\f{2\pi a}{L}\left(L^{z}_0+\overline{L}^{\overline{z}}_0\right)}=\lambda^{2h_{\mathcal{O}}}\mathcal{O}(\lambda z,\lambda\overline{z}),
\ee
where $a$ is a real parameter, and $\lambda$ is defined as
\be
\lambda := e^{\f{2\pi a}{L}}.
\ee
Define $(w^{\text{New}}_a,\overline{w}^{\text{New}}_a)$ as 
\be 
(e^{\f{2\pi w^{\text{New}}_a}{L}},e^{\f{2\pi \overline{w}^{\text{New}}_a}{L}}):=(\lambda z, \lambda \overline{z}).
\ee
By using the conformal map, during the Euclidean time evolution induced by $H_0$, the transformation of the primary operator reduces to
\be \label{eq:trf-H0}
e^{aH_0}\mathcal{O}(w,\overline{w})e^{-aH_0}=\left|\f{dw_a^{\text{New}}}{dw}\right|^{2h_{\mathcal{O}}}\mathcal{O}(w_a^{\text{New}},\overline{w}_a^{\text{New}})
\ee

Next, we take a closer look at $H_{\text{M\"obius}}$.
Starting from the complex coordinates, $(w, \overline{w})$, it is written as
\be
\begin{split}
    H_{\text{M\"obius}}=&\f{1}{2i\pi}\oint dw \left[T(w)-\f{\tanh{2\theta}}{2}\left(e^{\f{2\pi w}{L}}+e^{\f{-2\pi w}{L}}\right)T(w)\right]\\
    &+\f{1}{2i\pi} \oint d\overline{w} \left[\overline{T}(\overline{w})-\f{\tanh{2\theta}}{2}\left(e^{\f{2\pi \overline{w}}{L}}+e^{\f{-2\pi \overline{w}}{L}}\right)\overline{T}(\overline{w})\right].
\end{split}
\ee
We next map from $(w, \overline{w})$ to $(z, \overline{z})$ where the Hamiltonian is expressed as
\be
\begin{split}
    H_{\text{M\"obius}}&= \left(\f{2\pi}{L}\right)\times \f{1}{2i\pi} \oint dz \left[zT(z)-\f{\tanh{2\theta}}{2}(z^2+1)T(z)\right]\\
    &+\left(\f{2\pi}{L}\right)\times\f{1}{2i\pi} \oint d\overline{z} \left[\overline{z}\overline{T}(\overline{z})-\f{\tanh{2\theta}}{2}(\overline{z}^2+1)\overline{T}(\overline{z})\right]-\f{c}{12}\times \left(\f{2\pi}{L}\right).
\end{split}
\ee
Subsequently, we map $(z, \overline{z})$ to $(\tilde{z}, \overline{\tilde{z}})=\left(\f{-\cosh{\theta}z+\sinh{\theta}}{-\cosh{\theta}+\sinh{\theta}z},\f{-\cosh{\theta}\overline{z}+\sinh{\theta}}{-\cosh{\theta}+\sinh{\theta}\overline{z}}\right)$,
 and the M\"obius Hamiltonian reduces to 
\be
\begin{split}
    H_{\text{M\"obius}}&=\left(\f{2\pi}{L}\right)\times\f{1}{2i\pi} \left[\oint d\tilde{z}\tilde{z}T(\tilde{z})+\oint d\overline{\tilde{z}}\overline{\tilde{z}}\overline{T}(\overline{\tilde{z}})\right]-\f{c}{12}\times \left(\f{2\pi}{L}\right) \\
    &=\left(\f{2\pi}{L_{\text{eff}}}\right)\times\left[L^{\tilde{z}}_0+L^{\overline{\tilde{z}}}_0-\f{c}{12}\right],
\end{split}
\ee
where the Virasoro generators on $(\tilde{z},\overline{\tilde{z}})$ are defined as 
\be
L^{\tilde{z}}_n=\f{1}{2i\pi}\oint d\tilde{z} \tilde{z}^{n+1} T(\tilde{z}),~ \overline{L}^{\overline{\tilde{z}}}_n=\f{1}{2i\pi}\oint d\overline{\tilde{z}} \overline{\tilde{z}}^{n+1} \overline{T}(\overline{\tilde{z}}).
\ee
Thus, the time evolution induced by $H_{\text{M\"obius}}$ is equivalent to the evolution induced by the dilatation operator on $(\tilde{z},\overline{\tilde{z}})$.
Therefore, the transformation of the primary operator during the M\"obius time evolution is given by
\be
\mathcal{O}_{\text{H}}(\tilde{z},\overline{\tilde{z}},a):=e^{aH_{\text{M\"obius}}}\mathcal{O}(\tilde{z},\overline{\tilde{z}})e^{-aH_{\text{M\"obius}}}=\lambda_{\text{eff}}^{2h_{\mathcal{O}}}\mathcal{O}(\lambda_{\text{eff}} \tilde{z},\lambda_{\text{eff}}\overline{\tilde{z}}),
\ee
where $L_{\text{eff}}$ is defined by replacing $L$ of $\lambda$ with $L_{\text{eff}}$.
On a separate note, let us consider the Schr\"odinger equation of the local operator with respect to $a$.
It is given by
\be
\f{d\mathcal{O}_{\text{H}}(\tilde{z},\overline{\tilde{z}},a)}{da}=\f{2\pi}{L_{\text{eff}}} \left[2h_{\mathcal{O}}\mathcal{O}_{\text{H}}(\tilde{z},\overline{\tilde{z}},a)+\tilde{z}\partial_{\tilde{z}}\mathcal{O}_{\text{H}}(\tilde{z},\overline{\tilde{z}},a)+\overline{\tilde{z}}\partial_{\overline{\tilde{z}}}\mathcal{O}_{\text{H}}(\tilde{z},\overline{\tilde{z}},a)\right].
\ee
This is equivalent to the transformation, induced by the dilation operator in $(\tilde{z}, \overline{\tilde{z}})$, of $\mathcal{O}_{\text{H}}(\tilde{z},\overline{\tilde{z}},a)$,
\be \label{eq:Sch-eq-tz}
\f{d\mathcal{O}_{\text{H}}(\tilde{z},\overline{\tilde{z}},a)}{da}= \left[H_{\text{M\"obius}},\mathcal{O}_{\text{H}}(\tilde{z},\overline{\tilde{z}},a) \right]=\left[\left(\f{2\pi}{L_{\text{eff}}}\right)\cdot\left[L^{\tilde{z}}_0+L^{\overline{\tilde{z}}}_0-\f{c}{12}\right], \mathcal{O}_{\text{H}}(\tilde{z},\overline{\tilde{z}},a)\right].
\ee
 Since the M\"obius Hamiltonian is given by the linear combination of the Virasoro as in (\ref{eq:Hamiltonian-virasoro}), $\mathcal{O}_{\text{H}}(z,\overline{z},a)$ should follow the Schr\"odinger equation in the complex coordinates, $(z,\overline{z})$, 
 % \be\label{eq:Sch-eq-z}
 % \begin{split}
 %     \f{d\mathcal{O}_{\text{H}}(z,\overline{z},a)}{da}&= \left[H_{\text{M\"obius}},\mathcal{O}_{\text{H}}(z,\overline{z},a) \right]=\f{2\pi}{L}\left[L^z_0+\overline{L}^{\overline{z}}_0-\f{\tanh{2\theta}}{2}\left(L^z_{1}+L^z_{-1}+\overline{L}^{\overline{z}}_{1}+\overline{L}^{\overline{z}}_{-1}\right),\mathcal{O}_{\text{H}}(z,\overline{z},a)\right]\\
 %     &=\f{2\pi}{L_{\text{eff}}}\bigg{[}h_{\mathcal{O}}\left(\cosh{2\theta}-z \sinh{2\theta}\right)\mathcal{O}_{\text{H}}(z,\overline{z},a) +\left(z\cosh{2\theta}-\f{\sinh{2\theta}}{2}(1+z^2)\right)\partial_z\mathcal{O}_{\text{H}}(z,\overline{z},a)\bigg{]}\\
 %     &+\f{2\pi}{L_{\text{eff}}}\bigg{[}h_{\mathcal{O}}\left(\cosh{2\theta}-\overline{z} \sinh{2\theta}\right)\mathcal{O}_{\text{H}}(z,\overline{z},a) +\left(z\cosh{2\theta}-\f{\sinh{2\theta}}{2}(1+\overline{z} ^2)\right)\partial_{\overline{z} }\mathcal{O}_{\text{H}}(z,\overline{z},a)\bigg{]}.
 % \end{split}
 % \ee
\begin{align}\label{eq:Sch-eq-z}
    &\f{d\mathcal{O}_{\text{H}}(z,\overline{z},a)}{da} =\left[H_{\text{M\"obius}},\mathcal{O}_{\text{H}}(z,\overline{z},a) \right] \nonumber \\
    =&\f{2\pi}{L}\left[L^z_0+\overline{L}^{\overline{z}}_0-\f{\tanh{2\theta}}{2}\left(L^z_{1}+L^z_{-1}+\overline{L}^{\overline{z}}_{1}+\overline{L}^{\overline{z}}_{-1}\right),\mathcal{O}_{\text{H}}(z,\overline{z},a)\right]\nonumber\\
     =&\f{2\pi}{L_{\text{eff}}}\bigg{[}h_{\mathcal{O}}\left(\cosh{2\theta}-z \sinh{2\theta}\right)\mathcal{O}_{\text{H}}(z,\overline{z},a) +\left(z\cosh{2\theta}-\f{\sinh{2\theta}}{2}(1+z^2)\right)\partial_z\mathcal{O}_{\text{H}}(z,\overline{z},a)\bigg{]}\nonumber\\
     +&\f{2\pi}{L_{\text{eff}}}\bigg{[}h_{\mathcal{O}}\left(\cosh{2\theta}-\overline{z} \sinh{2\theta}\right)\mathcal{O}_{\text{H}}(z,\overline{z},a) +\left(\overline{z}\cosh{2\theta}-\f{\sinh{2\theta}}{2}(1+\overline{z} ^2)\right)\partial_{\overline{z} }\mathcal{O}_{\text{H}}(z,\overline{z},a)\bigg{]}.
\end{align}
%\textcolor{red}{(MT: In the last line of A.18, should it be $\overline{z}\cosh{2\theta}$?)}
 
By mapping from $(\tilde{z},\overline{\tilde{z}})$ to $(z,\overline{z})$, the equation (\ref{eq:Sch-eq-tz}) reduces to (\ref{eq:Sch-eq-z}).
Thus, (\ref{eq:Sch-eq-z}) is consistent with (\ref{eq:Sch-eq-tz}).

Now let us return to the main part of the analysis in this section.
Define the complex coordinates $(w^{\text{New}}_{a, \text{M\"obius}}, \overline{w}^{\text{New}}_{a, \text{M\"obius}}) $ as
% \be
% (w^{\text{New}}_{a, \text{M\"obius}}, \overline{w}^{\text{New}}_{a, \text{M\"obius}}):=\left(\lambda_{\text{eff}}\tilde{z}, \lambda_{\text{eff}}\overline{\tilde{z}}\right).
% \ee
\be
(e^{\frac{2\pi w^{\text{New}}_{a, \text{M\"obius}}}{L}}, e^{\frac{2\pi \overline{w}^{\text{New}}_{a, \text{M\"obius}}}{L}}):=\left(\lambda_{\text{eff}}\tilde{z}, \lambda_{\text{eff}}\overline{\tilde{z}}\right).
\ee
Consequently, the transformation of the primary operator, $e^{aH_{\text{M\"obius}}} \mathcal{O}(w,\overline{w}) e^{-aH_{\text{M\"obius}}}$, is given by replacing $w^{\text{New}}_{a}$ and $\overline{w}^{\text{New}}_{a}$ of (\ref{eq:trf-H0}) with $w^{\text{New}}_{a,\text{M\"obius}}$ and $\overline{w}^{\text{New}}_{a,\text{M\"obius}}$,
\be \label{eq:trf-HM}
e^{aH_{\text{M\"obius}}}\mathcal{O}(w,\overline{w})e^{-aH_{\text{M\"obius}}}=\left|\f{dw_{a,\text{M\"obius}}^{\text{New}}}{dw}\right|^{2h_{\mathcal{O}}}\mathcal{O}(w_{a,\text{M\"obius}}^{\text{New}},\overline{w}_{a,\text{M\"obius}}^{\text{New}})
\ee
The transformation in (\ref{eq: op-in-Heisenberg}) is given by the operation constructed of (\ref{eq:trf-H0}) and (\ref{eq:trf-HM}).
For example, for $i=1$ ($i=3$)%\textcolor{red}{$i=3$($i=1$)?}
, we perform the transformation, induced by $H_0$ ($H_{\text{M\"obius}}$), of the primary operator, and then we do the one induced by $H_{\text{M\"obius}}$ ($H_0$), while for $i=2$ ($i=4$)%\textcolor{red}{$i=4$($i=2$)?}
, we do the one induced by $H_{\text{M\"obius}}$ ($H_0$), and then we do the one induced by $H_0$ ($H_{\text{M\"obius}}$).

\if[0]
\be
\begin{split}
    &e^{-aH_0}\mathcal{O}(w,\overline{w})e^{aH_0}=e^{\f{-2a\pi}{L}\left(L^{z}_0+\overline{L}^{\overline{z}}_0\right)}\mathcal{O}(w,\overline{w})e^{\f{2a\pi}{L}\left(L^{z}_0+\overline{L}^{\overline{z}}_0\right)}\\
    &=\mathcal{O}(w,\overline{w})-\f{2a\pi}{L}\left[L^{z}_0+\overline{L}^{\overline{z}}_0,\mathcal{O}(w,\overline{w})\right]+\f{(-2a\pi)^2}{2!}\left[L^{z}_0+\overline{L}^{\overline{z}}_0, \left[L^{z}_0+\overline{L}^{\overline{z}}_0,\mathcal{O}(w,\overline{w})\right]\right]+\cdots,
\end{split}
\ee
where $a$ is a real parameter, and in the last line, we used Baker-Campbell-Hausdorff formula.
In the coordinates $(w,\overline{w})$, the commutators are given by
\be
\left[L^{z}_0+\overline{L}^{\overline{z}}_0,\mathcal{O}(w,\overline{w})\right]=2h_{\mathcal{O}}\mathcal{O}(w,\overline{w})+w\partial_w\mathcal{O}(w,\overline{w})+\overline{w}\partial_{\overline{w}}\mathcal{O}(w,\overline{w})\
\ee
\fi
%%%%%%%%%%%%%%%%%%%%%%%%
\subsection{The details of the trajectories }
%%%%%%%%%%%%%%%%%%%%%%%%
Here, we present the details of the local operator trajectories before the analytic continuation.
They are given by
\be
\begin{split}
      &w^{\text{New},1}_{\pm \epsilon}=\f{L}{2\pi}\log{\left[\f{\left[(1-\hat{\lambda}_{-\tau_1})\cosh{2\theta}-(\hat{\lambda}_{-\tau_1}+1)\right]\lambda_{\pm\epsilon}z+(\hat{\lambda}_{-\tau_1}-1)\sinh{2\theta}}{(1-\hat{\lambda}_{-\tau_1})\sinh{2\theta}\lambda_{\pm\epsilon}z+(\hat{\lambda}_{-\tau_1}-1)\cosh{2\theta}-(\hat{\lambda}_{-\tau_1}+1)}\right]},\\
    &\overline{w}^{\text{New},1}_{\pm \epsilon}=\f{L}{2\pi}\log{\left[\f{\left[(1-\hat{\lambda}_{-\tau_1})\cosh{2\theta}-(\hat{\lambda}_{-\tau_1}+1)\right]\lambda_{\pm\epsilon}\overline{z}+(\hat{\lambda}_{-\tau_1}-1)\sinh{2\theta}}{(1-\hat{\lambda}_{-\tau_1})\sinh{2\theta}\lambda_{\pm\epsilon}\overline{z}+(\hat{\lambda}_{-\tau_1}-1)\cosh{2\theta}-(\hat{\lambda}_{-\tau_1}+1)}\right]},\\
   % &w^{\text{New}}_{2}=\f{L}{2\pi}\log{\left[\f{\left[(1-\lambda_{-\tau_1})\cosh{2\theta}-(\lambda_{-\tau_1}+1)\right]\lambda_{-\epsilon}z+(\lambda_{-\tau_1}-1)\sinh{2\theta}}{(1-\lambda_{-\tau_1})\sinh{2\theta}\lambda_{-\epsilon}z+(\lambda_{-\tau_1}-1)\cosh{2\theta}-(\lambda_{-\tau_1}+1)}\right]},\\
  %  &\overline{w}^{\text{New}}_{2}=\f{L}{2\pi}\log{\left[\f{\left[(1-\lambda_{-\tau_1})\cosh{2\theta}-(\lambda_{-\tau_1}+1)\right]\lambda_{-\epsilon}\overline{z}+(\lambda_{-\tau_1}-1)\sinh{2\theta}}{(1-\lambda_{-\tau_1})\sinh{2\theta}\lambda_{-\epsilon}\overline{z}+(\lambda_{-\tau_1}-1)\cosh{2\theta}-(\lambda_{-\tau_1}+1)}\right]},\\
     &w^{\text{New},2}_{\pm\epsilon}=\pm\epsilon+\f{L}{2\pi}\log{\left[\f{\left[(1-\hat{\lambda}_{-\tau_1})\cosh{2\theta}-(\hat{\lambda}_{-\tau_1}+1)\right]z+(\hat{\lambda}_{-\tau_1}-1)\sinh{2\theta}}{(1-\hat{\lambda}_{-\tau_1})\sinh{2\theta}z+(\hat{\lambda}_{-\tau_1}-1)\cosh{2\theta}-(\hat{\lambda}_{-\tau_1}+1)}\right]},\\
    &\overline{w}^{\text{New},2}_{\pm\epsilon}=\pm\epsilon+\f{L}{2\pi}\log{\left[\f{\left[(1-\hat{\lambda}_{-\tau_1})\cosh{2\theta}-(\hat{\lambda}_{-\tau_1}+1)\right]\overline{z}+(\hat{\lambda}_{-\tau_1}-1)\sinh{2\theta}}{(1-\hat{\lambda}_{-\tau_1})\sinh{2\theta}\overline{z}+(\hat{\lambda}_{-\tau_1}-1)\cosh{2\theta}-(\hat{\lambda}_{-\tau_1}+1)}\right]},\\
   % &w^{\text{New}}_{4}=-\epsilon+\f{L}{2\pi}\log{\left[\f{\left[(1-\lambda_{-\tau_1})\cosh{2\theta}-(\lambda_{-\tau_1}+1)\right]z+(\lambda_{-\tau_1}-1)\sinh{2\theta}}{(1-\lambda_{-\tau_1})\sinh{2\theta}z+(\lambda_{-\tau_1}-1)\cosh{2\theta}-(\lambda_{-\tau_1}+1)}\right]},\\
   % &\overline{w}^{\text{New}}_{4}=-\epsilon+\f{L}{2\pi}\log{\left[\f{\left[(1-\lambda_{-\tau_1})\cosh{2\theta}-(\lambda_{-\tau_1}+1)\right]\overline{z}+(\lambda_{-\tau_1}-1)\sinh{2\theta}}{(1-\lambda_{-\tau_1})\sinh{2\theta}\overline{z}+(\lambda_{-\tau_1}-1)\cosh{2\theta}-(\lambda_{-\tau_1}+1)}\right]},
    &w^{\text{New},3}_{\pm\epsilon}=-\tau_0+\f{L}{2\pi}\log{\left[\f{\left[(1-\hat{\lambda}_{\pm \epsilon})\cosh{2\theta}-(\hat{\lambda}_{\pm \epsilon}+1)\right]z+(\hat{\lambda}_{\pm \epsilon}-1)\sinh{2\theta}}{(1-\hat{\lambda}_{\pm \epsilon})\sinh{2\theta}z+(\hat{\lambda}_{\pm \epsilon}-1)\cosh{2\theta}-(\hat{\lambda}_{\pm \epsilon}+1)}\right]},\\
    &\overline{w}^{\text{New},3}_{\pm\epsilon}=-\tau_0+\f{L}{2\pi}\log{\left[\f{\left[(1-\hat{\lambda}_{\pm \epsilon})\cosh{2\theta}-(\hat{\lambda}_{\pm \epsilon}+1)\right]\overline{z}+(\hat{\lambda}_{\pm \epsilon}-1)\sinh{2\theta}}{(1-\hat{\lambda}_{\pm \epsilon})\sinh{2\theta}\overline{z}+(\hat{\lambda}_{\pm \epsilon}-1)\cosh{2\theta}-(\hat{\lambda}_{\pm \epsilon}+1)}\right]},\\
    &w^{\text{New},4}_{\pm \epsilon}=\f{L}{2\pi}\log{\left[\f{\left[(1-\hat{\lambda}_{\pm \epsilon})\cosh{2\theta}-(\hat{\lambda}_{\pm \epsilon}+1)\right]\lambda_{-\tau_0}z+(\hat{\lambda}_{\pm \epsilon}-1)\sinh{2\theta}}{(1-\hat{\lambda}_{\pm \epsilon})\sinh{2\theta}\lambda_{-\tau_0}z+(\hat{\lambda}_{\pm \epsilon}-1)\cosh{2\theta}-(\hat{\lambda}_{\pm \epsilon}+1)}\right]},\\
    %%%%%%%%%%%%%%%%%%%%%%%%%%%%%%%%%%
     &\overline{w}^{\text{New},4}_{\pm \epsilon}=\f{L}{2\pi}\log{\left[\f{\left[(1-\hat{\lambda}_{\pm \epsilon})\cosh{2\theta}-(\hat{\lambda}_{\pm \epsilon}+1)\right]\lambda_{-\tau_0}\overline{z}+(\hat{\lambda}_{\pm \epsilon}-1)\sinh{2\theta}}{(1-\hat{\lambda}_{\pm \epsilon})\sinh{2\theta}\lambda_{-\tau_0}\overline{z}+(\hat{\lambda}_{\pm \epsilon}-1)\cosh{2\theta}-(\hat{\lambda}_{\pm \epsilon}+1)}\right]},\\
\end{split}
\ee
%\textcolor{red}{(MT: In $w^{\text{New},3}$, $\lambda_{\pm\epsilon}$ should be $\hat{\lambda}_{\pm\epsilon}$)}
where $z$, $\overline{z}$, $\hat{\lambda}_{\pm \epsilon}$, $\hat{\lambda}_{-\tau_1}$, $\lambda_{\pm \epsilon}$, and $\lambda_{-\tau_0}$ are defined by%\footnote{\textcolor{red}{\bf MN: I will add the expression in M\"obius case, later.}}
% \be
% z=e^{i\f{2\pi x}{L}},  \overline{z}=e^{-i\f{2\pi x}{L}}, \hat{\lambda}_{\pm \epsilon}=\exp{\left(\f{\pm2\pi \epsilon}{L \cosh{2\theta}}\right)}, \hat{\lambda}_{-\tau_1}=\exp{\left(\f{-2\pi \tau_{1}}{L \cosh{2\theta}}\right)}, \lambda_{\pm\epsilon}=e^{\f{\pm2\pi \epsilon}{L}}, \lambda_{-\tau_0}=e^{\left(\f{-2\pi \tau_0}{L}\right)}.
% \ee
\begin{align}
    &z=e^{i\f{2\pi x}{L}},  \overline{z}=e^{-i\f{2\pi x}{L}}, \hat{\lambda}_{\pm \epsilon}=\exp{\left(\f{\pm2\pi \epsilon}{L \cosh{2\theta}}\right)}, \hat{\lambda}_{-\tau_1}=\exp{\left(\f{-2\pi \tau_{1}}{L \cosh{2\theta}}\right)},
    \nonumber\\
    &\lambda_{\pm\epsilon}=e^{\f{\pm2\pi \epsilon}{L}}, \lambda_{-\tau_0}=e^{\left(\f{-2\pi \tau_0}{L}\right)}.
\end{align}
In SSD limit, $\theta \rightarrow \infty$, the trajectories of the local operators reduce to 
\be
\begin{split}
    &w^{\text{New},1}_{\pm \epsilon}=\f{L}{2\pi}\log{\left[\f{\pi \tau_1\left(1-e^{\f{\pm2\pi \epsilon}{L}}e^{\f{2i\pi x}{L}}\right)+Le^{\f{\pm2\pi \epsilon}{L}}e^{\f{2i\pi x}{L}}}{\pi \tau_1\left(1-e^{\f{\pm2\pi \epsilon}{L}}e^{\f{2i\pi x}{L}}\right)+L}\right]},\\
    %%%%%%%%%%%%%%%%%%%%%%%%%%%%%%%%%%
     &\overline{w}^{\text{New},1}_{\pm \epsilon}=\f{L}{2\pi}\log{\left[\f{\pi \tau_1\left(1-e^{\f{\pm2\pi \epsilon}{L}}e^{\f{-2i\pi x}{L}}\right)+Le^{\f{\pm2\pi \epsilon}{L}}e^{\f{-2i\pi x}{L}}}{\pi \tau_1\left(1-e^{\f{\pm2\pi \epsilon}{L}}e^{\f{-2i\pi x}{L}}\right)+L}\right]},\\
    %%%%%%%%%%%%%%%%%%%%%%%%%%%%%%%%%%
    % &w^{\text{New}}_2=\f{L}{2\pi}\log{\left[\f{\pi \tau_1\left(1-e^{\f{-2\pi \epsilon}{L}}e^{\f{2i\pi x}{L}}\right)+Le^{\f{-2\pi \epsilon}{L}}e^{\f{2i\pi x}{L}}}{\pi \tau_1\left(1-e^{\f{-2\pi \epsilon}{L}}e^{\f{2i\pi x}{L}}\right)+L}\right]},\\
    %%%%%%%%%%%%%%%%%%%%%%%%%%%%%%%%%%
   % &\overline{w}^{\text{New}}_2=\f{L}{2\pi}\log{\left[\f{\pi \tau_1\left(1-e^{\f{-2\pi \epsilon}{L}}e^{\f{-2i\pi x}{L}}\right)+Le^{\f{-2\pi \epsilon}{L}}e^{\f{-2i\pi x}{L}}}{\pi \tau_1\left(1-e^{\f{-2\pi \epsilon}{L}}e^{\f{-2i\pi x}{L}}\right)+L}\right]},\\
    %%%%%%%%%%%%%%%%%%%%%%%%%%%%%%%%%%
     &w^{\text{New},2}_{\pm\epsilon}=\pm\epsilon+\f{L}{2\pi}\log{\left[\f{\pi \tau_1\left(1-e^{\f{2i\pi x}{L}}\right)+Le^{\f{2i\pi x}{L}}}{\pi \tau_1\left(1-e^{\f{2i\pi x}{L}}\right)+L}\right]},\\
     &\overline{w}^{\text{New},2}_{\pm\epsilon}=\pm\epsilon+\f{L}{2\pi}\log{\left[\f{\pi \tau_1\left(1-e^{\f{-2i\pi x}{L}}\right)+Le^{\f{-2i\pi x}{L}}}{\pi \tau_1\left(1-e^{\f{-2i\pi x}{L}}\right)+L}\right]},\\
      %%%%%%%%%%%%%%%%%%%%%%%%%%%%%%%%%%
     &w^{\text{New},3}_{\pm \epsilon}=-\tau_0+\f{L}{2\pi}\log{\left[\f{\pm\pi \epsilon\left(1-e^{\f{2i\pi x}{L}}\right)+Le^{\f{2i\pi x}{L}}}{\pm\pi \epsilon\left(1-e^{\f{2i\pi x}{L}}\right)+L}\right]},\\
     %%%%%%%%%%%%%%%%%%%%%%%%%
& \overline{w}^{\text{New},3}_{\pm \epsilon}=-\tau_0+\f{L}{2\pi}\log{\left[\f{\pm\pi \epsilon\left(1-e^{\f{-2i\pi x}{L}}\right)+Le^{\f{-2i\pi x}{L}}}{\pm\pi \epsilon\left(1-e^{\f{-2i\pi x}{L}}\right)+L}\right]},\\
    %%%%%%%%%%%%%%%%%%%%%%%%%%%%%%%%%%
     &w^{\text{New},4}_{\pm \epsilon}=\f{L}{2\pi}\log{\left[\f{\pm\pi \epsilon\left(1-e^{\f{-2\pi \tau_0}{L}}e^{\f{2i\pi x}{L}}\right)+Le^{\f{-2\pi \tau_0}{L}}e^{\f{2i\pi x}{L}}}{\pm\pi \epsilon\left(1-e^{\f{-2\pi \tau_0}{L}}e^{\f{2i\pi x}{L}}\right)+L}\right]},\\
    %%%%%%%%%%%%%%%%%%%%%%%%%%%%%%%%%%
     &\overline{w}^{\text{New},4}_{\pm \epsilon}=\f{L}{2\pi}\log{\left[\f{\pm\pi \epsilon\left(1-e^{\f{-2\pi \tau_0}{L}}e^{\f{-2i\pi x}{L}}\right)+Le^{\f{-2\pi \tau_0}{L}}e^{\f{-2i\pi x}{L}}}{\pm\pi \epsilon\left(1-e^{\f{-2\pi \tau_0}{L}}e^{\f{-2i\pi x}{L}}\right)+L}\right]}.
     %&\overline{w}^{\text{New},2}_{\pm\epsilon}=\epsilon+\f{L}{2\pi}\log{\left[\f{\pi \tau_1\left(1-e^{\f{2i\pi x}{L}}\right)+Le^{\f{2i\pi x}{L}}}{\pi \tau_1\left(1-e^{\f{2i\pi x}{L}}\right)+L}\right]},\overline{w}^{\text{New}}_4=-\epsilon+\f{L}{2\pi}\log{\left[\f{\pi \tau_1\left(1-e^{\f{-2i\pi x}{L}}\right)+Le^{\f{-2i\pi x}{L}}}{\pi \tau_1\left(1-e^{\f{-2i\pi x}{L}}\right)+L}\right]},\\
\end{split}
\ee
%%%%%%%%%%%%%%%%%%%%%%%%

%%%%%%%%%%%%%%%%%%%%%%%%
\section{The details of energy densities without taking the small $\epsilon$ limit }\label{App:energy-densities-for-WO}
%%%%%%%%%%%%%%%%%%%%%%%%
Here, without taking the small $\epsilon$ limit, we present the details of energy densities during M\"obius/SSD time evolution.
%%%%%%%%%%%%%%%%%%%%%%%%
\subsection{M\"obius evolution}
%%%%%%%%%%%%%%%%%%%%%%%%

%The time evolution of the energy densities during the M\"obius time evolution is given by: \textcolor{red}{(MT: This sentence seems incomplete. Perhaps remove it?)}

First, we will closely look at the time dependence of the chiral and anti-chiral energy densities during the M\"obius time evolution.
After performing analytic continuations, $\tau_{i=0,1}=it_{0,1}$, without the small $\epsilon$ expansion, they are determined by
\be
\begin{split}
    &\left \langle T_{ww}\left(w_X\right) \right \rangle_{i=1,\cdot,4} \approx \left(\f{2\pi}{L}\right)^2\left[-\f{c}{24}+ \mathcal{T}^{\theta}_{i=1,\cdot,4}\right],
    %%%%%%%%%%%%%%%%%%%%%%%%%%%%%%%
    ~\left \langle T_{\overline{w}\overline{w}}\left(\overline{w}_X\right) \right \rangle_{i=1,\cdot,4}\approx \left(\f{2\pi}{L}\right)^2\left[-\f{c}{24}+\overline{\mathcal{T}}^{\theta}_{i=1,\cdot,4}\right],\\
\end{split}
\ee
where functions, $\mathcal{T}^{\theta}_{i=1,\cdots, 4}$, $\overline{\mathcal{T}}^{\theta}_{i=1,\cdots, 4}$, are defined as 
\be
\begin{split}
    &\mathcal{T}^{\theta}_{i=1}=\left( \frac{1}{2} \sinh \left(\frac{2 \pi  \epsilon }{L}\right)\right)^2\\
    &\text{\footnotesize{$\times\left( \frac{1}{\cos \left(\frac{\pi  t_1}{L_{\text{eff}}}\right) \sin \left(\frac{\pi  (x-X-i \epsilon )}{L}\right)-\sin \left(\frac{\pi  t_1}{L_{\text{eff}}}\right) \left(\cosh (2 \theta ) \cos \left(\frac{\pi  (x-X-i \epsilon )}{L}\right)-\sinh (2 \theta ) \cos \left(\frac{\pi  (x+X-i \epsilon )}{L}\right)\right)}\right)^2$}}\\
    &\text{\footnotesize{$\times\left(\frac{1}{\cos \left(\frac{\pi  t_1}{L_{\text{eff}}}\right) \sin \left(\frac{\pi  (x-X+i \epsilon )}{L}\right)-\sin \left(\frac{\pi  t_1}{L_{\text{eff}}}\right) \left(\cosh (2 \theta ) \cos \left(\frac{\pi  (x-X+i \epsilon )}{L}\right)-\sinh (2 \theta ) \cos \left(\frac{\pi  (x+X+i \epsilon )}{L}\right)\right)}\right)^2$}},\\
    &\overline{\mathcal{T}}^{\theta}_{i=1}=\left( \frac{1}{2} \sinh \left(\frac{2 \pi  \epsilon }{L}\right)\right)^2\\
    &\text{\footnotesize{$\times\left( \frac{1}{\sin \left(\frac{\pi  t_1}{L_{\text{eff}}}\right) \left(\cosh (2 \theta ) \cos \left(\frac{\pi  (x-X-i \epsilon )}{L}\right)-\sinh (2 \theta ) \cos \left(\frac{\pi  (x+X-i \epsilon )}{L}\right)\right)+\cos \left(\frac{\pi  t_1}{L_{\text{eff}}}\right) \sin \left(\frac{\pi  (x-X-i \epsilon )}{L}\right)}\right)^2$}}\\
    &\text{\footnotesize{$\times\left( \frac{1}{\sin \left(\frac{\pi  t_1}{L_{\text{eff}}}\right) \left(\cosh (2 \theta ) \cos \left(\frac{\pi  (x-X+i \epsilon )}{L}\right)-\sinh (2 \theta ) \cos \left(\frac{\pi  (x+X+i \epsilon )}{L}\right)\right)+\cos \left(\frac{\pi  t_1}{L_{\text{eff}}}\right) \sin \left(\frac{\pi  (x-X+i \epsilon )}{L}\right)}\right)^2$}}\\
    %%%%%%%%%%%%%%%%%%%%%%%%%%%%%%%%%%%%%%%%%%%%%%%%%%%%%
    &\mathcal{T}^{\theta}_{i=2}=\left( \frac{1}{2} \sinh \left(\frac{2 \pi  \epsilon }{L}\right)\right)^2\\
    &\text{\footnotesize{$\times\left(\sinh (2 \theta ) \sin \left(\frac{2 \pi  x}{L}\right) \sin \left(\frac{2 \pi  t_1}{L_{\text{eff}}}\right)+\sin ^2\left(\frac{\pi  t_1}{L_{\text{eff}}}\right) \left(\cosh (4 \theta )-\sinh (4 \theta ) \cos \left(\frac{2 \pi  x}{L}\right)\right)+\cos ^2\left(\frac{\pi  t_1}{L_{\text{eff}}}\right)\right)^2$}}\\
    &\text{\footnotesize{$\times\left( \frac{1}{\cos \left(\frac{\pi  t_1}{L_{\text{eff}}}\right) \sin \left(\frac{\pi  (x-X-i \epsilon )}{L}\right)-\sin \left(\frac{\pi  t_1}{L_{\text{eff}}}\right) \left(\cosh (2 \theta ) \cos \left(\frac{\pi  (x-X-i \epsilon )}{L}\right)-\sinh (2 \theta ) \cos \left(\frac{\pi  (x+X-i \epsilon )}{L}\right)\right)}\right)^2$}}\\
    &\text{\footnotesize{$\times\left(\frac{1}{\cos \left(\frac{\pi  t_1}{L_{\text{eff}}}\right) \sin \left(\frac{\pi  (x-X+i \epsilon )}{L}\right)-\sin \left(\frac{\pi  t_1}{L_{\text{eff}}}\right) \left(\cosh (2 \theta ) \cos \left(\frac{\pi  (x-X+i \epsilon )}{L}\right)-\sinh (2 \theta ) \cos \left(\frac{\pi  (x+X+i \epsilon )}{L}\right)\right)}\right)^2$}},\\
    &\overline{\mathcal{T}}^{\theta}_{i=2}=\left( \frac{1}{2} \sinh \left(\frac{2 \pi  \epsilon }{L}\right)\right)^2\\
    &\text{\footnotesize{$ \left(-\sinh (2 \theta ) \sin \left(\frac{2 \pi  x}{L}\right) \sin \left(\frac{2 \pi  t_1}{L_{\text{eff}}}\right)+\sin ^2\left(\frac{\pi  t_1}{L_{\text{eff}}}\right) \left(\cosh (4 \theta )-\sinh (4 \theta ) \cos \left(\frac{2 \pi  x}{L}\right)\right)+\cos ^2\left(\frac{\pi  t_1}{L_{\text{eff}}}\right)\right)^2$}}\\
    &\text{\footnotesize{$\times\left( \frac{1}{\sin \left(\frac{\pi  t_1}{L_{\text{eff}}}\right) \left(\cosh (2 \theta ) \cos \left(\frac{\pi  (x-X-i \epsilon )}{L}\right)-\sinh (2 \theta ) \cos \left(\frac{\pi  (x+X-i \epsilon )}{L}\right)\right)+\cos \left(\frac{\pi  t_1}{L_{\text{eff}}}\right) \sin \left(\frac{\pi  (x-X-i \epsilon )}{L}\right)}\right)^2$}}\\
    &\text{\footnotesize{$\times\left( \frac{1}{\sin \left(\frac{\pi  t_1}{L_{\text{eff}}}\right) \left(\cosh (2 \theta ) \cos \left(\frac{\pi  (x-X+i \epsilon )}{L}\right)-\sinh (2 \theta ) \cos \left(\frac{\pi  (x+X+i \epsilon )}{L}\right)\right)+\cos \left(\frac{\pi  t_1}{L_{\text{eff}}}\right) \sin \left(\frac{\pi  (x-X+i \epsilon )}{L}\right)}\right)^2$}}\\
    %%%%%%%%%%%%%%%%%%%%%%%%%%%%%%%%%%%%%%%%%%%%%%%%%%%%%%
    \end{split}
    \ee
    \be
    \begin{split}
    %%%%%%%%%%%%%%%%%%%%%%%%%%%%%%%%%%%%%%%%%%%%%%%%%%%%%
    &\mathcal{T}^{\theta}_{i=3}=\\
    &  \text{\footnotesize{$\left(\frac{\sinh \left(\frac{2 \pi  \epsilon }{L_{\text{eff}}}\right) \left(\sinh (2 \theta ) \cos \left(\frac{2 \pi  x}{L}\right)-\cosh (2 \theta )\right)}{2 \sinh ^2\left(\frac{\pi  \epsilon }{L_{\text{eff}}}\right) \left(\cosh (2 \theta ) \cos \left(\frac{\pi  \left(t_0-x+X\right)}{L}\right)-\sinh (2 \theta ) \cos \left(\frac{\pi  \left(t_0+x+X\right)}{L}\right)\right){}^2+2 \cosh ^2\left(\frac{\pi  \epsilon }{L_{\text{eff}}}\right) \sin ^2\left(\frac{\pi  \left(t_0-x+X\right)}{L}\right)}\right)^2$}},\\
    &\overline{\mathcal{T}}^{\theta}_{i=3}=\\
    &  \text{\footnotesize{$\left(\frac{\sinh \left(\frac{2 \pi  \epsilon }{L_{\text{eff}}}\right) \left(\sinh (2 \theta ) \cos \left(\frac{2 \pi  x}{L}\right)-\cosh (2 \theta )\right)}{2 \sinh ^2\left(\frac{\pi  \epsilon }{L_{\text{eff}}}\right) \left(\cosh (2 \theta ) \cos \left(\frac{\pi  \left(t_0+x-X\right)}{L}\right)-\sinh (2 \theta ) \cos \left(\frac{\pi  \left(-t_0+x+X\right)}{L}\right)\right){}^2+2 \cosh ^2\left(\frac{\pi  \epsilon }{L_{\text{eff}}}\right) \sin ^2\left(\frac{\pi  \left(t_0+x-X\right)}{L}\right)}\right)^2$}},\\
    %%%%%%%%%%%%%%%%%%%%%%%%%%%%%%%%%%%%%%%%%%%%%%%%%%%%%
    &\mathcal{T}^{\theta}_{i=4}=\\
    &  \text{\footnotesize{$\left(\frac{\sinh \left(\frac{2 \pi  \epsilon }{L_{\text{eff}}}\right) \left(\sinh (2 \theta ) \cos \left(\frac{2 \pi  \left(t_0-x\right)}{L}\right)-\cosh (2 \theta )\right)}{2 \sinh ^2\left(\frac{\pi  \epsilon }{L_{\text{eff}}}\right) \left(\cosh (2 \theta ) \cos \left(\frac{\pi  \left(t_0-x+X\right)}{L}\right)-\sinh (2 \theta ) \cos \left(\frac{\pi  \left(t_0-x-X\right)}{L}\right)\right){}^2+2 \cosh ^2\left(\frac{\pi  \epsilon }{L_{\text{eff}}}\right) \sin ^2\left(\frac{\pi  \left(t_0-x+X\right)}{L}\right)}\right)^2$}},\\
    &\overline{\mathcal{T}}^{\theta}_{i=4}=\\
    &  \text{\footnotesize{$\left(\frac{\sinh \left(\frac{2 \pi  \epsilon }{L_{\text{eff}}}\right) \left(\sinh (2 \theta ) \cos \left(\frac{2 \pi  \left(t_0+x\right)}{L}\right)-\cosh (2 \theta )\right)}{2 \sinh ^2\left(\frac{\pi  \epsilon }{L_{\text{eff}}}\right) \left(\cosh (2 \theta ) \cos \left(\frac{\pi  \left(t_0+x-X\right)}{L}\right)-\sinh (2 \theta ) \cos \left(\frac{\pi  \left(t_0+x+X\right)}{L}\right)\right){}^2+2 \cosh ^2\left(\frac{\pi  \epsilon }{L_{\text{eff}}}\right) \sin ^2\left(\frac{\pi  \left(t_0+x-X\right)}{L}\right)}\right)^2$}},\\
    %%%%%%%%%%%%%%%%%%%%%%%%%%%%%%%%%%%%%%%%%%%%%%%%%%%%%%
\end{split}
\ee
where the effective system size is defined as $L_{\text{eff}}=L \cosh{2\theta}$.

%%%%%%%%%%%%%%%%%%%%%%%%
\subsection{SSD evolution}
%%%%%%%%%%%%%%%%%%%%%%%%
The time evolution of the energy densities during the SSD evolution is given by
\be \label{eq:ee-without-small-epsilon-limit}
\begin{split}
    &\left \langle T_{ww}\left(w_X\right) \right \rangle_{1}=\left(\f{2\pi}{L}\right)^2 \Bigg[ -\f{c}{24}+\f{h_{\mathcal{O}}L^4\sinh^2{\left[\f{2\pi \epsilon}{L}\right]}}{4\left[2\pi t_1 \sin{\left[\f{\pi X}{L}\right]}\sin{\left[\f{\pi \left(x-i\epsilon \right)}{L}\right]}+L\sin{\left[\f{\pi\left((X-x)+i\epsilon\right)}{L}\right]}\right]^2}\\
    &\times\f{1}{\left[2\pi t_1 \sin{\left[\f{\pi X}{L}\right]}\sin{\left[\f{\pi (x+i\epsilon)}{L}\right]}+L\sin{\left[\f{\pi\left((X-x)-i\epsilon\right)}{L}\right]}\right]^2}\Bigg],\\
    %%%%%%%%%%%%%%%%%%%%%%%%%%%%%%%%%%%%%%
    &\left \langle T_{\overline{w}\overline{w}}\left(\overline{w}_X\right) \right \rangle_{1}=\left(\f{2\pi}{L}\right)^2 \Bigg[ -\f{c}{24}+\f{h_{\mathcal{O}}L^4\sinh^2{\left[\f{2\pi \epsilon}{L}\right]}}{4\left[2\pi t_1 \sin{\left[\f{\pi X}{L}\right]}\sin{\left[\f{\pi \left(x+i\epsilon \right)}{L}\right]}-L\sin{\left[\f{\pi\left((X-x)-i\epsilon\right)}{L}\right]}\right]^2}\\
    &\times\f{1}{\left[2\pi t_1 \sin{\left[\f{\pi X}{L}\right]}\sin{\left[\f{\pi (x-i\epsilon)}{L}\right]}-L\sin{\left[\f{\pi\left((X-x)+i\epsilon\right)}{L}\right]}\right]^2}\Bigg],\\
    %%%%%%%%%%%%%%%%%%%%%%%%%%%%%%%%%%%%%%%
    &\left \langle T_{ww}\left(w_X\right) \right \rangle_{2}=\left(\f{2\pi}{L}\right)^2 \Bigg[ -\f{c}{24}+\f{1}{\left[2\pi t_1 \sin{\left[\f{\pi (X-i\epsilon)}{L}\right]}\sin{\left[\f{\pi x}{L}\right]}+L\sin{\left[\f{\pi\left((X-x)-i\epsilon\right)}{L}\right]}\right]^2}\\
    &\times\f{h_{\mathcal{O}}\sinh^2{\left[\f{2\pi \epsilon}{L}\right]}\left[L^2+4\pi t_1 L\sin{\left(\f{\pi x}{L}\right)}\cos{\left(\f{\pi x}{L}\right)}+4\pi^2t_1^2\sin^2{\left(\f{\pi x}{L}\right)}\right]^2}{\left[2\pi t_1 \sin{\left[\f{\pi (X+i\epsilon)}{L}\right]}\sin{\left[\f{\pi x }{L}\right]}+L\sin{\left[\f{\pi\left((X-x)+i\epsilon\right)}{L}\right]}\right]^2}\Bigg],\\
    %%%%%%%%%%%%%%%%%%%%%%%%%%%%%%%%%%%%%%%
    &\left \langle T_{\overline{w}\overline{w}}\left(\overline{w}_X\right) \right \rangle_{2}=\left(\f{2\pi}{L}\right)^2 \Bigg[ -\f{c}{24}+\f{1}{\left[2\pi t_1 \sin{\left[\f{\pi (X+i\epsilon)}{L}\right]}\sin{\left[\f{\pi x}{L}\right]}-L\sin{\left[\f{\pi\left((X-x)+i\epsilon\right)}{L}\right]}\right]^2}\\
    &\times\f{h_{\mathcal{O}}\sinh^2{\left[\f{2\pi \epsilon}{L}\right]}\left[L^2-4\pi t_1 \sin{\left(\f{\pi x}{L}\right)}\cos{\left(\f{\pi x}{L}\right)}+4\pi^2t_1^2\sin^2{\left(\f{\pi x}{L}\right)}\right]^2}{\left[2\pi t_1 \sin{\left[\f{\pi (X-i\epsilon)}{L}\right]}\sin{\left[\f{\pi x }{L}\right]}-L\sin{\left[\f{\pi\left((X-x)-i\epsilon\right)}{L}\right]}\right]^2}\Bigg],\\
    %%%%%%%%%%%%%%%%%%%%%%%%%%%%%%%%%%%%%%%%
    &\left \langle T_{ww}\left(w_X\right) \right \rangle_{3}=\left(\f{2\pi}{L}\right)^2 \left[ -\f{c}{24}+\f{4h_{\mathcal{O}}\pi^2L^2\epsilon^2\sin^2{\left[\f{4\pi x}{L}\right]}}{\left[L^2\sin^2{\left[\f{\pi (t_0-x+X)}{L}\right]}+4\pi^2\epsilon^2\sin^2{\left[\f{\pi(t_0+X)}{L}\right]}\sin^2{\left[\f{\pi x}{L}\right]}\right]^2}\right],\\
    %%%%%%%%%%%%%%%%%%%%%%%%%%%%%%%%%%%%%%%%
    &\left \langle T_{\overline{w}\overline{w}}\left(\overline{w}_X\right) \right \rangle_{3}=\left(\f{2\pi}{L}\right)^2 \left[ -\f{c}{24}+\f{4h_{\mathcal{O}}\pi^2L^2\epsilon^2\sin^2{\left[\f{4\pi x}{L}\right]}}{\left[L^2\sin^2{\left[\f{\pi (t_0+x-X)}{L}\right]}+4\pi^2\epsilon^2\sin^2{\left[\f{\pi(t_0-X)}{L}\right]}\sin^2{\left[\f{\pi x}{L}\right]}\right]^2}\right],\\
    %%%%%%%%%%%%%%%%%%%%%%%%%%%%%%%%%%%%%%%%
     &\left \langle T_{ww}\left(w_X\right) \right \rangle_{4}=\left(\f{2\pi}{L}\right)^2 \left[ -\f{c}{24}+\f{4h_{\mathcal{O}}\pi^2L^2\epsilon^2\sin^2{\left[\f{\pi \pi (-x+t_0)}{L}\right]}}{\left[L^2\sin^2{\left[\f{\pi (t_0-x+X)}{L}\right]}+4\pi^2\epsilon^2\sin^2{\left[\f{\pi(t_0-X)}{L}\right]}\sin^2{\left[\f{\pi x}{L}\right]}\right]^2}\right],\\
    %%%%%%%%%%%%%%%%%%%%%%%%%%%%%%%%%%%%%%%%
    &\left \langle T_{\overline{w}\overline{w}}\left(\overline{w}_X\right) \right \rangle_{4}=\left(\f{2\pi}{L}\right)^2 \left[ -\f{c}{24}+\f{4h_{\mathcal{O}}\pi^2L^2\epsilon^2\sin^4{\left[\f{\pi (x+t_0)}{L}\right]}}{\left[L^2\sin^2{\left[\f{\pi (t_0+x-X)}{L}\right]}+4\pi^2\epsilon^2\sin^2{\left[\f{\pi(t_0+X)}{L}\right]}\sin^2{\left[\f{\pi x}{L}\right]}\right]^2}\right],\\
\end{split}
\ee

%%%%%%%%%%%%%%%%%%%%%%%%
\section{Results for the Integrable Theories}\label{IntegrableTheoryAppendix}%%%%%%%%%%%%%%%%%%%%%%%%

As mentioned in Section \ref{SummaryIntegrableTheories}, the calculation of the change in the second R\'{e}nyi entropy in the integrable theories is involved so we collect the relevant techincal details and results in this appendix.

\subsection{M\"{o}bius Case}
% \textcolor{red}{\bf MN: It seems to me that you have started explaining the behavior of EE using quasiparticle picture without explaining what is quasiparticle picture. My suggestion is (1) We will report on the time-dependence of EE. (2) Then, we explain what is the quasiparticle picture. (3) Subsequently, we explain the time dependence of EE in terms of quasiprticles. Or We put all the results analytically-driven on the appendix. We only explain the important time-dependence of EE in terms of quasiparticle picture. For example, behavior doesn't depend on the order of operations and so on.}

Let hatted quantities stand for both holomorphic/anti-holomorphic quantities and let $\sigma=+1/-1$ for holomorphic/anti-holomorphic quantities respectively. Insert the local operator at position $x$. Set the phase $\psi=\pi+\frac{\pi(X_1-X_2)}{L}$ to align the branch cut along the negative real axis in the $\zeta^n$ and $\bar{\zeta}^n$ planes. Then, perform the analytic continuation $\tau_1\rightarrow it_1$ for $i=1,2$ and $\tau_0\rightarrow it_0$ for $i=3,4$ and expand in powers of $\delta=\frac{2\pi \epsilon}{L}$ up to the second order to find:
\subsubsection*{\underline{$\mathbf{i=1}:$}}
\begin{align}\label{MobiusCoordsiEqual1}
&\hat{\zeta}^n_{\rho\epsilon}\nonumber\\ 
    =&  \frac{\sin{\frac{\pi t_1}{L\cosh{2\theta}}}\left(\cos{\frac{\pi(x+X_2)}{L}}\tanh{2\theta}-\cos{\frac{\pi(x-X_2)}{L}}\right)+\cos{\frac{\pi t_1}{L\cosh{2\theta}}}\sin{\frac{\pi \sigma(x-X_2)}{L}}\sech{2\theta}}{
\sin{\frac{\pi t_1}{L\cosh{2\theta}}}\left(\cos{\frac{\pi(x+X_1)}{L}}\tanh{2\theta}-\cos{\frac{\pi(x-X_1)}{L}}\right)+\cos{\frac{\pi t_1}{L\cosh{2\theta}}}\sin{\frac{\pi \sigma(x-X_1)}{L}}\sech{2\theta}}\Bigg(1\nonumber\\
+&\left(\frac{2\pi \epsilon}{L}\right)\frac{i\rho \sin{\frac{\pi\sigma(X_1-X_2)}{L}}\sech^2 2\theta}{2\left[\cos{\frac{\pi t_1}{L\cosh{2\theta}}} \sech2\theta\sin{\frac{\pi \sigma(x-X_1)}{L}}+\sin{\frac{\pi t_1}{L\cosh{2\theta}}}\left(\cos{\frac{\pi(x+X_1)}{L}}\tanh{2\theta}-\cos{\frac{\pi(x-X_1)}{L}}\right)\right] }\nonumber \\
\times&\frac{1}{\Bigg[\cos{\frac{\pi t_1}{L\cosh{2\theta}}} \sech2\theta\sin{\frac{\pi \sigma(x-X_2)}{L}}+\sin{\frac{\pi t_1}{L\cosh{2\theta}}}\left(\cos{\frac{\pi(x+X_2)}{L}}\tanh{2\theta}-\cos{\frac{\pi(x-X_2)}{L}}\right)\Bigg]}\nonumber \\
-&\left(\frac{\pi \epsilon}{L}\right)^2\frac{\sin{\frac{\pi\sigma(X_1-X_2)}{L}}\sech^2 2\theta}{\left[\cos{\frac{\pi t_1}{L\cosh{2\theta}}} \sech2\theta\sin{\frac{\pi \sigma(x-X_1)}{L}}+\sin{\frac{\pi t_1}{L\cosh{2\theta}}}\left(\cos{\frac{\pi(x+X_1)}{L}}\tanh{2\theta}-\cos{\frac{\pi(x-X_1)}{L}}\right)\right]^2}\nonumber \\
\times& 
\frac{\left[\cos{\frac{\pi t_1}{L\cosh{2\theta}}} \sech2\theta\cos{\frac{\pi(x-X_1)}{L}}+\sin{\frac{\pi t_1}{L\cosh{2\theta}}}\left(\sin{\frac{\pi\sigma(x-X_1)}{L}}-\sin{\frac{\pi\sigma(x+X_1)}{L}}\tanh{2\theta}\right)\right]}{\left[\cos{\frac{\pi t_1}{L\cosh{2\theta}}} \sech2\theta\sin{\frac{\pi \sigma(x-X_2)}{L}}+\sin{\frac{\pi t_1}{L\cosh{2\theta}}}\left(\cos{\frac{\pi(x+X_2)}{L}}\tanh{2\theta}-\cos{\frac{\pi(x-X_2)}{L}}\right)\right]}\Bigg)
\end{align}

\subsubsection*{\underline{$\mathbf{i=2}:$}}
\begin{align}\label{MobiusCoordsiEqual2}
    &\hat{\zeta}^n_{\rho\epsilon}\nonumber\\ 
    =&  \frac{\sin{\frac{\pi t_1}{L\cosh{2\theta}}}\left(\cos{\frac{\pi(x+X_2)}{L}}\tanh{2\theta}-\cos{\frac{\pi(x-X_2)}{L}}\right)+\cos{\frac{\pi t_1}{L\cosh{2\theta}}}\sin{\frac{\pi \sigma(x-X_2)}{L}}\sech{2\theta}}{
\sin{\frac{\pi t_1}{L\cosh{2\theta}}}\left(\cos{\frac{\pi(x+X_1)}{L}}\tanh{2\theta}-\cos{\frac{\pi(x-X_1)}{L}}\right)+\cos{\frac{\pi t_1}{L\cosh{2\theta}}}\sin{\frac{\pi \sigma(x-X_1)}{L}}\sech{2\theta}}\Bigg(1\nonumber\\
+&\left(\frac{2\pi \epsilon}{L}\right)\frac{i\rho \sin{\frac{\pi\sigma(X_1-X_2)}{L}}}{2\left[\cos{\frac{\pi t_1}{L\cosh{2\theta}}} \sech2\theta\sin{\frac{\pi \sigma(x-X_1)}{L}}+\sin{\frac{\pi t_1}{L\cosh{2\theta}}}\left(\cos{\frac{\pi(x+X_1)}{L}}\tanh{2\theta}-\cos{\frac{\pi(x-X_1)}{L}}\right)\right]}\nonumber\\
\times& \bigg[\sech^2 2\theta\cos^2\frac{\pi t_1}{L\cosh{2\theta}}+\sin^2\frac{\pi t_1}{L\cosh{2\theta}}\left(1+\tanh^22\theta-2\tanh{2\theta}\cos{\frac{2\pi x}{L}}\right)\nonumber \\
+&\sin{\frac{2\pi t_1}{L\cosh{2\theta}}}\tanh{2\theta}\sech2\theta \sin{\frac{2\pi\sigma x}{L}}\bigg]
\nonumber\\
\times&\frac{1}{\Bigg[\cos{\frac{\pi t_1}{L\cosh{2\theta}}} \sech2\theta\sin{\frac{\pi \sigma(x-X_2)}{L}}+\sin{\frac{\pi t_1}{L\cosh{2\theta}}}\left(\cos{\frac{\pi(x+X_2)}{L}}\tanh{2\theta}-\cos{\frac{\pi(x-X_2)}{L}}\right)\Bigg]} \nonumber \\ 
-&\left(\frac{\pi \epsilon}{L}\right)^2\frac{\sin{\frac{\pi\sigma(X_1-X_2)}{L}}}{\left[\cos{\frac{\pi t_1}{L\cosh{2\theta}}} \sech2\theta\sin{\frac{\pi \sigma(x-X_2)}{L}}+\sin{\frac{\pi t_1}{L\cosh{2\theta}}}\left(\cos{\frac{\pi(x+X_2)}{L}}\tanh{2\theta}-\cos{\frac{\pi(x-X_2)}{L}}\right)\right]}\nonumber\\
\times& \frac{\left[\cos{\frac{\pi t_1}{L\cosh{2\theta}}} \sech2\theta\cos{\frac{\pi(x-X_1)}{L}}+\sin{\frac{\pi t_1}{L\cosh{2\theta}}}\left(\sin{\frac{\pi\sigma(x-X_1)}{L}}+\sin{\frac{\pi\sigma(x+X_1)}{L}}\tanh{2\theta}\right)\right]}{\left[\cos{\frac{\pi t_1}{L\cosh{2\theta}}} \sech2\theta\sin{\frac{\pi \sigma(x-X_1)}{L}}+\sin{\frac{\pi t_1}{L\cosh{2\theta}}}\left(\cos{\frac{\pi(x+X_1)}{L}}\tanh{2\theta}-\cos{\frac{\pi(x-X_1)}{L}}\right)\right]^2}\nonumber \\
\times& \bigg[\sech^2 2\theta \cos^2\frac{\pi t_1}{L\cosh{2\theta}} +\sin^2\frac{\pi t_1}{L\cosh{2\theta}}\left(1+\tanh^2 2\theta-2\tanh2\theta\cos{\frac{2\pi x}{L}}\right) \nonumber \\
+&\sin{\frac{2\pi\sigma x}{L}}\sin{\frac{2\pi t_1}{L\cosh{2\theta}}}\tanh{2\theta}\sech2\theta\bigg]
\end{align}

\subsubsection*{\underline{$\mathbf{i=3}:$}}
\begin{align}\label{MobiusCoordsiEqual3}
    &\hat{\zeta}^n_{\rho\epsilon}\nonumber\\ 
    =&  \frac{\sin{\frac{\pi\left[t_0+\sigma(X_2-x)\right]}{L}}}{\sin{\frac{\pi\left[t_0+\sigma(X_1-x)\right]}{L}}} \Bigg\{1+\left(\frac{2\pi\epsilon}{L}\right)\frac{i\rho \sin{\frac{\pi\sigma(X_1-X_2)}{L}}\left(1-\cos{\frac{2\pi x}{L}}\tanh{2\theta}\right)}{2\sin{\frac{\pi\left[t_0+\sigma(X_1-x)\right]}{L}}\sin{\frac{\pi\left[t_0+\sigma(X_2-x)\right]}{L}}}\nonumber\\
+&\left(\frac{2\pi\epsilon}{L}\right)^2\frac{\sin{\frac{\pi\sigma(X_1-X_2)}{L}}\left(1-\cos{\frac{2\pi x}{L}}\tanh{2\theta}\right)\left[\cos{\frac{\pi\left[t_0+\sigma(X_1-x)\right]}{L}}-\cos{\frac{\pi\left[t_0+\sigma(X_1+x)\right]}{L}}\tanh{2\theta}\right]}{4\sin^2\frac{\pi\left[t_0+\sigma(X_1-x)\right]}{L}\sin{\frac{\pi\left[t_0+\sigma(X_2-x)\right]}{L}}}\Bigg\}
\end{align}

\subsubsection*{\underline{$\mathbf{i=4}:$}}
\begin{align}\label{MobiusCoordsiEqual4}
&\hat{\zeta}^n_{\rho\epsilon}\nonumber\\ 
    =&
    \frac{\sin{\frac{\pi\left[t_0+\sigma(X_2-x)\right]}{L}}}{\sin{\frac{\pi\left[t_0+\sigma(X_1-x)\right]}{L}}} \Bigg\{1+\left(\frac{2\pi\epsilon}{L}\right)\frac{i\rho \sin{\frac{\pi\sigma(X_1-X_2)}{L}}\left(1-\cos{\frac{2\pi (t_0-x\sigma)}{L}}\tanh{2\theta}\right)}{2\sin{\frac{\pi\left[t_0+\sigma(X_1-x)\right]}{L}}\sin{\frac{\pi\left[t_0+\sigma(X_2-x)\right]}{L}}}+\left(\frac{\pi\epsilon}{L}\right)^2\nonumber \\
\times&\frac{\sin{\frac{\pi\sigma(X_1-X_2)}{L}}\left(1-\cos{\frac{2\pi (t_0-\sigma x)}{L}}\tanh{2\theta}\right)\left[\cos{\frac{\pi\left[t_0+\sigma(X_1-x)\right]}{L}}-\cos{\frac{\pi\left[t_0-\sigma(X_1+x)\right]}{L}}\tanh{2\theta}\right]}{\sin^2\frac{\pi\left[t_0+\sigma(X_1-x)\right]}{L}\sin{\frac{\pi\left[t_0+\sigma(X_2-x)\right]}{L}}}\Bigg\}
\end{align}
In these formulas, $\rho=\pm 1$. Note that the signs always appear in the form $\rho\epsilon$, $\sigma x$, $\sigma X_1$ and $\sigma X_2$ so the anti-holomorphic coordinates can be obtained from the holomorphic ones by simply flipping the signs $x\rightarrow -x$, $X_1\rightarrow -X_1$ and $X_2\rightarrow -X_2$. For finite values of $\theta$, when $\sin{\frac{\pi t_1}{L\cosh{2\theta}}}\left(\cos{\frac{\pi(x+X_j)}{L}}\tanh{2\theta}-\cos{\frac{\pi(x-X_j)}{L}}\right)+\cos{\frac{\pi t_1}{L\cosh{2\theta}}}\sin{\frac{\pi \sigma(x-X_j)}{L}}\sech{2\theta} \neq0$ for $j=1,2$ and $i=1,2$, $\xi_0,\xi_1$ as defined in \eqref{UniformizationCoordinateExpansion} are non-zero and finite and $\xi_2$ is finite. The envelope functions for $i=3$ and $i=4$, $1-\cos{\frac{2\pi x}{L}}\tanh{2\theta}$ and $1-\cos{\frac{2\pi (t_0-\sigma x)}{L}}\tanh{2\theta}$ respectively, are never zero for finite $\theta$ so for finite $\theta$ and $\frac{t_0+\sigma(X_j-x)}{L}\not\in\mathbb{Z}$, $\xi_0,\xi_1\neq0$ and are finite while $\xi_2$ is finite. The cross ratios to second order in $\frac{\epsilon}{L}$ are 
\begin{align}\label{UnitaryMobiusQuenchCrossRatio}
    &\hat{\eta}_2 \nonumber\\=&\begin{cases}
        \frac{1}{2}\bigg(1-\text{sgn}\left[\frac{\sin{\frac{\pi t_1}{L\cosh{2\theta}}}\left(\cos{\frac{\pi(x+X_2)}{L}}\tanh{2\theta}-\cos{\frac{\pi(x-X_2)}{L}}\right)+\cos{\frac{\pi t_1}{L\cosh{2\theta}}}\sin{\frac{\pi \sigma(x-X_2)}{L}}\sech{2\theta}}{
\sin{\frac{\pi t_1}{L\cosh{2\theta}}}\left(\cos{\frac{\pi(x+X_1)}{L}}\tanh{2\theta}-\cos{\frac{\pi(x-X_1)}{L}}\right)+\cos{\frac{\pi t_1}{L\cosh{2\theta}}}\sin{\frac{\pi \sigma(x-X_1)}{L}}\sech{2\theta}}\right] \\
\times\bigg[1-\left(\frac{\pi \epsilon}{L}\right)^2 \frac{\sin^2\frac{\pi(X_1-X_2)}{L}\sech^4 2\theta}{2\left[\cos{\frac{\pi t_1}{L\cosh{2\theta}}} \sech2\theta\sin{\frac{\pi \sigma(x-X_1)}{L}}+\sin{\frac{\pi t_1}{L\cosh{2\theta}}}\left(\cos{\frac{\pi(x+X_1)}{L}}\tanh{2\theta}-\cos{\frac{\pi(x-X_1)}{L}}\right)\right]^2}\\
\times\frac{1}{\left[\cos{\frac{\pi t_1}{L\cosh{2\theta}}} \sech2\theta\sin{\frac{\pi \sigma(x-X_2)}{L}}+\sin{\frac{\pi t_1}{L\cosh{2\theta}}}\left(\cos{\frac{\pi(x+X_2)}{L}}\tanh{2\theta}-\cos{\frac{\pi(x-X_2)}{L}}\right)\right]^2}
\bigg]\bigg),& i=1\\
%%%%%%%%%%%%%%%%%%%%%
\frac{1}{2}\bigg(1-\text{sgn}\left[\frac{\sin{\frac{\pi t_1}{L\cosh{2\theta}}}\left(\cos{\frac{\pi(x+X_2)}{L}}\tanh{2\theta}-\cos{\frac{\pi(x-X_2)}{L}}\right)+\cos{\frac{\pi t_1}{L\cosh{2\theta}}}\sin{\frac{\pi \sigma(x-X_2)}{L}}\sech{2\theta}}{
\sin{\frac{\pi t_1}{L\cosh{2\theta}}}\left(\cos{\frac{\pi(x+X_1)}{L}}\tanh{2\theta}-\cos{\frac{\pi(x-X_1)}{L}}\right)+\cos{\frac{\pi t_1}{L\cosh{2\theta}}}\sin{\frac{\pi \sigma(x-X_1)}{L}}\sech{2\theta}}\right] \\
\times\bigg[1-\left(\frac{\pi \epsilon}{L}\right)^2 \frac{\sin^2\frac{\pi(X_1-X_2)}{L}}{2\left[\cos{\frac{\pi t_1}{L\cosh{2\theta}}} \sech2\theta\sin{\frac{\pi \sigma(x-X_1)}{L}}+\sin{\frac{\pi t_1}{L\cosh{2\theta}}}\left(\cos{\frac{\pi(x+X_1)}{L}}\tanh{2\theta}-\cos{\frac{\pi(x-X_1)}{L}}\right)\right]^2}\\ \frac{1}{\left[\cos{\frac{\pi t_1}{L\cosh{2\theta}}} \sech2\theta\sin{\frac{\pi \sigma(x-X_2)}{L}}+\sin{\frac{\pi t_1}{L\cosh{2\theta}}}\left(\cos{\frac{\pi(x+X_2)}{L}}\tanh{2\theta}-\cos{\frac{\pi(x-X_2)}{L}}\right)\right]^2}\\
\times \bigg\{
\sech^2 2\theta \cos^2\frac{\pi t_1}{L\cosh{2\theta}} 
+\sin^2\frac{\pi t_1}{L\cosh{2\theta}}\left(1+\tanh^2 2\theta-2\tanh2\theta\cos{\frac{2\pi x}{L}}\right) \\
+\sin{\frac{2\pi\sigma x}{L}}\sin{\frac{2\pi t_1}{L\cosh{2\theta}}}\tanh{2\theta}\sech2\theta\bigg\}^2\bigg]\bigg) ,& i=2\\
%%%%%%%%%%%%%%%%%%%%%%%
\frac{1}{2}\left\{1-\text{sgn}\left[\frac{\sin{\frac{\pi\left[t_0+\sigma(X_2-x)\right]}{L}}}{\sin{\frac{\pi\left[t_0+\sigma(X_1-x)\right]}{L}}}\right]\left[1-\left(\frac{2\pi\epsilon}{L}\right)^2\frac{\sin^2\frac{\pi(X_1-X_2)}{L}\left(1-\cos{\frac{2\pi x}{L}}\tanh{2\theta}\right)^2}{8\sin^2\frac{\pi\left[t_0+\sigma(X_1-x)\right]}{L}\sin^2\frac{\pi\left[t_0+\sigma(X_2-x)\right]}{L}}\right]
\right\},&i=3\\
%%%%%%%%%%%%%%%%%%%%%%%
\frac{1}{2}\left\{1-\text{sgn}\left[\frac{\sin{\frac{\pi\left[t_0+\sigma(X_2-x)\right]}{L}}}{\sin{\frac{\pi\left[t_0+\sigma(X_1-x)\right]}{L}}}\right]\left[1-\left(\frac{2\pi\epsilon}{L}\right)^2\frac{\sin^2\frac{\pi(X_1-X_2)}{L}\left(1-\cos{\frac{2\pi (t_0-\sigma x)}{L}}\tanh{2\theta}\right)^2}{8\sin^2\frac{\pi\left[t_0+\sigma(X_1-x)\right]}{L}\sin^2\frac{\pi\left[t_0+\sigma(X_2-x)\right]}{L}}\right]
\right\},&i=4.\\
\end{cases} 
\end{align}
Note that the cross ratios are periodic in time with period $L\cosh2\theta$ for $i=1,2$ and $L$ for $i=3$ and they are completely symmetric under an exchange of $X_1$ and $X_2$. These cross ratios reduce to \eqref{SSDCrossRatio} in the SSD $\theta\rightarrow\infty$ limit.

\subsubsection{Operator at $X_1^f$}
In the $\frac{\epsilon}{L}\rightarrow0$ limit, the cross ratios are approximately
\begin{equation}\label{MobiusCrossRatioOperatorAtOrigin}
    \hat{\eta}_2=\begin{cases}
    \frac{1}{2}\Bigg\{1-\text{sgn}\Bigg[\frac{\cos{\frac{\pi X_2}{L}}}{\cos{\frac{\pi X_1}{L}}}\frac{\tan\frac{\pi t_1}{L\cosh{2\theta}}+\sigma e^{2\theta}\tan\frac{\pi X_2}{L}}{\tan\frac{\pi t_1}{L\cosh{2\theta}}+\sigma e^{2\theta}\tan\frac{\pi X_1}{L}}\Bigg]\\
    \times\left[1-\left(\frac{\epsilon}{L}\right)^2 \times(\text{positive term})\right]
    \Bigg\},&   i=1,2,\cos{\frac{\pi t_1}{L\cosh{2\theta}}}\neq 0\\
\frac{1}{2}\left\{1-\text{sgn}\left[\frac{\cos{\frac{\pi X_2}{L}}}{\cos{\frac{\pi X_1}{L}}}\right]\left[1-\left(\frac{\epsilon}{L}\right)^2 \times(\text{positive term})\right]
    \right\},&   i=1,2,\cos{\frac{\pi t_1}{L\cosh{2\theta}}}= 0\\
%%%%%%%%%%%%%%%%%%%%
    \frac{1}{2}\left\{1-\text{sgn}\left[\frac{\sin{\frac{\pi\left(t_0+\sigma X_2\right)}{L}}}{\sin{\frac{\pi\left(t_0+\sigma X_1\right)}{L}}}\right]\left[1-\left(\frac{\epsilon}{L}\right)^2  \times(\text{positive term})\right]
\right\},&i=3\\
%%%%%%%%%%%%%%%%%%%%
    \frac{1}{2}\Bigg\{1-\text{sgn}\left[\frac{\sin{\frac{\pi\left(t_0+\sigma X_2\right)}{L}}}{\sin{\frac{\pi\left(t_0+\sigma X_1\right)}{L}}}\right]\\ \times\left[1-\left(\frac{\epsilon}{L}\right)^2\left(1-\cos{\frac{2\pi t_0}{L}}\tanh{2\theta}\right)^2  \times(\text{positive term})\right]
\Bigg\},&i=4
    \end{cases}
\end{equation}
where we used the fact that $X_j \neq \frac{L}{2}$ for $i=1,2$. 

\subsubsection*{Case (a)}
Consider the interval $A=[0,X_2]\cup [X_1,L]$ with $0<X_2<\frac{L}{2}<X_1<L$ and  $X_2>L-X_1>0$. The second R\'{e}nyi entanglement entropy is 
\begin{equation}\label{MobiusOperatorAtOriginCaseA}
   \Delta S_{A,i}^{(2)} (t_i)=\begin{cases}
    0,&i=1,2, m<\frac{t_1}{L\cosh{2\theta}}< m-\frac{\tan^{-1}\left(e^{2\theta}\tan\frac{\pi X_1}{L}\right)}{\pi}\\
    \log 2,&i=1,2, m-\frac{\tan^{-1}\left(e^{2\theta}\tan\frac{\pi X_1}{L}\right)}{\pi}<\frac{t_1}{L\cosh{2\theta}}<m+\frac{\tan^{-1}\left(e^{2\theta}\tan\frac{\pi X_2}{L}\right)}{\pi}\\
0,&i=1,2,m+\frac{\tan^{-1}\left(e^{2\theta}\tan\frac{\pi X_2}{L}\right)}{\pi}<\frac{t_1}{L\cosh{2\theta}}<m+1-\frac{\tan^{-1}\left(e^{2\theta}\tan\frac{\pi X_2}{L}\right)}{\pi} \\
    \log 2,&i=1,2, m+1-\frac{\tan^{-1}\left(e^{2\theta}\tan\frac{\pi X_2}{L}\right)}{\pi}<\frac{t_1}{L\cosh{2\theta}}< m+1+\frac{\tan^{-1}\left(e^{2\theta}\tan\frac{\pi X_1}{L}\right)}{\pi} \\
    0,&i=1,2, m+1+\frac{\tan^{-1}\left(e^{2\theta}\tan\frac{\pi X_1}{L}\right)}{\pi}<\frac{t_1}{L\cosh{2\theta}}< m+1\\
    %%%%%%%%%%%%
    0,&i=3,4, m<\frac{t_0}{L}<m+1-
    \frac{X_1}{L}\\
    \log 2,&i=3,4, m+1-\frac{X_1}{L}<\frac{t_0}{L}<m+\frac{X_2}{L}\\
    0,&i=3,4, m+\frac{X_2}{L}<\frac{t_0}{L}<m+1-\frac{X_2}{L}\\
    \log 2,&i=3,4, m+1-\frac{X_2}{L}<\frac{t_0}{L}<m+\frac{X_1}{L}\\
    0,&i=3,4, m+\frac{X_1}{L}<\frac{t_0}{L}<m+1\\
\end{cases} 
\end{equation}
for non-negative integers $m$. The principal branch of the arctangent is chosen, i.e. $\tan^{-1}(x)\in (-\frac{\pi}{2},\frac{\pi}{2})$. The $i=1,2$ results are fully explained by the M\"{o}bius quasiparticle picture with finite $\theta$. Initially, $\Delta S_{A,1}^{(2)}(t_1)=0$ because both quasiparticles are contained in subsystem $A$. At $\frac{t_1}{L\cosh{2\theta}}=m-\frac{\tan^{-1}\left(e^{2\theta}\tan\frac{\pi X_1}{L}\right)}{\pi}$, the left-moving quasiparticle hits $X_1$ and exits subsystem $A$ so the second R\'{e}nyi entropy jumps to $\Delta S_{A,1}^{(2)}(t_1)=\log 2$. At $t_1=L\cosh{2\theta}\left[m+\frac{\tan^{-1}\left(e^{2\theta}\tan\frac{\pi X_2}{L}\right)}{\pi}\right]$, the right-moving quasiparticle hits $X_2$ and also exits the subsystem so the second R\'{e}nyi entropy drops back down to $\Delta S_{A,1}^{(2)}(t_1)=0$. At $t_1=L\cosh{2\theta}\left[m+1-\frac{\tan^{-1}\left(e^{2\theta}\tan\frac{\pi X_2}{L}\right)}{\pi}\right]$, the left-moving quasiparticle hits $X_2$ and re-enters the subsystem so the second R\'{e}nyi entropy increases back up to $\Delta S_{A,1}^{(2)}(t_1)=\log 2$ until $t_1=L\cosh{2\theta}\left[m+1+\frac{\tan^{-1}\left(e^{2\theta}\tan\frac{\pi X_1}{L}\right)}{\pi}\right]$ when the right moving quasiparticle re-enters $A$ so the second R\'{e}nyi entropy drops back down to $\Delta S_{A,1}^{(2)}(t_1)=0$. Setting $\theta=0$ gives the obvious answer that generalizes the result in \cite{PhysRevD.90.041701} but for a finite spatial system instead. Note that $\Delta S_{A,i}^{(2)}(t_1)$ for $i=1,2$ is continuous even when $\cos{\frac{\pi t_1}{L \cosh{2\theta}}}=0$ to leading order in $\frac{\epsilon}{L}$.

On the other hand, the $i=3,4$ case is explained by the uniform $\theta=0$ quasiparticle picture.

\subsubsection*{Case (b)}
Consider the interval $A=[X_2,X_1]$ with $X_1>\frac{L}{2}>X_2>0$ and $\frac{L}{2}-X_2>X_1-\frac{L}{2}$. The second R\'{e}nyi entanglement entropy is 
\begin{equation}\label{MobiusOperatorAtOriginCaseb}
   \Delta S_{A,i}^{(2)} (t_i)=\begin{cases}
    0,&i=1,2, m<\frac{t_1}{L\cosh{2\theta}}< m+\frac{\tan^{-1}\left(e^{2\theta}\tan\frac{\pi X_2}{L}\right)}{\pi}\\
    \log 2,&i=1,2, m+\frac{\tan^{-1}\left(e^{2\theta}\tan\frac{\pi X_2}{L}\right)}{\pi}<\frac{t_1}{L\cosh{2\theta}}<m-\frac{\tan^{-1}\left(e^{2\theta}\tan\frac{\pi X_1}{L}\right)}{\pi}\\
0,&i=1,2,m-\frac{\tan^{-1}\left(e^{2\theta}\tan\frac{\pi X_1}{L}\right)}{\pi}<\frac{t_1}{L\cosh{2\theta}}<m+1+\frac{\tan^{-1}\left(e^{2\theta}\tan\frac{\pi X_1}{L}\right)}{\pi} \\
    \log 2,&i=1,2, m+1+\frac{\tan^{-1}\left(e^{2\theta}\tan\frac{\pi X_1}{L}\right)}{\pi}<\frac{t_1}{L\cosh{2\theta}}< m+1-\frac{\tan^{-1}\left(e^{2\theta}\tan\frac{\pi X_2}{L}\right)}{\pi} \\
    0,&i=1,2, m+1-\frac{\tan^{-1}\left(e^{2\theta}\tan\frac{\pi X_2}{L}\right)}{\pi}<\frac{t_1}{L\cosh{2\theta}}< m+1\\
%%%%%%%%%%%%
    0,&i=3,4, m<\frac{t_0}{L}<m+\frac{X_2}{L}\\
    \log 2,&i=3,4, m+\frac{X_2}{L}<\frac{t_0}{L}<m+1-\frac{X_1}{L}\\
    0,&i=3,4, m+1-\frac{X_1}{L}<\frac{t_0}{L}<m+\frac{X_1}{L}\\
    \log 2,&i=3,4, m+\frac{X_1}{L}<\frac{t_0}{L}<m+1-\frac{X_2}{L}\\
    0,&i=3,4, m+1-\frac{X_2}{L}<\frac{t_0}{L}<m+1\\
\end{cases} 
\end{equation}
for non-negative integers $m$. The $i=1,2$ case is explained by the M\"{o}bius quasiparticle picture. Initially, both quasiparticles are outside of $A$ so $\Delta S_{A,1}^{(2)}(t_1)=0$. At $t_1=L\cosh{2\theta}\left[m+\frac{\tan^{-1}\left(e^{2\theta}\tan\frac{\pi X_2}{L}\right)}{\pi}\right]$, the right-moving quasiparticle hits $X_2$ and enters subsystem $A$ so the second R\'{e}nyi entropy increases to $\Delta S_{A,1}^{(2)}(t_1)=\log 2$. The left-moving quasiparticle hits $X_1$ at $\frac{t_1}{L\cosh{2\theta}}=m-\frac{\tan^{-1}\left(e^{2\theta}\tan\frac{\pi X_1}{L}\right)}{\pi}$ and also enters subsystem $A$ so the second R\'{e}nyi entropy drops back down to $\Delta S_{A,1}^{(2)}(t_1)=0$. The right-moving quasiparticle subsequently arrives at $X_1$ at $\frac{t_1}{L\cosh{2\theta}}=m+1+\frac{\tan^{-1}\left(e^{2\theta}\tan\frac{\pi X_1}{L}\right)}{\pi}$ and leaves subsystem $A$ so the second R\'{e}nyi entanglement entropy goes back up to $\Delta S_{A,1}^{(2)}(t_1)=\log 2$. Finally, at $t_1=L\cosh{2\theta}\left[m+1-\frac{\tan^{-1}\left(e^{2\theta}\tan\frac{\pi X_2}{L}\right)}{\pi}\right]$, the left-moving quasiparticle reaches $X_2$ and also leaves the subsystem so the second R\'{e}nyi entropy drops back down to $\Delta S_{A,1}^{(2)}(t_1)=0$. The change in the second R\'{e}nyi entropy $\Delta S_{A,i}^{(2)}(t_1)$ for $i=1,2$ is continuous even when $\cos{\frac{\pi t_1}{L \cosh{2\theta}}}=0$ to leading order in $\frac{\epsilon}{L}$.

On the other hand, the $i=3,4$ case is described by the uniform $\theta=0$ quasiparticle picture.

\subsubsection*{Case (c)}
Consider the interval $A=[X_2,X_1]$ with $0<X_2<X_1<\frac{L}{2}$. The second R\'{e}nyi entanglement entropy is 
\begin{equation}\label{MobiusOperatorAtOriginCaseC}
   \Delta S_{A,i}^{(2)}(t_i) =\begin{cases}
    0,&i=1,2, m<\frac{t_1}{L\cosh{2\theta}}< m+\frac{\tan^{-1}\left(e^{2\theta}\tan\frac{\pi X_2}{L}\right)}{\pi} \\
    \log 2,&i=1,2, m+\frac{\tan^{-1}\left(e^{2\theta}\tan\frac{\pi X_2}{L}\right)}{\pi}<\frac{t_1}{L\cosh{2\theta}}<m+\frac{\tan^{-1}\left(e^{2\theta}\tan\frac{\pi X_1}{L}\right)}{\pi}\\
    0,&i=1,2,m+\frac{\tan^{-1}\left(e^{2\theta}\tan\frac{\pi X_1}{L}\right)}{\pi}<\frac{t_1}{L\cosh{2\theta}}<m+1-\frac{\tan^{-1}\left(e^{2\theta}\tan\frac{\pi X_1}{L}\right)}{\pi} \\
    \log 2,&i=1,2, m+1-\frac{\tan^{-1}\left(e^{2\theta}\tan\frac{\pi X_1}{L}\right)}{\pi}<\frac{t_1}{L\cosh{2\theta}}< m+1-\frac{\tan^{-1}\left(e^{2\theta}\tan\frac{\pi X_2}{L}\right)}{\pi} \\
    0,&i=1,2, m+1-\frac{\tan^{-1}\left(e^{2\theta}\tan\frac{\pi X_2}{L}\right)}{\pi}<\frac{t_1}{L\cosh{2\theta}}< m+1\\
 %%%%%%%%%%%%
    0,&i=3,4, m<\frac{t_0}{L}<m+\frac{X_2}{L}\\
    \log 2,&i=3,4, m+\frac{X_2}{L}<\frac{t_0}{L}<m+\frac{X_1}{L}\\
    0,&i=3,4, m+\frac{X_1}{L}<\frac{t_0}{L}<m+1-\frac{X_1}{L}\\
    \log 2,&i=3,4, m+1-\frac{X_1}{L}<\frac{t_0}{L}<m+1-\frac{X_2}{L}\\
    0,&i=3,4, m+1-\frac{X_2}{L}<\frac{t_0}{L}<m+1\\
\end{cases} 
\end{equation}
for non-negative integers $m$. The M\"{o}bius quasiparticle picture explains the $i=1,2$ cases. Initially, both quasiparticles are outside subsystem $A$ so $\Delta S_{A,1}^{(2)}(t_1)=0$. The right-moving quasiparticle arrives at $X_2$ and enters subsystem $A$ at $t_1=L\cosh{2\theta}\left[m+\frac{\tan^{-1}\left(e^{2\theta}\tan\frac{\pi X_2}{L}\right)}{\pi}\right]$ so the second R\'{e}nyi entanglement entropy increases to $\Delta S_{A,1}^{(2)}(t_1)=\log 2$. It then reaches $X_1$ at $\frac{t_1}{L\cosh{2\theta}}=m+\frac{\tan^{-1}\left(e^{2\theta}\tan\frac{\pi X_1}{L}\right)}{\pi}$ at exits the subsystem $A$ so the second R\'{e}nyi entanglement entropy drops back down to $\Delta S_{A,1}^{(2)}(t_1)=0$. Subsequently, the left-moving quasiparticle enters subsystem $A$ through $X_1$ at $t_1=L\cosh{2\theta}\left[m+1-\frac{\tan^{-1}\left(e^{2\theta}\tan\frac{\pi X_1}{L}\right)}{\pi}\right]$ so the second R\'{e}nyi entanglement entropy goes back up to $\Delta S_{A,1}^{(2)}(t_1)=\log 2$. Finally, at $t_1=L\cosh{2\theta}\left[m+1-\frac{\tan^{-1}\left(e^{2\theta}\tan\frac{\pi X_2}{L}\right)}{\pi}\right]$, the left-moving quasiparticle hits $X_2$ and exits subsystem $A$ so the second R\'{e}nyi entanglement entropy drops back down to $\Delta S_{A,1}^{(2)}(t_1)=0$. As before, the change in the second R\'{e}nyi entropy $\Delta S_{A,i}^{(2)}(t_1)$ for $i=1,2$ is continuous even when $\cos{\frac{\pi t_1}{L \cosh{2\theta}}}=0$ to leading order in $\frac{\epsilon}{L}$.

On the other hand, the $i=3,4$ case is described by the uniform $\theta=0$ quasiparticle picture.

\subsubsection{Operator at $X_2^f$}
The trajectory of quasiparticles that begin at $x = \frac{L}{2}$ is given by sending $x\rightarrow \frac{L}{2}$ in \eqref{QuasiparticleTrajectoryGeneralTheta} and can be rewritten as
\begin{equation}
    t_1 = L\cosh{2\theta}\left[k- \frac{\mu\tan^{-1}\left(e^{-2\theta}\cot{\frac{\pi X}{L}}\right)}{\pi}\right]
\end{equation}
where $k\in\mathbb{Z}$ and $\mu=+1/-1$  for right/left moving quasiparticles.

When the local operator is placed at the second fixed point $x=X_2^f$, the cross ratio \eqref{UnitaryMobiusQuenchCrossRatio} simplifies to 
\begin{equation}
    \hat{\eta}_2=
    \begin{cases}
    \frac{1}{2}\Bigg\{1-\text{sgn}\left[\frac{\tan\frac{\pi t_1}{L\cosh{2\theta}}-\sigma e^{-2\theta}\cot\frac{\pi X_2}{L}}{\tan\frac{\pi t_1}{L\cosh{2\theta}}-\sigma e^{-2\theta}\cot\frac{\pi X_1}{L}}\right]\\
    \times\left[1-\left(\frac{\epsilon}{L}\right)^2 \times(\text{positive term})\right]
    \Bigg\},&i=1,2,  \cos{\frac{\pi t_1}{L\cosh{2\theta}}}\neq 0\\
    \mathcal{O}\left(\left(\frac{\epsilon}{L}\right)^2\right),&i=1,2,  \cos{\frac{\pi t_1}{L\cosh{2\theta}}}= 0\\
    %%%%%%%%%%%%%%%%%%%%%%%%%%%%%%%%%%%%%
\frac{1}{2}\left\{1-\text{sgn}\left[\frac{\cos{\frac{\pi(t_0+\sigma X_2)}{L}}}{\cos{\frac{\pi(t_0+\sigma X_1)}{L}}}\right]\left[1-\left(\frac{\epsilon}{L}\right)^2\times \text{positive number}\right]\right\},&i=3\\
%%%%%%%%%%%%%%%%%%%%
    \frac{1}{2}\Bigg\{1-\text{sgn}\left[\frac{\cos{\frac{\pi\left(t_0+\sigma X_2\right)}{L}}}{\cos{\frac{\pi\left(t_0+\sigma X_1\right)}{L}}}\right]\\ \times\left[1-\left(\frac{\epsilon}{L}\right)^2\left(1+\cos{\frac{2\pi t_0}{L}}\tanh{2\theta}\right)^2  \times\text{positive term}\right]
\Bigg\},&i=4
    \end{cases}
\end{equation}
where we used the fact that $0<X_2,X_1<L$. 

\subsubsection*{Case (a)}
Consider the interval $A=[0,X_2]\cup [X_1,L]$ with $X_2>L-X_1>0$. The second R\'{e}nyi entanglement entropy is 
\begin{equation}\label{MobiusOperatorAtMidpointCaseA}
   \Delta S_{A,i}^{(2)}(t_i) =\begin{cases}
    0,&i=1,2, m<\frac{t_1}{L\cosh{2\theta}}< m+\frac{\tan^{-1}\left(e^{-2\theta}\cot\frac{\pi X_2}{L}\right)}{\pi} \\
    \log 2,&i=1,2, m+\frac{\tan^{-1}\left(e^{-2\theta}\cot\frac{\pi X_2}{L}\right)}{\pi}<\frac{t_1}{L\cosh{2\theta}}<m-\frac{\tan^{-1}\left(e^{-2\theta}\cot\frac{\pi X_1}{L}\right)}{\pi} \\
    0,&i=1,2,m-\frac{\tan^{-1}\left(e^{-2\theta}\cot\frac{\pi X_1}{L}\right)}{\pi}<\frac{t_1}{L\cosh{2\theta}}<m+1+\frac{\tan^{-1}\left(e^{-2\theta}\cot\frac{\pi X_1}{L}\right)}{\pi} \\
    \log 2,&i=1,2, m+1+\frac{\tan^{-1}\left(e^{-2\theta}\cot\frac{\pi X_1}{L}\right)}{\pi}<\frac{t_1}{L\cosh{2\theta}}< m+1-\frac{\tan^{-1}\left(e^{-2\theta}\cot\frac{\pi X_2}{L}\right)}{\pi} \\
    0,&i=1,2, m+1-\frac{\tan^{-1}\left(e^{-2\theta}\cot\frac{\pi X_2}{L}\right)}{\pi}<\frac{t_1}{L\cosh{2\theta}}< m+1\\
%%%%%%%%%%%%%%%%%%%%%%%%%%%%%%%
        0,&i=3,4, m<\frac{t_0}{L}<m+\frac{1}{2}-\frac{X_2}{L}\\
        \log 2,&i=3,4, m+\frac{1}{2}-\frac{X_2}{L}<\frac{t_0}{L}<m+\frac{X_1}{L}-\frac{1}{2}\\
        0,&i=3,4, m+\frac{X_1}{L}-\frac{1}{2} <\frac{t_0}{L}< m +\frac{3}{2}-\frac{X_1}{L}\\
        \log 2,&i=3,4,  m+\frac{3}{2}-\frac{X_1}{L} <\frac{t_0}{L}<m+\frac{1}{2}+\frac{X_2}{L} \\
        0,&i=3,4,  m+\frac{1}{2}+\frac{X_2}{L} <\frac{t_0}{L}<(m+1)    \end{cases} 
\end{equation}
for non-negative integers $m$. 
\begin{sloppypar}
For $i=1,2$, the R\'{e}nyi entropy is continuous at $t_1 = \frac{2k+1}{2}L \cosh{2\theta}$ at leading order in $\frac{\epsilon}{L}$. Taking the $\theta=0$ limit for $i=1,2$ gives the obvious generalization of the result in \cite{PhysRevD.90.041701}. Initially, both quasiparticles are outside of $A$ so $\Delta S_{A,1}^{(2)}(t_1)=0$. At $t_1=L\cosh{2\theta}\left[m+\frac{\tan^{-1}\left(e^{-2\theta}\cot\frac{\pi X_2}{L}\right)}{\pi}\right]$, the left-moving quasiparticle hits $X_2$ and enters $A$ so $\Delta S_{A,1}^{(2)}(t_1)=\log 2$. The right-moving quasiparticle hits $X_1$ at $\frac{t_1}{L\cosh{2\theta}}=m-\frac{\tan^{-1}\left(e^{-2\theta}\cot\frac{\pi X_1}{L}\right)}{\pi}$ and enters $A$ as well so the second R\'{e}nyi entanglement entropy drops back down to $\Delta S_{A,1}^{(2)}(t_1)=0$. At $t_1=L\cosh{2\theta}\left[m+1+\frac{\tan^{-1}\left(e^{-2\theta}\cot\frac{\pi X_1}{L}\right)}{\pi}\right]$, the left moving quasiparticle arrives at $X_1$ and exits subsystem $A$ so the second R\'{e}nyi entropy goes back up to $\Delta S_{A,1}^{(2)}(t_1)=\log 2$. The right moving quasiparticle also leaves the subsystem at $t_1=L\cosh{2\theta}\left[m+1-\frac{\tan^{-1}\left(e^{-2\theta}\cot\frac{\pi X_2}{L}\right)}{\pi}\right]$ so the second R\'{e}nyi entanglement entropy returns to its original value of $\Delta S_{A,1}^{(2)}(t_1)=0$. Note that the change in the second R\'{e}nyi entropy $\Delta S_{A,i}^{(2)}(t_1)$ for $i=1,2$ is continuous even when $\cos{\frac{\pi t_1}{L \cosh{2\theta}}}=0$ to leading order in $\frac{\epsilon}{L}$.
\end{sloppypar}

On the other hand, the $i=3,4$ R\'{e}nyi entropy is described by the uniform $\theta=0$ quasiparticle picture.

\subsubsection*{Case (b)}
Consider the interval $A=[X_2,X_1]$ with $X_1>\frac{L}{2}>X_2>0$ and $\frac{L}{2}-X_2>X_1-\frac{L}{2}$. The second R\'{e}nyi entanglement entropy is 
\begin{equation}\label{MobiusOperatorAtMidpointCaseB}
   \Delta S_{A,i}^{(2)}(t_i) =\begin{cases}
    0,&i=1,2, m<\frac{t_1}{L\cosh{2\theta}}< m-\frac{\tan^{-1}\left(e^{-2\theta}\cot\frac{\pi X_1}{L}\right)}{\pi} \\
    \log 2,&i=1,2,m-\frac{\tan^{-1}\left(e^{-2\theta}\cot\frac{\pi X_1}{L}\right)}{\pi}<\frac{t_1}{L\cosh{2\theta}}<m+\frac{\tan^{-1}\left(e^{-2\theta}\cot\frac{\pi X_2}{L}\right)}{\pi}\\
    0,&i=1,2, m+\frac{\tan^{-1}\left(e^{-2\theta}\cot\frac{\pi X_2}{L}\right)}{\pi}<\frac{t_1}{L\cosh{2\theta}}<m+1-\frac{\tan^{-1}\left(e^{-2\theta}\cot\frac{\pi X_2}{L}\right)}{\pi} \\
    \log 2,&i=1,2,  m+1-\frac{\tan^{-1}\left(e^{-2\theta}\cot\frac{\pi X_2}{L}\right)}{\pi}<\frac{t_1}{L\cosh{2\theta}}< m+1+\frac{\tan^{-1}\left(e^{-2\theta}\cot\frac{\pi X_1}{L}\right)}{\pi}\\
    0,& i=1,2, m+1+\frac{\tan^{-1}\left(e^{-2\theta}\cot\frac{\pi X_1}{L}\right)}{\pi}<\frac{t_1}{L\cosh{2\theta}}<m+1\\
     %%%%%%%%%%%%%%%%%%%%%%%%%%%%%%%
        0,&i=3,4, m<\frac{t_0}{L}<m+\frac{X_1}{L}-\frac{1}{2}
        \\
        \log 2,&i=3,4, m+\frac{X_1}{L}-\frac{1}{2}<\frac{t_0}{L}<m+\frac{1}{2}-\frac{X_2}{L}\\
        0,&i=3,4, m+\frac{1}{2}-\frac{X_2}{L} <\frac{t_0}{L}<m+\frac{1}{2}+\frac{X_2}{L} \\
        \log 2,&i=3,4,  m+\frac{1}{2}+\frac{X_2}{L} <\frac{t_0}{L}<m+\frac{3}{2}-\frac{X_1}{L} \\
        0,&i=3,4,  m+\frac{3}{2}-\frac{X_1}{L} <\frac{t_0}{L}<m+1
\end{cases} 
\end{equation}
for non-negative integers $m$. For $i=1,2$, the R\'{e}nyi entropy is continuous at $t_1 = \frac{2k+1}{2}L \cosh{2\theta}$ for integer $k$ at leading order in $\frac{\epsilon}{L}$. For $i=1,2$, the uniform $\theta=0$ case gives the obvious sensible answer. For finite $\theta$, both particles begin in subsystem $A$ so $\Delta S_{A,1}^{(2)}(t_1)=0$. The right moving quasiparticle reaches $X_1$ at $t_1=L\cosh{2\theta}\left[m-\frac{\tan^{-1}\left(e^{-2\theta}\cot\frac{\pi X_1}{L}\right)}{\pi}\right]$ and exits the subsystem $A$ so the second R\'{e}nyi entropy goes up to $\Delta S_{A,1}^{(2)}(t_1)=\log 2$. The left moving quasiparticle also exits the subsystem at $t_1=L\cosh{2\theta}\left[m+\frac{\tan^{-1}\left(e^{-2\theta}\cot\frac{\pi X_2}{L}\right)}{\pi}\right]$ so the second R\'{e}nyi entropy drops back down to $\Delta S_{A,1}^{(2)}(t_1)=0$. The right-moving quasiparticle subsequently re-enters subsystem $A$ at $t_1=L\cosh{2\theta}\left[m+1-\frac{\tan^{-1}\left(e^{-2\theta}\cot\frac{\pi X_2}{L}\right)}{\pi}\right]$ so the second R\'{e}nyi entropy goes back up to $\Delta S_{A,1}^{(2)}(t_1)=\log 2$. Finally, the left moving quasiparticle re-enters $A$ at $\frac{t_1}{L\cosh{2\theta}}=m+1+\frac{\tan^{-1}\left(e^{-2\theta}\cot\frac{\pi X_1}{L}\right)}{\pi}$ so the second R\'{e}nyi entropy returns to $\Delta S_{A,1}^{(2)}(t_1)=0$. As before, the change in the second R\'{e}nyi entropy $\Delta S_{A,i}^{(2)}(t_1)$ for $i=1,2$ is continuous even when $\cos{\frac{\pi t_1}{L \cosh{2\theta}}}=0$ to leading order in $\frac{\epsilon}{L}$.

On the other hand, the $i=3,4$ cases are explained by the $\theta=0$ uniform quasiparticle picture.

\subsubsection*{Case (c)}
Consider the interval $A=[X_2,X_1]$ with $0<X_2<X_1<\frac{L}{2}$. The second R\'{e}nyi entanglement entropy is 
\begin{equation}\label{MobiusOperatorAtMidpointCaseC}
   \Delta S_{A,i}^{(2)}(t_i) =\begin{cases}
    0,&i=1,2, m<\frac{t_1}{L\cosh{2\theta}}< m+\frac{\tan^{-1}\left(e^{-2\theta}\cot\frac{\pi X_1}{L}\right)}{\pi} \\
    \log 2,&i=1,2, m+\frac{\tan^{-1}\left(e^{-2\theta}\cot\frac{\pi X_1}{L}\right)}{\pi}<\frac{t_1}{L\cosh{2\theta}}<m+\frac{\tan^{-1}\left(e^{-2\theta}\cot\frac{\pi X_2}{L}\right)}{\pi} \\
    0,&i=1,2,m+\frac{\tan^{-1}\left(e^{-2\theta}\cot\frac{\pi X_2}{L}\right)}{\pi}<\frac{t_1}{L\cosh{2\theta}}<m+1-\frac{\tan^{-1}\left(e^{-2\theta}\cot\frac{\pi X_2}{L}\right)}{\pi} \\
    \log 2,&i=1,2, m+1-\frac{\tan^{-1}\left(e^{-2\theta}\cot\frac{\pi X_2}{L}\right)}{\pi}<\frac{t_1}{L\cosh{2\theta}}< m+1-\frac{\tan^{-1}\left(e^{-2\theta}\cot\frac{\pi X_1}{L}\right)}{\pi} \\
    0,&i=1,2, m+1-\frac{\tan^{-1}\left(e^{-2\theta}\cot\frac{\pi X_1}{L}\right)}{\pi}<\frac{t_1}{L\cosh{2\theta}}< m+1\\
    %%%%%%%%%%%%%%%%%%%%%%%%%%%%%%%
        0,&i=3,4, m<\frac{t_0}{L}<m+\frac{1}{2}-\frac{X_1}{L}
        \\
        \log 2,&i=3,4, m+\frac{1}{2}-\frac{X_1}{L}<\frac{t_0}{L}<m+\frac{1}{2}-\frac{X_2}{L}\\
        0,&i=3,4, m+\frac{1}{2}-\frac{X_2}{L} <\frac{t_0}{L}<m+\frac{1}{2}+\frac{X_2}{L} \\
        \log 2,&i=3,4,  m+\frac{1}{2}+\frac{X_2}{L} <\frac{t_0}{L}<m+\frac{1}{2}+\frac{X_1}{L} \\
        0,&i=3,4,  m+\frac{1}{2}+\frac{X_1}{L} <\frac{t_0}{L}<m+1
\end{cases} 
\end{equation}
for non-negative integers $m$. For $i=1,2$, the R\'{e}nyi entropy is continuous at $t_1 = \frac{2k+1}{2}L \cosh{2\theta}$ for integer $k$ at leading order in $\frac{\epsilon}{L}$. For $i=1,2$, the $\theta=0$ case gives the sensible answer.  For finite $\theta$, initially, both quasiparticles are located outside $A$ so $\Delta S_{A,1}^{(2)}(t_1)=0$. At $t_1=L\cosh{2\theta}\left[m+\frac{\tan^{-1}\left(e^{-2\theta}\cot\frac{\pi X_1}{L}\right)}{\pi}\right]$, the left moving quasiparticle reaches $X_1$ and enters subsystem $A$ so $\Delta S_{A,1}^{(2)}(t_1)=\log 2$. It then travels to $X_2$ at $t_1=L\cosh{2\theta}\left[m+\frac{\tan^{-1}\left(e^{-2\theta}\cot\frac{\pi X_2}{L}\right)}{\pi}\right]$ and exits the subsystem so the second R\'{e}nyi entropy drops back down to $\Delta S_{A,1}^{(2)}(t_1)=0$. The right moving quasiparticle arrives at $X_2$ at $t_1=L\cosh{2\theta}\left[m+1-\frac{\tan^{-1}\left(e^{-2\theta}\cot\frac{\pi X_2}{L}\right)}{\pi}\right]$ and enters subsystem $A$ so the second R\'{e}nyi entropy jumps to $\Delta S_{A,1}^{(2)}(t_1)=\log 2$. Finally, at $t_1=L\cosh{2\theta}\left[m+1-\frac{\tan^{-1}\left(e^{-2\theta}\cot\frac{\pi X_1}{L}\right)}{\pi}\right]$, the right moving quasiparticle reaches $X_1$ and exits the subsystem so the second R\'{e}nyi entropy drops back down to $\Delta S_{A,1}^{(2)}(t_1)=0$. As before, the change in the second R\'{e}nyi entropy $\Delta S_{A,i}^{(2)}(t_1)$ for $i=1,2$ is continuous even when $\cos{\frac{\pi t_1}{L \cosh{2\theta}}}=0$ to leading order in $\frac{\epsilon}{L}$.

As usual, the $i=3,4$ cases are explained by the $\theta=0$ uniform quasiparticle picture.

\subsection{SSD Quench}

Putting the  $w^{\text{New,i}}, \overline{w}^{\text{New,i}}$ coordinates that correspond to SSD evolution into the uniformization map \eqref{UniformizationMap}, aligning the branch cut of $\zeta^n$, $\overline{\zeta}^n$ along the negative real axis by setting $\psi = \pi + \frac{\pi(X_1-X_2)}{L}$ and analytically continuing to real time $\tau_1 = it_1$ for $i=1,2$ and $\tau_0 = it_0$ for $i=3,4$, the coordinates on the $n$-sheeted Riemann surface to second order in 
$\frac{2\pi\epsilon}{L}$ are
\begin{align}\label{SSDCoordinates}
    \hat{\zeta}^n_{\rho\epsilon}=\begin{cases}
    \frac{t_1\sin{\frac{\pi x}{L}}\sin{\frac{\pi X_2}{L}}+\frac{L}{2\pi}\sin{\frac{\pi\sigma(X_2-x)}{L}}}{t_1\sin{\frac{\pi x}{L}}\sin{\frac{\pi X_1}{L}}+\frac{L}{2\pi}\sin{\frac{\pi\sigma(X_1-x)}{L}}}\Bigg\{1\\
    +\left(\frac{2\pi\epsilon}{L}\right)\frac{i\rho L^2 \sin{\frac{\pi\sigma(X_1-X_2)}{L}}}{8\pi^2\left[t_1\sin{\frac{\pi x}{L}}\sin{\frac{\pi X_1}{L}}+\frac{L}{2\pi}\sin{\frac{\pi \sigma(X_1-x)}{L}}\right]\left[t_1\sin{\frac{\pi x}{L}}\sin{\frac{\pi X_2}{L}}+\frac{L}{2\pi}\sin{\frac{\pi \sigma(X_2-x)}{L}}\right]}\\
    -\left(\frac{2\pi\epsilon}{L}\right)^2 
    \frac{L^2 \sin{\frac{\pi\sigma(X_1-X_2)}{L}}\left[t_1\sin{\frac{\pi\sigma X_1}{L}}\cos{\frac{\pi x}{L}}-\frac{L}{2\pi}\cos{\frac{\pi (x-X_1)}{L}}\right]
    }{16\pi^2\left[t_1\sin{\frac{\pi x}{L}}\sin{\frac{\pi X_1}{L}}+\frac{L}{2\pi}\sin{\frac{\pi \sigma(X_1-x)}{L}}\right]^2\left[t_1\sin{\frac{\pi x}{L}}\sin{\frac{\pi X_2}{L}}+\frac{L}{2\pi}\sin{\frac{\pi \sigma(X_2-x)}{L}}\right]}\Bigg\},& i=1\\ 
%%%%%%%%%%%%%%%%%%%%%%%%%%%%%%%   
\frac{t_1\sin{\frac{\pi x}{L}}\sin{\frac{\pi X_2}{L}}+\frac{L}{2\pi}\sin{\frac{\pi\sigma(X_2-x)}{L}}}{t_1\sin{\frac{\pi x}{L}}\sin{\frac{\pi X_1}{L}}+\frac{L}{2\pi}\sin{\frac{\pi\sigma(X_1-x)}{L}}}\Bigg\{1\\
    +\left(\frac{2\pi\epsilon}{L}\right)\frac{i\rho  \sin{\frac{\pi\sigma(X_1-X_2)}{L}}\left(L^2+4\pi^2t_1^2 \sin^2 \frac{\pi x}{L}+2\pi t_1 L\sin{\frac{2\pi x\sigma}{L}}\right)}{8\pi^2\left[t_1\sin{\frac{\pi x}{L}}\sin{\frac{\pi X_1}{L}}+\frac{L}{2\pi}\sin{\frac{\pi \sigma(X_1-x)}{L}}\right]\left[t_1\sin{\frac{\pi x}{L}}\sin{\frac{\pi X_2}{L}}+\frac{L}{2\pi}\sin{\frac{\pi \sigma(X_2-x)}{L}}\right]}\\
    +\left(\frac{2\pi\epsilon}{L}\right)^2 
    \frac{\sin{\frac{\pi\sigma(X_1-X_2)}{L}}\left[t_1\sin{\frac{\pi\sigma x}{L}}\cos{\frac{\pi X_1}{L}}+\frac{L}{2\pi}\cos{\frac{\pi (x-X_1)}{L}}\right]\left(L^2+4\pi^2t_1^2 \sin^2 \frac{\pi x}{L}+2\pi t_1 L\sin{\frac{2\pi x\sigma}{L}}\right)}{16\pi^2\left[t_1\sin{\frac{\pi x}{L}}\sin{\frac{\pi X_1}{L}}+\frac{L}{2\pi}\sin{\frac{\pi \sigma(X_1-x)}{L}}\right]^2\left[t_1\sin{\frac{\pi x}{L}}\sin{\frac{\pi X_2}{L}}+\frac{L}{2\pi}\sin{\frac{\pi \sigma(X_2-x)}{L}}\right]}\Bigg\},& i=2\\
%%%%%%%%%%%%%%%%%%%%%%%%%%%%%%%%%%%%%%
\frac{\sin{\frac{\pi\left[t_0+\sigma(X_2-x)\right]}{L}}}{\sin{\frac{\pi\left[t_0+\sigma(X_1-x)\right]}{L}}}\Bigg\{1
-\left(\frac{2\pi\epsilon}{L}\right)\frac{i\rho \sin^2\frac{\pi x}{L}\sin{\frac{\pi\sigma(X_1-X_2)}{L}}}{\sin{\frac{\pi\left[t_0+\sigma(X_1-x)\right]}{L}}\sin{\frac{\pi\left[t_0+\sigma(X_2-x)\right]}{L}}}\\
+\left(\frac{2\pi\epsilon}{L}\right)^2\frac{\sin^3\frac{\pi x\sigma}{L}\sin{\frac{\pi\sigma(X_1-X_2)}{L}}\sin{\frac{\pi(t_0+\sigma X_1)}{L}}}{\sin^2\frac{\pi\left[t_0+\sigma(X_1-x)\right]}{L}\sin{\frac{\pi\left[t_0+\sigma(X_2-x)\right]}{L}}}\Bigg\},&i=3 \\
%%%%%%%%%%%%%%%%%%%%%%%%%%%%%%%%%%%
\frac{\sin{\frac{\pi\left[t_0+\sigma(X_2-x)\right]}{L}}}{\sin{\frac{\pi\left[t_0+\sigma(X_1-x)\right]}{L}}}\Bigg\{1
-\left(\frac{2\pi\epsilon}{L}\right)\frac{i\rho \sin^2\frac{\pi (t_0-\sigma x)}{L}\sin{\frac{\pi\sigma(X_1-X_2)}{L}}}{\sin{\frac{\pi\left[t_0+\sigma(X_1-x)\right]}{L}}\sin{\frac{\pi\left[t_0+\sigma(X_2-x)\right]}{L}}}\\
-\left(\frac{2\pi\epsilon}{L}\right)^2\frac{\sin^3\frac{\pi (t_0-\sigma x)}{L}\sin{\frac{\pi\sigma(X_1-X_2)}{L}}\sin{\frac{\pi\sigma X_1}{L}}}{\sin^2\frac{\pi\left[t_0+\sigma(X_1-x)\right]}{L}\sin{\frac{\pi\left[t_0+\sigma(X_2-x)\right]}{L}}}\Bigg\},&i=4
\end{cases}  
\end{align}
where $x$ is the position where the local operator was inserted. Taking the $\theta\rightarrow\infty$ limit of the M\"{o}bius coordinates (\ref{MobiusCoordsiEqual1},\ref{MobiusCoordsiEqual2},\ref{MobiusCoordsiEqual3},\ref{MobiusCoordsiEqual4}) reproduces the SSD coordinates \eqref{SSDCoordinates}. When $x=0$ for $i=3$, the coordinates of the two local operators are
\begin{equation}
    \hat{\zeta}_{\pm\epsilon}=\frac{\sin{\frac{\pi(t_0+\sigma X_2)}{L}}}{\sin{\frac{\pi(t_0+\sigma X_1)}{L}}}
\end{equation}
which is equivalent to setting the regulator $\epsilon=0$. The two local operators are coincident and hence the correlator is ill-defined. Naively, if 
\begin{equation}
    \frac{\sin{\frac{\pi(t_0+\sigma X_2)}{L}}}{\sin{\frac{\pi(t_0+\sigma X_1)}{L}}}<0,
\end{equation}
the coordinates of the two operators $\hat{\zeta}_{\pm\epsilon}^2$ sit on the branch cut on the negative real axis and the rotation that we have been doing would be ill-defined. Similarly, for $i=4$, when $t_0 = \sigma x+nL$ for $n\in\mathbb{Z}$, the regulator $\epsilon$ drops out of the exact coordinates and the two operators become co-incident.

When $t_1 \sin{\frac{\pi x}{L}}\sin{\frac{\pi X_j}{L}}+\frac{L}{2\pi}\sin{\frac{\pi\sigma(X_j-x)}{L}}\neq0$ for $j=1,2$ for $i=1,2$ and $x\neq0$, $\frac{t_0+\sigma(X_j-x)}{L}\not\in \mathbb{Z}$ for $j=1,2$ for $i=3$, and $\frac{t_0-\sigma x}{L}\not\in\mathbb{Z}$ and $\frac{t_0+\sigma(X_j-x)}{L}\not\in \mathbb{Z}$ for $j=1,2$ for $i=4$, both $\xi_0,\xi_1$ as defined in \eqref{UniformizationCoordinateExpansion} are non-zero and are finite and $\xi_2$ is finite so the cross ratios to second order in $\frac{\epsilon}{L}$ \eqref{CrossRatioSecondOrder} are
\begin{align}
\hat{\eta}_2=&\begin{cases}
\frac{1}{2}\Bigg\{1-\text{sgn}\left[\frac{t_1\sin{\frac{\pi x}{L}}\sin{\frac{\pi X_2}{L}}+\frac{L}{2\pi}\sin{\frac{\pi\sigma(X_2-x)}{L}}}{t_1\sin{\frac{\pi x}{L}}\sin{\frac{\pi X_1}{L}}+\frac{L}{2\pi}\sin{\frac{\pi\sigma(X_1-x)}{L}}} \right]
\Bigg[1\\
-\left(\frac{2\pi \epsilon}{L}\right)^2 
\frac{L^4 \sin^2\frac{\pi(X_1-X_2)}{L}}{128\pi^4  \left[t_1\sin{\frac{\pi x}{L}}\sin{\frac{\pi X_1}{L}}+\frac{L}{2\pi}\sin{\frac{\pi \sigma(X_1-x)}{L}}\right]^2\left[t_1\sin{\frac{\pi x}{L}}\sin{\frac{\pi X_2}{L}}+\frac{L}{2\pi}\sin{\frac{\pi \sigma(X_2-x)}{L}}\right]^2}
\Bigg]\Bigg\},&i=1\\
%%%%%%%%%%%%%%%%%%%%%%%%%%%%%%%%%
\frac{1}{2}\Bigg\{1-\text{sgn}\left[\frac{t_1\sin{\frac{\pi x}{L}}\sin{\frac{\pi X_2}{L}}+\frac{L}{2\pi}\sin{\frac{\pi\sigma(X_2-x)}{L}}}{t_1\sin{\frac{\pi x}{L}}\sin{\frac{\pi X_1}{L}}+\frac{L}{2\pi}\sin{\frac{\pi\sigma(X_1-x)}{L}}} \right]
\Bigg[1\\
-\left(\frac{2\pi \epsilon}{L}\right)^2 
\frac{\sin^2\frac{\pi(X_1-X_2)}{L}\left[L^2+4\pi^2 t_1^2 \sin^2\frac{\pi x}{L}+2\pi t_1 L\sin{\frac{2\pi x\sigma}{L}}\right]^2}{128\pi^4  \left[t_1\sin{\frac{\pi x}{L}}\sin{\frac{\pi X_1}{L}}+\frac{L}{2\pi}\sin{\frac{\pi \sigma(X_1-x)}{L}}\right]^2\left[t_1\sin{\frac{\pi x}{L}}\sin{\frac{\pi X_2}{L}}+\frac{L}{2\pi}\sin{\frac{\pi \sigma(X_2-x)}{L}}\right]^2}
\Bigg]\Bigg\},&i=2\\
%%%%%%%%%%%%%%%%%%%%%%%%%%%%%%%
\frac{1}{2}\Bigg\{1-\text{sgn}\left[\frac{\sin{\frac{\pi\left[t_0+\sigma(X_2-x)\right]}{L}}}{\sin{\frac{\pi\left[t_0+\sigma(X_1-x)\right]}{L}}}\right]\Bigg[1-\left(\frac{2\pi\epsilon}{L}\right)^2\frac{\sin^4 \frac{\pi x}{L} \sin^2 \frac{\pi(X_1-X_2)}{L}}{2\sin^2 \frac{\pi \left[t_0+\sigma(X_1-x)\right]}{L}\sin^2 \frac{\pi \left[t_0+\sigma(X_2-x)\right]}{L}}\Bigg]\Bigg\},&i=3\\
%%%%%%%%%%%%%%%%%%%%%%%%%%%%%%
\frac{1}{2}\Bigg\{1-\text{sgn}\left[\frac{\sin{\frac{\pi\left[t_0+\sigma(X_2-x)\right]}{L}}}{\sin{\frac{\pi\left[t_0+\sigma(X_1-x)\right]}{L}}}\right]\Bigg[1-\left(\frac{2\pi\epsilon}{L}\right)^2\frac{\sin^4 \frac{\pi (t_0 -\sigma x)}{L} \sin^2 \frac{\pi(X_1-X_2)}{L}}{2\sin^2 \frac{\pi \left[t_0+\sigma(X_1-x)\right]}{L}\sin^2 \frac{\pi \left[t_0+\sigma(X_2-x)\right]}{L}}\Bigg]\Bigg\},&i=4
\end{cases}\label{SSDCrossRatio}
\end{align}

\subsubsection{Operator at $X_1^f$}
Setting $x=0$ in \eqref{SSDCrossRatio} gives
\begin{align}\label{SSDCrossRatiox0}
    \hat{\eta}_2(x=0)=\begin{cases}
    \mathcal{O}\left(\frac{\epsilon^2}{L^2}\right),i=1,2, \forall t_1>0 \\
%%%%%%%%%%%%%%%%%%%%%%%%%%%%%%%% 
\frac{1}{2}\Bigg\{1-\text{sgn}\left[\frac{\sin{\frac{\pi(t_0+\sigma X_2)}{L}}}{\sin{\frac{\pi\left(t_0+\sigma X_1\right)}{L}}}\right]\Bigg[1-\left(\frac{\epsilon}{L}\right)^2\sin^4 \frac{\pi t_0 }{L}\times\text{positive number}\Bigg]\Bigg\},i=4
    \end{cases}
\end{align}
for $0<X_1,X_2<L$. The second order term in the cross ratio for the $i=4$ case in \eqref{SSDCrossRatiox0} vanishes when $\frac{t_0}{L}\in \mathbb{Z}$. As explained previously, setting $x=0$ for the $i=3$ case as well as $t_0 = nL$ for $n\in \mathbb{Z}$ for $i=4$ leads to correlators that are ill-defined because $w_\epsilon^{\text{New,i}}=w_{-\epsilon}^{\text{New,i}}$. That is, the operators become coincident and the cross ratios $\eta_2=\overline{\eta}_2=0$ vanish identically to all orders in $\epsilon$. The following results hold away from these special cases. 

Contrast the $i=4$ cross ratio with the corresponding M\"{o}bius cross ratio \eqref{MobiusCrossRatioOperatorAtOrigin} where the regulator term does not vanish for finite $\theta$ for all time $t_0\in\mathbb{R}$. The time-dependence of the cross ratio $\hat{\eta}_2$ is the same to $\mathcal{O}\left(\frac{\epsilon}{L}\right)$ as in the M\"{o}bius case so $\Delta S_{A,4}^{(2)}$ is the same for the M\"{o}bius case as it is for the SSD case with the exception that the correlation functions and hence the R\'{e}nyi entropy remains well-defined even when $\frac{t_0}{L}\in\mathbb{Z}$ for the M\"{o}bius evolution. Note also that for $i=4$, the cross ratio is periodic in time $t_0$ with period $L$.

\subsubsection*{Case (a)}
The change in second R\'{e}nyi entropy is 
\begin{equation}\label{UnitarySSDquenchOperatorAtOriginCaseA}
    \Delta S_{A,i}^{(2)}(t_i)=\begin{cases}
        0,& i=1,2 \\
        0,&i=4, mL<t_0<mL+L-X_1 \\
        \log 2,&i=4,  mL+L-X_1<t_0<mL+X_2\\
        0,&i=4, mL+X_2<t_0< mL+L-X_2 \\
        \log 2,&i=4,  mL+L-X_2<t_0< mL+X_1 \\
        0,&i=4, mL+X_1 <t_0<(m+1)L 
    \end{cases}
\end{equation}
for non-negative integers $m$. Taking the SSD limit of \eqref{MobiusOperatorAtOriginCaseA} for $i=1,2$ reproduces the corresponding result in \eqref{UnitarySSDquenchOperatorAtOriginCaseA}. The $i=4$ case is explained by the $\theta=0$ uniform quasiparticle picture. Keep in mind that for $i=4$, the local operator correlation function and hence the second R\'{e}nyi entropy is ill-defined at $\frac{t_0}{L}\in \mathbb{Z}$.

\subsubsection*{Case (b)}
The change in second R\'{e}nyi entropy is 
\begin{equation}\label{UnitarySSDquenchOperatorAtOriginCaseB}
    \Delta S_{A,i}^{(2)}(t_i)=\begin{cases}
        0,& i=1,2 \\
        %%%%%%%%%%%%%%%%
        0,&i=4, mL<t_0<mL+X_2 \\
        \log 2,&i=4,  mL+X_2<t_0<mL+L-X_1\\
        0,&i=4, mL+L-X_1<t_0< mL+X_1 \\
        \log 2,&i=4,  mL+X_1<t_0< mL+L-X_2 \\
        0,&i=4, mL+L-X_2 <t_0<(m+1)L 
    \end{cases}
\end{equation}
for non-negative integers $m$. Taking the $\theta\rightarrow\infty$ SSD limit in \eqref{MobiusOperatorAtOriginCaseb} reproduces the first line of \eqref{UnitarySSDquenchOperatorAtOriginCaseB}.
The $i=4$ case is explained by the uniform $\theta=0$ quasiparticle picture.

\subsubsection*{Case (c)}
The change in second R\'{e}nyi entropy is 
\begin{equation}\label{UnitarySSDquenchOperatorAtOriginCaseC}
    \Delta S_{A,i}^{(2)}(t_i)=\begin{cases}
        0,& i=1,2 \\
        %%%%%%%%%%%%%%%%%%%%%%%%
        %%%%%%%%%%%%%%%%
        0,&i=4, mL<t_0<mL+X_2 \\
        \log 2,&i=4,  mL+X_2<t_0<mL+X_1\\
        0,&i=4, mL+X_1<t_0< mL+L-X_1\\
        \log 2,&i=4,  mL+L-X_1<t_0< mL+L-X_2 \\
        0,&i=4, mL+L-X_2 <t_0<(m+1)L 
    \end{cases}
\end{equation}
for non-negative integers $m$. Taking the $\theta\rightarrow\infty$ SSD limit in \eqref{MobiusOperatorAtOriginCaseC} reproduces the first line of \eqref{UnitarySSDquenchOperatorAtOriginCaseC}. As before, the $i=4$ case is explained by the uniform $\theta=0$ quasiparticle picture.

\subsubsection{Operator at $X_2^f$}
Setting $x=\frac{L}{2}$ in \eqref{SSDCrossRatio} gives
\begin{align}
    &\hat{\eta}_2\left(x=\frac{L}{2}\right)\nonumber \\
    =&\begin{cases}
        \frac{1}{2}\left\{1-\text{sgn}\left[\frac{t_1-\frac{\sigma L}{2\pi}\cot{\frac{\pi X_2}{L}}}{t_1-\frac{\sigma L}{2\pi}\cot{\frac{\pi X_1}{L}}}\right]\left[1-\left(\frac{\epsilon}{L}\right)^2\times \text{positive number}\right]\right\},&i=1,2 \\
        \frac{1}{2}\left\{1-\text{sgn}\left[\frac{\cos{\frac{\pi(t_0+\sigma X_2)}{L}}}{\cos{\frac{\pi(t_0+\sigma X_1)}{L}}}\right]\left[1-\left(\frac{\epsilon}{L}\right)^2\times \text{positive number}\right]\right\},&i=3 \\
        \frac{1}{2}\left\{1-\text{sgn}\left[\frac{\cos{\frac{\pi(t_0+\sigma X_2)}{L}}}{\cos{\frac{\pi(t_0+\sigma X_1)}{L}}}\right]\left[1-\left(\frac{\epsilon}{L}\right)^2\cos^4\frac{\pi t_0}{L}\times \text{positive number}\right]\right\},&i=4 
    \end{cases}
\end{align}
where the cross ratio is periodic in time with $t_0\sim t_0+L$ for $i=3$. The cross ratio for $i=3$ is the same for both M\"{o}bius and SSD quenches so for $i=3$, $\Delta S_{A,i}^{(2)}$ is the same for both M\"{o}bius and SSD when the local operator is inserted at $x=\frac{L}{2}$.

As before, when $t_0 = \frac{2k+1}{2}L$ for $k\in \mathbb{Z}$, the regulator term vanishes for the $i=4$ case and the correlator and hence the second R\'{e}nyi entropy is ill-defined since $w_\epsilon^{\text{New,4}}=w_{-\epsilon}^{\text{New,4}}$ which means that the operators become coincident and the cross ratios $\eta_2=\overline{\eta}_2=0$ vanish identically to all orders in $\epsilon$. The second R\'{e}nyi entropy is ill-defined at these times for $i=4$ for holographic CFTs as well. Contrast this with the M\"{o}bius case where the regulator term does not vanish for finite $\theta$ and the cross ratios are identical to those of the SSD quench for $\Delta S_{A,4}^{(2)}(t_0)$ with $x=\frac{L}{2}$ to order $\mathcal{O}\left(\frac{\epsilon}{L}\right)$ so the R\'{e}nyi entropy is the same for both M\"{o}bius and SSD for $i=4$ with the exception that the correlators and hence the R\'{e}nyi entropy are well defined even at $t_0 = \frac{2k+1}{2}L$ for integers $k$ in the M\"{o}bius quench.

\subsubsection*{Case (a)}
The change in second R\'{e}nyi entropy is
\begin{equation}\label{UnitarySSDquenchCaseA}
    \Delta S_{A,i}^{(2)}(t_i) = \begin{cases}
        \log 2, &i=1,2, \frac{L}{2\pi}\cot{\frac{\pi X_2}{L}}<t_1< \frac{L}{2\pi}|\cot{\frac{\pi X_1}{L}}|\\
        0&i=1,2,0<t_1< \frac{L}{2\pi}\cot{\frac{\pi X_2}{L}}\quad \text{and}\quad \frac{L}{2\pi}|\cot{\frac{\pi X_1}{L}}|<t_1\\
%%%%%%%%%%%%%%%%%%%%%%%%%%%%%%%
        0,&i=3,4, mL<t_0<mL+\frac{L}{2}-X_2\\
        \log 2,&i=3,4, mL+\frac{L}{2}-X_2<t_0<mL+X_1-\frac{L}{2}\\
        0,&i=3,4, mL+X_1-\frac{L}{2} <t_0< mL +\frac{3L}{2}-X_1\\
        \log 2,&i=3,4,  mL+\frac{3L}{2}-X_1 <t_0<mL+\frac{L}{2}+X_2 \\
        0,&i=3,4,  mL+\frac{L}{2}+X_2 <t_0<(m+1)L
    \end{cases}
\end{equation}
for non-negative integers $m$. For $i=1,2$, both quasiparticles are initially outside the subsystem so $\Delta S_{A,1}^{(2)}(t_1)=0$. The left-moving quasiparticle enters the subsystem at $\frac{L}{2\pi}\cot{\frac{\pi X_2}{L}}$ and $\Delta S_{A,1}^{(2)}(t_1)$ jumps to $\log 2$ before the right-moving quasiparticle also enters the subsystem at $\frac{L}{2\pi}|\cot{\frac{\pi X_1}{L}}|$ and $\Delta S_{A,1}^{(2)}(t_1)$ drops back down to 0. Taking the $\theta\rightarrow\infty$ SSD limit of \eqref{MobiusOperatorAtMidpointCaseA}, where only the $m=0$ period matters, gives the $i=1,2$ result in \eqref{UnitarySSDquenchCaseA}.

The $i=3,4$ case is explained by the uniform $\theta=0$ quasiparticle picture.

\subsubsection*{Case (b)}
The change in second R\'{e}nyi entropy is 
\begin{equation}\label{UnitarySSDquenchCaseB}
    \Delta S_{A,i}^{(2)}(t_i) = \begin{cases}
        \log 2, &i=1,2, \frac{L}{2\pi}|\cot{\frac{\pi X_1}{L}}|< t_1<\frac{L}{2\pi}\cot{\frac{\pi X_2}{L}} \\
        0&i=1,2,0<t_1<\frac{L}{2\pi}|\cot{\frac{\pi X_1}{L}}| \quad\text{and}\quad \frac{L}{2\pi}\cot{\frac{\pi X_2}{L}}<t_1\\
        %%%%%%%%%%%%%%%%%%%%%%%%%%%%%%%
        0,&i=3,4, mL<t_0<mL+X_1-\frac{L}{2}
        \\
        \log 2,&i=3,4, mL+X_1-\frac{L}{2}<t_0<mL+\frac{L}{2}-X_2\\
        0,&i=3,4, mL+\frac{L}{2}-X_2 <t_0<mL+\frac{L}{2}+X_2 \\
        \log 2,&i=3,4,  mL+\frac{L}{2}+X_2 <t_0<mL+\frac{3L}{2}-X_1 \\
        0,&i=3,4,  mL+\frac{3L}{2}-X_1 <t_0<(m+1)L
    \end{cases}
\end{equation}
for non-negative integers $m$. For $i=1,2$, both quasiparticles begin in subsystem $A$ so $\Delta S_{A,1}^{(2)}(t_1)=0$. At time $\frac{L}{2\pi}|\cot{\frac{\pi X_1}{L}}|$, the right-moving quasiparticle leaves subsystem $A$ so $\Delta S_{A,1}^{(2)}(t_1)$ jumps to $\log 2$. Subsequently, the left-moving quasiparticle leaves the subsystem $A$ at $\frac{L}{2\pi}\cot{\frac{\pi X_2}{L}}$ and $\Delta S_{A,1}^{(2)}(t_1)$ drops back down to 0. Taking the $\theta\rightarrow\infty$ SSD limit in \eqref{MobiusOperatorAtMidpointCaseB} for $i=1,2$ reproduces the corresponding result in \eqref{UnitarySSDquenchCaseB}.

The uniform $\theta=0$ quasiparticle picture fully explains the $i=3,4$ case.

\subsubsection*{Case (c)}
The change in second R\'{e}nyi entropy entropy is 
\begin{equation}
\label{UnitarySSDquenchCaseC}
    \Delta S_{A,i}^{(2)}(t_i) = \begin{cases}
        \log 2, &i=1,2, \frac{L}{2\pi}\cot{\frac{\pi X_1}{L}}< t_1<\frac{L}{2\pi}\cot{\frac{\pi X_2}{L}} \\
        0&i=1,2,0<t_1<\frac{L}{2\pi}\cot{\frac{\pi X_1}{L}} \quad \text{and}\quad \frac{L}{2\pi}\cot{\frac{\pi X_2}{L}}<t_1\\
        %%%%%%%%%%%%%%%%%%%%%%%%%%%%%%%
        0,&i=3,4, mL<t_0<mL+\frac{L}{2}-X_1
        \\
        \log 2,&i=3,4, mL+\frac{L}{2}-X_1<t_0<mL+\frac{L}{2}-X_2\\
        0,&i=3,4, mL+\frac{L}{2}-X_2 <t_0<mL+\frac{L}{2}+X_2 \\
        \log 2,&i=3,4,  mL+\frac{L}{2}+X_2 <t_0<mL+\frac{L}{2}+X_1 \\
        0,&i=3,4,  mL+\frac{L}{2}+X_1 <t_0<(m+1)L
    \end{cases}
\end{equation}
for non-negative integers $m$. For $i=1,2$, both quasiparticles are initially outside subsystem $A$. The left-moving quasiparticle enters the subsystem at $\frac{L}{2\pi}\cot{\frac{\pi X_1}{L}}$, causing $\Delta S_{A,1}^{(2)}(t_1)$ to jump to $\log 2$ before exiting the subsystem at $\frac{L}{2\pi}\cot{\frac{\pi X_2}{L}}$ at which time $\Delta S_{A,1}^{(2)}(t_1)$ drops back down to 0. Taking the $\theta\rightarrow\infty$ SSD limit of \eqref{MobiusOperatorAtMidpointCaseC} for $i=1,2$ reproduces the corresponding result in \eqref{UnitarySSDquenchCaseC}.

The $i=3,4$ case is explained by the uniform $\theta=0$ quasiparticle picture.

\subsubsection{Information survival in integrable theories}
As we have seen in the previous subsections, when the time evolution operator is the SSD Hamiltonian, the late time second R\'{e}nyi entropy for the local operator quenched state returns to the vacuum value as the subsystems are situated away from the fixed point $X_1^f$ where the quasiparticles accumulate. To have a non-vacuum second R\'{e}nyi entropy at late times, place the local operator at $X_2^f$ and consider a subsystem $A = [X_2,X_1]$ where $X_2=0$ and $0<X_1<\frac{L}{2}$.

The coordinates are
\begin{align}\label{UniformizationCoordinatesEndpointOrigin}
  \hat{\zeta}_{\rho\epsilon}^n =
  \begin{cases}
  \frac{1}{\cos{\frac{\pi X_1}{L}}-\frac{2\pi t_1}{L}\sin{\frac{\pi \sigma X_1}{L}}}  \Bigg\{1+\left(\frac{2\pi \epsilon}{L}\right)\frac{i\rho \sin{\frac{\pi \sigma X_1}{L}}}{2\left(\cos{\frac{\pi X_1}{L}}-\frac{2\pi t_1}{L}\sin{\frac{\pi\sigma X_1}{L}}\right)}\\-\left(\frac{2\pi \epsilon}{L}\right)^2 \frac{\sin^2\frac{\pi X_1}{L}}{4\left(\cos{\frac{\pi X_1}{L}}-\frac{2\pi t_1}{L}\sin{\frac{\pi\sigma X_1}{L}}\right)^2}\Bigg\} ,&i=1    \\
  \frac{1}{\cos{\frac{\pi X_1}{L}}-\frac{2\pi t_1}{L}\sin{\frac{\pi \sigma X_1}{L}}}  \Bigg\{1+\left(\frac{2\pi \epsilon}{L}\right)\frac{i\rho \sin{\frac{\pi \sigma X_1}{L}}\left(L^2+4\pi^2 t_1^2\right)}{2L^2 \left(\cos{\frac{\pi X_1}{L}}-\frac{2\pi t_1}{L}\sin{\frac{\pi\sigma X_1}{L}}\right)}\\-\left(\frac{2\pi \epsilon}{L}\right)^2 \frac{\sin\frac{\pi\sigma X_1}{L}\left(\frac{2\pi t_1}{L}\cos{\frac{\pi X_1}{L}}+\sin{\frac{\pi \sigma X_1}{L}}\right)\left(L^2+4\pi^2 t_1^2\right)}{4L^2\left(\cos{\frac{\pi X_1}{L}}-\frac{2\pi t_1}{L}\sin{\frac{\pi\sigma X_1}{L}}\right)^2}\Bigg\} ,&i=2  \\ %%%%%%%%%%%%%%%%%%%%%%%%%%
  \frac{\cos{\frac{\pi t_0}{L}}}{\cos{\frac{\pi (t_0+\sigma X_1)}{L}}} \Bigg\{ 1-\left(\frac{2\pi \epsilon}{L}\right)\frac{i\rho \sin{\frac{\pi \sigma X_1}{L}}}{\cos{\frac{\pi (t_0+\sigma X_1)}{L}}\cos{\frac{\pi t_0}{L}}}\\
  -\left(\frac{2\pi\epsilon}{L}\right)^2 \frac{\sin{\frac{\pi\sigma X_1}{L}}\sin{\frac{\pi(t_0+\sigma X_1)}{L}}}{\cos^2\frac{\pi(t_0+\sigma X_1)}{L}\cos{\frac{\pi t_0}{L}}}\Bigg\},&i=3\\
  %%%%%%%%%%%%%%%%%%%%%%%%%%%%%%%%%
  \frac{\cos{\frac{\pi t_0}{L}}}{\cos{\frac{\pi (t_0+\sigma X_1)}{L}}} \Bigg\{ 1-\left(\frac{2\pi \epsilon}{L}\right)\frac{i\rho \sin{\frac{\pi \sigma X_1}{L}}\cos{\frac{\pi t_0}{L}}}{\cos{\frac{\pi (t_0+\sigma X_1)}{L}}}\\
  -\left(\frac{2\pi\epsilon}{L}\right)^2 \frac{\sin^2\frac{\pi X_1}{L}\cos^2\frac{\pi t_0}{L}}{\cos^2\frac{\pi(t_0+\sigma X_1)}{L}}\Bigg\},&i=4
  \end{cases}
\end{align}
These coordinates \eqref{UniformizationCoordinatesEndpointOrigin} can be obtained either by setting $X_2=0$ in \eqref{UniformizationMap} and then performing a Taylor expansion in $\frac{2\pi\epsilon}{L}$ or by sending $X_2\rightarrow0$ in the SSD coordinates \eqref{SSDCoordinates}. Since $0<X_1<\frac{L}{2}$, when $t_1\neq \frac{L \sigma}{2\pi}\cot{\frac{\pi X_1}{L}}$, $\xi_0$, $\xi_1$ and $\xi_2$ as defined in \eqref{UniformizationCoordinateExpansion} are finite and non-zero for $i=1,2$ while for $i=3,4$, when $\cos{\frac{\pi t_0}{L}},\cos{\frac{\pi (t_0+\sigma X_1)}{L}}\neq 0$, $\xi_0$ and $\xi_1$ are non-zero and finite while $\xi_2$ does not diverge. $t_1=\frac{L \sigma}{2\pi}\cot{\frac{\pi X_1}{L}}$ for $i=1,2$ and $t_0=(m+\frac{1}{2})L-\sigma X_1$ with integers $m$ for $i=3,4$ correspond to the discontinuities in the second R\'{e}nyi entropy anyway so away from these points the second R\'{e}nyi entropy is well defined. When $t_0=(m+\frac{1}{2})L$ for integers $m$, the change in the second R\'{e}nyi entropy becomes ill-defined for $i=3,4$. Applying \eqref{CrossRatioSecondOrder}, the cross ratios are
\begin{align}
    \hat{\eta}_2 = \begin{cases}
        \frac{1}{2}\Bigg\{1-\text{sgn}\left[\cos{\frac{\pi X_1}{L}}-\frac{2\pi t_1}{L}\sin{\frac{\pi \sigma X_1}{L}}\right]\left[1-\left(\frac{2\pi\epsilon}{L}\right)^2\frac{\sin^2\frac{\pi X_1}{L}}{8\left(\cos{\frac{\pi X_1}{L}}-\frac{2\pi t_1}{L}\sin{\frac{\pi \sigma X_1}{L}}\right)^2}\right]
        \Bigg\},&i=1\\
        \frac{1}{2}\Bigg\{1-\text{sgn}\left[\cos{\frac{\pi X_1}{L}}-\frac{2\pi t_1}{L}\sin{\frac{\pi \sigma X_1}{L}}\right]\left[1-\left(\frac{2\pi\epsilon}{L}\right)^2\frac{\sin^2\frac{\pi X_1}{L}\left(L^2+4\pi^2 t_1^2\right)^2}{8L^4\left(\cos{\frac{\pi X_1}{L}}-\frac{2\pi t_1}{L}\sin{\frac{\pi \sigma X_1}{L}}\right)^2}\right]
        \Bigg\},&i=2\\
        %%%%%%%%%%%%%%%%%%%%%%%%%%
        \frac{1}{2}\Bigg\{1-\text{sgn}\left[\frac{\cos{\frac{\pi t_0}{L}}}{\cos{\frac{\pi (t_0+\sigma X_1)}{L}}}\right]\left[1-\left(\frac{2\pi\epsilon}{L}\right)^2\frac{\sin^2\frac{\pi X_1}{L}}{2\cos^2\frac{\pi (t_0+\sigma X_1)}{L}\cos^2\frac{\pi t_0}{L}}\right]
        \Bigg\},&i=3\\
        %%%%%%%%%%%%%%%%%%%%%%%%%%
        \frac{1}{2}\Bigg\{1-\text{sgn}\left[\frac{\cos{\frac{\pi t_0}{L}}}{\cos{\frac{\pi (t_0+\sigma X_1)}{L}}}\right]\left[1-\left(\frac{2\pi\epsilon}{L}\right)^2\frac{\sin^2\frac{\pi X_1}{L}\cos^2\frac{\pi t_0}{L}}{2\cos^2\frac{\pi (t_0+\sigma X_1)}{L}}\right]
        \Bigg\},&i=4
    \end{cases}
\end{align}
which can also be directly obtained by setting $X_2=0$, $x=\frac{L}{2}$ in \eqref{SSDCrossRatio} for all cases $i=1,2,3,4$. Since the cross ratios are periodic with $t_0 \sim t_0 +L$ for $i=3,4$, the corresponding change in second R\'{e}nyi entropy will have the same periodicity as well. The change in the second R\'{e}nyi entropy for the local operator \eqref{SumVertexOperator} in the free boson theory is \begin{equation}
    \Delta S_{A,i}^{(2)}=\begin{cases}
        0,& i=1,2,0<t_1<\frac{L}{2\pi} \cot{\frac{\pi X_1}{L}}\\
        \log 2,&i=1,2, \frac{L}{2\pi} \cot{\frac{\pi X_1}{L}}<t_1 \\
        0,&i=3,4, mL<t_0<mL+\frac{L}{2}-X_1 \\
        \log 2,&i=3,4, mL+\frac{L}{2}-X_1<t_0< mL+\frac{L}{2}+X_1 \\
        0,&i=3,4, mL+\frac{L}{2}+X_1<t_0 <(m+1)L
    \end{cases}
\end{equation}
for non-negative integers $m$. For $i=1,2$, the time $t_1=\frac{L}{2\pi} \cot{\frac{\pi X_1}{L}}$ is the time it takes for a left moving quasiparticle starting at $X_2^f$ to arrive at the rightmost boundary of the subsystem $X_1$. On the other hand, the $i=3,4$ cases are well-defined by the $\theta=0$ quasiparticle picture. For $i=4$, the R\'{e}nyi is ill-defined for $t_0=(k+\frac{1}{2})L$ for integers $k$. We get the same answer by sending $X_2\rightarrow 0^+$ in \eqref{UnitarySSDquenchCaseC} for all cases $i=1,2,3,4$.

%%%%%%%%%%%%%%%%%%%%%%%%%%%%%%%%%%%%%%%%%%%%%%%%%%%%%%%%%%%%%%%%%%%%%%%
\section{The details of the cross ratios \label{App:crossratiosX1}}
%%%%%%%%%%%%%%%%%%%%%%%%%%%%%%%%%%%%%%%%%%%%%%%%%%%%%%%%%%%%%%%%%%%%%%%
Let us present the details of the cross ratio after we perform the analytic continuation to real time.
%%%%%%%%%%%%%%%%%%%%%%%%%%%%%%%%%%%%%%%
\subsection{When the local operator is inserted at $x=X^f_1=0$.\label{App:crossratiosX1}}
%%%%%%%%%%%%%%%%%%%%%%%%%%%%%%%%%%%%%%%
By the second order in the small $\epsilon$ expansion, the details of the cross ratios are given by 
\be
\begin{split}
&z_{c,1} \approx 1-\frac{i2\pi \epsilon  \sin{\left[\f{\pi (X_1-X_2)}{L}\right]}}{L e^{-2 \theta }\prod_{i=1,2}\left(\cosh{\theta } \sin{\left[\f{\pi (t_1+X_i\cosh{2\theta})}{L\cosh{2\theta}}\right]}-\sinh{\theta}\sin{\left[\f{\pi (t_1-X_i\cosh{2\theta})}{L\cosh{2\theta}}\right]}\right) } +\mathcal{O}(\epsilon^2),\\
&\overline{z}_{c,1} \approx 1+\frac{i2\pi \epsilon  \sin{\left[\f{\pi (X_1-X_2)}{L}\right]}}{L e^{-2 \theta } \prod_{i=1,2}\left(\cosh{\theta } \sin{\left[\f{\pi (t_1-X_i\cosh{2\theta})}{L\cosh{2\theta}}\right]}-\sinh{\theta}\sin{\left[\f{\pi (t_1+X_i\cosh{2\theta})}{L\cosh{2\theta}}\right]}\right) } +\mathcal{O}(\epsilon^2),\\
&z_{c,2} \approx 1-\frac{i2\pi \epsilon  \sin{\left[\f{\pi (X_1-X_2)}{L}\right]}\left(e^{2\theta}\cos^2{\left(\f{\pi t_1}{L\cosh{2\theta}}\right)}+e^{-2\theta}\sin^2{\left(\f{\pi t_1}{L\cosh{2\theta}}\right)}\right)}{L \prod_{i=1,2}\left(\cosh{\theta } \sin{\left[\f{\pi (t_1+X_i\cosh{2\theta})}{L\cosh{2\theta}}\right]}-\sinh{\theta}\sin{\left[\f{\pi (t_1-X_i\cosh{2\theta})}{L\cosh{2\theta}}\right]}\right) } +\mathcal{O}(\epsilon^2),\\
&\overline{z}_{c,2} \approx 1+\frac{i2\pi \epsilon  \sin{\left[\f{\pi (X_1-X_2)}{L}\right]}\left(e^{2\theta}\cos^2{\left(\f{\pi t_1}{L\cosh{2\theta}}\right)}+e^{-2\theta}\sin^2{\left(\f{\pi t_1}{L\cosh{2\theta}}\right)}\right)}{L \prod_{i=1,2}\left(-\cosh{\theta } \sin{\left[\f{\pi (t_1-X_i\cosh{2\theta})}{L\cosh{2\theta}}\right]}+\sinh{\theta}\sin{\left[\f{\pi (t_1+X_i\cosh{2\theta})}{L\cosh{2\theta}}\right]}\right) } +\mathcal{O}(\epsilon^2),\\
&z_{c,3} \approx 1+\frac{2i \epsilon \pi  (\tanh (2 \theta )-1) \sin{\left[\f{\pi(X_1-X_2)}{L}\right]}}{L \prod_{i=1,2}\sin{\left[\f{\pi(t_0+X_i)}{L}\right]}}+\mathcal{O}(\epsilon^2),\\
&\overline{z}_{c,3} \approx 1-\frac{2i \epsilon \pi  (\tanh (2 \theta )-1) \sin{\left[\f{\pi(X_1-X_2)}{L}\right]}}{L \prod_{i=1,2}\sin{\left[\f{\pi(t_0-X_i)}{L}\right]}}+\mathcal{O}(\epsilon^2),\\
&z_{c,4}\approx 1+\frac{2i \pi  \epsilon  \sin{\left[\f{\pi(X_1-X_2)}{L}\right]}\left(\cos{\left(\f{2\pi t_0}{L}\right)}\tanh{2\theta}-1\right)}{L \prod_{i=1,2}\sin{\left[\f{\pi(t_0+X_i)}{L}\right]}}+\mathcal{O}(\epsilon^2),\\
&\overline{z}_{c,4} \approx 1-\frac{2i \pi  \epsilon  \sin{\left[\f{\pi(X_1-X_2)}{L}\right]}\left(\cos{\left(\f{2\pi t_0}{L}\right)}\tanh{2\theta}-1\right)}{L \prod_{i=1,2}\sin{\left[\f{\pi(t_0-X_i)}{L}\right]}}+\mathcal{O}(\epsilon^2),
\end{split}
\ee
where around $t_1=\f{L \cosh{2\theta}}{\pi}\tan^{-1}{\left[\pm e^{2\theta}\tan{\left(\f{\pi X_{i}}{L}\right)}\right]}+n L \cosh{2\theta}$, the small $\epsilon$ expansion of $z_{c,i=1,2}$ or $\overline{z}_{c,i=1,2}$ breaks down, while around $t_0=nL \pm X_{i=1,2}$, that of $z_{c,i=3,4}$ or $\overline{z}_{c,i=3,4}$ breaks down. Here, we assume that $0<\tan^{-1}{\left[\pm e^{2\theta}\tan{\left(\f{\pi X_{i}}{L}\right)}\right]}<\pi$ and $n$ is an integer greater than or equal to zero.
Let us consider the time dependence of $S_{A,i}$ in cases, (a), (b), and (c).

%%%%%%%%%%%%%%%%%%%%%%%%%%%%%%%%%%%%%%%
\subsection{When the local operator is inserted at $x=\f{L}{2}$.\label{App:crossratiosX2}}
%%%%%%%%%%%%%%%%%%%%%%%%%%%%%%%%%%%%%%%

When the local operator is inserted at $x=\f{L}{2}$, the small $\epsilon$ expansion of the cross ratios analytically-continued to real time is approximated by 
\be
\begin{split}
    &z_{c,1} \approx 1-\f{2i\pi e^{-2\theta}\epsilon \sin{\left[\f{\pi (X_1-X_2)}{L}\right]}}{L\prod_{i=1,2}\left[\cos{\left(\f{\pi\left(X_i \cosh{2\theta}+t_1\right)}{L\cosh{2\theta}}\right)}\cosh{\theta}-\cos{\left(\f{\pi\left(X_i \cosh{2\theta}-t_1\right)}{L\cosh{2\theta}}\right)}\sinh{\theta}\right]}+\mathcal{O}(\epsilon^2),\\
    &\overline{z}_{c,1} \approx 1+\f{2i\pi e^{-2\theta}\epsilon \sin{\left[\f{\pi (X_1-X_2)}{L}\right]}}{L\prod_{i=1,2}\left[\cos{\left(\f{\pi\left(X_i \cosh{2\theta}-t_1\right)}{L\cosh{2\theta}}\right)}\cosh{\theta}-\cos{\left(\f{\pi\left(X_i \cosh{2\theta}+t_1\right)}{L\cosh{2\theta}}\right)}\sinh{\theta}\right]}+\mathcal{O}(\epsilon^2),\\
     &z_{c,2} \approx 1-\f{2i\pi \epsilon \sin{\left[\f{\pi (X_1-X_2)}{L}\right]}\left(\cosh{2\theta}-\sinh{2\theta}\cos{\left(\f{2\pi t_1}{L\cosh{2\theta}}\right)}\right)}{L\prod_{i=1,2}\left[\cos{\left(\f{\pi\left(X_i \cosh{2\theta}+t_1\right)}{L\cosh{2\theta}}\right)}\cosh{\theta}-\cos{\left(\f{\pi\left(X_i \cosh{2\theta}-t_1\right)}{L\cosh{2\theta}}\right)}\sinh{\theta}\right]}+\mathcal{O}(\epsilon^2),\\
    &\overline{z}_{c,2} \approx 1+\f{2i\pi \epsilon \sin{\left[\f{\pi (X_1-X_2)}{L}\right]}\left(\cosh{2\theta}-\sinh{2\theta}\cos{\left(\f{2\pi t_1}{L\cosh{2\theta}}\right)}\right)}{L\prod_{i=1,2}\left[\cos{\left(\f{\pi\left(X_i \cosh{2\theta}-t_1\right)}{L\cosh{2\theta}}\right)}\cosh{\theta}-\cos{\left(\f{\pi\left(X_i \cosh{2\theta}+t_1\right)}{L\cosh{2\theta}}\right)}\sinh{\theta}\right]}+\mathcal{O}(\epsilon^2),\\
    &z_{c,3} \approx 1-i \f{2\pi \epsilon (1+\tanh{2\theta})\sin{\left[\f{\pi (X_1-X_2)}{L}\right]}}{L\prod_{i=1,2}\cos{\left[\f{\pi(t_0+X_i)}{L}\right]}}+\mathcal{O}(\epsilon^2),\\
    &\overline{z}_{c,3} \approx 1+i \f{2\pi \epsilon (1+\tanh{2\theta})\sin{\left[\f{\pi (X_1-X_2)}{L}\right]}}{L\prod_{i=1,2}\cos{\left[\f{\pi(t_0-X_i)}{L}\right]}}+\mathcal{O}(\epsilon^2),\\
    &z_{c,4} \approx 1-i \f{2\pi \epsilon \left(1+\cos{\left(\f{2\pi t_0}{L}\right)}\tanh{2\theta}\right)\sin{\left[\f{\pi (X_1-X_2)}{L}\right]}}{L\prod_{i=1,2}\cos{\left[\f{\pi(t_0+X_i)}{L}\right]}}+\mathcal{O}(\epsilon^2),\\
    &\overline{z}_{c,4} \approx 1+i \f{2\pi \epsilon \left(1+\cos{\left(\f{2\pi t_0}{L}\right)}\tanh{2\theta}\right)\sin{\left[\f{\pi (X_1-X_2)}{L}\right]}}{L\prod_{i=1,2}\cos{\left[\f{\pi(t_0-X_i)}{L}\right]}}+\mathcal{O}(\epsilon^2),\\
\end{split}
\ee
where around $t_1=\f{L_{\text{eff}}}{\pi}\tan^{-1}{\left[\pm e^{-2\theta}\cot{\left(\f{\pi X_i}{L}\right)}\right]}+n L_{\text{eff}}$, the small $\epsilon$ expansion of $z_{c,i=1,2}$, $\overline{z}_{c,i=1,2}$, breaks down, while around $t_0= \pm X_i+L\left(\f{1}{2}+n\right)$, the small $\epsilon$ expansion of $z_{c,i=3,4}$, $\overline{z}_{c,i=3,4}$ breaks down. Let us define the characteristic times, $0<\hat{t}<L_{\text{eff}}$, as 
\be
\begin{split}
 &\hat{t}_{\f{L}{2}-X_2}=\text{\footnotesize{$\f{L_{\text{eff}}}{\pi }\tan^{-1}{\left(e^{-2\theta}\tan{\left[\f{\pi}{L}\left(\f{L}{2}-X_2\right)\right]}\right)}$}}, \hat{t}_{\f{L}{2}+X_2}=\text{\footnotesize{$\f{L_{\text{eff}}}{\pi }\tan^{-1}{\left(e^{-2\theta}\tan{\left[\f{\pi}{L}\left(\f{L}{2}+X_2\right)\right]}\right)}$}},\\
&\hat{t}_{X_1-\f{L}{2}}=\text{\footnotesize{$\f{L_{\text{eff}}}{\pi }\tan^{-1}{\left(e^{-2\theta}\tan{\left[\f{\pi}{L}\left(X_1-\f{L}{2}\right)\right]}\right)}$}}, \hat{t}_{\f{3L}{2}-X_1}=\text{\footnotesize{$\f{L_{\text{eff}}}{\pi }\tan^{-1}{\left(e^{-2\theta}\tan{\left[\f{\pi}{L}\left(\f{3L}{2}-X_1\right)\right]}\right)}$}},\\
\end{split}
\ee
where $L_{\text{eff}}>\hat{t}_{\f{L}{2}+X_2}>\hat{t}_{\f{3L}{2}-X_1}>\hat{t}_{X_1-\f{L}{2}}>\hat{t}_{\f{L}{2}-X_2}>0$.

%%%%%%%%%%%%%%%%%%%%%%%%%%%%%%%%%%%%%%%
\subsection{When the local operator is inserted at general $x$.\label{App:crossratiosgeneralx}}
%%%%%%%%%%%%%%%%%%%%%%%%%%%%%%%%%%%%%%%

To second order in the small $\epsilon$ expansion, the cross ratios are given by 
\be\label{eq:general-x-mo-crossratio}
\begin{split}
&z_{c,1} \approx \mathcal{O}(\epsilon^2) + \\& \text{\footnotesize{$ 1-\frac{2 i \pi  \epsilon  \sin \left[\frac{\pi  \left(X_1-X_2\right)}{L}\right]}{L \prod_{i=1,2} \left[\sin \left(\frac{\pi  t_1}{L_{\text{eff}}}\right) \left(\cosh (2 \theta ) \cos \left(\frac{\pi  \left(x-X_i\right)}{L}\right)-\sinh (2 \theta ) \cos \left(\frac{\pi  \left(X_i+x\right)}{L}\right)\right)-\cos \left(\frac{\pi  t_1}{L_{\text{eff}}}\right) \sin \left(\frac{\pi  \left(x-X_i\right)}{L}\right)\right]} $}},\\
&\overline{z}_{c,1} \approx \mathcal{O}(\epsilon^2) + \\&\text{\footnotesize{$ 1+\frac{2 i \pi  \epsilon  \sin \left[\frac{\pi  \left(X_1-X_2\right)}{L}\right]}{L \prod_{i=1,2}  \left[\sin \left(\frac{\pi  t_1}{L_{\text{eff}}}\right) \left(\cosh (2 \theta ) \cos \left(\frac{\pi  \left(x-X_i\right)}{L}\right)-\sinh (2 \theta ) \cos \left(\frac{\pi  \left(X_i+x\right)}{L}\right)\right)+\cos \left(\frac{\pi  t_1}{L_{\text{eff}}}\right) \sin \left(\frac{\pi  \left(x-X_i\right)}{L}\right)\right]}$}},\\
&z_{c,2} \text{\footnotesize{$ \approx \mathcal{O}(\epsilon^2) + 1-  2 i \pi  \epsilon  \sin \left(\frac{\pi  \left(X_1-X_2\right)}{L}\right) \left(\cos \left(\frac{\pi  t_1}{L_{\text{eff}}}\right)-i \sin \left(\frac{\pi  t_1}{L_{\text{eff}}}\right) \left(\cosh (2 \theta )-\sinh (2 \theta ) e^{-\frac{2 i \pi  x}{L}}\right)\right)$}}\\& \times \text{\footnotesize{$\frac{\left.\cos \left(\frac{\pi  t_1}{L_{\text{eff}}}\right)-i \sin \left(\frac{\pi  t_1}{L_{\text{eff}}}\right) \left(-\cosh (2 \theta )+\sinh (2 \theta ) e^{\frac{2 i \pi  x}{L}}\right)\right)}{L \prod_{i=1,2} \left[\sin \left(\frac{\pi  t_1}{L_{\text{eff}}}\right) \left(\cosh (2 \theta ) \cos \left(\frac{\pi  \left(x-X_i\right)}{L}\right)-\sinh (2 \theta ) \cos \left(\frac{\pi  \left(X_i+x\right)}{L}\right)\right)-\cos \left(\frac{\pi  t_1}{L_{\text{eff}}}\right) \sin \left(\frac{\pi  \left(x-X_i\right)}{L}\right)\right]}$}},\\
&\overline{z}_{c,2} \approx \text{\footnotesize{$\mathcal{O}(\epsilon^2) + 1+ 2 i \pi  \epsilon  \sin \left(\frac{\pi  \left(X_1-X_2\right)}{L}\right) \left(\cos \left(\frac{\pi  t_1}{L_{\text{eff}}}\right)-i \sin \left(\frac{\pi  t_1}{L_{\text{eff}}}\right) \left(\cosh (2 \theta )-\sinh (2 \theta ) e^{\frac{2 i \pi  x}{L}}\right)\right) $}} \\&\times \text{\footnotesize{$\frac{\cos \left(\frac{\pi  t_1}{L_{\text{eff}}}\right)-i \sin \left(\frac{\pi  t_1}{L_{\text{eff}}}\right) \left(-\cosh (2 \theta )+\sinh (2 \theta ) e^{-\frac{2 i \pi  x}{L}}\right)}{L \prod_{i=1,2}  \left[\sin \left(\frac{\pi  t_1}{L_{\text{eff}}}\right) \left(\cosh (2 \theta ) \cos \left(\frac{\pi  \left(x-X_i\right)}{L}\right)-\sinh (2 \theta ) \cos \left(\frac{\pi  \left(X_i+x\right)}{L}\right)\right)+\cos \left(\frac{\pi  t_1}{L_{\text{eff}}}\right) \sin \left(\frac{\pi  \left(x-X_i\right)}{L}\right)\right]}$}},\\
    &z_{c,3} \approx 1+\frac{2 i \pi  \epsilon  \sin \left[\frac{\pi  \left(X_1-X_2\right)}{L}\right] \left(\tanh (2 \theta ) \cos \left(\frac{2 \pi  x}{L}\right)-1\right)}{L \prod_{i=1,2} \sin \left[\frac{\pi  \left(t_0-x+X_i\right)}{L}\right]}+\mathcal{O}(\epsilon^2),\\
&\overline{z}_{c,3} \approx 1-\frac{2 i \pi  \epsilon  \sin \left[\frac{\pi  \left(X_1-X_2\right)}{L}\right]\left(\tanh (2 \theta ) \cos \left(\frac{2 \pi  x}{L}\right)-1\right)}{L \prod_{i=1,2} \sin \left[\frac{\pi  \left(t_0+x-X_i\right)}{L}\right]}+\mathcal{O}(\epsilon^2),\\
&z_{c,4}\approx 1+\frac{2 i \pi  \epsilon  \sin \left[\frac{\pi  \left(X_1-X_2\right)}{L}\right]\left(\tanh (2 \theta ) \cos \left(\frac{2 \pi  \left(t_0-x\right)}{L}\right)-1\right)}{L \prod_{i=1,2} \sin \left[\frac{\pi  \left(t_0-x+X_i\right)}{L}\right] }+\mathcal{O}(\epsilon^2),\\
&\overline{z}_{c,4} \approx 1-\frac{2 i \pi  \epsilon  \sin \left[\frac{\pi  \left(X_1-X_2\right)}{L}\right] \left(\tanh (2 \theta ) \cos \left(\frac{2 \pi  \left(t_0+x\right)}{L}\right)-1\right)}{L \prod_{i=1,2} \sin \left[\frac{\pi  \left(t_0+x-X_i\right)}{L}\right] }+\mathcal{O}(\epsilon^2),
\end{split}
\ee
where around $t_1=\f{L_{\text{eff}}}{\pi}\tan^{-1}{\left[\left| \frac{\sin \left(\frac{\pi  \left(x-X_i\right)}{L}\right)}{\cosh (2 \theta ) \cos \left(\frac{\pi  \left(x-X_i\right)}{L}\right)-\sinh (2 \theta ) \cos \left(\frac{\pi  \left(X_i+x\right)}{L}\right)}\right|\right]}+n L_{\text{eff}}$  , the small $\epsilon$ expansion of $z_{c,i=1,2}$, $\overline{z}_{c,i=1,2}$, breaks down, while around $t_0= \pm \left(x-X_i\right)+n L$, the small $\epsilon$ expansion of $z_{c,i=3,4}$, $\overline{z}_{c,i=3,4}$ breaks down.

%%%%%%%%%%%%%%%%%%%%%%%%%%%%%%%%%%%%%%%%%%%%%%%%%%%%%%%%%%%%%%%%%%%%%%%

%%%%%%%%%%%%%%%%%%%%%%%%%%%%%%%%%%%%%%%%%%%%%%%%%%%%%
\section{The time dependence of $S_{A,i}$ in $2$d holographic CFTs}
%%%%%%%%%%%%%%%%%%%%%%%%%%%%%%%%%%%%%%%%%%%%%%%%%%%%%
Here, we present the time dependence of $S_{A,i}$ for M\"obius/SSD cases.

%%%%%%%%%%%%%%%%%%%%%%%%%%%%%%%%%%%%%%%%%%%%%%%%%%%%%
\subsection{M\"obius case}
%%%%%%%%%%%%%%%%%%%%%%%%%%%%%%%%%%%%%%%%%%%%%%%%%%%%%
First, we report on the time dependence of $S_{A,i}$ for M\"obius case.
Here, we assume that the local operator is inserted at $x=0,\f{L}{2}$.

%%%%%%%%%%%%%%%%%%%%%%%%%%%%%%%%%%%%%%%%%%%%%%%%%%%%%
\subsubsection{$x=X_1^f$\label{App:HEE-MO-x0}}
%%%%%%%%%%%%%%%%%%%%%%%%%%%%%%%%%%%%%%%%%%%%%%%%%%%%%
We present the time dependence of $S_{A,i}$ for the M\"obius case when the local operator is inserted at $x=0$.
The subsystems considered here are (b) and (c).

%%%%%%%%%%%%%%%%%%%%%%%%%%
 For case (b), the time dependence of $S_{A,i}$ is given by
\be
\begin{split}\nonumber
%%%%%%%%%%%%%%%%%%%%%%%%%%%%%%%%%
     &S_{A,1} \approx \frac{c}{3}  \log \left[\frac{L }{\pi } \sin \left(\frac{\pi  \left(X_1-X_2\right)}{L}\right)\right]   
    \\ &+ \begin{cases}
0 &  t_{X_2}>t_1>0\\
\frac{c}{6} \log \left[\frac{2 \sin \left[\pi \alpha_{\mathcal{O}}\right]}{ \alpha_{\mathcal{O}}}\right]-\frac{c}{6} \log \left[- \epsilon  g_1(t_1,X^{f}_1,\theta)\right] &   nL_{\text{eff}}+t_{1,+}^t>t_1>nL_{\text{eff}}+t_{X_2}\\
\frac{c}{6} \log \left[\frac{2 \sin \left[\pi \alpha_{\mathcal{O}}\right]}{ \alpha_{\mathcal{O}}}\right]-\frac{c}{6} \log \left[- \epsilon  f_1(t_1,X^{f}_1,\theta)\right]  &   nL_{\text{eff}}+t_{L-X_1}>t_1>nL_{\text{eff}}+t_{1,+}^t\\
0 &   nL_{\text{eff}}+t_{X_1}>t_1>nL_{\text{eff}}+t_{L-X_1}\\
\frac{c}{6} \log \left[\frac{2 \sin \left[\pi \alpha_{\mathcal{O}}\right]}{ \alpha_{\mathcal{O}}}\right]-\frac{c}{6} \log \left[\epsilon  g_1(t_1,X^{f}_1,\theta)\right] &  nL_{\text{eff}}+ t_{1,-}^t>t_1>nL_{\text{eff}}+t_{X_1}\\
\frac{c}{6} \log \left[\frac{2 \sin \left[\pi \alpha_{\mathcal{O}}\right]}{ \alpha_{\mathcal{O}}}\right]-\frac{c}{6} \log \left[\epsilon  f_1(t_1,X^{f}_1,\theta)\right] &   nL_{\text{eff}}+t_{L-X_2}>t_1>nL_{\text{eff}}+ t_{1,-}^t\\
0 & (n+1) L_{\text{eff}}+t_{X_2}>t_1>nL_{\text{eff}}+t_{L-X_2}\\
     \end{cases},\\
%%%%%%%%%%%%%%%%%%%%%%%%%%%%%%%%%
\end{split}
\ee
     \be
\begin{split}
     &S_{A,2} \approx  \frac{c}{3}  \log \left[\frac{L }{\pi } \sin \left(\frac{\pi  \left(X_1-X_2\right)}{L}\right)\right]   
    \\ &+ \begin{cases}
0 &  t_{X_2}>t_1>0\\
\frac{c}{6} \log \left[\frac{2 \sin \left[\pi \alpha_{\mathcal{O}}\right]}{ \alpha_{\mathcal{O}}}\right]-\frac{c}{6} \log \left[- \epsilon  g_2(t_1,X^{f}_1,\theta)\right] &   nL_{\text{eff}}+t_{1,+}^t>t_1>nL_{\text{eff}}+t_{X_2}\\
\frac{c}{6} \log \left[\frac{2 \sin \left[\pi \alpha_{\mathcal{O}}\right]}{ \alpha_{\mathcal{O}}}\right]-\frac{c}{6} \log \left[- \epsilon  f_2(t_1,X^{f}_1,\theta)\right]  &   nL_{\text{eff}}+t_{L-X_1}>t_1>nL_{\text{eff}}+t_{1,+}^t\\
0 &   nL_{\text{eff}}+t_{X_1}>t_1>nL_{\text{eff}}+t_{L-X_1}\\
\frac{c}{6} \log \left[\frac{2 \sin \left[\pi \alpha_{\mathcal{O}}\right]}{ \alpha_{\mathcal{O}}}\right]-\frac{c}{6} \log \left[\epsilon  g_2(t_1,X^{f}_1,\theta)\right] &  nL_{\text{eff}}+ t_{1,-}^t>t_1>nL_{\text{eff}}+t_{X_1}\\
\frac{c}{6} \log \left[\frac{2 \sin \left[\pi \alpha_{\mathcal{O}}\right]}{ \alpha_{\mathcal{O}}}\right]-\frac{c}{6} \log \left[\epsilon  f_2(t_1,X^{f}_1,\theta)\right] &   nL_{\text{eff}}+t_{L-X_2}>t_1>nL_{\text{eff}}+ t_{1,-}^t\\
0 & (n+1) L_{\text{eff}}+t_{X_2}>t_1>nL_{\text{eff}}+t_{L-X_2}\\
     \end{cases},\\
%%%%%%%%%%%%%%%%%%%%%%%%%%%%%%%%%
      &S_{A,3} \approx \frac{c}{3}  \log \left[\frac{L }{\pi } \sin \left(\frac{\pi  \left(X_1-X_2\right)}{L}\right)\right]  
    \\ &+ \begin{cases}
0 & X_2>t_0>0\\
\frac{c}{6} \log \left[\frac{2 \sin \left[\pi \alpha_{\mathcal{O}}\right]}{ \alpha_{\mathcal{O}}}\right]-\frac{c}{6} \log \left[ -\epsilon g_3(t_0,X^{f}_1,\theta)\right] &nL+t^{t}_{0, +}>t_0> nL+X_2\\
\frac{c}{6} \log \left[\frac{2 \sin \left[\pi \alpha_{\mathcal{O}}\right]}{ \alpha_{\mathcal{O}}}\right]-\frac{c}{6} \log \left[ -\epsilon  f_3(t_0,X^{f}_1,\theta)\right] &(n+1)L-X_1>t_0>nL+t^{t}_{0, +}\\ 
0&nL+X_1>t_0>(n+1)L-X_1\\
\frac{c}{6} \log \left[\frac{2 \sin \left[\pi \alpha_{\mathcal{O}}\right]}{ \alpha_{\mathcal{O}}}\right]-\frac{c}{6} \log \left[ \epsilon  g_3(t_0,X^{f}_1,\theta)\right] &nL+t^{t}_{0, -}>t_0>nL+X_1\\
\frac{c}{6} \log \left[\frac{2 \sin \left[\pi \alpha_{\mathcal{O}}\right]}{ \alpha_{\mathcal{O}}}\right]-\frac{c}{6} \log \left[ \epsilon  f_3(t_0,X^{f}_1,\theta)\right] &(n+1)L-X_2>t_0>nL+t^{t}_{0, -}\\
0 &(n+1)L+X_2>t_0>(n+1)L-X_2\\
    \end{cases},\\
%%%%%%%%%%%%%%%%%%%%%%%%%%%%%%%%
    &S_{A,4}\approx \frac{c}{3}  \log \left[\frac{L }{\pi } \sin \left(\frac{\pi  \left(X_1-X_2\right)}{L}\right)\right]  
       \\ &+ \begin{cases}
0 & X_2>t_0>0\\
\frac{c}{6} \log \left[\frac{2 \sin \left[\pi \alpha_{\mathcal{O}}\right]}{ \alpha_{\mathcal{O}}}\right]-\frac{c}{6} \log \left[ -\epsilon g_4(t_0,X^{f}_1,\theta)\right] &nL+t^{t}_{0, +}>t_0> nL+X_2\\
\frac{c}{6} \log \left[\frac{2 \sin \left[\pi \alpha_{\mathcal{O}}\right]}{ \alpha_{\mathcal{O}}}\right]-\frac{c}{6} \log \left[ -\epsilon  f_4(t_0,X^{f}_1,\theta)\right] &(n+1)L-X_1>t_0>nL+t^{t}_{0, +}\\ 
0&nL+X_1>t_0>(n+1)L-X_1\\
\frac{c}{6} \log \left[\frac{2 \sin \left[\pi \alpha_{\mathcal{O}}\right]}{ \alpha_{\mathcal{O}}}\right]-\frac{c}{6} \log \left[ \epsilon  g_4(t_0,X^{f}_1,\theta)\right] &nL+t^{t}_{0, -}>t_0>nL+X_1\\
\frac{c}{6} \log \left[\frac{2 \sin \left[\pi \alpha_{\mathcal{O}}\right]}{ \alpha_{\mathcal{O}}}\right]-\frac{c}{6} \log \left[ \epsilon  f_4(t_0,X^{f}_1,\theta)\right] &(n+1)L-X_2>t_0>nL+t^{t}_{0, -}\\
0 &(n+1)L+X_2>t_0>(n+1)L-X_2\\
    \end{cases}.
\end{split}
\ee
For case (c), the time dependence of $S_{A,i}$ is given by
\be\label{EE-for-Mobius-0-c}
\begin{split}
%%%%%%%%%%%%%%%%%%%%%%%%%%%%%%%%%
     &S_{A,1} \approx \frac{c}{3}  \log \left[\frac{L }{\pi } \sin \left(\frac{\pi  \left(X_1-X_2\right)}{L}\right)\right]   
    \\ &+ \begin{cases}
0 &  t_{X_2}>t_1>0\\
\frac{c}{6} \log \left[\frac{2 \sin \left[\pi \alpha_{\mathcal{O}}\right]}{ \alpha_{\mathcal{O}}}\right]-\frac{c}{6} \log \left[- \epsilon  g_1(t_1,X^{f}_1,\theta)\right] &   nL_{\text{eff}}+t_{X_1}>t_1>nL_{\text{eff}}+t_{X_2}\\
0 &  nL_{\text{eff}}+ t_{L-X_1}>t_1>nL_{\text{eff}}+t_{X_1}\\
\frac{c}{6} \log \left[\frac{2 \sin \left[\pi \alpha_{\mathcal{O}}\right]}{ \alpha_{\mathcal{O}}}\right]-\frac{c}{6} \log \left[ \epsilon  f_1(t_1,X^{f}_1,\theta)\right] &  nL_{\text{eff}}+ t_{L-X_2}>t_1>nL_{\text{eff}}+t_{L-X_1}\\
0 & (n+1) L_{\text{eff}}+t_{X_2}>t_1>nL_{\text{eff}}+t_{L-X_2}\\
     \end{cases},\\
%%%%%%%%%%%%%%%%%%%%%%%%%%%%%%%%%
 &S_{A,2} \approx  \frac{c}{3}  \log \left[\frac{L }{\pi } \sin \left(\frac{\pi  \left(X_1-X_2\right)}{L}\right)\right]   
      \\ &+ \begin{cases}
0 &  t_{X_2}>t_1>0\\
\frac{c}{6} \log \left[\frac{2 \sin \left[\pi \alpha_{\mathcal{O}}\right]}{ \alpha_{\mathcal{O}}}\right]-\frac{c}{6} \log \left[ -\epsilon  g_2(t_1,X^{f}_1,\theta)\right] &   nL_{\text{eff}}+t_{X_1}>t_1>nL_{\text{eff}}+t_{X_2}\\
0 &  nL_{\text{eff}}+ t_{L-X_1}>t_1>nL_{\text{eff}}+t_{X_1}\\
\frac{c}{6} \log \left[\frac{2 \sin \left[\pi \alpha_{\mathcal{O}}\right]}{ \alpha_{\mathcal{O}}}\right]-\frac{c}{6} \log \left[\epsilon  f_2(t_1,X^{f}_1,\theta)\right] &  nL_{\text{eff}}+ t_{L-X_2}>t_1>nL_{\text{eff}}+t_{L-X_1}\\
0 & (n+1) L_{\text{eff}}+t_{X_2}>t_1>nL_{\text{eff}}+t_{L-X_2}\\
     \end{cases},\\
%%%%%%%%%%%%%%%%%%%%%%%%%%%%%%%%%
      &S_{A,3} \approx \frac{c}{3}  \log \left[\frac{L }{\pi } \sin \left(\frac{\pi  \left(X_1-X_2\right)}{L}\right)\right]  
    \\ &+ \begin{cases}
0 & X_2>t_0>0\\
\frac{c}{6} \log \left[\frac{2 \sin \left[\pi \alpha_{\mathcal{O}}\right]}{ \alpha_{\mathcal{O}}}\right]-\frac{c}{6} \log \left[ -\epsilon  g_3(t_0,X^{f}_1,\theta)\right] &nL+X_1>t_0>nL+X_2\\
0&(n+1)L-X_1>t_0>nL+X_1\\
\frac{c}{6} \log \left[\frac{2 \sin \left[\pi \alpha_{\mathcal{O}}\right]}{ \alpha_{\mathcal{O}}}\right]-\frac{c}{6} \log \left[ \epsilon  f_3(t_0,X^{f}_1,\theta)\right] &(n+1)L-X_2>t_0>(n+1)L-X_1\\
0 &(n+1)L+X_2>t_0>(n+1)L-X_2\\
    \end{cases},\\
%%%%%%%%%%%%%%%%%%%%%%%%%%%%%%%%
      &S_{A,4} \approx \frac{c}{3}  \log \left[\frac{L }{\pi } \sin \left(\frac{\pi  \left(X_1-X_2\right)}{L}\right)\right]  
    \\ &+ \begin{cases}
0 & X_2>t_0>0\\
\frac{c}{6} \log \left[\frac{2 \sin \left[\pi \alpha_{\mathcal{O}}\right]}{ \alpha_{\mathcal{O}}}\right]-\frac{c}{6} \log \left[ -\epsilon  g_4(t_0,X^{f}_1,\theta)\right] &nL+X_1>t_0>nL+X_2\\
0&(n+1)L-X_1>t_0>nL+X_1\\
\frac{c}{6} \log \left[\frac{2 \sin \left[\pi \alpha_{\mathcal{O}}\right]}{ \alpha_{\mathcal{O}}}\right]-\frac{c}{6} \log \left[ \epsilon  f_4(t_0,X^{f}_1,\theta)\right] &(n+1)L-X_2>t_0>(n+1)L-X_1\\
0 &(n+1)L+X_2>t_0>(n+1)L-X_2\\
    \end{cases}.
\end{split}
\ee

%%%%%%%%%%%%%%%%%
%%%%%%%%%%%%%%%%%%%%%%%%%%%%%%%%%%%%%%%
\subsubsection{ $x=X_2^f$.\label{App:HEE-MO-xL2}}
%%%%%%%%%%%%%%%%%%%%%%%%%%%%%%%%%%%%%%%
Next, we consider the time dependence of entanglement entropy for the states with the insertion at $x=\f{L}{2}$ of the local operator.
To second order in the small $\epsilon$ expansion, the cross ratios are approximately given by (\ref{eq:small-cross-Xf1}).
The details of the cross ratios are reported in Appendix \ref{App:crossratiosX2}.
The time dependence of the cross ratios follows the propagation of quasiparticles.

Let us define $\hat{t}_x$ by
\be
\hat{t}_x=\f{L_{\text{eff}}}{\pi} \tan^{-1}{\left[e^{-2\theta} \tan{\left(\f{\pi x}{L}\right)}\right]}.
\ee
The time dependence of cross ratios for $i=3,4$ with the insertion at $x=X_2^f$ of the local operator is similar to that with the insertion at $x=X_1^f$ of the local operator. 
Therefore, we present here the time dependence of the cross ratios for $i=1,2$.
In the cases of (a) and (b) for $i=1,2$, $f_{i=1,2}(T_{i=1,2},x=X_2^f,\theta)$ ($g_{i=1,2}(T_{i=1,2},X_2^f,\theta)$) is positive in the time intervals, $(n+1)L_{\text{eff}}+\hat{t}_{\f{L}{2}-X_2}>t_1>n L_{\text{eff}}+\hat{t}_{\f{3 }{2} L-X_1}$ ($nL_{\text{eff}}+\hat{t}_{\f{L}{2}+X_2}>t_1>n L_{\text{eff}}+\hat{t}_{X_1-\f{L}{2}}$), where $n$ is an integer, while it is negative in the time intervals, $nL_{\text{eff}}+\hat{t}_{\f{3}{2}L-X_1}>t_1>n L_{\text{eff}}+\hat{t}_{\f{L}{2}-X_2}$ ($(n+1)L_{\text{eff}}+\hat{t}_{X_1-\f{L}{2}}>t_1>n L_{\text{eff}}+\hat{t}_{\f{L}{2}+X_2}$). 

In the case of (c), $f_{i=1,2}(T_{i=1,2},X_2^f,\theta)$ ($g_{i=1,2}(T_{i=1,2},X_2^f,\theta)$)  is positive in the time intervals, $nL_{\text{eff}}+\hat{t}_{\f{L}{2}-X_2}>t_1>n L_{\text{eff}}+\hat{t}_{\f{L}{2}-X_1}$ ($(n+1)L_{\text{eff}}+\hat{t}_{X_2+\f{L}{2}}>t_1>n L_{\text{eff}}+\hat{t}_{X_1+\f{L}{2}}$), while it is negative in the time intervals, $(n+1) L_{\text{eff}}+t_{\f{L}{2}-X_1}>t_1>n L_{\text{eff}}+t_{\f{L}{2}-X_2}$ ($nL_{\text{eff}}+\hat{t}_{\f{L}{2}+X_1}>t_1>nL_{\text{eff}}+\hat{t}_{\f{L}{2}+X_2}$).
Since the denominators of $z_{c,i}$ ($\overline{z}_{c,i}$) vanishes around the times, $t_1=nL_{\text{eff}} +t_{X_1}$ or $t_1=nL_{\text{eff}} +t_{X_2}$ ($t_1=nL_{\text{eff}} +t_{L-X_1}$ or $t_1=nL_{\text{eff}} +t_{L-X_2}$), the small $\epsilon$ expansion in (\ref{eq:small-cross-Xf1}) breaks down.

After choosing the branches correctly, the time dependence of entanglement entropy for (a), (b), and (c) is determined as follows. For case (a), the time dependence of $S_{A,i}$ is given by
\be\nonumber
\begin{split}
   & S_{A,1}\approx \f{c}{3}\log{\left[\f{L}{\pi}\sin{\left[\f{\pi(X_1-X_2)}{L}\right]}\right]} \\ + & \begin{cases}
         0 &  \hat{t}_{\f{L}{2}-X_2}>t_1>0\\
        %%%%%%%%%%%%%%%%%%%%%%%%%%%%%%%%
\frac{c}{6} \log \left[\frac{2 \sin \left[\pi \alpha_{\mathcal{O}}\right]}{ \alpha_{\mathcal{O}}}\right]-\frac{c}{6} \log \left[-\epsilon  f_1(t_1,X^{f}_2,\theta)\right]  & n L_{\text{eff}}+\hat{t}_{1,+}^t>t_1>n L_{\text{eff}}+ \hat{t}_{\f{L}{2}-X_2}\\
%%%%%%%%%%%%%%%%%%%%%%%%%%%%%
\frac{c}{6} \log \left[\frac{2 \sin \left[\pi \alpha_{\mathcal{O}}\right]}{ \alpha_{\mathcal{O}}}\right]-\frac{c}{6} \log \left[-\epsilon  g_1(t_1,X^{f}_2,\theta)\right]  & n L_{\text{eff}}+\hat{t}_{X_1-\f{L}{2}}>t_1>n L_{\text{eff}}+\hat{t}_{1,+}^t\\
%%%%%%%%%%%%%%%%%%%%%%%%%%%%%
0 & n L_{\text{eff}}+\hat{t}_{\frac{3}{2}L-X_1}>t_1>n L_{\text{eff}}+\hat{t}_{X_1-\f{L}{2}} \\
%%%%%%%%%%%%%%%%%%%%%%%%%%%%%%
\frac{c}{6} \log \left[\frac{2 \sin \left[\pi \alpha_{\mathcal{O}}\right]}{ \alpha_{\mathcal{O}}}\right]-\frac{c}{6} \log \left[\epsilon  f_1(t_1,X^{f}_2,\theta)\right]  & n L_{\text{eff}}+\hat{t}_{1,-}^t>t_1>n L_{\text{eff}}+\hat{t}_{\frac{3}{2}L-X_1}\\
\frac{c}{6} \log \left[\frac{2 \sin \left[\pi \alpha_{\mathcal{O}}\right]}{ \alpha_{\mathcal{O}}}\right]-\frac{c}{6} \log \left[\epsilon  g_1(t_1,X^{f}_2,\theta)\right]  &n L_{\text{eff}}+ \hat{t}_{\f{L}{2}+X_2}>t_1>n L_{\text{eff}}+\hat{t}_{1,-}^t\\
0 & (n+1) L_{\text{eff}}+ \hat{t}_{\f{L}{2}-X_2}>t_1>n L_{\text{eff}}+\hat{t}_{\f{L}{2}+X_2}\\
    \end{cases},\\
    %%%%%%%%%%%%%%%%%%%%%
    %%%%%%%%%%%%%%%%%%%%%
      & S_{A,2}\approx \f{c}{3}\log{\left[\f{L}{\pi}\sin{\left[\f{\pi(X_1-X_2)}{L}\right]}\right]} \\ + & \begin{cases}
         0 &  \hat{t}_{\f{L}{2}-X_2}>t_1>0\\
        %%%%%%%%%%%%%%%%%%%%%%%%%%%%%%%%
\frac{c}{6} \log \left[\frac{2 \sin \left[\pi \alpha_{\mathcal{O}}\right]}{ \alpha_{\mathcal{O}}}\right]-\frac{c}{6} \log \left[-\epsilon  f_2(t_1,X^{f}_2,\theta)\right]  & n L_{\text{eff}}+\hat{t}_{1,+}^t>t_1>n L_{\text{eff}}+ \hat{t}_{\f{L}{2}-X_2}\\
%%%%%%%%%%%%%%%%%%%%%%%%%%%%%
\frac{c}{6} \log \left[\frac{2 \sin \left[\pi \alpha_{\mathcal{O}}\right]}{ \alpha_{\mathcal{O}}}\right]-\frac{c}{6} \log \left[-\epsilon  g_2(t_1,X^{f}_2,\theta)\right]  & n L_{\text{eff}}+\hat{t}_{X_1-\f{L}{2}}>t_1>n L_{\text{eff}}+\hat{t}_{1,+}^t\\
%%%%%%%%%%%%%%%%%%%%%%%%%%%%%
0 & n L_{\text{eff}}+\hat{t}_{\frac{3}{2}L-X_1}>t_1>n L_{\text{eff}}+\hat{t}_{X_1-\f{L}{2}} \\
%%%%%%%%%%%%%%%%%%%%%%%%%%%%%%
\frac{c}{6} \log \left[\frac{2 \sin \left[\pi \alpha_{\mathcal{O}}\right]}{ \alpha_{\mathcal{O}}}\right]-\frac{c}{6} \log \left[\epsilon  f_2(t_1,X^{f}_2,\theta)\right]  & n L_{\text{eff}}+\hat{t}_{1,-}^t>t_1>n L_{\text{eff}}+\hat{t}_{\frac{3}{2}L-X_1}\\
\frac{c}{6} \log \left[\frac{2 \sin \left[\pi \alpha_{\mathcal{O}}\right]}{ \alpha_{\mathcal{O}}}\right]-\frac{c}{6} \log \left[\epsilon  g_2(t_1,X^{f}_2,\theta)\right]  &n L_{\text{eff}}+ \hat{t}_{\f{L}{2}+X_2}>t_1>n L_{\text{eff}}+\hat{t}_{1,-}^t\\
0 & (n+1) L_{\text{eff}}+ \hat{t}_{\f{L}{2}-X_2}>t_1>n L_{\text{eff}}+\hat{t}_{\f{L}{2}+X_2}\\
    \end{cases},\\
    %%%%%%%%%%%%%%%%%%%%%
    %%%%%%%%%%%%%%%%%%%%%
      & S_{A,3}\approx \f{c}{3}\log{\left[\f{L}{\pi}\sin{\left[\f{\pi(X_1-X_2)}{L}\right]}\right]}\\ &+\begin{cases}
         0 &\f{L}{2}-X_2>t_0>0\\
        %%%%%%%%%%%%%%%%%%%%%%%%%%%%%%%%
\frac{c}{6} \log \left[\frac{2 \sin \left[\pi \alpha_{\mathcal{O}}\right]}{ \alpha_{\mathcal{O}}}\right]-\frac{c}{6} \log \left[-\epsilon  f_3(t_0,X^{f}_2,\theta)\right] &nL+ \tilde{t}^{t}_{0,+}>t_1>\left(n+\f{1}{2}\right)L-X_2\\
%%%%%%%%%%%%%%%%%%%%%%%%%%%%%
\frac{c}{6} \log \left[\frac{2 \sin \left[\pi \alpha_{\mathcal{O}}\right]}{ \alpha_{\mathcal{O}}}\right]-\frac{c}{6} \log \left[-\epsilon  g_3(t_0,X^{f}_2,\theta)\right] &\left(n-\f{1}{2}\right)L+X_1>t_1>nL+ \tilde{t}^{t}_{0,+}\\
%%%%%%%%%%%%%%%%%%%%%%%%%%%%%
0 &\left(n+\f{3}{2}\right)L-X_1>t_0>\left(n-\f{1}{2}\right)L+X_1\\
%%%%%%%%%%%%%%%%%%%%%%%%%%%%%%
\frac{c}{6} \log \left[\frac{2 \sin \left[\pi \alpha_{\mathcal{O}}\right]}{ \alpha_{\mathcal{O}}}\right]-\frac{c}{6} \log \left[\epsilon  f_3(t_0,X^{f}_2,\theta)\right] & nL+\tilde{t}^t_{0,-}>t_0>\left(n+\f{3}{2}\right)L-X_1\\
%%%%%%%%%%%%%%%%%%%%%%%%%%%%%
\frac{c}{6} \log \left[\frac{2 \sin \left[\pi \alpha_{\mathcal{O}}\right]}{ \alpha_{\mathcal{O}}}\right]-\frac{c}{6} \log \left[\epsilon  g_3(t_0,X^{f}_2,\theta)\right] & \left(n+\f{1}{2}\right)L+X_2>t_0>nL+\tilde{t}^t_{0,-}\\
0 &\left(n+\f{3}{2}\right)L-X_2>t_0> \left(n+\f{1}{2}\right)L+X_2\\
%%%%%%%%%%%%%%%%%%%%%%%%%%%%%
    \end{cases},\\
\end{split}
\ee
\be
\begin{split}
 & S_{A,4}\approx \f{c}{3}\log{\left[\f{L}{\pi}\sin{\left[\f{\pi(X_1-X_2)}{L}\right]}\right]}\\ &+\begin{cases}
         0 &\f{L}{2}-X_2>t_0>0\\
        %%%%%%%%%%%%%%%%%%%%%%%%%%%%%%%%
\frac{c}{6} \log \left[\frac{2 \sin \left[\pi \alpha_{\mathcal{O}}\right]}{ \alpha_{\mathcal{O}}}\right]-\frac{c}{6} \log \left[-\epsilon  f_4(t_0,X^{f}_2,\theta)\right] &nL+ \tilde{t}^{t}_{0,+}>t_1>\left(n+\f{1}{2}\right)L-X_2\\
%%%%%%%%%%%%%%%%%%%%%%%%%%%%%
\frac{c}{6} \log \left[\frac{2 \sin \left[\pi \alpha_{\mathcal{O}}\right]}{ \alpha_{\mathcal{O}}}\right]-\frac{c}{6} \log \left[-\epsilon  g_4(t_0,X^{f}_2,\theta)\right] &\left(n-\f{1}{2}\right)L+X_1>t_1>nL+ \tilde{t}^{t}_{0,+}\\
%%%%%%%%%%%%%%%%%%%%%%%%%%%%%
0 &\left(n+\f{3}{2}\right)L-X_1>t_0>\left(n-\f{1}{2}\right)L+X_1\\
%%%%%%%%%%%%%%%%%%%%%%%%%%%%%%
\frac{c}{6} \log \left[\frac{2 \sin \left[\pi \alpha_{\mathcal{O}}\right]}{ \alpha_{\mathcal{O}}}\right]-\frac{c}{6} \log \left[\epsilon  f_4(t_0,X^{f}_2,\theta)\right] & nL+\tilde{t}^t_{0,-}>t_0>\left(n+\f{3}{2}\right)L-X_1\\
%%%%%%%%%%%%%%%%%%%%%%%%%%%%%
\frac{c}{6} \log \left[\frac{2 \sin \left[\pi \alpha_{\mathcal{O}}\right]}{ \alpha_{\mathcal{O}}}\right]-\frac{c}{6} \log \left[\epsilon  g_4(t_0,X^{f}_2,\theta)\right] & \left(n+\f{1}{2}\right)L+X_2>t_0>nL+\tilde{t}^t_{0,-}\\
0 &\left(n+\f{3}{2}\right)L-X_2>t_0> \left(n+\f{1}{2}\right)L+X_2\\
%%%%%%%%%%%%%%%%%%%%%%%%%%%%%
    \end{cases},\\
\end{split}
\ee
where $n$ is an integer greater than or equal to $0$, and $\hat{t}_{i,\pm}$ are positive. The characteristic parameters are given by
\be\label{eq:ttpm-mobius-L/2}
\begin{split}
&\hat{t}_{1,\pm}^t= \f{L_{\text{eff}}}{\pi} \tan^{-1} \left[\pm \sqrt{e^{-4 \theta } \tan \left(\frac{\pi  \left(X_1-\frac{L}{2}\right)}{L}\right) \tan \left(\frac{\pi  \left(\frac{L}{2}-X_2\right)}{L}\right) }\right] 
, \\
&\cos \left(\frac{\pi \tilde{t}^{t}_{0, \pm}}{L}\right)= \pm \sqrt{\frac{-\prod_{i=1}^2 \sin \left[\frac{\pi X_i}{L}\right]}{\cos \left[\frac{\pi\left(X_1+X_2\right)}{L}\right]}}.
\end{split}
\ee

Note that the computation on $S_{A,4}$ breaks down at $t_0=L\left(n+\f{1}{2}\right)$ because $z_{c,4}$ and $\overline{z}_{c,4}$ exactly becomes unity. 
Since the Hamiltonian density at $x=X_1^f$ of $H_{\text{SSD}}$ is zero, the damping factor, $e^{-\epsilon H_{\text{SSD}}}$, can not tame the high-energy mode there adequately.
At $t_0=L\left(n+\f{1}{2}\right)$, the excitation created by the local operator at $x=X_2^f$ could arrive at $x=X_1^f$, so that $S_{A,4}$ becomes diverge.
For case (b), the time dependence of $S_{A,i}$ is given by

\be\nonumber
\begin{split}
	& S_{A,1}\approx \f{c}{3}\log{\left[\f{L}{\pi}\sin{\left[\f{\pi(X_1-X_2)}{L}\right]}\right]} \\ + & \begin{cases}
	 0  & \hat{t}_{X_1-\f{L}{2}}>t_1>0\\
\frac{c}{6} \log \left[\frac{2 \sin \left[\pi \alpha_{\mathcal{O}}\right]}{ \alpha_{\mathcal{O}}}\right]-\frac{c}{6} \log \left[\epsilon  g_1(t_1,X^{f}_2,\theta)\right] & nL_{\text{eff}}+\hat{t}_{1,+}^t>t_1>nL_{\text{eff}}+\hat{t}_{X_1-\f{L}{2}} \\
\frac{c}{6} \log \left[\frac{2 \sin \left[\pi \alpha_{\mathcal{O}}\right]}{ \alpha_{\mathcal{O}}}\right]-\frac{c}{6} \log \left[\epsilon  f_1(t_1,X^{f}_2,\theta)\right] & nL_{\text{eff}}+\hat{t}_{\f{L}{2}-X_2}>t_1>nL_{\text{eff}}+\hat{t}_{1,+}^t \\
0 & nL_{\text{eff}}+\hat{t}_{\f{L}{2}+X_2}>t_1>nL_{\text{eff}}+\hat{t}_{\f{L}{2}-X_2}\\
\frac{c}{6} \log \left[\frac{2 \sin \left[\pi \alpha_{\mathcal{O}}\right]}{ \alpha_{\mathcal{O}}}\right]-\frac{c}{6} \log \left[-\epsilon  g_1(t_1,X^{f}_2,\theta)\right] & nL_{\text{eff}}+\hat{t}_{1,-}^t>t_1>nL_{\text{eff}}+\hat{t}_{\f{L}{2}+X_2}\\
\frac{c}{6} \log \left[\frac{2 \sin \left[\pi \alpha_{\mathcal{O}}\right]}{ \alpha_{\mathcal{O}}}\right]-\frac{c}{6} \log \left[-\epsilon  f_1(t_1,X^{f}_2,\theta)\right] & nL_{\text{eff}}+\hat{t}_{\frac{3}{2}L-X_1}>t_1>nL_{\text{eff}}+\hat{t}_{1,-}^t\\
0  & (n+1) L_{\text{eff}}+\hat{t}_{X_1-\f{L}{2}}+>t_1>n L_{\text{eff}}+\hat{t}_{\frac{3}{2}L-X_1}\\
	\end{cases},\\
	%%%%%%%%%%%%%%%%%%%%%
	%%%%%%%%%%%%%%%%%%%%%
	& S_{A,2}\approx \f{c}{3}\log{\left[\f{L}{\pi}\sin{\left[\f{\pi(X_1-X_2)}{L}\right]}\right]} \\ + & \begin{cases}
	 0  & \hat{t}_{X_1-\f{L}{2}}>t_1>0\\
\frac{c}{6} \log \left[\frac{2 \sin \left[\pi \alpha_{\mathcal{O}}\right]}{ \alpha_{\mathcal{O}}}\right]-\frac{c}{6} \log \left[\epsilon  g_2(t_1,X^{f}_2,\theta)\right] & nL_{\text{eff}}+\hat{t}_{1,+}^t>t_1>nL_{\text{eff}}+\hat{t}_{X_1-\f{L}{2}} \\
\frac{c}{6} \log \left[\frac{2 \sin \left[\pi \alpha_{\mathcal{O}}\right]}{ \alpha_{\mathcal{O}}}\right]-\frac{c}{6} \log \left[\epsilon  f_2(t_1,X^{f}_2,\theta)\right] & nL_{\text{eff}}+\hat{t}_{\f{L}{2}-X_2}>t_1>nL_{\text{eff}}+\hat{t}_{1,+}^t \\
0 & nL_{\text{eff}}+\hat{t}_{\f{L}{2}+X_2}>t_1>nL_{\text{eff}}+\hat{t}_{\f{L}{2}-X_2}\\
\frac{c}{6} \log \left[\frac{2 \sin \left[\pi \alpha_{\mathcal{O}}\right]}{ \alpha_{\mathcal{O}}}\right]-\frac{c}{6} \log \left[-\epsilon  g_2(t_1,X^{f}_2,\theta)\right] & nL_{\text{eff}}+\hat{t}_{1,-}^t>t_1>nL_{\text{eff}}+\hat{t}_{\f{L}{2}+X_2}\\
\frac{c}{6} \log \left[\frac{2 \sin \left[\pi \alpha_{\mathcal{O}}\right]}{ \alpha_{\mathcal{O}}}\right]-\frac{c}{6} \log \left[-\epsilon  f_2(t_1,X^{f}_2,\theta)\right] & nL_{\text{eff}}+\hat{t}_{\frac{3}{2}L-X_1}>t_1>nL_{\text{eff}}+\hat{t}_{1,-}^t\\
0  & (n+1) L_{\text{eff}}+\hat{t}_{X_1-\f{L}{2}}+>t_1>n L_{\text{eff}}+\hat{t}_{\frac{3}{2}L-X_1}\\
	\end{cases},\\
	%%%%%%%%%%%%%%%%%%%%%
	%%%%%%%%%%%%%%%%%%%%%
	& S_{A,3}\approx \f{c}{3}\log{\left[\f{L}{\pi}\sin{\left[\f{\pi(X_1-X_2)}{L}\right]}\right]}\\ &+\begin{cases}
		0 &X_1-\f{L}{2}>t_0>0\\
		%%%%%%%%%%%%%%%%%%%%%%%%%%%%%%%%
		\frac{c}{6} \log \left[\frac{2 \sin \left[\pi \alpha_{\mathcal{O}}\right]}{ \alpha_{\mathcal{O}}}\right]-\frac{c}{6} \log \left[\epsilon  g_3(t_0,X^{f}_2,\theta)\right] &n L+ \tilde{t}^{t}_{0,+}>t_0>(n-\f{1}{2})L+X_1\\
		%%%%%%%%%%%%%%%%%%%%%%%%%%%%%
		\frac{c}{6} \log \left[\frac{2 \sin \left[\pi \alpha_{\mathcal{O}}\right]}{ \alpha_{\mathcal{O}}}\right]-\frac{c}{6} \log \left[\epsilon  f_3(t_0,X^{f}_2,\theta)\right] &\left(n+\f{1}{2}\right)L-X_2>t_0>n L+ \tilde{t}^{t}_{0,+}\\
		%%%%%%%%%%%%%%%%%%%%%%%%%%%%%
		0 &\left(n+\f{1}{2}\right)L+X_2>t_0>\left(n+\f{1}{2}\right)L-X_2\\
		%%%%%%%%%%%%%%%%%%%%%%%%%%%%%%
		\frac{c}{6} \log \left[\frac{2 \sin \left[\pi \alpha_{\mathcal{O}}\right]}{ \alpha_{\mathcal{O}}}\right]-\frac{c}{6} \log \left[-\epsilon  g_3(t_0,X^{f}_2,\theta)\right] & nL+\tilde{t}^t_{0,-}>t_0>\left(n+\f{1}{2}\right)L+X_2\\
		%%%%%%%%%%%%%%%%%%%%%%%%%%%%%
		\frac{c}{6} \log \left[\frac{2 \sin \left[\pi \alpha_{\mathcal{O}}\right]}{ \alpha_{\mathcal{O}}}\right]-\frac{c}{6} \log \left[-\epsilon  f_3(t_0,X^{f}_2,\theta)\right] & \left(n+\f{3}{2}\right)L-X_1>t_0>nL+\tilde{t}^t_{0,-}\\
		%%%%%%%%%%%%%%%%%%%%%%%%%%%%%
  0 &(n+\f{1}{2})L+X_1>t_0>\left(n+\f{3}{2}\right)L-X_1\\
	\end{cases},\\
\end{split}
\ee
\be
\begin{split}
 & S_{A,4}\approx \f{c}{3}\log{\left[\f{L}{\pi}\sin{\left[\f{\pi(X_1-X_2)}{L}\right]}\right]}\\ &+\begin{cases}
		0 &X_1-\f{L}{2}>t_0>0\\
		%%%%%%%%%%%%%%%%%%%%%%%%%%%%%%%%
		\frac{c}{6} \log \left[\frac{2 \sin \left[\pi \alpha_{\mathcal{O}}\right]}{ \alpha_{\mathcal{O}}}\right]-\frac{c}{6} \log \left[\epsilon  g_4(t_0,X^{f}_2,\theta)\right] &n L+ \tilde{t}^{t}_{0,+}>t_0>(n-\f{1}{2})L+X_1\\
		%%%%%%%%%%%%%%%%%%%%%%%%%%%%%
		\frac{c}{6} \log \left[\frac{2 \sin \left[\pi \alpha_{\mathcal{O}}\right]}{ \alpha_{\mathcal{O}}}\right]-\frac{c}{6} \log \left[\epsilon  f_4(t_0,X^{f}_2,\theta)\right] &\left(n+\f{1}{2}\right)L-X_2>t_0>n L+ \tilde{t}^{t}_{0,+}\\
		%%%%%%%%%%%%%%%%%%%%%%%%%%%%%
		0 &\left(n+\f{1}{2}\right)L+X_2>t_0>\left(n+\f{1}{2}\right)L-X_2\\
		%%%%%%%%%%%%%%%%%%%%%%%%%%%%%%
		\frac{c}{6} \log \left[\frac{2 \sin \left[\pi \alpha_{\mathcal{O}}\right]}{ \alpha_{\mathcal{O}}}\right]-\frac{c}{6} \log \left[-\epsilon  g_4(t_0,X^{f}_2,\theta)\right] & nL+\tilde{t}^t_{0,-}>t_0>\left(n+\f{1}{2}\right)L+X_2\\
		%%%%%%%%%%%%%%%%%%%%%%%%%%%%%
		\frac{c}{6} \log \left[\frac{2 \sin \left[\pi \alpha_{\mathcal{O}}\right]}{ \alpha_{\mathcal{O}}}\right]-\frac{c}{6} \log \left[-\epsilon  f_4(t_0,X^{f}_2,\theta)\right] & \left(n+\f{3}{2}\right)L-X_1>t_0>nL+\tilde{t}^t_{0,-}\\
		%%%%%%%%%%%%%%%%%%%%%%%%%%%%%
  0 &(n+\f{1}{2})L+X_1>t_0>\left(n+\f{3}{2}\right)L-X_1\\
	\end{cases},\\
\end{split}
\ee
For case (c), the time dependence of $S_{A,i}$ is given by
\be\nonumber
\begin{split}
	& S_{A,1}\approx \f{c}{3}\log{\left[\f{L}{\pi}\sin{\left[\f{\pi(X_1-X_2)}{L}\right]}\right]} \\ + & \begin{cases}
0 & \hat{t}_{\f{L}{2}-X_1}>t_1>0\\
\frac{c}{6} \log \left[\frac{2 \sin \left[\pi \alpha_{\mathcal{O}}\right]}{ \alpha_{\mathcal{O}}}\right]-\frac{c}{6} \log \left[\epsilon  f_1(t_1,X^{f}_2,\theta)\right] &   nL_{\text{eff}}+\hat{t}_{\f{L}{2}-X_2}>t_1> nL_{\text{eff}}+\hat{t}_{\f{L}{2}-X_1} \\
0 &  nL_{\text{eff}}+\hat{t}_{\f{L}{2}+X_2} >t_1> nL_{\text{eff}}+\hat{t}_{\f{L}{2}-X_2} \\
\frac{c}{6} \log \left[\frac{2 \sin \left[\pi \alpha_{\mathcal{O}}\right]}{ \alpha_{\mathcal{O}}}\right]-\frac{c}{6} \log \left[- \epsilon  g_1(t_1,X^{f}_2,\theta)\right] &  nL_{\text{eff}}+\hat{t}_{\f{L}{2}+X_1} >t_1> nL_{\text{eff}}+\hat{t}_{\f{L}{2}+X_2} \\
0 & (n+1) L_{\text{eff}}+ \hat{t}_{\f{L}{2}-X_1}>t_1> nL_{\text{eff}}+\hat{t}_{\f{L}{2}+X_1}\\
	\end{cases},\\
	%%%%%%%%%%%%%%%%%%%%%
	%%%%%%%%%%%%%%%%%%%%%
	& S_{A,2}\approx \f{c}{3}\log{\left[\f{L}{\pi}\sin{\left[\f{\pi(X_1-X_2)}{L}\right]}\right]} \\ + & \begin{cases}
0 & \hat{t}_{\f{L}{2}-X_1}>t_1>0\\
\frac{c}{6} \log \left[\frac{2 \sin \left[\pi \alpha_{\mathcal{O}}\right]}{ \alpha_{\mathcal{O}}}\right]-\frac{c}{6} \log \left[\epsilon  f_2(t_1,X^{f}_2,\theta)\right] &   nL_{\text{eff}}+\hat{t}_{\f{L}{2}-X_2}>t_1> nL_{\text{eff}}+\hat{t}_{\f{L}{2}-X_1} \\
0 &  nL_{\text{eff}}+\hat{t}_{\f{L}{2}+X_2} >t_1> nL_{\text{eff}}+\hat{t}_{\f{L}{2}-X_2} \\
\frac{c}{6} \log \left[\frac{2 \sin \left[\pi \alpha_{\mathcal{O}}\right]}{ \alpha_{\mathcal{O}}}\right]-\frac{c}{6} \log \left[- \epsilon  g_2(t_1,X^{f}_2,\theta)\right] &  nL_{\text{eff}}+\hat{t}_{\f{L}{2}+X_1} >t_1> nL_{\text{eff}}+\hat{t}_{\f{L}{2}+X_2} \\
0 & (n+1) L_{\text{eff}}+ \hat{t}_{\f{L}{2}-X_1}>t_1> nL_{\text{eff}}+\hat{t}_{\f{L}{2}+X_1}\\
	\end{cases},\\
	%%%%%%%%%%%%%%%%%%%%%
	%%%%%%%%%%%%%%%%%%%%%
	& S_{A,3}\approx \f{c}{3}\log{\left[\f{L}{\pi}\sin{\left[\f{\pi(X_1-X_2)}{L}\right]}\right]}\\ &+\begin{cases}
			0 & \f{L}{2} -X_1>t_1>0 \\
	\frac{c}{6} \log \left[\frac{2 \sin \left[\pi \alpha_{\mathcal{O}}\right]}{ \alpha_{\mathcal{O}}}\right]-\frac{c}{6} \log \left[\epsilon  f_3(t_0,X^{f}_2,\theta)\right] & (n+\f{1}{2})L-X_2>t_1>(n+\f{1}{2}) L-X_1 \\
	0 & (n+\f{1}{2})L+X_2>t_1>(n+\f{1}{2})L-X_2\\
	\frac{c}{6} \log \left[\frac{2 \sin \left[\pi \alpha_{\mathcal{O}}\right]}{ \alpha_{\mathcal{O}}}\right]-\frac{c}{6} \log \left[-\epsilon  g_3(t_0,X^{f}_2,\theta)\right] & (n+\f{1}{2})L+X_1>t_1>(n+\f{1}{2})L+X_2\\
 0 & (n+\f{3}{2})-X_1>t_1>(n+\f{1}{2}) L+X_1\\
	\end{cases},\\
\end{split}
\ee
\be
\begin{split}
 	%%%%%%%%%%%%%%%%%%%%%
	& S_{A,4}\approx \f{c}{3}\log{\left[\f{L}{\pi}\sin{\left[\f{\pi(X_1-X_2)}{L}\right]}\right]}\\ &+\begin{cases}
			0 & \f{L}{2} -X_1>t_1>0 \\
	\frac{c}{6} \log \left[\frac{2 \sin \left[\pi \alpha_{\mathcal{O}}\right]}{ \alpha_{\mathcal{O}}}\right]-\frac{c}{6} \log \left[\epsilon  f_4(t_0,X^{f}_2,\theta)\right] & (n+\f{1}{2})L-X_2>t_1>(n+\f{1}{2}) L-X_1 \\
	0 & (n+\f{1}{2})L+X_2>t_1>(n+\f{1}{2})L-X_2\\
	\frac{c}{6} \log \left[\frac{2 \sin \left[\pi \alpha_{\mathcal{O}}\right]}{ \alpha_{\mathcal{O}}}\right]-\frac{c}{6} \log \left[-\epsilon  g_4(t_0,X^{f}_2,\theta)\right] & (n+\f{1}{2})L+X_1>t_1>(n+\f{1}{2})L+X_2\\
 0 & (n+\f{3}{2})-X_1>t_1>(n+\f{1}{2}) L+X_1\\
	\end{cases},\\
\end{split}
\ee
The time dependence of $S_{A,i}$ is periodic with the periodicity $L_{\text{eff}}$ for $i=1,2$ and $L$ for $i=3,4$.
The time dependence of $S_{A,i}$ follows the propagation of quasiparticles. 
%%%%%%%%%%%%%%%%%%%%%%%%%%%%%%%%%%%%%%%%%%%%%%%%%%%%%
\subsubsection{General $x$ \label{App:HEE-MO-generalx}}
The details of the small $\epsilon$ expansion of cross ratios in general $x$ during the M\"obius time evolution are reported in Appendix \ref{eq:general-x-mo-crossratio} .
Define $\grave{t}_{y}$ as
\be
\grave{t}_{y}=\f{L_{\text{eff}}}{\pi}\tan^{-1}{\left[\left| \frac{\sin \left(\frac{\pi  \left(x-y\right)}{L}\right)}{\cosh (2 \theta ) \cos \left(\frac{\pi  \left(x-y\right)}{L}\right)-\sinh (2 \theta ) \cos \left(\frac{\pi  \left(y+x\right)}{L}\right)}\right|\right]}
\ee
% One can easily check that for $0<x<y_1<y_2<L$ and any finite or infinity $\theta$, we  have $\grave{t}^{x}_{y_1}<\grave{t}^{x}_{y_2}$.
For any given general $x, X_{1},X_{2}$, where $0<X_2<X_1<L$, when $i=1,2$ we observe several different behaviors when $x$ is inserted in different regions on the circle. 
Then, we define some characteristic parameters:
 \be
 \begin{split}
     &\grave{t}^{t}_{1,\pm}=\f{L_{\text{eff}}}{\pi}\tan^{-1}\left[\pm\sqrt{\f{2 L \sin \left(\frac{\pi  \left(X_1-x\right)}{L}\right) \sin \left(\frac{\pi  \left(x-X_2\right)}{L}\right)}{\delta}}\right],\\
     &\cos \left(\frac{\pi  \check{t}^{t}_{0,\pm}}{L}\right)=\pm\sqrt{ \frac{\cos \left(\frac{\pi  \left(2 x-X_1-X_2\right)}{L}\right)+\cos \left(\frac{\pi  \left(X_1-X_2\right)}{L}\right)}{2 \cos \left(\frac{\pi  \left(2 x-X_1-X_2\right)}{L}\right)} }.
 \end{split}
 \ee
 
Here, $\delta$ is defined as
 \be
 \begin{split}
 &\delta=\\
 &\text{\footnotesize{$ 2  L \left(\sinh ^2(2 \theta ) \cos \left(\frac{\pi  \left(x+X_1\right)}{L}\right) \cos \left(\frac{\pi  \left(x+X_2\right)}{L}\right)+\cosh ^2(2 \theta ) \cos \left(\frac{\pi  \left(x-X_1\right)}{L}\right) \cos \left(\frac{\pi  \left(x-X_2\right)}{L}\right)\right) $}}\\
 &\text{\footnotesize{$ -L \sinh (4 \theta ) \left(\cos \left(\frac{\pi  \left(x+X_1\right)}{L}\right) \cos \left(\frac{\pi  \left(x-X_2\right)}{L}\right)+\cos \left(\frac{\pi  \left(x-X_1\right)}{L}\right) \cos \left(\frac{\pi  \left(x+X_2\right)}{L}\right)\right)$}}
 \end{split}
 \ee
 
Notice that in the general $x$ case, even if we insert a local excitation in the middle of the interval ($X_1 - x = x - X_2 > 0$) at $t=0$, we still have $t_{X_2} \neq t_{X_1}$ in general.

We select four typical cases to consider the time-dependence of $S_{A,i}$ here (See Fig.\ \ref{Fig:generalsubsystems} ):
\be\label{eq:generalsubsystems}
\begin{split}
    A=\begin{cases}
        \left\{y\big{|}X_2 \le y\le X_1\right\},~\text{where}~,\\ \begin{cases}
              ~\text{when}~  i=1,2 , \\  
X_{2}<x<X_{1}, \grave{t}_{X_2}<\grave{t}_{X_1}, L_{\text{eff}}-\grave{t}_{X_1}> \grave{t}_{X_1} ~\text{and}~ L_{\text{eff}}-\grave{t}_{X_2} > \grave{t}_{X_2}   \\  ~\text{when}~ i=3,4,\\  X_{2}<x<X_{1} ,  \left|x-X_2\right| < \left| X_1-x\right| ,  \left|L+x-X_1\right| > \left| X_1-x\right|  \\~\text{and}~  \left|L+x-X_2\right| > \left| X_2-x\right| 
         \end{cases} & ~\text{for}~\text{(\romannumeral1)}~\\
        \left\{y\big{|}X_2 \le y\le X_1\right\}, ~\text{where}~ ,\\ \begin{cases}
              ~\text{when}~  i=1,2 , \\  X_{1}<x<L, \grave{t}_{X_1}<\grave{t}_{X_2}, L_{\text{eff}}-\grave{t}_{X_1}> \grave{t}_{X_1} ~\text{and}~ L_{\text{eff}}-\grave{t}_{X_2} > \grave{t}_{X_2}    \\  ~\text{when}~ i=3,4,\\ X_{1}<x<L,  \left| X_1-x\right|<\left|X_2+L-x\right|, \left|L+x-X_1\right| > \left| X_1-x\right|\\ ~\text{and}~ \left|L+x-X_2\right| > \left| X_2-x\right|.
         \end{cases}  &~\text{for}~\text{(\romannumeral2)}~\\
         \left\{y\big{|}X_2 \le y\le X_1\right\}, ~\text{where}~ ,\\\begin{cases}
              ~\text{when}~  i=1,2 , \\ 0<x<X_{2} ,  \grave{t}_{X_2}<\grave{t}_{X_1} ,  L_{\text{eff}}-\grave{t}_{X_1}> \grave{t}_{X_1}  ~\text{and}~ L_{\text{eff}}-\grave{t}_{X_2} > \grave{t}_{X_2}    \\  ~\text{when}~ i=3,4,\\ 0<x<X_{2},  \left| X_2-x\right|<\left|X_1-x\right| , \\ \left|L+x-X_1\right| > \left| X_1-x\right|  ~\text{and}~ \left|L+x-X_2\right| > \left| X_2-x\right|.
         \end{cases} &~\text{for}~\text{(\romannumeral3)}~\\
         \left\{y\big{|}X_2 \le y\le X_1\right\} ,  ~\text{where}~ ,\\ \begin{cases}
              ~\text{when}~ i=1,2: \grave{t}_{X_2}=\grave{t}_{X_1} ~\text{or}~   \grave{t}_{X_2}= L_{\text{eff}}-\grave{t}_{X_1}  \\  ~\text{when}~ i=3,4: \left|x-X_2\right| = \left| X_1-x\right| ~\text{or}~   \left|X_2+L-x\right|=\left| X_1-x\right|
         \end{cases}
        & ~\text{for}~\text{(\romannumeral4)}~\\
    \end{cases}
\end{split}
\ee
\begin{center} 
\begin{tikzpicture}
    \begin{scope}[shift={(0,0)}]
        % Figure a
        \draw (0,0) circle (1cm);
        \filldraw (00:1) circle (2pt) node[above right] {$X_2$};
        \filldraw (240:1) circle (2pt) node[below left] {$X_1$};
        \draw[line width=2pt] (00:1) arc[start angle=0, end angle=240, radius=1] node[midway, above] {A};
        \node at (0,-1.7) {Case (\romannumeral1)};
        \filldraw (0,1) circle (1.5pt) node[below] {$x $};
    \end{scope}
    
    \begin{scope}[shift={(4,0)}] 
        % Figure b
        \draw (0,0) circle (1cm);
        \filldraw (180:1) circle (2pt) node[above left] {$X_1$};
        \filldraw (20:1) circle (2pt) node[below right] {$X_2$};
        \draw[line width=2pt] (180:1) arc[start angle=180, end angle=20, radius=1] node[midway, below] {A};
        \node at (0,-1.7) {Case (\romannumeral2)};
         \filldraw (240:1) circle (1pt) node[below] {$x $};
    \end{scope}
    
    \begin{scope}[shift={(8,0)}]
        \draw (0,0) circle (1cm);
        \filldraw (60:1) circle (2pt) node[above right] {$X_1$};
        \filldraw (-30:1) circle (2pt) node[below right] {$X_2$};
        \draw[line width=2pt] (-30:1) arc[start angle=-30, end angle=60, radius=1] node[midway, left] {A};
        \node at (0,-1.7) {Case (\romannumeral3)};
         \filldraw (-70:1) circle (1pt) node[below] {$x $};
    \end{scope}

    \begin{scope}[shift={(12,0)}]
        \draw (0,0) circle (1cm);
        \filldraw (00:1) circle (2pt) node[above right] {$X_2$};
        \filldraw (180:1) circle (2pt) node[below left] {$X_1$};
        \draw[line width=2pt] (00:1) arc[start angle=0, end angle=180, radius=1] node[midway, above] {A};
        \node at (0,-1.7) {Case (\romannumeral4)};
        \filldraw (0,1) circle (1.5pt) node[below] {$x $};
    \end{scope}
\end{tikzpicture}

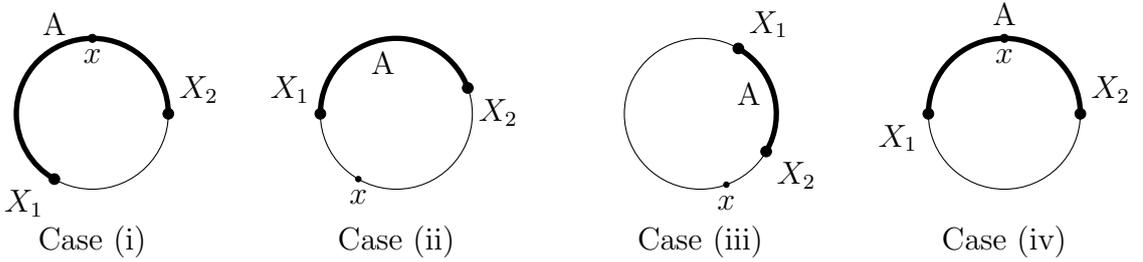
\captionof{figure}{Four typical cases. The details of the subsystems illustrated here are in (\ref{eq:generalsubsystems}). \label{Fig:generalsubsystems}}
\end{center}

The time dependence of $S_{A,i}$ considered in the case of (\romannumeral1) is similar to inserting $x$ at $X^f_2$ ,  specifically case (a) for SSD cases, as shown in equation  (\ref{EE-for-SSD-L/2-a}). The time-dependence of $S_{A,i}$ is given by
\be
\begin{split}\nonumber
& S_{A,1}\approx \f{c}{3}\log{\left[\f{L}{\pi}\sin{\left[\f{\pi(X_1-X_2)}{L}\right]}\right]}\\ &+\begin{cases}
			0 & \grave{t}_{X_2} >t_1> 0\\
	 \frac{c}{6} \log \left[\frac{2 \sin \left[\pi \alpha_{\mathcal{O}}\right]}{ \alpha_{\mathcal{O}}}\right]-\frac{c}{6} \log \left[ \epsilon \left| f_1(t_1,x,\theta)\right|\right]&nL_{\text{eff}}+\grave{t}^{t}_{0,+}>t_1> nL_{\text{eff}}+ \grave{t}_{X_2}\\
   \frac{c}{6} \log \left[\frac{2 \sin \left[\pi \alpha_{\mathcal{O}}\right]}{ \alpha_{\mathcal{O}}}\right]-\frac{c}{6} \log \left[ \epsilon \left| g_1(t_1,x,\theta)\right|\right]  & nL_{\text{eff}}+\grave{t}_{X_1}>t_1> nL_{\text{eff}}+ \grave{t}^{t}_{0,+}\\
	0 & (n+1) L_{\text{eff}}-\grave{t}_{X_1}>t_1> nL_{\text{eff}}+\grave{t}_{X_1}\\
\frac{c}{6} \log \left[\frac{2 \sin \left[\pi \alpha_{\mathcal{O}}\right]}{ \alpha_{\mathcal{O}}}\right]-\frac{c}{6} \log \left[ \epsilon \left| f_1(t_1,x,\theta)\right|\right]& n  L_{\text{eff}}+\grave{t}^{t}_{0,-}>t_1> (n+1) L_{\text{eff}}-\grave{t}_{X_1}\\
   \frac{c}{6} \log \left[\frac{2 \sin \left[\pi \alpha_{\mathcal{O}}\right]}{ \alpha_{\mathcal{O}}}\right]-\frac{c}{6} \log \left[ \epsilon \left| g_1(t_1,x,\theta)\right|\right] & (n+1) L_{\text{eff}}-\grave{t}_{X_2}>t_1> n  L_{\text{eff}}+\grave{t}^{t}_{0,-}\\
   0 & (n+1) L_{\text{eff}}+\grave{t}_{X_2}>t_1> (n+1) L_{\text{eff}}-\grave{t}_{X_2}\\
\end{cases},\\
& S_{A,2}\approx \f{c}{3}\log{\left[\f{L}{\pi}\sin{\left[\f{\pi(X_1-X_2)}{L}\right]}\right]}\\ &+\begin{cases}
			0 & \grave{t}_{X_2} >t_1> 0\\
	 \frac{c}{6} \log \left[\frac{2 \sin \left[\pi \alpha_{\mathcal{O}}\right]}{ \alpha_{\mathcal{O}}}\right]-\frac{c}{6} \log \left[ \epsilon \left| f_2(t_1,x,\theta)\right|\right]&nL_{\text{eff}}+\grave{t}^{t}_{0,+}>t_1> nL_{\text{eff}}+ \grave{t}_{X_2}\\
   \frac{c}{6} \log \left[\frac{2 \sin \left[\pi \alpha_{\mathcal{O}}\right]}{ \alpha_{\mathcal{O}}}\right]-\frac{c}{6} \log \left[ \epsilon \left| g_2(t_1,x,\theta)\right|\right]  & nL_{\text{eff}}+\grave{t}_{X_1}>t_1> nL_{\text{eff}}+ \grave{t}^{t}_{0,+}\\
	0 & (n+1) L_{\text{eff}}-\grave{t}_{X_1}>t_1> nL_{\text{eff}}+\grave{t}_{X_1}\\
\frac{c}{6} \log \left[\frac{2 \sin \left[\pi \alpha_{\mathcal{O}}\right]}{ \alpha_{\mathcal{O}}}\right]-\frac{c}{6} \log \left[ \epsilon \left| f_2(t_1,x,\theta)\right|\right]& n  L_{\text{eff}}+\grave{t}^{t}_{0,-}>t_1> (n+1) L_{\text{eff}}-\grave{t}_{X_1}\\
   \frac{c}{6} \log \left[\frac{2 \sin \left[\pi \alpha_{\mathcal{O}}\right]}{ \alpha_{\mathcal{O}}}\right]-\frac{c}{6} \log \left[ \epsilon \left| g_2(t_1,x,\theta)\right|\right] & (n+1) L_{\text{eff}}-\grave{t}_{X_2}>t_1> n  L_{\text{eff}}+\grave{t}^{t}_{0,-}\\
   0 & (n+1) L_{\text{eff}}+\grave{t}_{X_2}>t_1> (n+1) L_{\text{eff}}-\grave{t}_{X_2}\\
\end{cases},\\
	& S_{A,3}\approx \f{c}{3}\log{\left[\f{L}{\pi}\sin{\left[\f{\pi(X_1-X_2)}{L}\right]}\right]}\\ &+\begin{cases}
			0 & \ x-X_2 >t_0> 0\\
	 \frac{c}{6} \log \left[\frac{2 \sin \left[\pi \alpha_{\mathcal{O}}\right]}{ \alpha_{\mathcal{O}}}\right]-\frac{c}{6} \log \left[ \epsilon \left| f_3(t_0,x,\theta)\right|\right]& n L+\check{t}^{t}_{0,+}>t_0> nL+ x-X_2\\
   \frac{c}{6} \log \left[\frac{2 \sin \left[\pi \alpha_{\mathcal{O}}\right]}{ \alpha_{\mathcal{O}}}\right]-\frac{c}{6} \log \left[ \epsilon \left| g_3(t_0,x,\theta)\right|\right]  & n L+X_1-x>t_0> nL+ \check{t}^{t}_{0,+}\\
	0 & (n+1)L+x-X_1>t_0> n L+X_1-x\\
\frac{c}{6} \log \left[\frac{2 \sin \left[\pi \alpha_{\mathcal{O}}\right]}{ \alpha_{\mathcal{O}}}\right]-\frac{c}{6} \log \left[ \epsilon \left| f_3(t_0,x,\theta)\right|\right]& n L+\check{t}^{t}_{0,-}>t_0> (n+1)L+x-X_1\\
   \frac{c}{6} \log \left[\frac{2 \sin \left[\pi \alpha_{\mathcal{O}}\right]}{ \alpha_{\mathcal{O}}}\right]-\frac{c}{6} \log \left[ \epsilon \left| g_3(t_0,x,\theta)\right|\right] & (n+1)L+X_2-x>t_0> n L+\check{t}^{t}_{0,-}\\
   0 & (n+1)L+ x-X_2>t_0> (n+1)L+ X_2-x\\
\end{cases},\\
\end{split}
\ee
\be
\begin{split}
  & S_{A,4}\approx \f{c}{3}\log{\left[\f{L}{\pi}\sin{\left[\f{\pi(X_1-X_2)}{L}\right]}\right]}\\ &+\begin{cases}
			0 & \ x-X_2 >t_0> 0\\
	 \frac{c}{6} \log \left[\frac{2 \sin \left[\pi \alpha_{\mathcal{O}}\right]}{ \alpha_{\mathcal{O}}}\right]-\frac{c}{6} \log \left[ \epsilon \left| f_4(t_0,x,\theta)\right|\right]& n L+\check{t}^{t}_{0,+}>t_0> nL+ x-X_2\\
   \frac{c}{6} \log \left[\frac{2 \sin \left[\pi \alpha_{\mathcal{O}}\right]}{ \alpha_{\mathcal{O}}}\right]-\frac{c}{6} \log \left[ \epsilon \left| g_4(t_0,x,\theta)\right|\right]  & n L+X_1-x>t_0> nL+ \check{t}^{t}_{0,+}\\
	0 & (n+1)L+x-X_1>t_0> n L+X_1-x\\
\frac{c}{6} \log \left[\frac{2 \sin \left[\pi \alpha_{\mathcal{O}}\right]}{ \alpha_{\mathcal{O}}}\right]-\frac{c}{6} \log \left[ \epsilon \left| f_4(t_0,x,\theta)\right|\right]& n L+\check{t}^{t}_{0,-}>t_0> (n+1)L+x-X_1\\
   \frac{c}{6} \log \left[\frac{2 \sin \left[\pi \alpha_{\mathcal{O}}\right]}{ \alpha_{\mathcal{O}}}\right]-\frac{c}{6} \log \left[ \epsilon \left| g_4(t_0,x,\theta)\right|\right] & (n+1)L+X_2-x>t_0> n L+\check{t}^{t}_{0,-}\\
   0 & (n+1)L+ x-X_2>t_0> (n+1)L+ X_2-x\\
\end{cases}.\\
\end{split}
\ee

In the case of (\romannumeral2), as in (\ref{EE-for-Mobius-0-a}), the time dependence of $S_{A,i}$ is similar to that in the case (a)  with  $\theta \neq 0$ and the insertion of the local operator at $X^f_1$. The time-dependence of $S_{A,i}$ is given by
\be \nonumber
\begin{split}
& S_{A,1}\approx \f{c}{3}\log{\left[\f{L}{\pi}\sin{\left[\f{\pi(X_1-X_2)}{L}\right]}\right]}\\ &+\begin{cases}
			0 & \grave{t}_{X_1} >t_1> 0\\
	 \frac{c}{6} \log \left[\frac{2 \sin \left[\pi \alpha_{\mathcal{O}}\right]}{ \alpha_{\mathcal{O}}}\right]-\frac{c}{6} \log \left[ \epsilon \left| f_1(t_1,x,\theta)\right|\right]&nL_{\text{eff}}+\grave{t}^{t}_{0,+}>t_1> nL_{\text{eff}}+ \grave{t}_{X_1}\\
   \frac{c}{6} \log \left[\frac{2 \sin \left[\pi \alpha_{\mathcal{O}}\right]}{ \alpha_{\mathcal{O}}}\right]-\frac{c}{6} \log \left[ \epsilon \left| g_1(t_1,x,\theta)\right|\right]  & nL_{\text{eff}}+\grave{t}_{X_2}>t_1> nL_{\text{eff}}+ \grave{t}^{t}_{0,+}\\
	0 & (n+1) L_{\text{eff}}-\grave{t}_{X_2}>t_1> nL_{\text{eff}}+\grave{t}_{X_2}\\
\frac{c}{6} \log \left[\frac{2 \sin \left[\pi \alpha_{\mathcal{O}}\right]}{ \alpha_{\mathcal{O}}}\right]-\frac{c}{6} \log \left[ \epsilon \left| f_1(t_1,x,\theta)\right|\right]& n  L_{\text{eff}}+\grave{t}^{t}_{0,-}>t_1> (n+1) L_{\text{eff}}-\grave{t}_{X_2}\\
   \frac{c}{6} \log \left[\frac{2 \sin \left[\pi \alpha_{\mathcal{O}}\right]}{ \alpha_{\mathcal{O}}}\right]-\frac{c}{6} \log \left[ \epsilon \left| g_1(t_1,x,\theta)\right|\right] & (n+1) L_{\text{eff}}-\grave{t}_{X_1}>t_1> n  L_{\text{eff}}+\grave{t}^{t}_{0,-}\\
   0 & (n+1) L_{\text{eff}}+\grave{t}_{X_1}>t_1> (n+1) L_{\text{eff}}-\grave{t}_{X_1}\\
\end{cases},\\
& S_{A,2}\approx \f{c}{3}\log{\left[\f{L}{\pi}\sin{\left[\f{\pi(X_1-X_2)}{L}\right]}\right]}\\ &+\begin{cases}
			0 & \grave{t}_{X_1} >t_1> 0\\
	 \frac{c}{6} \log \left[\frac{2 \sin \left[\pi \alpha_{\mathcal{O}}\right]}{ \alpha_{\mathcal{O}}}\right]-\frac{c}{6} \log \left[ \epsilon \left| f_2(t_1,x,\theta)\right|\right]&nL_{\text{eff}}+\grave{t}^{t}_{0,+}>t_1> nL_{\text{eff}}+ \grave{t}_{X_1}\\
   \frac{c}{6} \log \left[\frac{2 \sin \left[\pi \alpha_{\mathcal{O}}\right]}{ \alpha_{\mathcal{O}}}\right]-\frac{c}{6} \log \left[ \epsilon \left| g_2(t_1,x,\theta)\right|\right]  & nL_{\text{eff}}+\grave{t}_{X_2}>t_1> nL_{\text{eff}}+ \grave{t}^{t}_{0,+}\\
	0 & (n+1) L_{\text{eff}}-\grave{t}_{X_2}>t_1> nL_{\text{eff}}+\grave{t}_{X_2}\\
\frac{c}{6} \log \left[\frac{2 \sin \left[\pi \alpha_{\mathcal{O}}\right]}{ \alpha_{\mathcal{O}}}\right]-\frac{c}{6} \log \left[ \epsilon \left| f_2(t_1,x,\theta)\right|\right]& n  L_{\text{eff}}+\grave{t}^{t}_{0,-}>t_1> (n+1) L_{\text{eff}}-\grave{t}_{X_2}\\
   \frac{c}{6} \log \left[\frac{2 \sin \left[\pi \alpha_{\mathcal{O}}\right]}{ \alpha_{\mathcal{O}}}\right]-\frac{c}{6} \log \left[ \epsilon \left| g_2(t_1,x,\theta)\right|\right] & (n+1) L_{\text{eff}}-\grave{t}_{X_1}>t_1> n  L_{\text{eff}}+\grave{t}^{t}_{0,-}\\
   0 & (n+1) L_{\text{eff}}+\grave{t}_{X_1}>t_1> (n+1) L_{\text{eff}}-\grave{t}_{X_1}\\
\end{cases},\\
\end{split}
\ee
\be
\begin{split}
	& S_{A,3}\approx \f{c}{3}\log{\left[\f{L}{\pi}\sin{\left[\f{\pi(X_1-X_2)}{L}\right]}\right]}\\ &+\begin{cases}
			0 & \ x-X_1 >t_0> 0\\
	\frac{c}{6} \log \left[\frac{2 \sin \left[\pi \alpha_{\mathcal{O}}\right]}{ \alpha_{\mathcal{O}}}\right]-\frac{c}{6} \log \left[ \epsilon \left| f_3(t_0,x,\theta)\right|\right] & nL+\check{t}^{t}_{0,+}>t_0> nL+ x-X_1\\
   \frac{c}{6} \log \left[\frac{2 \sin \left[\pi \alpha_{\mathcal{O}}\right]}{ \alpha_{\mathcal{O}}}\right]-\frac{c}{6} \log \left[ \epsilon \left| g_3(t_0,x,\theta)\right|\right] & (n+1) L+X_2-x>t_0> nL+\check{t}^{t}_{0,+}\\
	0 & nL+x-X_2>t_0>(n+1) L+X_2-x\\
\frac{c}{6} \log \left[\frac{2 \sin \left[\pi \alpha_{\mathcal{O}}\right]}{ \alpha_{\mathcal{O}}}\right]-\frac{c}{6} \log \left[ \epsilon \left| f_3(t_0,x,\theta)\right|\right]& nL+\check{t}^{t}_{0,-}>t_0> nL+x-X_2\\
   \frac{c}{6} \log \left[\frac{2 \sin \left[\pi \alpha_{\mathcal{O}}\right]}{ \alpha_{\mathcal{O}}}\right]-\frac{c}{6} \log \left[ \epsilon \left| g_3(t_0,x,\theta)\right|\right] & (n+1)L+X_1-x>t_0> nL+\check{t}^{t}_{0,-}\\
   0 & (n+1)L+ x-X_1>t_0> (n+1)L+X_1-x\\
\end{cases},\\
 & S_{A,4}\approx \f{c}{3}\log{\left[\f{L}{\pi}\sin{\left[\f{\pi(X_1-X_2)}{L}\right]}\right]}\\ &+\begin{cases}
			0 & \ x-X_1 >t_0> 0\\
	\frac{c}{6} \log \left[\frac{2 \sin \left[\pi \alpha_{\mathcal{O}}\right]}{ \alpha_{\mathcal{O}}}\right]-\frac{c}{6} \log \left[ \epsilon \left| f_4(t_0,x,\theta)\right|\right] & nL+\check{t}^{t}_{0,+}>t_0> nL+ x-X_1\\
   \frac{c}{6} \log \left[\frac{2 \sin \left[\pi \alpha_{\mathcal{O}}\right]}{ \alpha_{\mathcal{O}}}\right]-\frac{c}{6} \log \left[ \epsilon \left| g_4(t_0,x,\theta)\right|\right] & (n+1) L+X_2-x>t_0> nL+\check{t}^{t}_{0,+}\\
	0 & nL+x-X_2>t_0>(n+1) L+X_2-x\\
\frac{c}{6} \log \left[\frac{2 \sin \left[\pi \alpha_{\mathcal{O}}\right]}{ \alpha_{\mathcal{O}}}\right]-\frac{c}{6} \log \left[ \epsilon \left| f_4(t_0,x,\theta)\right|\right]& nL+\check{t}^{t}_{0,-}>t_0> nL+x-X_2\\
   \frac{c}{6} \log \left[\frac{2 \sin \left[\pi \alpha_{\mathcal{O}}\right]}{ \alpha_{\mathcal{O}}}\right]-\frac{c}{6} \log \left[ \epsilon \left| g_4(t_0,x,\theta)\right|\right] & (n+1)L+X_1-x>t_0> nL+\check{t}^{t}_{0,-}\\
   0 & (n+1)L+ x-X_1>t_0> (n+1)L+X_1-x\\
\end{cases}.\\
\end{split}
\ee

As shown in equation (\ref{EE-for-Mobius-0-c}), the time dependence of entanglement entropy in case (\romannumeral3) is similar to that with $\theta \neq \infty$ and the insertion of the local operator at $X^f_1$,  specifically case (c).
The time-dependence of $S_{A,i}$ is given by

\be
\begin{split}
& S_{A,1}\approx \f{c}{3}\log{\left[\f{L}{\pi}\sin{\left[\f{\pi(X_1-X_2)}{L}\right]}\right]}\\ &+\begin{cases}
			0 & \grave{t}_{X_2} >t_1> 0\\
	 \frac{c}{6} \log \left[\frac{2 \sin \left[\pi \alpha_{\mathcal{O}}\right]}{ \alpha_{\mathcal{O}}}\right]-\frac{c}{6} \log \left[ \epsilon \left| g_1(t_1,x,\theta)\right|\right]&nL_{\text{eff}}+\grave{t}_{X_1}>t_1> nL_{\text{eff}}+ \grave{t}_{X_2}\\
	0 & (n+1) L_{\text{eff}}-\grave{t}_{X_1}>t_1> nL_{\text{eff}}+\grave{t}_{X_1}\\
\frac{c}{6} \log \left[\frac{2 \sin \left[\pi \alpha_{\mathcal{O}}\right]}{ \alpha_{\mathcal{O}}}\right]-\frac{c}{6} \log \left[ \epsilon \left| f_1(t_1,x,\theta)\right|\right]& (n+1) L_{\text{eff}}-\grave{t}_{X_2}>t_1> (n+1) L_{\text{eff}}-\grave{t}_{X_1}\\
   0 & (n+1) L_{\text{eff}}+\grave{t}_{X_2}>t_1> (n+1) L_{\text{eff}}-\grave{t}_{X_2}\\
\end{cases},\\
& S_{A,2}\approx \f{c}{3}\log{\left[\f{L}{\pi}\sin{\left[\f{\pi(X_1-X_2)}{L}\right]}\right]}\\ &+\begin{cases}
			0 & \grave{t}_{X_2} >t_1> 0\\
	 \frac{c}{6} \log \left[\frac{2 \sin \left[\pi \alpha_{\mathcal{O}}\right]}{ \alpha_{\mathcal{O}}}\right]-\frac{c}{6} \log \left[ \epsilon \left| g_2(t_1,x,\theta)\right|\right]&nL_{\text{eff}}+\grave{t}_{X_1}>t_1> nL_{\text{eff}}+ \grave{t}_{X_2}\\
	0 & (n+1) L_{\text{eff}}-\grave{t}_{X_1}>t_1> nL_{\text{eff}}+\grave{t}_{X_1}\\
\frac{c}{6} \log \left[\frac{2 \sin \left[\pi \alpha_{\mathcal{O}}\right]}{ \alpha_{\mathcal{O}}}\right]-\frac{c}{6} \log \left[ \epsilon \left| f_2(t_1,x,\theta)\right|\right]& (n+1) L_{\text{eff}}-\grave{t}_{X_2}>t_1> (n+1) L_{\text{eff}}-\grave{t}_{X_1}\\
   0 & (n+1) L_{\text{eff}}+\grave{t}_{X_2}>t_1> (n+1) L_{\text{eff}}-\grave{t}_{X_2}\\
\end{cases},\\
	& S_{A,3}\approx \f{c}{3}\log{\left[\f{L}{\pi}\sin{\left[\f{\pi(X_1-X_2)}{L}\right]}\right]}\\ &+\begin{cases}
			0 & \ X_2-x>t_0> 0\\
	 \frac{c}{6} \log \left[\frac{2 \sin \left[\pi \alpha_{\mathcal{O}}\right]}{ \alpha_{\mathcal{O}}}\right]-\frac{c}{6} \log \left[ \epsilon \left| g_3(t_0,x,\theta)\right|\right]  & n L+X_1-x>t_0> nL+ X_2-x\\
	0 & (n+1)L+x-X_1>t_0>n L+X_1-x\\
	\frac{c}{6} \log \left[\frac{2 \sin \left[\pi \alpha_{\mathcal{O}}\right]}{ \alpha_{\mathcal{O}}}\right]-\frac{c}{6} \log \left[ \epsilon \left| f_3(t_0,x,\theta)\right|\right]  & (n+1)L+x-X_2>t_0> (n+1)L+x-X_1\\
   0 & (n+1)L+ X_2-x>t_0> (n+1)L+x-X_2\\
\end{cases},\\
 & S_{A,4}\approx \f{c}{3}\log{\left[\f{L}{\pi}\sin{\left[\f{\pi(X_1-X_2)}{L}\right]}\right]}\\ &+\begin{cases}
						0 & \ X_2-x>t_0> 0\\
	 \frac{c}{6} \log \left[\frac{2 \sin \left[\pi \alpha_{\mathcal{O}}\right]}{ \alpha_{\mathcal{O}}}\right]-\frac{c}{6} \log \left[ \epsilon \left| g_4(t_0,x,\theta)\right|\right]  & n L+X_1-x>t_0> nL+ X_2-x\\
	0 & (n+1)L+x-X_1>t_0>n L+X_1-x\\
	\frac{c}{6} \log \left[\frac{2 \sin \left[\pi \alpha_{\mathcal{O}}\right]}{ \alpha_{\mathcal{O}}}\right]-\frac{c}{6} \log \left[ \epsilon \left| f_4(t_0,x,\theta)\right|\right]  & (n+1)L+x-X_2>t_0> (n+1)L+x-X_1\\
   0 & (n+1)L+ X_2-x>t_0> (n+1)L+x-X_2\\
\end{cases}.\\
\end{split}
\ee

In the case of (\romannumeral4), the time-dependence of $S_{A,i}$ is given by
\be
\begin{split}
& S_{1}\approx \f{c}{3}\log{\left[\f{L}{\pi}\sin{\left[\f{\pi(X_1-X_2)}{L}\right]}\right]} \ \ \  t_1 >0\\
    %%%%%%%%%%%%%%%%%%%%%
& S_{2}\approx \f{c}{3}\log{\left[\f{L}{\pi}\sin{\left[\f{\pi(X_1-X_2)}{L}\right]}\right]} \ \ \  t_1 >0\\
        %%%%%%%%%%%%%%%%%%%%%%%%%%%%%%%%
& S_{3}\approx \f{c}{3}\log{\left[\f{L}{\pi}\sin{\left[\f{\pi(X_1-X_2)}{L}\right]}\right]} \ \ \  t_0 >0\\
    %%%%%%%%%%%%%%%%%%%%%
& S_{4}\approx \f{c}{3}\log{\left[\f{L}{\pi}\sin{\left[\f{\pi(X_1-X_2)}{L}\right]}\right]} \ \ \  t_0 >0\\
        %%%%%%%%%%%%%%%%%%%%%%%%%%%%%%%%
\end{split}
\ee

where $n$ is an integer greater than or equal to $0$. 
%%%%%%%%%%%%%%%%%%%%%%%%%%%%%%%%%%%%%%%%%%%%%%%%%%%%%
\subsection{SSD case}
%%%%%%%%%%%%%%%%%%%%%%%%%%%%%%%%%%%%%%%%%%%%%%%%%%%%%
Here we report on the time dependence of $S_{A,i}$ for (b) and (c)  during the SSD time evolution when the local operator is inserted at $x=X_1^f$ or $x=X_2^f$.
%%%%%%%%%%%%%%%%%%%%%%%%%%%%%%%%%%%%%%%%%%%%%%%%%%%%%
\subsubsection{$x=X_1^f$ \label{app:EE-SSD-Xf1}}
%%%%%%%%%%%%%%%%%%%%%%%%%%%%%%%%%%%%%%%%%%%%%%%%%%%%%
Here, we assume that the local operator is inserted at $X_1^f$.
Now, let us present the time dependence of $S_{A,i=4}$ for case (b).
In this case, the time dependence of $S_{A,i=4}$ is determined by
\be
\begin{split}
    &S_{A,4}\approx \frac{c}{3}  \log \left[\frac{L }{\pi } \sin \left(\frac{\pi  \left(X_1-X_2\right)}{L}\right)\right]  
       \\ &+ \begin{cases}
0 & X_2>t_0>0\\
\frac{c}{6} \log \left[\frac{2 \sin \left[\pi \alpha_{\mathcal{O}}\right]}{ \alpha_{\mathcal{O}}}\right]-\frac{c}{6} \log \left[ -\epsilon g_4(t_0,X^{f}_1,\infty)\right] &nL+t^{t}_{0, +}>t_0> nL+X_2\\
\frac{c}{6} \log \left[\frac{2 \sin \left[\pi \alpha_{\mathcal{O}}\right]}{ \alpha_{\mathcal{O}}}\right]-\frac{c}{6} \log \left[ -\epsilon  f_4(t_0,X^{f}_1,\infty)\right] &(n+1)L-X_1>t_0>nL+t^{t}_{0, +}\\ 
0&nL+X_1>t_0>(n+1)L-X_1\\
\frac{c}{6} \log \left[\frac{2 \sin \left[\pi \alpha_{\mathcal{O}}\right]}{ \alpha_{\mathcal{O}}}\right]-\frac{c}{6} \log \left[ \epsilon  g_4(t_0,X^{f}_1,\infty)\right] &nL+t^{t}_{0, -}>t_0>nL+X_1\\
\frac{c}{6} \log \left[\frac{2 \sin \left[\pi \alpha_{\mathcal{O}}\right]}{ \alpha_{\mathcal{O}}}\right]-\frac{c}{6} \log \left[ \epsilon  f_4(t_0,X^{f}_1,\infty)\right] &(n+1)L-X_2>t_0>nL+t^{t}_{0, -}\\
0 &(n+1)L+X_2>t_0>(n+1)L-X_2\\
    \end{cases}
\end{split}
\ee
Subsequently, we present the time dependence of $S_{A,i=4}$ for case (c).
It is determined by
\be
\begin{split}
    &S_{A,4} \approx \frac{c}{3}  \log \left[\frac{L }{\pi } \sin \left(\frac{\pi  \left(X_1-X_2\right)}{L}\right)\right]  
    \\ &+ \begin{cases}
0 & X_2>t_0>0\\
\frac{c}{6} \log \left[\frac{2 \sin \left[\pi \alpha_{\mathcal{O}}\right]}{ \alpha_{\mathcal{O}}}\right]-\frac{c}{6} \log \left[ -\epsilon  g_4(t_0,X^{f}_1,\infty)\right] &nL+X_1>t_0>nL+X_2\\
0&(n+1)L-X_1>t_0>nL+X_1\\
\frac{c}{6} \log \left[\frac{2 \sin \left[\pi \alpha_{\mathcal{O}}\right]}{ \alpha_{\mathcal{O}}}\right]-\frac{c}{6} \log \left[ \epsilon  f_4(t_0,X^{f}_1,\infty)\right] &(n+1)L-X_2>t_0>(n+1)L-X_1\\
0 &(n+1)L+X_2>t_0>(n+1)L-X_2\\
    \end{cases}
\end{split}
\ee
%%%%%%%%%%%%%%%%%%%%%%%%%%%%%%%%%%%%%%%%%%%%%%%%%%%%%
\subsubsection{$x=X_2^f$  \label{app:EE-SSD-Xf2}}
%%%%%%%%%%%%%%%%%%%%%%%%%%%%%%%%%%%%%%%%%%%%%%%%%%%%%
Here, we assume that the local operator is inserted at $X_2^f$. Now, let us study the time dependence of $S_{A,i}$ as follows. The time dependence of $S_{A,i}$ in case (b) is given by 
\be
\begin{split}\nonumber
& S_{A,1}\approx \f{c}{3}\log{\left[\f{L}{\pi}\sin{\left[\f{\pi(X_1-X_2)}{L}\right]}\right]} \\&+\begin{cases}
         0 &\tilde{t}_{1,+}>t_1>0\\
        %%%%%%%%%%%%%%%%%%%%%%%%%%%%%%%%
\frac{c}{6} \log \left[\frac{2 \sin \left[\pi \alpha_{\mathcal{O}}\right]}{ \alpha_{\mathcal{O}}}\right]-\frac{c}{6} \log \left[ \epsilon  g_1(t_1,X^{f}_2,\infty)\right] & \tilde{t}^{t}_{1}>t_1>\tilde{t}_{1,+}\\
%%%%%%%%%%%%%%%%%%%%%%%%%%%%%%
\frac{c}{6} \log \left[\frac{2 \sin \left[\pi \alpha_{\mathcal{O}}\right]}{ \alpha_{\mathcal{O}}}\right]-\frac{c}{6} \log \left[ \epsilon  f_1(t_1,X^{f}_2,\infty)\right]   & \tilde{t}_{2,-}>t_1>\tilde{t}^t_{1}\\
    %%%%%%%%%%%%%%%%%%%%%
0  & t_1>\tilde{t}_{2,-}\\
        \end{cases},\\
%%%%%%%%%%%%%%%%%%%%%%%%%%%%%%%%%%%%%%%%%%
%%%%%%%%%%%%%%%%%%%%%%%%%%%%%%%%%%%%%%%%%%
      & S_{A,2}\approx \f{c}{3}\log{\left[\f{L}{\pi}\sin{\left[\f{\pi(X_1-X_2)}{L}\right]}\right]}\\& +\begin{cases}
         0 &\tilde{t}_{1,+}>t_1>0\\
        %%%%%%%%%%%%%%%%%%%%%%%%%%%%%%%%
\frac{c}{6} \log \left[\frac{2 \sin \left[\pi \alpha_{\mathcal{O}}\right]}{ \alpha_{\mathcal{O}}}\right]-\frac{c}{6} \log \left[\epsilon  g_2(t_1,X^{f}_2,\infty)\right]  & \tilde{t}^{t}_{1}>t_1>\tilde{t}_{1,+}\\
%%%%%%%%%%%%%%%%%%%%%%%%%%%%%%
\frac{c}{6} \log \left[\frac{2 \sin \left[\pi \alpha_{\mathcal{O}}\right]}{ \alpha_{\mathcal{O}}}\right]-\frac{c}{6} \log \left[\epsilon  f_2(t_1,X^{f}_2,\infty)\right]   & \tilde{t}_{2,-}>t_1>\tilde{t}^t_{1}\\
    %%%%%%%%%%%%%%%%%%%%%
0  & t_1>\tilde{t}_{2,-}\\
        \end{cases},\\
%%%%%%%%%%%%%%%%%%%%%%%%%%%%%%
\end{split}
\ee
\be
\begin{split}
%%%%%%%%%%%%%%%%%%%%%%%%%%%%%%
& S_{A,3}\approx \f{c}{3}\log{\left[\f{L}{\pi}\sin{\left[\f{\pi(X_1-X_2)}{L}\right]}\right]} \\
%%%%%%%%%%%%%%%%%%%%%%%%%%%%%%
&+\begin{cases}
         0 &X_1-\f{L}{2}>t_0>0\\
        %%%%%%%%%%%%%%%%%%%%%%%%%%%%%%%%
\frac{c}{6} \log \left[\frac{2 \sin \left[\pi \alpha_{\mathcal{O}}\right]}{ \alpha_{\mathcal{O}}}\right]-\frac{c}{6} \log \left[ \epsilon  g_3(t_0,X^{f}_2,\infty)\right]  & nL+\tilde{t}^t_{0,+}>t_0>\left(n-\f{1}{2}\right)L+X_1\\
%%%%%%%%%%%%%%%%%%%%%%%%%%%%%%
\frac{c}{6} \log \left[\frac{2 \sin \left[\pi \alpha_{\mathcal{O}}\right]}{ \alpha_{\mathcal{O}}}\right]-\frac{c}{6} \log \left[ \epsilon f_3(t_0,X^{f}_2,\infty)\right]   & \left(n+\f{1}{2}\right)L-X_2>t_0>nL+\tilde{t}^t_{0,+}\\
    %%%%%%%%%%%%%%%%%%%%%
0  & \left(\f{1}{2}+n\right)L+X_2>t_0>\left(\f{1}{2}+n\right)L-X_2\\
%%%%%%%%%%%%%%%%%%%%%%%%%%%%%
\frac{c}{6} \log \left[\frac{2 \sin \left[\pi \alpha_{\mathcal{O}}\right]}{ \alpha_{\mathcal{O}}}\right]-\frac{c}{6} \log \left[ -\epsilon  g_3(t_0,X^{f}_2,\infty)\right]  & nL+\tilde{t}^{t}_{0,-}>t_0>\left(\f{1}{2}+n\right)L +X_2\\
\frac{c}{6} \log \left[\frac{2 \sin \left[\pi \alpha_{\mathcal{O}}\right]}{ \alpha_{\mathcal{O}}}\right]-\frac{c}{6} \log \left[- \epsilon  f_3(t_0,X^{f}_2,\infty)\right]  & \left(\f{3}{2}+n\right)L-X_1>t_0>nL+\tilde{t}^t_{0,-}\\
0 & \left(\f{1}{2}+n\right)L +X_1 >t_0>\left(\f{3}{2}+n\right)L-X_1
        \end{cases},\\
%%%%%%%%%%%%%%%%%%%%%%%%%%%%%%%%%%%%%%%%%%
%%%%%%%%%%%%%%%%%%%%%%%%%%%%%%%%%%%%%%%%%%
&S_{A,4} \approx \f{c}{3}\log{\left[\f{L}{\pi}\sin{\left[\f{\pi(X_1-X_2)}{L}\right]}\right]} \\
%%%%%%%%%%%%%%%%%%%%%%%%%%%%%%
&+\begin{cases}
         0 &X_1-\f{L}{2}>t_0>0\\
        %%%%%%%%%%%%%%%%%%%%%%%%%%%%%%%%
\frac{c}{6} \log \left[\frac{2 \sin \left[\pi \alpha_{\mathcal{O}}\right]}{ \alpha_{\mathcal{O}}}\right]-\frac{c}{6} \log \left[ \epsilon  g_4(t_0,X^{f}_2,\infty)\right]  & nL+\tilde{t}^t_{0,+}>t_0>\left(n-\f{1}{2}\right)L+X_1\\
%%%%%%%%%%%%%%%%%%%%%%%%%%%%%%
\frac{c}{6} \log \left[\frac{2 \sin \left[\pi \alpha_{\mathcal{O}}\right]}{ \alpha_{\mathcal{O}}}\right]-\frac{c}{6} \log \left[ \epsilon  f_4(t_0,X^{f}_2,\infty)\right]   & \left(n+\f{1}{2}\right)L-X_2>t_0>nL+\tilde{t}^t_{0,+}\\
    %%%%%%%%%%%%%%%%%%%%%
0  & \left(\f{1}{2}+n\right)L+X_2>t_0>\left(\f{1}{2}+n\right)L-X_2\\
%%%%%%%%%%%%%%%%%%%%%%%%%%%%%
\frac{c}{6} \log \left[\frac{2 \sin \left[\pi \alpha_{\mathcal{O}}\right]}{ \alpha_{\mathcal{O}}}\right]-\frac{c}{6} \log \left[ -\epsilon  g_4(t_0,X^{f}_2,\infty)\right]  & nL+\tilde{t}^{t}_{0,-}>t_0>\left(\f{1}{2}+n\right)L +X_2\\
\frac{c}{6} \log \left[\frac{2 \sin \left[\pi \alpha_{\mathcal{O}}\right]}{ \alpha_{\mathcal{O}}}\right]-\frac{c}{6} \log \left[ -\epsilon  f_4(t_0,X^{f}_2,\infty)\right]  & \left(\f{3}{2}+n\right)L-X_1>t_0>nL+\tilde{t}^t_{0,-}\\
0 & \left(\f{1}{2}+n\right)L +X_1 >t_0>\left(\f{3}{2}+n\right)L-X_1
        \end{cases},\\
\end{split}
\ee
where $n$ is a positive integer and $t_0$ is a positive number.
%For large $t_1$, $S_{A,i=1,2}$ are approximated by (\ref{eq:asymp-ee1}).

The time dependence of $S_{A,i}$ in case (c) is given by 
\be
\begin{split}
    & S_{A,1}\approx \f{c}{3}\log{\left[\f{L}{\pi}\sin{\left[\f{\pi(X_1-X_2)}{L}\right]}\right]} \\&+\begin{cases}
         0 &t_{1,-}>t_1>0\\
        %%%%%%%%%%%%%%%%%%%%%%%%%%%%%%%%
\frac{c}{6} \log \left[\frac{2 \sin \left[\pi \alpha_{\mathcal{O}}\right]}{ \alpha_{\mathcal{O}}}\right]-\frac{c}{6} \log \left[ \epsilon  f_1(t_1,X^{f}_2,\infty)\right] & t_{2,-}>t_1>t_{1,-}\\
%%%%%%%%%%%%%%%%%%%%%%%%%%%%%%
0 & t_1>t_{2,-}\\
    \end{cases},\\
    %%%%%%%%%%%%%%%%%%%%%
    %%%%%%%%%%%%%%%%%%%%%
       & S_{A,2}\approx \f{c}{3}\log{\left[\f{L}{\pi}\sin{\left[\f{\pi(X_1-X_2)}{L}\right]}\right]} \\&+\begin{cases}
         0 &t_{1,-}>t_1>0\\
        %%%%%%%%%%%%%%%%%%%%%%%%%%%%%%%%
\frac{c}{6} \log \left[\frac{2 \sin \left[\pi \alpha_{\mathcal{O}}\right]}{ \alpha_{\mathcal{O}}}\right]-\frac{c}{6} \log \left[ \epsilon  f_2(t_1,X^{f}_2,\infty)\right] & t_{2,-}>t_1>t_{1,-}\\
%%%%%%%%%%%%%%%%%%%%%%%%%%%%%
%\f{c}{6}\log{\left[\f{\sin{\left[\pi \alpha_{\mathcal{O}}\right]}}{ \pi\alpha_{\mathcal{O}}}\right]}+\f{c}{6}\log{\left[g(X_1,X_2,t_1)\right]} & \tilde{t}_{1,+}>t_1>\tilde{t}^{t}_{1}\\
%%%%%%%%%%%%%%%%%%%%%%%%%%%%%%
0 & t_1>t_{2,-}\\
    \end{cases},\\
%%%%%%%%%%%%%%%%%%%%%%%%%%%%%%
%%%%%%%%%%%%%%%%%%%%%%%%%%%%%%
& S_{A,3}\approx \f{c}{3}\log{\left[\f{L}{\pi}\sin{\left[\f{\pi(X_1-X_2)}{L}\right]}\right]} \\
%%%%%%%%%%%%%%%%%%%%%%%%%%%%%%
&+\begin{cases}
         0 &\f{L}{2}-X_1>t_0>0\\
        %%%%%%%%%%%%%%%%%%%%%%%%%%%%%%%%
\frac{c}{6} \log \left[\frac{2 \sin \left[\pi \alpha_{\mathcal{O}}\right]}{ \alpha_{\mathcal{O}}}\right]-\frac{c}{6} \log \left[ \epsilon  f_3(t_0,X^{f}_2,\infty)\right] & \left(\f{1}{2}+n\right)L-X_2>t_0>\left(n+\f{1}{2}\right)L-X_1\\
%%%%%%%%%%%%%%%%%%%%%%%%%%%%%
0  & \left(n+\f{1}{2}\right)L+X_2>t_0>\left(n+\f{1}{2}\right)L-X_2\\
    %%%%%%%%%%%%%%%%%%%%%
\frac{c}{6} \log \left[\frac{2 \sin \left[\pi \alpha_{\mathcal{O}}\right]}{ \alpha_{\mathcal{O}}}\right]-\frac{c}{6} \log \left[- \epsilon  g_3(t_0,X^{f}_2,\infty)\right] & \left(\f{1}{2}+n\right)L+X_1>t_0>\left(\f{1}{2}+n\right)L +X_2\\
%%%%%%%%%%%%%%%%%%%%%%%%
0 & \left(\f{3}{2}+n\right)L -X_1 >t_0>\left(\f{1}{2}+n\right)L+X_1
        \end{cases},\\
%%%%%%%%%%%%%%%%%%%%%%%%%%%%%%%%%%%%%%%%%%
%%%%%%%%%%%%%%%%%%%%%%%%%%%%%%%%%%%%%%%%%%
    & S_{A,4}\approx \f{c}{3}\log{\left[\f{L}{\pi}\sin{\left[\f{\pi(X_1-X_2)}{L}\right]}\right]} \\
%%%%%%%%%%%%%%%%%%%%%%%%%%%%%%
&+\begin{cases}
         0 &\f{L}{2}-X_1>t_0>0\\
        %%%%%%%%%%%%%%%%%%%%%%%%%%%%%%%%
\frac{c}{6} \log \left[\frac{2 \sin \left[\pi \alpha_{\mathcal{O}}\right]}{ \alpha_{\mathcal{O}}}\right]-\frac{c}{6} \log \left[ \epsilon  f_4(t_0,X^{f}_2,\infty)\right] & \left(\f{1}{2}+n\right)L-X_2>t_0>\left(n+\f{1}{2}\right)L-X_1\\
%%%%%%%%%%%%%%%%%%%%%%%%%%%%%
0  & \left(n+\f{1}{2}\right)L+X_2>t_0>\left(n+\f{1}{2}\right)L-X_2\\
    %%%%%%%%%%%%%%%%%%%%%
\frac{c}{6} \log \left[\frac{2 \sin \left[\pi \alpha_{\mathcal{O}}\right]}{ \alpha_{\mathcal{O}}}\right]-\frac{c}{6} \log \left[- \epsilon  g_4(t_0,X^{f}_2,\infty)\right] & \left(\f{1}{2}+n\right)L+X_1>t_0>\left(\f{1}{2}+n\right)L +X_2\\
%%%%%%%%%%%%%%%%%%%%%%%%
0 & \left(\f{3}{2}+n\right)L -X_1 >t_0>\left(\f{1}{2}+n\right)L+X_1
        \end{cases}\\
\end{split}
\ee
\subsubsection{General $x$ \label{App:HEE-SSD-generalx}}
Here are the details of the small $\epsilon$ expansion of cross ratios for general $x$ during the SSD time evolution:
\be
\begin{split}
&z_{c,1} \approx 1-\frac{2 i \pi L \epsilon  \sin \left[\frac{\pi  \left(X_1-X_2\right)}{L}\right]}{ \prod_{i=1,2} \left[L \sin \left(\frac{\pi  \left(x-X_i\right)}{L}\right)-2 \pi  t_1 \sin \left(\frac{\pi  x}{L}\right) \sin \left(\frac{\pi  X_i}{L}\right)\right]} +\mathcal{O}(\epsilon^2) ,\\
&\overline{z}_{c,1} \approx 1+\frac{2 i \pi L \epsilon  \sin \left[\frac{\pi  \left(X_1-X_2\right)}{L}\right]}{\prod_{i=1,2}  \left[L \sin \left(\frac{\pi  \left(x-X_i\right)}{L}\right)+2 \pi  t_1 \sin \left(\frac{\pi  x}{L}\right) \sin \left(\frac{\pi  X_i}{L}\right) \right]} +\mathcal{O}(\epsilon^2) ,\\
&z_{c,2} \approx 1-\frac{ 2 i \pi   \epsilon  \sin \left[\frac{\pi  \left(X_1-X_2\right)}{L}\right] \left[ L^2+4 \pi ^2 t_1^2 \sin ^2\left(\frac{\pi  x}{L}\right)+2 \pi  L t_1 \sin \left(\frac{2 \pi  x}{L}\right)\right]
}{L \prod_{i=1,2} \left[L \sin \left(\frac{\pi  \left(x-X_i\right)}{L}\right)-2 \pi  t_1 \sin \left(\frac{\pi  x}{L}\right) \sin \left(\frac{\pi  X_i}{L}\right)\right]} +\mathcal{O}(\epsilon^2) ,\\
&\overline{z}_{c,2} \approx 1+\frac{2 i \pi  \epsilon  \sin \left[\frac{\pi  \left(X_1-X_2\right)}{L}\right]\left[L^2+4 \pi ^2 t_1^2 \sin ^2\left(\frac{\pi  x}{L}\right)-2 \pi  L t_1 \sin \left(\frac{2 \pi  x}{L}\right)\right]}{L  \prod_{i=1,2}  \left[ L \sin \left(\frac{\pi  \left(x-X_i\right)}{L}\right)+2 \pi  t_1 \sin \left(\frac{\pi  x}{L}\right) \sin \left(\frac{\pi  X_i}{L}\right) \right]} +\mathcal{O}(\epsilon^2) ,\\
    &z_{c,3} \approx 1+\frac{2 i \pi  \epsilon  \sin \left[\frac{\pi  \left(X_1-X_2\right)}{L}\right] \left( \cos \left(\frac{2 \pi  x}{L}\right)-1\right)}{L \prod_{i=1,2} \sin \left[\frac{\pi  \left(t_0-x+X_i\right)}{L}\right]}+\mathcal{O}(\epsilon^2),\\
&\overline{z}_{c,3} \approx 1-\frac{2 i \pi  \epsilon  \sin \left[\frac{\pi  \left(X_1-X_2\right)}{L}\right]\left( \cos \left(\frac{2 \pi  x}{L}\right)-1\right)}{L \prod_{i=1,2} \sin \left[\frac{\pi  \left(t_0+x-X_i\right)}{L}\right]}+\mathcal{O}(\epsilon^2),\\
&z_{c,4}\approx 1+\frac{2 i \pi  \epsilon  \sin \left[\frac{\pi  \left(X_1-X_2\right)}{L}\right]\left( \cos \left(\frac{2 \pi  \left(t_0-x\right)}{L}\right)-1\right)}{L \prod_{i=1,2} \sin \left[\frac{\pi  \left(t_0-x+X_i\right)}{L}\right] }+\mathcal{O}(\epsilon^2),\\
&\overline{z}_{c,4} \approx 1-\frac{2 i \pi  \epsilon  \sin \left[\frac{\pi  \left(X_1-X_2\right)}{L}\right] \left( \cos \left(\frac{2 \pi  \left(t_0+x\right)}{L}\right)-1\right)}{L \prod_{i=1,2} \sin \left[\frac{\pi  \left(t_0+x-X_i\right)}{L}\right] }+\mathcal{O}(\epsilon^2),
\end{split}
\ee
Define $\check{t}_{y}$ as
\be
\check{t}_{y}=\left|\f{L \sin\left[ \f{\pi(y-x)}{L}\right] }{2 \pi \sin\left( \f{\pi x}{L}\right)\sin\left( \f{\pi y}{L}\right)}  \right|
\ee
One can easily check that for $0<x<y_1<y_2<L$ , we  have $\check{t}_{y_1}<\check{t}_{y_2}$.

Around $t_1=\check{t}_{X_i}+n L_{\text{eff}}$, the small $\epsilon$ expansion of $z_{c,i=1,2}$, $\overline{z}_{c,i=1,2}$, breaks down, while around $t_0= \pm \left(x-X_i\right)+n L$, the small $\epsilon$ expansion of $z_{c,i=3,4}$, $\overline{z}_{c,i=3,4}$ breaks down. 
 For any given $x, X_{1},X_{2}$, where $0<X_2<X_1<L$, when $i=1,2$ we observe several different behaviors as $x$ is inserted in different regions on the circle. 
 Here, we define the special point $X_{\text{spe}}$ as the one satisfying 
  $X_2<X_{\text{spe}}<X_1$  which obeys the following equation for  $i=1,2$,
 \be
 \sin \left(\frac{\pi  X_1}{L}\right) \sin \left[\frac{\pi  \left(x_{\text{spe}}-X_2\right)}{L}\right]=\sin \left(\frac{\pi  X_2}{L}\right) \sin \left[\frac{\pi  \left(X_1-x_{\text{spe}}\right)}{L}\right].
 \ee
The parameters are given by
 \be
 \begin{split}
     &\check{t}^{t}_{1}=\frac{\sqrt{ L^2 \left(\cot \left(\frac{\pi  x}{L}\right)-\cot \left(\frac{\pi  X_1}{L}\right)\right) \left(\cot \left(\frac{\pi  X_2}{L}\right)-\cot \left(\frac{\pi  x}{L}\right)\right)}}{2 \pi },\\
     &\cos \left(\frac{\pi  \check{t}^{t}_{0,\pm}}{L}\right)=\sqrt{\pm \frac{\cos \left(\frac{\pi  \left(2 x-X_1-X_2\right)}{L}\right)+\cos \left(\frac{\pi  \left(X_1-X_2\right)}{L}\right)}{2 \cos \left(\frac{\pi  \left(2 x-X_1-X_2\right)}{L}\right)} }.
 \end{split}
 \ee
% In the following, we also define $\check{t}_{X_1-x}, \check{t}_{X_2-x}\ $ by
% \be
% \check{t}_{X_1-x}=\left|\f{L \sin\left[ \f{\pi(X_1-x)}{L}\right] }{2 \pi \sin\left( \f{\pi x}{L}\right)\sin\left( \f{\pi X_1}{L}\right)}  \right|
% \ee
% \be
% \check{t}_{X_2-x}=\left|\f{L \sin\left[ \f{\pi(X_1-x)}{L}\right] }{2 \pi \sin\left( \f{\pi x}{L}\right)\sin\left( \f{\pi X_2}{L}\right)}  \right|
% \ee
In general, there exist five cases which correspond to the different behaviors of time-dependence of $S_{A,i}$ (See Fig.\ \ref{Fig:SSDgeneralsubsystems} ):
\be\label{eq:SSDgeneralsubsystems}
\begin{split}
    A=\begin{cases}
         \left\{y\big{|}X_2 \le y\le X_1\right\}, ~\text{where}~  0<x<X_{2} 
           &~\text{for}~(\text{\uppercase\expandafter{\romannumeral1}})\\
         \left\{y\big{|}X_2 \le y\le X_1\right\}, ~\text{where}~  X_{2}<x<X_{spe} &~\text{for}~(\text{\uppercase\expandafter{\romannumeral2}})\\        
         \left\{y\big{|}X_2 \le y\le X_1\right\}, ~\text{where}~    x = X_{spe}  &~\text{for}~(\text{\uppercase\expandafter{\romannumeral3}})\\
         \left\{y\big{|}X_2 \le y\le X_1\right\}, ~\text{where}~    X_{spe}<x<X_{1} &~\text{for}~(\text{\uppercase\expandafter{\romannumeral4}})\\
         \left\{y\big{|}X_2 \le y\le X_1\right\} ,  ~\text{where}~ X_{1}<x<L & ~\text{for}~  (\text{\uppercase\expandafter{\romannumeral5}})\\
    \end{cases}
\end{split}
\ee
\begin{center} 
\begin{tikzpicture}
    \begin{scope}[shift={(0,0)}]
        \draw (0,0) circle (1cm);
        \filldraw (60:1) circle (2pt) node[above right] {$X_2$};
        \filldraw (240:1) circle (2pt) node[below left] {$X_1$};
        \draw[line width=2pt] (60:1) arc[start angle=60, end angle=240, radius=1] node[midway, left] {A};
        \node at (0,-1.7) {Case (\uppercase\expandafter{\romannumeral1})};
        \filldraw (00:1) circle (1.5pt) node[right] {$x $};
    \end{scope}
    
    \begin{scope}[shift={(3,0)}]
        \draw (0,0) circle (1cm);
        \filldraw (60:1) circle (2pt) node[above right] {$X_2$};
        \filldraw (240:1) circle (2pt) node[below left] {$X_1$};
        \draw[line width=2pt] (60:1) arc[start angle=60, end angle=240, radius=1] node[midway, left] {A};
        \node at (0,-1.7) {Case (\uppercase\expandafter{\romannumeral2})};
        \filldraw (90:1) circle (1.5pt) node[above] {$x $};
    \end{scope}
    
    \begin{scope}[shift={(6,0)}]
        \draw (0,0) circle (1cm);
        \filldraw (60:1) circle (2pt) node[above right] {$X_2$};
        \filldraw (240:1) circle (2pt) node[below left] {$X_1$};
        \draw[line width=2pt] (60:1) arc[start angle=60, end angle=240, radius=1] node[midway, left] {A};
        \node at (0,-1.7) {Case (\uppercase\expandafter{\romannumeral3})};
        \filldraw (170:1) circle (1.5pt) node[right] {$x $};
    \end{scope}

    \begin{scope}[shift={(9,0)}]
        \draw (0,0) circle (1cm);
        \filldraw (60:1) circle (2pt) node[above right] {$X_2$};
        \filldraw (240:1) circle (2pt) node[below left] {$X_1$};
        \draw[line width=2pt] (60:1) arc[start angle=60, end angle=240, radius=1] node[midway, left] {A};
        \node at (0,-1.7) {Case (\uppercase\expandafter{\romannumeral4})};
        \filldraw (220:1) circle (1.5pt) node[left] {$x $};
    \end{scope}

       \begin{scope}[shift={(12,0)}]
        \draw (0,0) circle (1cm);
        \filldraw (60:1) circle (2pt) node[above right] {$X_2$};
        \filldraw (240:1) circle (2pt) node[below left] {$X_1$};
        \draw[line width=2pt] (60:1) arc[start angle=60, end angle=240, radius=1] node[midway, left] {A};
        \node at (0,-1.7) {Case (\uppercase\expandafter{\romannumeral5})};
        \filldraw (260:1) circle (1.5pt) node[below] {$x $};
    \end{scope}
\end{tikzpicture}

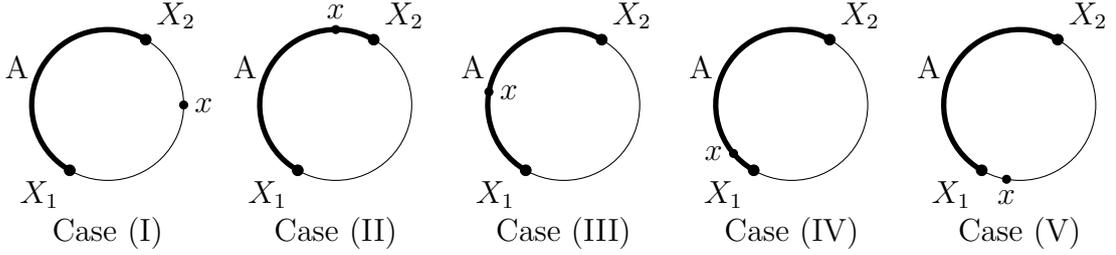
\captionof{figure}{Five general cases. The details of subsystem illustrated here are reported in (\ref{eq:SSDgeneralsubsystems}). \label{Fig:SSDgeneralsubsystems}}
\end{center}

In the case of (\uppercase\expandafter{\romannumeral1}), the time dependence of $S_{A,i}$ is given by
\be
\begin{split}
     & S_{1}\approx \f{c}{3}\log{\left[\f{L}{\pi}\sin{\left[\f{\pi(X_1-X_2)}{L}\right]}\right]}\\& +\begin{cases}
         0 &\check{t}_{X_2}>t_1>0\\
        %%%%%%%%%%%%%%%%%%%%%%%%%%%%%%%%
    \frac{c}{6} \log \left[\frac{2 \sin \left[\pi \alpha_{\mathcal{O}}\right]}{ \alpha_{\mathcal{O}}}\right]-\frac{c}{6} \log \left[ \epsilon \left| g_1(t_1,x,\infty)\right|\right] & \check{t}_{X_1}>t_1>\check{t}_{X_2}\\
    %%%%%%%%%%%%%%%%%%%%%%
0 & \check{t}_{X_1}\
    \end{cases},\\
    %%%%%%%%%%%%%%%%%%%%%
    %%%%%%%%%%%%%%%%%%%%%
       & S_{2}\approx \f{c}{3}\log{\left[\f{L}{\pi}\sin{\left[\f{\pi(X_1-X_2)}{L}\right]}\right]}\\& +\begin{cases}
         0 &\check{t}_{X_2}>t_1>0\\
        %%%%%%%%%%%%%%%%%%%%%%%%%%%%%%%%
 \frac{c}{6} \log \left[\frac{2 \sin \left[\pi \alpha_{\mathcal{O}}\right]}{ \alpha_{\mathcal{O}}}\right]-\frac{c}{6} \log \left[ \epsilon \left| g_2(t_1,x,\infty)\right|\right]  & \check{t}_{X_1}>t_1>\check{t}_{X_2}\\
%%%%%%%%%%%%%%%%%%%%%%%%%%%%%
%\f{c}{6}\log{\left[\f{\sin{\left[\pi \alpha_{\mathcal{O}}\right]}}{ \pi\alpha_{\mathcal{O}}}\right]}+\f{c}{6}\log{\left[g(X_1,X_2,t_1)\right]} & \check{t}_{1,+}>t_1>\check{t}^{t}_{1}\\
%%%%%%%%%%%%%%%%%%%%%%%%%%%%%%
0 & t_1>\check{t}_{X_1}\\
    \end{cases},\\
%%%%%%%%%%%%%%%%%%%%%%%%%%%%%%
\end{split}
\ee

In the case of (\uppercase\expandafter{\romannumeral2}), the time dependence of $S_{A,i}$ is given by
\be
\begin{split}
     & S_{1}\approx \f{c}{3}\log{\left[\f{L}{\pi}\sin{\left[\f{\pi(X_1-X_2)}{L}\right]}\right]} \\& +\begin{cases}
         0 &\check{t}_{X_2}>t_1>0\\
        %%%%%%%%%%%%%%%%%%%%%%%%%%%%%%%%
   \frac{c}{6} \log \left[\frac{2 \sin \left[\pi \alpha_{\mathcal{O}}\right]}{ \alpha_{\mathcal{O}}}\right]-\frac{c}{6} \log \left[ \epsilon \left| f_1(t_1,x,\infty)\right|\right]  & \check{t}^{t}_{1}>t_1>\check{t}_{X_2}\\
   \frac{c}{6} \log \left[\frac{2 \sin \left[\pi \alpha_{\mathcal{O}}\right]}{ \alpha_{\mathcal{O}}}\right]-\frac{c}{6} \log \left[ \epsilon \left| g_1(t_1,x,\infty)\right|\right] & \check{t}_{X_1}>t_1>\check{t}^{t}_{1}\\
    %%%%%%%%%%%%%%%%%%%%%%
0 & t_1>\check{t}_{X_1}\
    \end{cases},\\
    %%%%%%%%%%%%%%%%%%%%%
    %%%%%%%%%%%%%%%%%%%%%
       & S_{2}\approx \f{c}{3}\log{\left[\f{L}{\pi}\sin{\left[\f{\pi(X_1-X_2)}{L}\right]}\right]}\\& +\begin{cases}
         0 &\check{t}_{X_2}>t_1>0\\
        %%%%%%%%%%%%%%%%%%%%%%%%%%%%%%%%
   \frac{c}{6} \log \left[\frac{2 \sin \left[\pi \alpha_{\mathcal{O}}\right]}{ \alpha_{\mathcal{O}}}\right]-\frac{c}{6} \log \left[ \epsilon \left| f_2(t_1,x,\infty)\right|\right]  & \check{t}^{t}_{1}>t_1>\check{t}_{X_2}\\
   \frac{c}{6} \log \left[\frac{2 \sin \left[\pi \alpha_{\mathcal{O}}\right]}{ \alpha_{\mathcal{O}}}\right]-\frac{c}{6} \log \left[ \epsilon \left| g_2(t_1,x,\infty)\right|\right] & \check{t}_{X_1}>t_1>\check{t}^{t}_{1}\\
    %%%%%%%%%%%%%%%%%%%%%%
0 & t_1>\check{t}_{X_1}\
    \end{cases}.\\
%%%%%%%%%%%%%%%%%%%%%%%%%%%%%%
\end{split}
\ee
In the case of (\uppercase\expandafter{\romannumeral3}), the time dependence of $S_{A,i}$ is given by
\be
\begin{split}
& S_{1}\approx \f{c}{3}\log{\left[\f{L}{\pi}\sin{\left[\f{\pi(X_1-X_2)}{L}\right]}\right]} \ \ \  t_1 >0\\
    %%%%%%%%%%%%%%%%%%%%%
& S_{2}\approx \f{c}{3}\log{\left[\f{L}{\pi}\sin{\left[\f{\pi(X_1-X_2)}{L}\right]}\right]} \ \ \  t_1 >0\\
        %%%%%%%%%%%%%%%%%%%%%%%%%%%%%%%%
\end{split}
\ee
In the case of (\uppercase\expandafter{\romannumeral4}), the time dependence of $S_{A,i}$ is given by
\be
\begin{split}
     & S_{1}\approx \f{c}{3}\log{\left[\f{L}{\pi}\sin{\left[\f{\pi(X_1-X_2)}{L}\right]}\right]} \\& +\begin{cases}
         0 &\check{t}_{X_1}>t_1>0\\
        %%%%%%%%%%%%%%%%%%%%%%%%%%%%%%%%
   \frac{c}{6} \log \left[\frac{2 \sin \left[\pi \alpha_{\mathcal{O}}\right]}{ \alpha_{\mathcal{O}}}\right]-\frac{c}{6} \log \left[ \epsilon \left| g_1(t_1,x,\infty)\right|\right]  & \check{t}^{t}_{1}>t_1>\check{t}_{X_1}\\
   \frac{c}{6} \log \left[\frac{2 \sin \left[\pi \alpha_{\mathcal{O}}\right]}{ \alpha_{\mathcal{O}}}\right]-\frac{c}{6} \log \left[ \epsilon \left| f_1(t_1,x,\infty)\right|\right] & \check{t}_{X_2}>t_1>\check{t}^{t}_{1}\\
    %%%%%%%%%%%%%%%%%%%%%%
0 & t_1>\check{t}_{X_2}\
    \end{cases},\\
    %%%%%%%%%%%%%%%%%%%%%
    %%%%%%%%%%%%%%%%%%%%%
       & S_{2}\approx \f{c}{3}\log{\left[\f{L}{\pi}\sin{\left[\f{\pi(X_1-X_2)}{L}\right]}\right]}\\& +\begin{cases}
         0 &\check{t}_{X_1}>t_1>0\\
        %%%%%%%%%%%%%%%%%%%%%%%%%%%%%%%%
   \frac{c}{6} \log \left[\frac{2 \sin \left[\pi \alpha_{\mathcal{O}}\right]}{ \alpha_{\mathcal{O}}}\right]-\frac{c}{6} \log \left[ \epsilon \left| g_2(t_1,x,\infty)\right|\right]  & \check{t}^{t}_{1}>t_1>\check{t}_{X_1}\\
   \frac{c}{6} \log \left[\frac{2 \sin \left[\pi \alpha_{\mathcal{O}}\right]}{ \alpha_{\mathcal{O}}}\right]-\frac{c}{6} \log \left[ \epsilon \left| f_2(t_1,x,\infty)\right|\right] & \check{t}_{X_2}>t_1>\check{t}^{t}_{1}\\
    %%%%%%%%%%%%%%%%%%%%%%
0 &t_1> \check{t}_{X_2}\
    \end{cases}.\\
%%%%%%%%%%%%%%%%%%%%%%%%%%%%%%
\end{split}
\ee

In the case of (\uppercase\expandafter{\romannumeral5}), the time dependence of $S_{A,i}$ is given by
\be
\begin{split}
     & S_{1}\approx \f{c}{3}\log{\left[\f{L}{\pi}\sin{\left[\f{\pi(X_1-X_2)}{L}\right]}\right]} \\& +\begin{cases}
         0 &\check{t}_{X_1}>t_1>0\\
        %%%%%%%%%%%%%%%%%%%%%%%%%%%%%%%%
    \frac{c}{6} \log \left[\frac{2 \sin \left[\pi \alpha_{\mathcal{O}}\right]}{ \alpha_{\mathcal{O}}}\right]-\frac{c}{6} \log \left[ \epsilon \left| f_1(t_1,x,\infty)\right|\right] & \check{t}_{X_2}>t_1>\check{t}_{X_1}\\
    %%%%%%%%%%%%%%%%%%%%%%
0 & t_1>\check{t}_{X_2}\
    \end{cases},\\
    %%%%%%%%%%%%%%%%%%%%%
    %%%%%%%%%%%%%%%%%%%%%
       & S_{2}\approx \f{c}{3}\log{\left[\f{L}{\pi}\sin{\left[\f{\pi(X_1-X_2)}{L}\right]}\right]} \\&+\begin{cases}
         0 &\check{t}_{X_1}>t_1>0\\
        %%%%%%%%%%%%%%%%%%%%%%%%%%%%%%%%
\frac{c}{6} \log \left[\frac{2 \sin \left[\pi \alpha_{\mathcal{O}}\right]}{ \alpha_{\mathcal{O}}}\right]-\frac{c}{6} \log \left[ \epsilon \left| f_2(t_1,x,\infty)\right|\right]& \check{t}_{X_2}>t_1>\check{t}_{X_1}\\
%%%%%%%%%%%%%%%%%%%%%%%%%%%%%%
0 & t_1>\check{t}_{X_2}\\
    \end{cases},\\
%%%%%%%%%%%%%%%%%%%%%%%%%%%%%%
\end{split}
\ee

For any given general $x, X_{1},X_{2}$, where  $0<X_2<X_1<L)$, for $i=3,4$, we select four typical cases, according to (\ref{eq:generalsubsystems}).
As shown in equation  (\ref{EE-for-SSD-L/2-a}), the time dependence of $S_{A,i}$ considered in the case of (\romannumeral1) is similar to that in case (a) with the insertion of the local operator at $X^f_2$ and $\theta=\infty$. The time dependence of $S_{A,i}$ is given by
\be
\begin{split}
	& S_{A,3}\approx \f{c}{3}\log{\left[\f{L}{\pi}\sin{\left[\f{\pi(X_1-X_2)}{L}\right]}\right]}\\ &+\begin{cases}
			0 & \ x-X_2 >t_0> 0\\
	 \frac{c}{6} \log \left[\frac{2 \sin \left[\pi \alpha_{\mathcal{O}}\right]}{ \alpha_{\mathcal{O}}}\right]-\frac{c}{6} \log \left[ \epsilon \left| f_3(t_0,x,\infty)\right|\right]& n L+\check{t}^{t}_{0,+}>t_0> nL+ x-X_2\\
   \frac{c}{6} \log \left[\frac{2 \sin \left[\pi \alpha_{\mathcal{O}}\right]}{ \alpha_{\mathcal{O}}}\right]-\frac{c}{6} \log \left[ \epsilon \left| g_3(t_0,x,\infty)\right|\right]  & n L+X_1-x>t_0> nL+ \check{t}^{t}_{0,+}\\
	0 & (n+1)L+x-X_1>t_0> n L+X_1-x\\
\frac{c}{6} \log \left[\frac{2 \sin \left[\pi \alpha_{\mathcal{O}}\right]}{ \alpha_{\mathcal{O}}}\right]-\frac{c}{6} \log \left[ \epsilon \left| f_3(t_0,x,\infty)\right|\right]& n L+\check{t}^{t}_{0,-}>t_0> (n+1)L+x-X_1\\
   \frac{c}{6} \log \left[\frac{2 \sin \left[\pi \alpha_{\mathcal{O}}\right]}{ \alpha_{\mathcal{O}}}\right]-\frac{c}{6} \log \left[ \epsilon \left| g_3(t_0,x,\infty)\right|\right] & (n+1)L+X_2-x>t_0> n L+\check{t}^{t}_{0,-}\\
   0 & (n+1)L+ x-X_2>t_0> (n+1)L+ X_2-x\\
\end{cases},\\
 & S_{A,4}\approx \f{c}{3}\log{\left[\f{L}{\pi}\sin{\left[\f{\pi(X_1-X_2)}{L}\right]}\right]}\\ &+\begin{cases}
			0 & \ x-X_2 >t_0> 0\\
	 \frac{c}{6} \log \left[\frac{2 \sin \left[\pi \alpha_{\mathcal{O}}\right]}{ \alpha_{\mathcal{O}}}\right]-\frac{c}{6} \log \left[ \epsilon \left| f_4(t_0,x,\infty)\right|\right]& n L+\check{t}^{t}_{0,+}>t_0> nL+ x-X_2\\
   \frac{c}{6} \log \left[\frac{2 \sin \left[\pi \alpha_{\mathcal{O}}\right]}{ \alpha_{\mathcal{O}}}\right]-\frac{c}{6} \log \left[ \epsilon \left| g_4(t_0,x,\infty)\right|\right]  & n L+X_1-x>t_0> nL+ \check{t}^{t}_{0,+}\\
	0 & (n+1)L+x-X_1>t_0> n L+X_1-x\\
\frac{c}{6} \log \left[\frac{2 \sin \left[\pi \alpha_{\mathcal{O}}\right]}{ \alpha_{\mathcal{O}}}\right]-\frac{c}{6} \log \left[ \epsilon \left| f_4(t_0,x,\infty)\right|\right]& n L+\check{t}^{t}_{0,-}>t_0> (n+1)L+x-X_1\\
   \frac{c}{6} \log \left[\frac{2 \sin \left[\pi \alpha_{\mathcal{O}}\right]}{ \alpha_{\mathcal{O}}}\right]-\frac{c}{6} \log \left[ \epsilon \left| g_4(t_0,x,\infty)\right|\right] & (n+1)L+X_2-x>t_0> n L+\check{t}^{t}_{0,-}\\
   0 & (n+1)L+ x-X_2>t_0> (n+1)L+ X_2-x\\
\end{cases}.\\
\end{split}
\ee

In the case of (\romannumeral2), the time dependence of $S_{A,i}$ is similar to case (a) when inserting $x$ at $X^f_1$ of the M\"obius cases,  as shown in equation  (\ref{EE-for-Mobius-0-a}). The time dependence of $S_{A,i}$ is given by
\be
\begin{split}
	& S_{A,3}\approx \f{c}{3}\log{\left[\f{L}{\pi}\sin{\left[\f{\pi(X_1-X_2)}{L}\right]}\right]}\\ &+\begin{cases}
			0 & \ x-X_1 >t_0> 0\\
	\frac{c}{6} \log \left[\frac{2 \sin \left[\pi \alpha_{\mathcal{O}}\right]}{ \alpha_{\mathcal{O}}}\right]-\frac{c}{6} \log \left[ \epsilon \left| f_3(t_0,x,\infty)\right|\right] & nL+\check{t}^{t}_{0,+}>t_0> nL+ x-X_1\\
   \frac{c}{6} \log \left[\frac{2 \sin \left[\pi \alpha_{\mathcal{O}}\right]}{ \alpha_{\mathcal{O}}}\right]-\frac{c}{6} \log \left[ \epsilon \left| g_3(t_0,x,\infty)\right|\right] & (n+1) L+X_2-x>t_0> nL+\check{t}^{t}_{0,+}\\
	0 & nL+x-X_2>t_0>(n+1) L+X_2-x\\
\frac{c}{6} \log \left[\frac{2 \sin \left[\pi \alpha_{\mathcal{O}}\right]}{ \alpha_{\mathcal{O}}}\right]-\frac{c}{6} \log \left[ \epsilon \left| f_3(t_0,x,\infty)\right|\right]& nL+\check{t}^{t}_{0,-}>t_0> nL+x-X_2\\
   \frac{c}{6} \log \left[\frac{2 \sin \left[\pi \alpha_{\mathcal{O}}\right]}{ \alpha_{\mathcal{O}}}\right]-\frac{c}{6} \log \left[ \epsilon \left| g_3(t_0,x,\infty)\right|\right] & (n+1)L+X_1-x>t_0> nL+\check{t}^{t}_{0,-}\\
   0 & (n+1)L+ x-X_1>t_0> (n+1)L+X_1-x\\
\end{cases},\\
 & S_{A,4}\approx \f{c}{3}\log{\left[\f{L}{\pi}\sin{\left[\f{\pi(X_1-X_2)}{L}\right]}\right]}\\ &+\begin{cases}
			0 & \ x-X_1 >t_0> 0\\
	\frac{c}{6} \log \left[\frac{2 \sin \left[\pi \alpha_{\mathcal{O}}\right]}{ \alpha_{\mathcal{O}}}\right]-\frac{c}{6} \log \left[ \epsilon \left| f_4(t_0,x,\infty)\right|\right] & nL+\check{t}^{t}_{0,+}>t_0> nL+ x-X_1\\
   \frac{c}{6} \log \left[\frac{2 \sin \left[\pi \alpha_{\mathcal{O}}\right]}{ \alpha_{\mathcal{O}}}\right]-\frac{c}{6} \log \left[ \epsilon \left| g_4(t_0,x,\infty)\right|\right] & (n+1) L+X_2-x>t_0> nL+\check{t}^{t}_{0,+}\\
	0 & nL+x-X_2>t_0>(n+1) L+X_2-x\\
\frac{c}{6} \log \left[\frac{2 \sin \left[\pi \alpha_{\mathcal{O}}\right]}{ \alpha_{\mathcal{O}}}\right]-\frac{c}{6} \log \left[ \epsilon \left| f_4(t_0,x,\infty)\right|\right]& nL+\check{t}^{t}_{0,-}>t_0> nL+x-X_2\\
   \frac{c}{6} \log \left[\frac{2 \sin \left[\pi \alpha_{\mathcal{O}}\right]}{ \alpha_{\mathcal{O}}}\right]-\frac{c}{6} \log \left[ \epsilon \left| g_4(t_0,x,\infty)\right|\right] & (n+1)L+X_1-x>t_0> nL+\check{t}^{t}_{0,-}\\
   0 & (n+1)L+ x-X_1>t_0> (n+1)L+X_1-x\\
\end{cases}.\\
\end{split}
\ee
As shown in equation  (\ref{EE-for-Mobius-0-c}), the time dependence of entanglement entropy in case (\romannumeral3) is similar to that with the insertion of the local operator at $X^f_1$  and $\theta \neq \infty$,  specifically case (c). The time dependence of $S_{A,i}$ is given by

\be
\begin{split}
	& S_{A,3}\approx \f{c}{3}\log{\left[\f{L}{\pi}\sin{\left[\f{\pi(X_1-X_2)}{L}\right]}\right]}\\ &+\begin{cases}
			0 & \ X_2-x>t_0> 0\\
	 \frac{c}{6} \log \left[\frac{2 \sin \left[\pi \alpha_{\mathcal{O}}\right]}{ \alpha_{\mathcal{O}}}\right]-\frac{c}{6} \log \left[ \epsilon \left| g_3(t_0,x,\infty)\right|\right]  & n L+X_1-x>t_0> nL+ X_2-x\\
	0 & (n+1)L+x-X_1>t_0>n L+X_1-x\\
	\frac{c}{6} \log \left[\frac{2 \sin \left[\pi \alpha_{\mathcal{O}}\right]}{ \alpha_{\mathcal{O}}}\right]-\frac{c}{6} \log \left[ \epsilon \left| f_3(t_0,x,\infty)\right|\right]  & (n+1)L+x-X_2>t_0> (n+1)L+x-X_1\\
   0 & (n+1)L+ X_2-x>t_0> (n+1)L+x-X_2\\
\end{cases},\\
 & S_{A,4}\approx \f{c}{3}\log{\left[\f{L}{\pi}\sin{\left[\f{\pi(X_1-X_2)}{L}\right]}\right]}\\ &+\begin{cases}
						0 & \ X_2-x>t_0> 0\\
	 \frac{c}{6} \log \left[\frac{2 \sin \left[\pi \alpha_{\mathcal{O}}\right]}{ \alpha_{\mathcal{O}}}\right]-\frac{c}{6} \log \left[ \epsilon \left| g_4(t_0,x,\infty)\right|\right]  & n L+X_1-x>t_0> nL+ X_2-x\\
	0 & (n+1)L+x-X_1>t_0>n L+X_1-x\\
	\frac{c}{6} \log \left[\frac{2 \sin \left[\pi \alpha_{\mathcal{O}}\right]}{ \alpha_{\mathcal{O}}}\right]-\frac{c}{6} \log \left[ \epsilon \left| f_4(t_0,x,\infty)\right|\right]  & (n+1)L+x-X_2>t_0> (n+1)L+x-X_1\\
   0 & (n+1)L+ X_2-x>t_0> (n+1)L+x-X_2\\
\end{cases}.\\
\end{split}
\ee

In the case of (\romannumeral4), the time dependence of $S_{A,i}$ is given by
\be
\begin{split}
& S_{3}\approx \f{c}{3}\log{\left[\f{L}{\pi}\sin{\left[\f{\pi(X_1-X_2)}{L}\right]}\right]} \ \ \  t_0 >0\\
    %%%%%%%%%%%%%%%%%%%%%
& S_{4}\approx \f{c}{3}\log{\left[\f{L}{\pi}\sin{\left[\f{\pi(X_1-X_2)}{L}\right]}\right]} \ \ \  t_0 >0\\
        %%%%%%%%%%%%%%%%%%%%%%%%%%%%%%%%
\end{split}
\ee

where $n$ is an integer greater than or equal to $0$.

\bibliographystyle{ieeetr}
\bibliography{reference}
\end{document}